\documentclass[a4paper,fleqn,usenatbib]{mnras}

\usepackage[T1]{fontenc}

\usepackage[english]{babel}
\usepackage{graphicx}
\usepackage{fleqn}
\usepackage{amsmath}
\usepackage{amssymb}
\usepackage{booktabs}
\usepackage{hyperref}
\usepackage{etoolbox} % for '\AtBeginEnvironment' macro
\AtBeginEnvironment{pmatrix}{\everymath{\displaystyle}}
\usepackage{multirow}
\usepackage[flushleft]{threeparttable}
\usepackage{txfonts} %older fonts, for compatibility with arXiv

\newcommand{\Bm}{\overline{\mathrm{B}}}

\newcommand{\codename}{SecularMultiple}
\newcommand{\complexi}{\mathrm{i}}
\renewcommand{\S}{Section}
\newcommand{\F}{Figure}
\newcommand{\unit}[1]{\hat{\boldsymbol{#1}}}
\newcommand{\ve}[1]{\boldsymbol{#1}}

\title[Hierarchical $N$-tuple systems]{Secular dynamics of hierarchical multiple systems composed of nested binaries, with an arbitrary number of bodies and arbitrary hierarchical structure. First applications to multiplanet and multistar systems}

\author[Adrian S. Hamers and Simon F. Portegies Zwart]{Adrian S. Hamers$^{1}$\thanks{E-mail: hamers@strw.leidenuniv.nl} and Simon F. Portegies Zwart$^{1}$\\
$^{1}$Leiden Observatory, Leiden University, PO Box 9513, NL-2300 RA Leiden, The Netherlands}

\date{}

\pubyear{2016}

\begin{document}
\label{firstpage}
\pagerange{\pageref{firstpage}--\pageref{lastpage}}
\maketitle

\begin{abstract}
We present a method for studying the secular gravitational dynamics of hierarchical multiple systems consisting of nested binaries, which is valid for an arbitrary number of bodies and arbitrary hierarchical structure. We derive the Hamiltonian of the system and expand it in terms of the -- assumed to be -- small ratios $x_i$ of binary separations. At the lowest nontrivial expansion order (quadrupole order, second order in $x_i$), the Hamiltonian consists of terms which, individually, depend on binary pairs. At higher orders, in addition to terms depending on binary pairs, we also find terms which, individually, depend on more than two binaries. In general, at order $n$ in $x_i$, individual terms depend on at most $n-1$ binaries. We explicitly derive the Hamiltonian including all terms up and including third order in $x_i$ (octupole order), and including the binary pairwise terms up and including fifth order in $x_i$. These terms are orbit averaged, and we present a new algorithm for efficiently solving the equations of motion. This algorithm is highly suitable for studying the secular evolution of hierarchical systems with complex hierarchies, making long-term integrations of such systems feasible. We show that accurate results are obtained for multiplanet systems with semimajor axis ratios as large as $\approx 0.4$, provided that high-order terms are included. In addition to multiplanet systems with a single star, we apply our results to multistar systems with multiple planets.
\end{abstract}


\section{Introduction}
\label{sect:introduction}
Stellar multiple systems are often arranged in a hierarchical configuration composed of binary orbits \citep{1968QJRAS...9..388E,1981CeMec..25...51N}. For most systems, these orbits are nearly Keplerian on short time-scales. The simplest and most common configuration is a hierarchical triple, in which the centre of mass of a binary is orbited by a tertiary companion. Observed triples are, by necessity, dynamically stable on time-scales of the order of the orbital periods. However, on time-scales much longer than the orbital periods, exchanges of torques can give rise to secular oscillations in the eccentricities of the orbits, i.e. Lidov-Kozai (LK) oscillations \citep{1962P&SS....9..719L,1962AJ.....67..591K}. These oscillations have important implications for e.g. short-period binaries \citep{1979A&A....77..145M,1998MNRAS.300..292K,2001ApJ...562.1012E,2006Ap&SS.304...75E,2007ApJ...669.1298F,2014ApJ...793..137N}. 

Similar oscillations can occur in quadruple systems, which are observed in two hierarchical configurations. In the `2+2' or `binary-binary' configuration, two binaries orbit each other's barycentre (e.g. \citealt{2013MNRAS.435..943P}, who studied the secular dynamics using $N$-body integration methods). In the `3+1' or `triple-single' configuration, a single body orbits the centre of mass of a hierarchical triple \citep{2015MNRAS.449.4221H}. Quintuple and even higher-order systems are also observed, although with lower frequency (e.g. \citealt{2014AJ....147...86T,2014AJ....147...87T}). The number of possible configurations of hierarchical $N$-body systems composed of nested binaries increases with $N$, and is $N-2$ for $N\geq 3$  \citep{1981CeMec..25...51N}. This suggests a large variety of high-$N$ systems (e.g. \citealt{2007MNRAS.379..111V}).

Even more complex hierarchies arise when considering exoplanets. Since the past decade, planets are being discovered at an exponential rate, and an increasing number of exoplanets are observed in binary, triple and even quadruple systems. In the simplest case of stellar binaries, planets have been observed orbiting an individual star (S-type orbits, in the nomenclature of \citealt{1982OAWMN.191..423D}), and orbiting a stellar binary (P-type orbits). Approximately 70 S-type planets in binaries have been found sofar \citep{2012&A...542A..92R}. Fewer circumbinary (or P-type) planets have been found; e.g., nine {\it Kepler} transiting circumbinary planets have been found sofar (e.g. \citealt{2015ARA&A..53..409W}). However, detecting such planets is generally more difficult compared to S-type planets \citep{2014A&A...570A..91M}; it has been estimated that the occurrence rate of P-type planets could be as high as the occurrence rate of exoplanets around single stars \citep{2014MNRAS.444.1873A}. 

Although their detection is still severely limited by observational biases, exoplanets have also been found in multistellar systems with more complex hierarchies. Examples include Gliese 667, a triple system with at least two planets orbiting the tertiary star (\citealt{2013A&A...549A.109B}; \citealt{2014MNRAS.437.3540F} and references therein), and 30 Arietis, a quadruple system with an S-type planet around one of the stars \citep{2006A&A...450..681T,2009A&A...507.1659G,2015ApJ...799....4R,2015AJ....149..118R}.

These observations indicate that planet formation in stellar multiple systems is not only possible, but may also be a common phenomenon. Multiplanet systems in multistellar systems give rise to complex hierarchies with many layers of binaries, starting from the largest binary, typically a wide stellar binary with periods as long as thousands of years, and ending with tight planetary orbits with periods as short as a few days. Moons enrich the complexity of this picture even more. 

So far, the secular, i.e. long-term, dynamics of high-order multiple systems have hardly been explored. This is largely due to the computational complexity of integrating, with sufficient precision, dynamical systems with orbital periods differing by many orders of magnitude \citep{2015ComAC...2....2B}. An approximate, but nevertheless useful, approach is to average the Hamiltonian over individual orbits, effectively smearing out the point mass moving in each orbit into a ring. The resulting dynamics are that of interacting rings rather than point masses. The typical required time-step of the 
orbit-averaged equations of motion is a (sufficiently) small fraction of the secular time-scales, which are much longer than the orbital periods. This allows for much faster integration. 

This method of averaging has a long history, dating back to the seminal works of Laplace and Lagrange (\citealt{lagrange_1781,laplace_1784}; see \citealt{2012arXiv1209.5996L} for an historic overview). In Laplace-Lagrange (LL) theory, bodies with similar masses are assumed to orbit a central massive body. The Hamiltonian is expanded in terms of the orbital eccentricities and inclinations which are assumed to be small, and non-secular terms are dropped. This method works well for planetary systems, provided that the eccentricities and inclinations are close to zero. However, on long time-scales (after thousands or more secular oscillations), secular chaotic diffusion can cause high eccentricities and inclinations, at which point the expansion is no longer accurate. For example, in the Solar system, the orbit of Mercury is unstable due to secular chaos on a time-scale of $\sim 5 \, \mathrm{Gyr}$ relative to the current Solar system \citep{1994A&A...287L...9L,2002MNRAS.336..483I,2008Icar..196....1L,2009Natur.459..817L}. Furthermore, secular chaos can lead to the production of hot Jupiters, and, more generally, play a crucial role in the long-term evolution of planetary systems \citep{2011ApJ...739...31L,2011ApJ...735..109W,2014PNAS..11112610L}. Also, in LL theory, the hierarchy is fixed, and more complex configurations such as moons around the planets, or one or more stellar companions to the central star, are not possible. 

An alternative method has been developed and applied by \citet{1962P&SS....9..719L,1962AJ.....67..591K} and later by \citet{1968AJ.....73..190H,2013MNRAS.431.2155N} and others. In this method, which has been applied to hierarchical triples, the Hamiltonian is expanded in terms of the ratio of the inner to the outer binary separations, which is assumed to be small. A major advantage of this method is that it is valid for arbitrary eccentricities and inclinations, provided that sufficient hierarchy is maintained (i.e. the bodies in the inner and outer binaries are sufficiently separated even at the closest mutual points in their orbits). 

\begin{figure}
\center
\includegraphics[scale = 0.47, trim = 0mm 15mm 0mm 10mm]{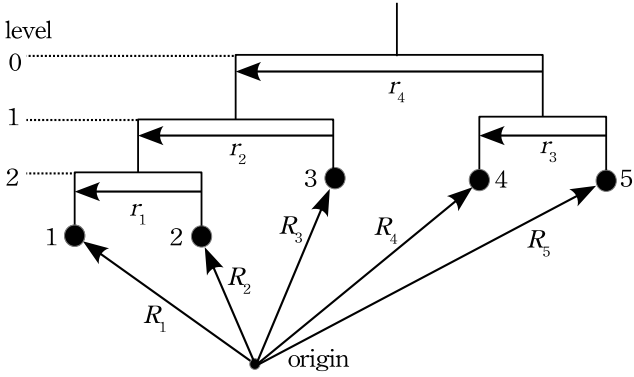}
\caption{\small A schematic representation of a hierarchical quintuple in a mobile diagram \citep{1968QJRAS...9..388E}. Absolute position vectors of bodies with respect to an arbitrary and fixed origin are denoted with capital letters, i.e. with $\ve{R}_i$. Binary separation vectors are denoted with small letters, i.e. with $\ve{r}_i$. Note that other choices for the directions of the relative vectors are also possible; e.g., $\ve{r}_1$ can also be defined as $\ve{r}_1=\ve{R}_2-\ve{R}_1$, i.e. pointing from body 1 towards body 2. Consequently, some of the quantities $B_{ijk}$ will change sign (cf. Appendix\,\ref{app:der:struc}). }
\label{fig:example_quintuple}
\end{figure}

Until recently, this approach has been limited to hierarchical triples. An extension to hierarchical quadruples was presented by \citet{2015MNRAS.449.4221H}, who derived the expanded and orbit-averaged Hamiltonian for both the `3+1' and `2+2' configurations. Here, we present a generalization to hierarchical multiple systems composed of binary orbits, with an arbitrary number of bodies in an arbitrary hierarchy. The main assumptions are as follows.
\begin{itemize}
\item The system is composed of nested binary orbits (see \F\,\ref{fig:example_quintuple} for an example quintuple system). Each binary has two `children', which can be either bodies (i.e. point particles) or binaries. If a child is a binary, then its position vector is the centre of mass position vector of all bodies contained within that child. This excludes trapezium-type systems (e.g. \citealt{ambartsumian_54}). 
\item The system is sufficiently hierarchical (see \S\,\ref{sect:der} for a quantitative definition). 
\item On short time-scales, the binary orbits are well approximated by bound Kepler orbits (i.e. the time-scales for angular momentum-exchange are much longer than the orbital periods).
\end{itemize}

Although the above assumptions may appear restrictive, it turns out that our method applies to a wide range of astrophysically relevant systems. Integration of the averaged and expanded equations of motion is computationally much less intensive compared to solving the full $N$-body problem. Therefore, our new method opens up the possibility for studying the long-term dynamics of complex hierarchical systems, without having to resort to costly $N$-body integrations. Such fast integration is particularly valuable, given the large parameter space involved in these systems,  

Moreover, the analytic perturbation method allows for specific effects to be associated with individual terms in the expansion. This allows for a clearer and deeper understanding of the gravitational dynamics compared to direct $N$-body integration, in which all effects are combined and cannot be disentangled. For example, in hierarchical triples, the octupole-order term, in contrast to the quadrupole-order term, can give rise to orbital flips between prograde and retrograde orbits, potentially leading to very high eccentricities \citep{2011ApJ...742...94L,2011PhRvL.107r1101K,2013MNRAS.431.2155N,2013ApJ...779..166T,2014ApJ...785..116L,2014ApJ...791...86L}.

This paper is structured as follows. In \S\,\ref{sect:der}, we summarize our derivation and our results of the generalized Hamiltonian for hierarchical multiple systems. A detailed and self-contained derivation is given in Appendix \ref{app:der}. We describe a new numerical algorithm to solve the orbit-averaged equations of motion. In the subsequent sections, \S s\,\ref{sect:multiplanet} and \ref{sect:observed}, we demonstrate the validity of the method and the algorithm and the potential of their applications, focusing on multiplanet systems in stellar multiple systems. We discuss our method in \S\,\ref{sect:discussion}, and we conclude in \S\,\ref{sect:conclusions}.

\section{The generalized Hamiltonian for hierarchical multiple systems}
\label{sect:der}

\subsection{Definition and description of the system}
\label{sect:der:def}
We consider hierarchical multiple systems that are composed of nested binaries and which can be represented by `simplex mobile' diagrams, introduced by \citet{1968QJRAS...9..388E}. We note that a similar hierarchical decomposition is made in the \textsc{Kira} integrator in \textsc{Starlab} (appendix B of \citealt{2001MNRAS.321..199P}). An example for a hierarchical quintuple system ($N=5$) is given in \F\,\ref{fig:example_quintuple}. The binaries have two `children', which can be either bodies, or binaries themselves. Conversely, each binary or body has a `parent', with the exception of the unique binary from which all other binaries and bodies originate, i.e. the `top' binary.  Each binary can be assigned a `level', defined as the number of binaries that need to be traversed in the mobile diagram to reach that binary, counting from the top binary. 

Here, we consider only multiple systems with the `simplex' structure, and not the more general `multiplex' structure, in which members of the system are allowed to have more than two children \citep{1968QJRAS...9..388E}. Note, however, that the structure of most observed hierarchical multiple systems is of the simplex type (e.g. \citealt{2014AJ....147...86T,2014AJ....147...87T}). This is also expected on the basis of the requirement of dynamical stability on long time-scales \citep{1981CeMec..25...51N}.

Mathematically, the structure of the system can be specified in terms of a `mass ratio matrix' $\bf{A}$. The components of this $N\times N$-matrix are mass ratios of bodies within the system. The matrix $\bf{A}$ is defined such that it gives the relation between the absolute position vectors $\ve{R}_i$ of bodies in the system, and the relative binary separation vectors $\ve{r}_i$ (see also \F\,\ref{fig:example_quintuple}). This description is similar to that used by \citet{1981CeMec..25...51N,1983CeMec..29..149W,1988CeMec..41..333A}. For a precise definition of $\bf{A}$ and examples, we refer to Appendix\,\ref{app:der:struc}.

\subsection{The expanded Hamiltonian}
\label{sect:der:ex}
In Appendix\,\ref{app:der}, we derive, from first principles, the general (Newtonian) Hamiltonian for hierarchical multiple systems as defined above, and we expand it in terms of ratios $x_i$ of separations of binaries at different levels. These ratios are assumed to be small, i.e. we assume that the system is hierarchical. The general expression is comprised of summations over both binaries and bodies, and is therefore not very useful for our purposes. Fortunately, it is possible to rewrite these summations over both binaries and bodies to summations over only binaries, and we carry out this rewriting up to and including fifth order in $x_i$, where for the highest two orders, $x_i^4$ and $x_i^5$, we only include binary pair terms. Our main results, at the various orders, are as follows.
\begin{enumerate}
\item The terms in the Hamiltonian at the lowest order, i.e. first order in $x_i$ (`dipole order'), vanish identically for {\it any} hierarchical configuration. 
\item The Hamiltonian at second order in $x_i$ (`quadrupole order') consists of terms which, individually, depend on binary {\it pairs}. For each pair, the binaries must be on different levels, and they must also be connected to each other through their children, i.e. the binary with the highest level must be a descendant of the binary with the lowest level. For example, in \F\,\ref{fig:example_quintuple}, the included binary pairs are $(r_1,r_4)$, $(r_2,r_4)$, $(r_3,r_4)$ and $(r_1,r_2)$, whereas pairs $(r_2,r_3)$ and $(r_1,r_3)$ are excluded. The individual binary pair terms are mathematically equivalent to the quadrupole-order terms of the hierarchical three-body Hamiltonian. 
\item At the `octupole order', i.e. third order in $x_i$, the Hamiltonian consists of two types of terms.
    \begin{enumerate}
    \item Terms which, individually, depend on binary {\it pairs}, with the binaries at different levels. The pairing occurs similarly as for the quadrupole-order terms. Mathematically, the pairwise terms are equivalent to the octupole-order terms of the hierarchical three-body Hamiltonian. 
    \item Terms which, individually, depend on binary {\it triplets}, with the binaries at different levels. The triplets to be included are those for which the binary $p$ with the highest level is a `descendent' of the binary with the intermediate level, $u$, and for which $u$ is a descendent of the binary $k$ with the lowest level (cf. equation~\ref{eq:S3_rewr}). In other words, all three binaries must be part of the same branch. In \F\,\ref{fig:example_quintuple}, there is only one triplet term, which applies to the binary triplet $(r_1,r_2,r_4)$. 
    
    The mathematical form of the individual triplet terms is equivalent to the term depending on three binaries at the octupole order in the hierarchical four-body Hamiltonian (`3+1' configuration; this term was first derived by \citealt{2015MNRAS.449.4221H}, where it was referred to as a `cross term'). In practice, at least for the type of systems considered in this paper, the triplet term can be safely neglected compared to the binary pair terms, even those of higher orders.
    \end{enumerate}
\item At the `hexadecupole order', i.e. fourth order in $x_i$, the Hamiltonian consists of three types of terms which depend on binary pairs, triplets and quadlets, respectively. Here, we have restricted to explicitly deriving the binary pair terms only. 
\item At the `dotriacontupole order', i.e. fifth order in $x_i$, the Hamiltonian consists of four types of terms which depend on binary pairs, triplets, quadlets and quintlets, respectively. As for the hexadecupole order, we have exclusively derived the binary pair terms explicitly. 
\item Generally, at order $n$ in $x_i$, there are $n-1$ types of terms. The individual terms depend on at most $n$ binaries. Whether all these terms appear is contingent on the system: if the system does not contain a sufficiently complex hierarchy, then not all different binary terms may appear. For example, in the case of a hierarchical triple, only the binary pair terms appear; a summation with three or more different binaries at high orders does not apply.  
\end{enumerate}

In summary, the general hierarchical $N$-body Hamiltonian can be constructed from the Hamiltonians of smaller subsystems. The higher the order, the larger the number of binaries within these subsystems. To lowest nontrivial order, i.e. the quadrupole order, the $N$-body Hamiltonian consists entirely of combinations of three-body (i.e. binary pair) Hamiltonians. In other words, at this order, only the interactions between binary pairs are included. 

As we show below (cf. \S\,\ref{sect:multiplanet}), the inclusion of high-order binary pair terms is necessary particularly in planetary systems, where the separation ratios may not be $\ll 1$. For example, in the Solar system (including the planets plus Pluto), the mean of the adjacent ratios of semimajor axes (i.e. $a_\mathrm{Mercury}/a_\mathrm{Venus}$, $a_\mathrm{Venus}/a_\mathrm{Earth}$, etc.) is $\approx 0.58$. Good results for this system are obtained only when terms of high orders are included.

\subsection{Orbit averaging}
\label{sect:der:oa}
After expanding the Hamiltonian in terms of the $x_i$, we orbit average the Hamiltonian assuming that the motion in each binary on suborbital time-scales is exactly Keplerian (cf. Appendix\,\ref{app:der:av}). We employ a vector formalism, where each binary orbit is described by its eccentricity vector, $\ve{e}_k$, and $\ve{j}_k$. Here, $\ve{j}_k$ is the specific angular momentum vector $\ve{h}_k$ normalized to the angular momentum of a circular orbit $\Lambda_k$, i.e.
\begin{align}
\ve{j}_k \equiv \ve{h}_k/\Lambda_k,
\end{align}
where 
\begin{align}
\ve{h}_k \equiv \ve{r}_k \times \dot{\ve{r}}_k
\end{align}
and 
\begin{align}
\label{eq:Lambda_main}
\Lambda_k = M_{k.\mathrm{C1}} M_{k.\mathrm{C2}} \, \sqrt{ \frac{G a_k}{M_k}}.
\end{align}
We define the mass of binary $k$, $M_k$, as the sum of the masses of {\it all bodies} contained within the hierarchy of binary $k$. The masses of the two children of binary $k$ are denoted with $M_{k.\mathrm{C1}}$ and $M_{k.\mathrm{C2}}$. Evidently, the latter satisfy
\begin{align}
M_{k.\mathrm{C1}} + M_{k.\mathrm{C2}} = M_k.
\end{align}
For example, in \F\,\ref{fig:example_quintuple}, $M_2 = m_1+m_2+m_3$, $M_{2.\mathrm{C1}} = m_1+m_2$ and $M_{2.\mathrm{C2}} = m_3$. Of course, the choice of which child of binary 2 is `child 1' or `child 2' is arbitrary; the alternative, and equally valid, choice is  $M_{2.\mathrm{C1}} = m_3$ and $M_{2.\mathrm{C2}} = m_1+m_2$.

As described in more detail in Appendix\,\ref{app:der:av}, the averaging procedure (cf. equation~\ref{eq:app:avdef}) is not a canonical transformation. However, a transformation of the `coordinates' $\ve{e}_k$ and $\ve{j}_k$ can be found, which leads to a transformed Hamiltonian that is equivalent to the old Hamiltonian, and amounts to averaging the Hamiltonian over the orbits \citep{2013MNRAS.431.2155N}. The transformed coordinates differ from the original ones.  However, as noted by \citet{2013MNRAS.431.2155N}, the untransformed and the transformed coordinates differ by order $x_i^2$. Therefore, this difference can usually be neglected. This is borne out by tests of our algorithm with other methods (cf. \S\,\ref{sect:multiplanet}).

Nevertheless, it is important to keep in mind that the orbit-averaged approximation can break down in situations where the time-scale for changes in the (secular) orbital parameters (in particular, the eccentricities) is shorter than the orbital period \citep{2012ApJ...757...27A,2014ApJ...781...45A,2014MNRAS.439.1079A}. In our numerical algorithm (cf. \S\,\ref{sect:der:alg}), we check for this condition during the integrations. In the context of planetary systems, the orbit-averaged approximation breaks down when the planets are sufficiently closely spaced for mean motion resonances to be important, which are not taken into account.

We have explicitly derived all orbit-averaged terms up and including octupole order, and including the pairwise hexadecupole- and dotriacontupole-order terms (cf. Appendix \ref{app:der:av:HO}). Furthermore, using complex integration techniques, we have derived a general expression for the averaged pairwise Hamiltonian to order $n$ in terms of quickly-evaluated derivatives of polynomial functions, rather than integrals (cf. Appendix \ref{app:der:av:gen_pair}).

\subsection{General implications}
\label{sect:der:impl}

\subsubsection{Quadrupole order}
\label{sect:der:impl:quad}
An immediate implication of \S\,\ref{sect:der:ex} is that to quadrupole order, the qualitative behaviour of the system can be characterized using simple arguments based on the LK time-scales of binary pairs, analogously to the ratio of LK time-scales that was considered for quadruple systems in \citet{2015MNRAS.449.4221H}. 

An order-of-magnitude estimate of the LK time-scale $P_{\mathrm{LK},pk}$, $\tau_{pk}$, for binaries $p$ and $k$ with $p$ a child of $k$, can be obtained by dividing $\Lambda_p$ by the associated order-of-magnitude quadrupole-order term in the orbit-averaged Hamiltonian (cf. equations~\ref{eq:EOM} and \ref{eq:S2_av}). This gives
\begin{align}
\nonumber \tau_{pk} &\equiv \Lambda_p \frac{M_p}{M_{p.\mathrm{C1}}M_{p.\mathrm{C2}}} \frac{a_k}{G M_{k.\mathrm{CS}(p)}} \left ( \frac{a_k}{a_p} \right )^2 \left (1-e_k^2 \right )^{3/2} \\
&= \frac{1}{2\pi} \frac{P_{\mathrm{orb},k}^2}{P_{\mathrm{orb},p}} \frac{M_k}{M_{k.\mathrm{CS}(p)}} \left (1-e_k^2 \right )^{3/2},
\end{align}
where we used equation~(\ref{eq:Lambda_main}), and the orbital periods $P_{\mathrm{orb},k}$ are defined through Kepler's law, 
\begin{align}
\label{eq:P_orb}
P_{\mathrm{orb},k} = 2\pi \sqrt{\frac{a_k^3}{GM_k}}.
\end{align}
In the case of isolated triples, the precise LK time-scale to quadrupole order (and with the further assumption of highly hierarchical systems) also depends on the initial $e_p$ and $\omega_p$ (i.e. the inner orbit eccentricity and argument of pericentre), as well as the relative inclination $i_{pk}$ between the inner and outer orbits \citep{2007CeMDA..98...67K}. Formally, we write the LK time-scale for hierarchical $N$-body systems as
\begin{align}
\label{eq:P_LK}
P_{\mathrm{LK},pk} = C(e_p,\omega_p,i_{pk}) \frac{P_{\mathrm{orb},k}^2}{P_{\mathrm{orb},p}} \frac{M_k}{M_{k.\mathrm{CS}(p)}} \left (1-e_k^2 \right )^{3/2}.
\end{align}
Here, the dimensionless factor $C(e_p,\omega_p,i_{pk})$ captures any dependence on $e_p$, $\omega_p$ and $i_{pk}$; generally, $C$ is of order unity. For example, for a hierarchical triple system with inner binary masses $m_1$ and $m_2$ and tertiary mass $m_3$, equation~(\ref{eq:P_LK}) reduces to the familiar expression (e.g. \citealt{1997AJ....113.1915I,1998MNRAS.300..292K,2015MNRAS.452.3610A})
\begin{align}
P_\mathrm{LK;in,out} = C(e_\mathrm{in},\omega_\mathrm{in},i_\mathrm{rel}) \frac{P_{\mathrm{orb},\mathrm{out}}^2}{P_{\mathrm{orb},\mathrm{in}}} \frac{m_1+m_2+m_3}{m_3} \left (1-e_\mathrm{out}^2 \right )^{3/2}.
\end{align}

The LK time-scale $P_{\mathrm{LK},pk}$ is a proxy for the strength of the secular torque of orbit $k$ on orbit $p$. Therefore, ratios of LK time-scales applied to different pairs of binaries give a measure for the relative importance of the secular torques. 

\begin{figure}
\center
\includegraphics[scale = 0.42, trim = 0mm 15mm 0mm 10mm]{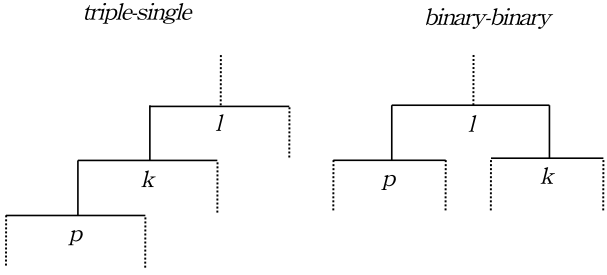}
\caption{\small Left: mobile diagram of a `fully nested' subsystem consisting of three binaries $p$, $k$ and $l$ for which $p$ is a child of $k$, and $k$ is a child of $l$, used to define the quantity $\mathcal{R}_{pkl}^{(\mathrm{fn})}$ (cf. equation~\ref{eq:R_def_fn}). Right: mobile diagram of a `binary-binary' subsystem consisting of three binaries $p$, $k$ and $l$ for which $p$ is a child of $l$, and $k$ is a child of $l$, used to define the quantity $\mathcal{R}_{pkl}^{(\mathrm{bb})}$ (cf. equation~\ref{eq:R_def_bb}). }
\label{fig:mobile_fn_bb}
\end{figure}

In the case of three adjacent binaries, $p$, $k$ and $l$, for which $p$ is a child of $k$, and $k$ is a child of $l$ (i.e. a `fully nested' subsystem, see the left part of \F\,\ref{fig:mobile_fn_bb}), we consider the ratio
\begin{align}
\label{eq:R_def_fn}
\nonumber &\mathcal{R}_{pkl}^{(\mathrm{fn})} \equiv \frac{P_{\mathrm{LK},{pk}}}{P_{\mathrm{LK},{kl}}} \\
\nonumber &= \frac{ C(e_p,\omega_p,i_{pk})}{ C(e_k,\omega_k,i_{kl})} \frac{P_{\mathrm{orb},k}^3}{P_{\mathrm{orb},p} P_{\mathrm{orb},l}^2} \frac{M_k}{M_l} \frac{M_{l.\mathrm{CS}(k)}}{M_{k.\mathrm{CS}(p)}} \left ( \frac{1-e_k^2}{1-e_l^2} \right )^{3/2} \\
&= \frac{ C(e_p,\omega_p,i_{pk})}{ C(e_k,\omega_k,i_{kl})} \left ( \frac{a_k^3}{a_p a_l^2} \right )^{3/2} \left ( \frac{M_p}{M_k} \right )^{1/2} \frac{M_{l.\mathrm{CS}(k)}}{M_{k.\mathrm{CS}(p)}} \left ( \frac{1-e_k^2}{1-e_l^2} \right )^{3/2},
\end{align}
where in the last line, the expression was written in terms of semimajor axes. Note that for quadruple systems, equation~(\ref{eq:R_def_fn}) reduces to equation (13) of \citet{2015MNRAS.449.4221H}.

Analogously to the case of fully nested hierarchical quadruple systems \citep{2015MNRAS.449.4221H}, the following qualitative behaviour applies.
\begin{enumerate}
\item $\mathcal{R}_{pkl}^{(\mathrm{fn})} \ll 1$: the torque of $k$ on $p$ dominates compared to the torque of $l$ on $k$. LK oscillations can be induced by $k$ on $p$. Furthermore, induced precession on $k$ from $p$ can quench LK oscillations in $k$ otherwise induced by $l$.
\item $\mathcal{R}_{pkl}^{(\mathrm{fn})} \gg 1$: the torque of $l$ on $k$ dominates compared to the torque of $k$ on $p$. LK oscillations can be induced by $l$ on $k$. Effectively, $p$ can be approximated as a point mass.
\item $\mathcal{R}_{pkl}^{(\mathrm{fn})} \sim 1$: complex, nonregular LK cycles can be induced in $p$; chaotic evolution is possible.
\end{enumerate}

In the case of three binaries, $p$, $k$ and $l$, for which $p$ is a child of $l$, and $k$ is a child of $l$ (i.e. a `binary-binary' subsystem, see the right part of \F\,\ref{fig:mobile_fn_bb}), we consider the ratio
\begin{align}
\label{eq:R_def_bb}
\nonumber \mathcal{R}_{pkl}^{(\mathrm{bb})} &\equiv \frac{P_{\mathrm{LK},{pl}}}{P_{\mathrm{LK},{kl}}} = \frac{ C(e_p,\omega_p,i_{pl})}{ C(e_k,\omega_k,i_{kl})} \frac{P_{\mathrm{orb},k}}{P_{\mathrm{orb},p}} \frac{M_p}{M_k} \\
&= \frac{ C(e_p,\omega_p,i_{pl})}{ C(e_k,\omega_k,i_{kl})} \left ( \frac{a_k}{a_p} \right )^{3/2} \left ( \frac{M_p}{M_k} \right )^{3/2}
\end{align}
(note that $M_{l.\mathrm{CS}(k)} = M_p$ and $M_{l.\mathrm{CS}(p)} = M_k$). 

In this case, the following characteristics apply.
\begin{enumerate}
\item $\mathcal{R}_{pkl}^{(\mathrm{bb})} \ll 1$: the torque of $l$ on $p$ dominates compared to the torque of $l$ on $k$. For sufficiently small values of $\mathcal{R}_{pkl}^{(\mathrm{bb})}$, $k$ can effectively be treated as a point mass. 
\item $\mathcal{R}_{pkl}^{(\mathrm{bb})} \gg 1$: the torque of $l$ on $k$ dominates compared to the torque of $l$ on $p$. For sufficiently large values of $\mathcal{R}_{pkl}^{(\mathrm{bb})}$, $p$ can effectively be treated as a point mass. 
\item $\mathcal{R}_{pkl}^{(\mathrm{bb})} \sim 1$: complex, nonregular LK cycles can be induced in $p$ and $k$ \citep{2013MNRAS.435..943P}.
\end{enumerate}

\subsubsection{Higher orders}
\label{sect:der:impl:higher}
In hierarchical triple systems, the octupole-order term gives rise to generally more complex, and potentially chaotic eccentricity oscillations compared to when including only the quadrupole-order term, especially if the semimajor axis ratio is small and/or the outer orbit eccentricity is high. Orbital flips can occur between prograde and retrograde orbits, potentially leading to very high eccentricities \citep{2011ApJ...742...94L,2011PhRvL.107r1101K,2013MNRAS.431.2155N,2013ApJ...779..166T,2014ApJ...785..116L,2014ApJ...791...86L}. The importance of these effects can be evaluated using the octupole parameter, which is essentially the ratio of the order-of-magnitude estimate of the orbit-averaged octupole-order term to the quadrupole-order term. It is given by \citep{2013MNRAS.431.2155N}
\begin{align}
\label{eq:eps_oct_triple}
\epsilon_\mathrm{oct,triple} = \frac{\left |m_1-m_2 \right |}{m_1+m_2} \frac{a_\mathrm{in}}{a_\mathrm{out}} \frac{e_\mathrm{out}}{1-e_\mathrm{out}^2}.
\end{align}
In the case of hierarchical $N$-body systems, the situation is more complicated because there will generally be more than one pair of binaries, and because at higher orders (starting with the octupole order), terms appear which individually depend on more than two binaries (cf. \S\,\ref{sect:der:ex}). 

In the applications considered in this paper, the octupole-order triplet terms are found to be negligible compared to the octupole-order binary pair terms, and binary pair terms at higher orders (e.g. \S\,\ref{sect:multiplanet:single}). This can be understood by estimating the ratio of the orbit-averaged octupole-order binary triplet term, $\overline{S}'_{3;3}$, to the orbit-averaged octupole-order binary pair term, $\overline{S}'_{3;2}$ (cf. equation~\ref{eq:S3_av}). Consider three binaries $p$, $u$ and $k$, where $p \in \{k.\mathrm{C}\}$, $p \in \{u.\mathrm{C} \}$ and $u \in \{k.\mathrm{C} \}$ (i.e. a connected binary triplet in which each binary is on a different level). A distinction should be made when evaluating $\overline{S}'_{3;2}$: we apply it to $(p,u)$ and $(u,k)$ (we do not consider $(p,k)$ because by assumption, $a_p/a_u \gg a_p/a_k$). In the case of $(p,u)$, we find
\begin{align}
\label{eq:ratio_S32_S33_p_u}
\left | \frac{ \overline{S}'_{3;3} }{ \overline{S}'_{3;2} } \right |_{(p,u)} \sim \frac{e_k}{e_p} \frac{M_{k.\mathrm{CS}(p)}}{\left |M_{p.\mathrm{C1}} - M_{p.\mathrm{C2}} \right |} \frac{M_p}{M_u} \left ( \frac{a_u}{a_k} \right )^4 \left (\frac{a_u}{a_p} \right ) \left ( \frac{1-e_u^2}{1-e_k^2} \right )^{7/2}.
\end{align}

In the case of $(u,k)$,
\begin{align}
\label{eq:ratio_S32_S33_u_k}
\left | \frac{ \overline{S}'_{3;3} }{ \overline{S}'_{3;2} } \right |_{(u,k)} \sim \frac{M_{u.\mathrm{CS}(p)}}{\left | M_{u.\mathrm{C1}}-M_{u.\mathrm{C2}} \right |} \frac{M_{p.\mathrm{C1}}M_{p.\mathrm{C2}}}{M_{u.\mathrm{C1}}M_{u.\mathrm{C2}}} \frac{M_u}{M_p} \left ( \frac{a_p}{a_u} \right )^2.
\end{align}
Note that equations~(\ref{eq:ratio_S32_S33_p_u}) and (\ref{eq:ratio_S32_S33_u_k}) diverge as $M_{p.\mathrm{C1}} \rightarrow M_{p.\mathrm{C2}}$, $M_{u.\mathrm{C1}} \rightarrow M_{u.\mathrm{C2}}$ and/or $e_p\rightarrow 0$. In those cases, the corresponding term $\overline{S}'_{3;2}$ vanishes. However, additionally, the binary pair hexadecupole-order term does not vanish (cf. equation~\ref{eq:S4_av}). Therefore, one should compare to $\overline{S}'_{4;2}$ (this is not done here). 

In the case of a planetary system with order Jupiter-mass planets around an order solar-mass star (with the star in the innermost orbit $p$), $M_{k.\mathrm{CS}(p)} \sim 10^{-3} \, \mathrm{M}_\odot$, $|M_{p.\mathrm{C1}}-M_{p.\mathrm{C2}}| \sim 1\, \mathrm{M}_\odot$, $M_p \sim M_u \sim 1\, \mathrm{M}_\odot$, $a_u \ll a_k$ and $a_u \gg a_p$ with $a_p/a_u \sim a_u/a_k$. Therefore, the mass ratios in equation~(\ref{eq:ratio_S32_S33_p_u}) are of the order of $10^{-3}$, whereas the semimajor axis ratio is $\ll 1$. Assuming eccentricities close to zero, this shows that the binary triplet term is negligible compared to the binary pair term in the case of $(p,u)$. Similarly, $M_{u.\mathrm{CS}(p)} \sim 10^{-3}\, \mathrm{M}_\odot$, $|M_{u.\mathrm{C1}}-M_{u.\mathrm{C2}}| \sim 1 \, \mathrm{M}_\odot$ and $M_{p.\mathrm{C1}} M_{p.\mathrm{C2}} \sim M_{u.\mathrm{C1}} M_{u.\mathrm{C2}} \sim 10^{-3} \, \mathrm{M}^2_\odot$, showing that a similar conclusion applies in the case of $(u,k)$.

In Appendix~\ref{app:der:est_nested}, we generalize these estimates for nested planetary systems of the importance of terms other than binary pair terms to arbitrary order $n$. 

When the non-pairwise binary terms are negligible, it suffices to consider the ratios of binary pair terms at sequential orders. We consider the ratio of the order-of-magnitude orbit-averaged binary pair term at order $n+1$, to that of order $n$. Using the general expression equation~(\ref{eq:S_n_pair_gen_final}), we find
\begin{align}
\label{eq:ratio_S_n}
\left | \frac{\overline{S}'_{n+1;2}}{\overline{S}'_{n;2}} \right |_{(p,k)} \sim \left | \frac{ M^n_{p.\mathrm{C2}} - (-1)^n M^n_{p.\mathrm{C1}} }{ M_p M^{n-1}_{p.\mathrm{C2}} + (-1)^n M_p M^{n-1}_{p.\mathrm{C1}} } \right | \frac{a_p}{a_k} \frac{f_n(e_k)}{1-e_k^2},
\end{align}
where $f_n(e_k)$ is a function of $e_k$. Based on the expressions in equations~(\ref{eq:S2_av}), (\ref{eq:S3_av}), (\ref{eq:S4_av}) and (\ref{eq:S5_av}), i.e. for $n \in \{2,3,4,5\}$,
\begin{align}
f_n(e_k) = \left \{
\begin{array}{cc}
e_k, & n \, \mathrm{even}; \\
1/e_k, & n \, \mathrm{odd},
\end{array} \right.
\end{align}
and we expect that this applies to any $n\geq 2$. For $n=2$, equation~(\ref{eq:ratio_S_n}) reads
\begin{align}
\left | \frac{\overline{S}'_{3;2}}{\overline{S}'_{2;2}} \right | \sim \frac{ \left | M_{p.\mathrm{C1}} - M_{p.\mathrm{C2}} \right | }{ M_{p.\mathrm{C1}} + M_{p.\mathrm{C2}} } \frac{a_p}{a_k} \frac{e_k}{1-e_k^2},
\end{align}
which reduces to equation~(\ref{eq:eps_oct_triple}) in the case of a hierarchical triple. 

Binary pair terms to high order turn out to be important in multiplanet systems. This is investigated in detail in \S\,\ref{sect:multiplanet:single}.

\subsection{Numerical algorithm}
\label{sect:der:alg}
We have developed a numerical algorithm written in \textsc{C++}, \textsc{\codename}, to efficiently integrate the equations of motion for hierarchical multiple systems based on the formalism described in \S\,\ref{sect:der}, and presented in detail in Appendix \ref{app:der}\footnote{This algorithm is a generalization of the algorithm that was presented in \citet{2015MNRAS.449.4221H}.}. The equations of motion are derived by taking gradients of the orbit-averaged Hamiltonian in terms of the orbital vectors, $\overline{H}$. Explicitly, the equations of motion are given by the Milankovitch equations (\citealt{milankovitch_39}, e.g. \citealt{1961JGR....66.2797M,1963PCPS...59..669A,1964RSPSA.280...97A,2005MNRAS.364.1222B,2009AJ....137.3706T}; see \citealt{2014CeMDA.118..197R} for a recent overview), i.e.
\begin{subequations}
\label{eq:EOM}
\begin{align}
\label{eq:EOM:j}
\frac{\mathrm{d} \boldsymbol{j}_k}{\mathrm{d} t} &= - \frac{1}{\Lambda_k} \left [ \, \boldsymbol{j}_k \times \nabla_{\boldsymbol{j}_k} \overline{H} + \boldsymbol{e}_k \times \nabla_{\boldsymbol{e}_k} \overline{H} \, \right ]; \\
\label{eq:EOM:e}
\frac{\mathrm{d} \boldsymbol{e}_k}{\mathrm{d} t} &= - \frac{1}{\Lambda_k} \left [ \, \boldsymbol{e}_k \times \nabla_{\boldsymbol{j}_k} \overline{H} + \boldsymbol{j}_k \times \nabla_{\boldsymbol{e}_k} \overline{H} \, \right ].
\end{align}
\end{subequations}
Here, $\Lambda_k$ is the angular momentum of a circular orbit given by equation~(\ref{eq:Lambda_main}). The system of first-order differential equations for the sets $(\ve{j}_k,\ve{e}_k)$ for all binaries is solved using the \textsc{CVODE} library \citep{1996ComPh..10..138C}.

Our algorithm is interfaced in the \textsc{AMUSE} framework \citep{2013CoPhC.183..456P,2013A&A...557A..84P}, and uses the \textsc{AMUSE} particle datamodel for the user to specify the hierarchical structure and associated parameters in the high-level \textsc{Python} language. Binaries and bodies are both members of a single particle set; the type of particle (body or binary) is specified with a Boolean parameter. The system structure is defined by linking the two children of each binary to other members of the same particle set. We have added support for easily switching between the secular approach and direct $N$-body integration with any of the direct $N$-body codes available in \textsc{AMUSE}.

The \textsc{\codename} algorithm also supports the inclusion of additional forces, including post-Newtonian (PN) terms to the 1PN and 2.5PN orders, and tidal bulges and tidal friction. The PN terms are implemented by treating the binaries in the system as being isolated, and adding the relevant equations for $\dot{\ve{e}}_k$ and $\dot{\ve{h}}_k$. Here, we neglect the contribution of any additional terms which may apply. For example, \citet{2013ApJ...773..187N} showed that for hierarchical triples and at the 1PN order, an additional term (an `interaction term') appears that is associated with both inner and outer orbits. Here, we only include the 1PN terms associated with individual binaries (cf. Appendix\,\ref{app:der:1PN}). 

To model the effects of tidal bulges and tidal friction, we adopt the equilibrium tide model of \citet{1998ApJ...499..853E} with a constant tidal time lag $\tau$ (or, equivalently, a constant viscous time-scale $t_\mathrm{V}$). For tidal friction, we adopted the equations derived by \citet{2009MNRAS.395.2268B} which are well defined in the limit $e_k \rightarrow 0$. We implement the equations in binaries where at least one of the children is a body. Here, we treat the `companion' as a point mass, even if the companion is, in reality, a binary. A self-consistent treatment of tides in hierarchical multiple systems is beyond the scope of this work. 

As mentioned in \S\,\ref{sect:der:oa}, the orbit-averaged approximation breaks down if the time-scale for changes of the angular momentum $j_k$ is smaller than the orbital time-scale \citep{2012ApJ...757...27A,2014ApJ...781...45A}. In \textsc{\codename}, we allow the user to check for this condition in any binary $k$ at any time in the integration. Using a root-finding algorithm, the integration is stopped whenever $t_{j,k} \leq P_{\mathrm{orb},k}$, where $P_{\mathrm{orb},k}$ is the orbital period of binary $k$ (cf. equation~\ref{eq:P_orb}) and $t_{j,k}$ is the time-scale for the angular momentum of binary $k$ to change by order itself. The latter is given by
\begin{align}
t_{j,k} = \left | \frac{1}{j_k} \frac{\mathrm{d} j_k}{\mathrm{d} t} \right |^{-1} = \left | \frac{e_k}{1-e_k^2} \frac{\mathrm{d} e_k}{\mathrm{d} t} \right |^{-1},
\label{eq:t_AM}
\end{align}
where the time derivates are obtained numerically from the equations of motion (cf. equations~\ref{eq:EOM}).

\section{Tests: S-type multiplanet systems in single and multiple stellar systems}
\label{sect:multiplanet}

\begin{figure}
\center
\includegraphics[scale = 0.47, trim = 0mm 15mm 0mm 10mm]{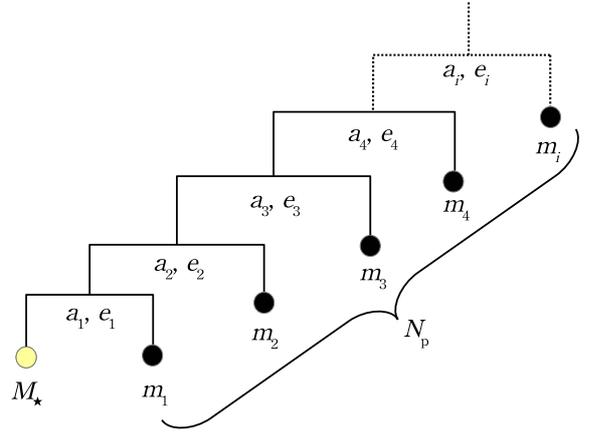}
\caption{\small A mobile diagram of a hierarchical multiplanet system with a single star (mass $M_\star$). In the framework of the formalism presented in Appendix \ref{app:der}, a multiplanet system can be represented as a `fully nested' system. Each planet is assumed to orbit around the centre of mass position of the central star plus the planets contained within its orbit. In practice, the latter position nearly coincides with that of the central star. The planetary masses are $m_i$, $1\leq i \leq N_\mathrm{p}$, where $N_\mathrm{p}$ is the number of planets. The same subindices are used to denote the orbital parameters. }
\label{fig:planetary_system}
\end{figure}

\subsection{Single-star systems}
\label{sect:multiplanet:single}
As a first test and application, we consider single-star multiplanet systems with planets in orbits with small eccentricities and inclinations. This allows for comparison of \textsc{\codename} with LL theory at relatively low order, i.e. to second order in the eccentricities and the inclinations. The hierarchy of the system is a `fully nested' configuration, and is depicted schematically in \F\,\ref{fig:planetary_system}. Note that in the framework of the formalism presented in Appendix \ref{app:der}, a multiplanet system can be represented as system in which each planet is assumed to orbit around the centre of mass of the central star plus the planets contained within its orbit. We consider various planet numbers $N_\mathrm{p}$. The mass of the star is assumed to be $M_\star = 1\, \mathrm{M}_\odot$. For simplicity, the planetary masses, $m_i$ with $1\leq i \leq N_\mathrm{p}$, are set equal to $m_i = 1 \, M_\mathrm{J}$.

\begin{figure}
\center
\includegraphics[scale = 0.5, trim = 15mm 0mm 0mm 0mm]{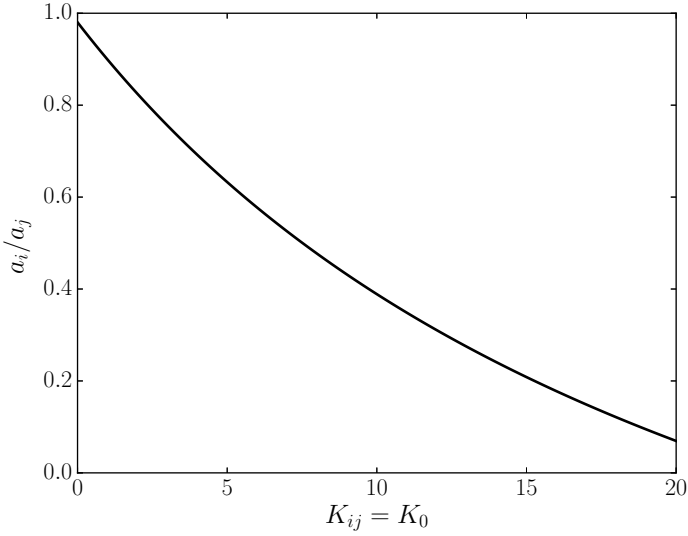}
\caption{\small The relation between the semimajor axis ratio $a_i/a_j$ and $K_{ij}=K_0$ according to equation~(\ref{eq:Kdef}). Assumed parameters are $M_\star = 1\, \mathrm{M}_\odot$, $m_i = 1 \, M_\mathrm{J}$ and $e_i = 0.01$.}
\label{fig:K_alpha_paper.eps}
\end{figure}

We consider a sequence of systems with different spacing between the planets. The spacing is quantified in terms of $K_{ij}$, which applies to pairs $(i,j)=(i,i+1)$ of adjacent planets, and is defined as
\begin{align}
\label{eq:Kdef}
K_{ij} \equiv \frac{a_j(1-e_j) - a_i(1+e_i)}{R_{\mathrm{H};ij}},
\end{align}
where $R_{\mathrm{H};ij}$ is the mutual Hill radius, defined by
\begin{align}
\label{eq:RH_def}
R_{\mathrm{H};ij} &\equiv \frac{a_i+a_j}{2C_{\mathrm{H};ij}}; \quad \quad \quad C_{\mathrm{H};ij} \equiv \left ( \frac{3M_\star}{m_i+m_j} \right )^{1/3}.
\end{align}
For simplicity, we assume a single initial spacing $K_{ij} = K_0$ for adjacent planets. Furthermore, we choose a single initial eccentricity of $e_i = 0.01$. With these assumptions and with the innermost semimajor axis $a_1$ specified, equation~(\ref{eq:Kdef}) defines the semimajor axes of an $N_\mathrm{p}$-planet system with constant semimajor axis ratios. For reference, we show in \F\,\ref{fig:K_alpha_paper.eps} the relation between the semimajor axis ratio $a_i/a_j$ and $K_0$ according to equation~(\ref{eq:Kdef}).

\begin{figure}
\center
\includegraphics[scale = 0.45, trim = 10mm 0mm 0mm 0mm]{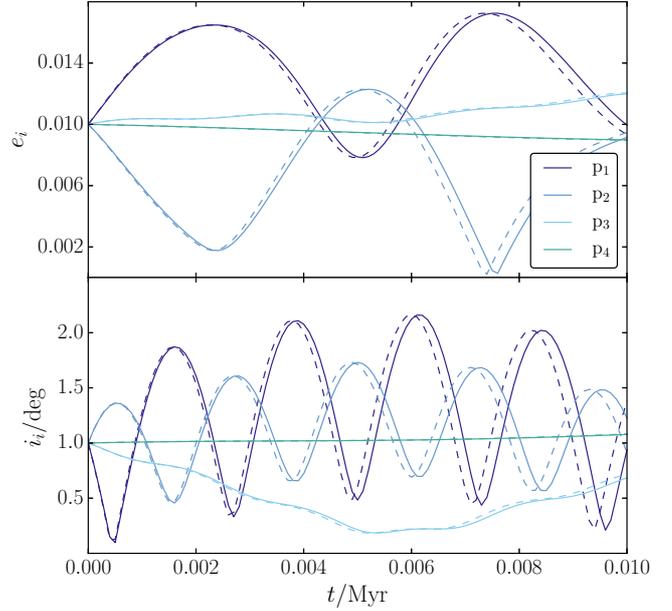}
\caption{\small The eccentricities (top panel) and inclinations (bottom panel) as a function of time for a four-planet system with spacing $K_0 = 12$ (cf. equation~\ref{eq:Kdef}). The initial eccentricities are $e_i = 0.01$ and inclinations $i_i=1^\circ$. Solid lines are according to \textsc{\codename} with all terms up and including octupole order, plus the hexadecupole and dotriacontupole binary pair terms. Dashed lines are according to second-order LL theory. Lines with different colours correspond to the four planets (see the legend). }
\label{fig:LL_example_all_planets_paper_Np_4e_example}
\end{figure}

\begin{figure}
\center
\includegraphics[scale = 0.45, trim = 10mm 0mm 0mm 0mm]{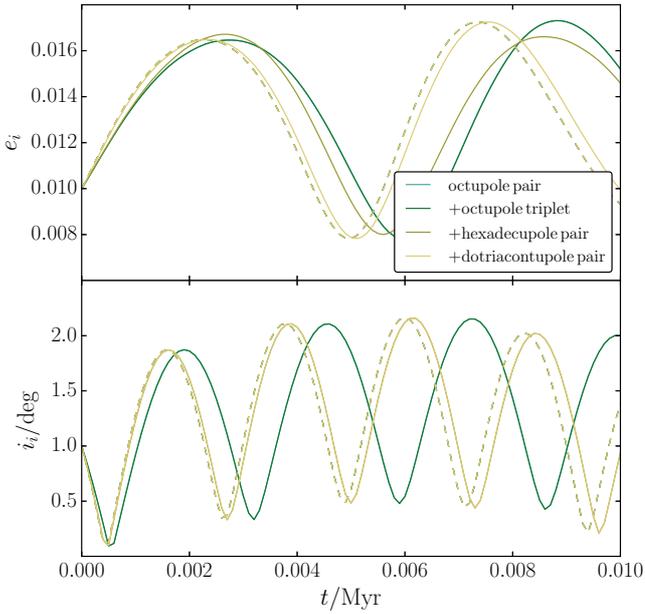}
\caption{\small As in \F\,\ref{fig:LL_example_all_planets_paper_Np_4e_example}, now showing the innermost planet's eccentricity and inclination according to LL theory (dashed lines), and \textsc{\codename} with inclusion of various terms (see the legend). Agreement of \textsc{\codename} with LL theory increases as higher order terms are included. }
\label{fig:LL_example_all_orders_paper_Np_4e_example}
\end{figure}

\begin{figure}
\center
\includegraphics[scale = 0.45, trim = 10mm 0mm 0mm 10mm]{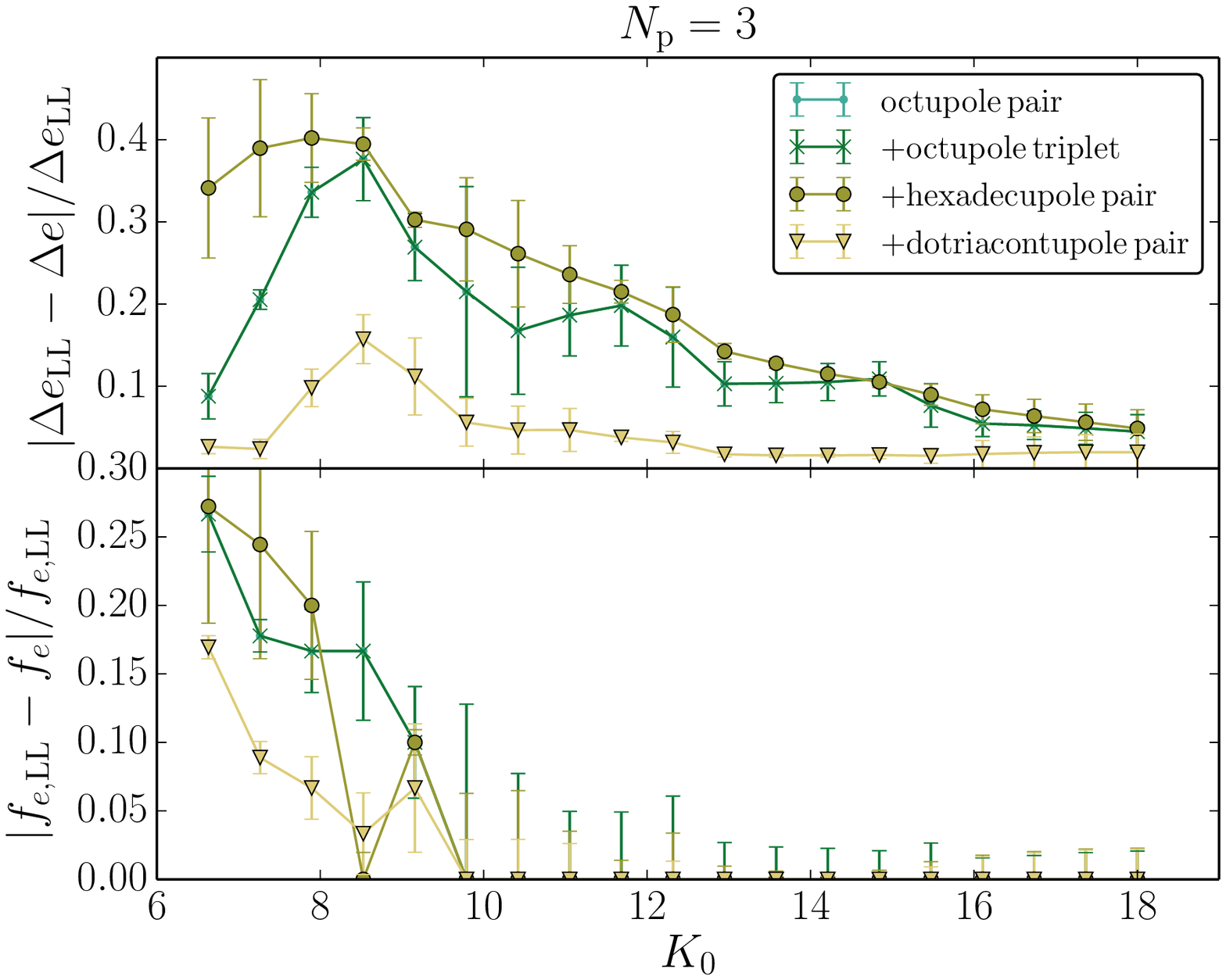}
\includegraphics[scale = 0.45, trim = 10mm 0mm 0mm 0mm]{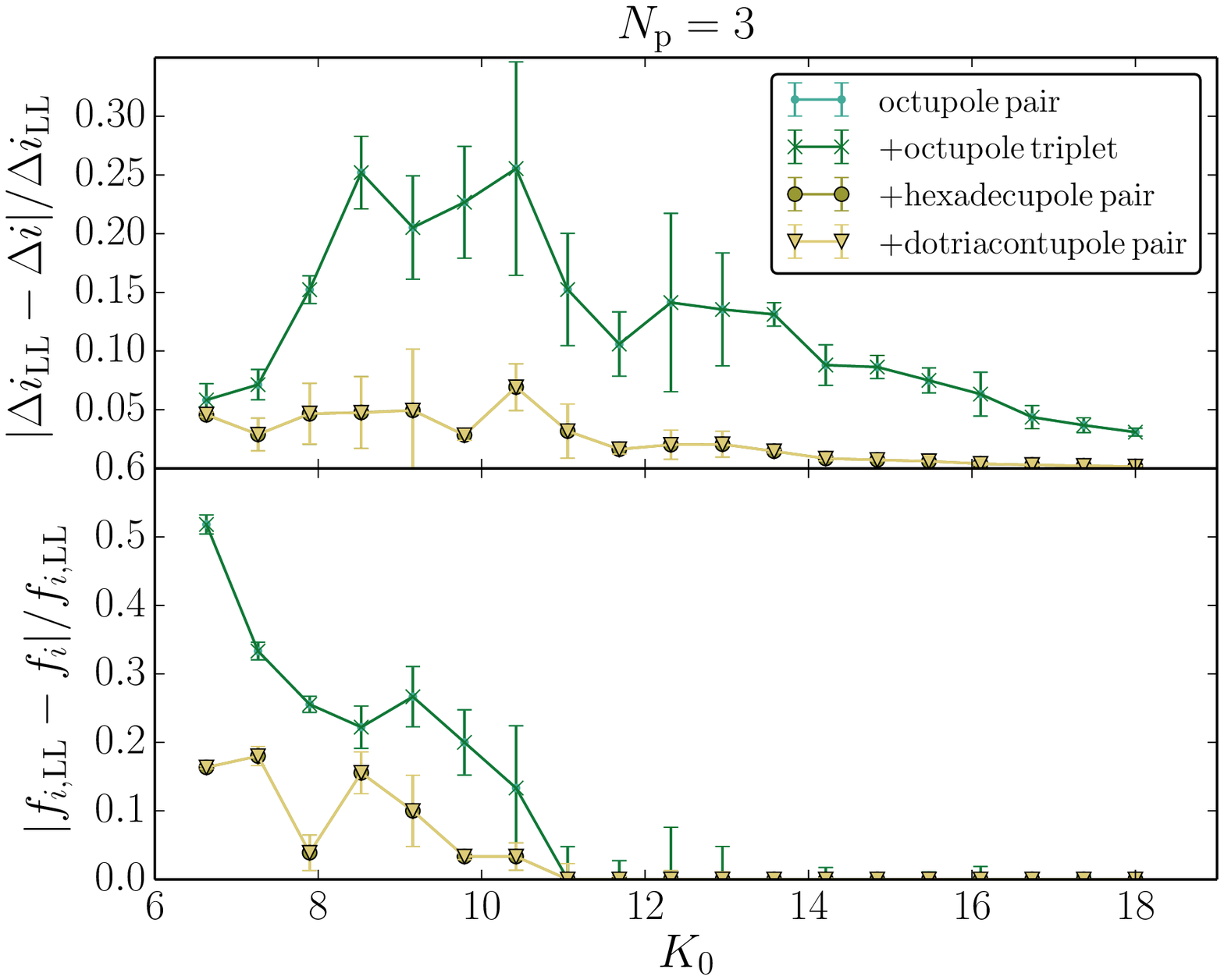}
\includegraphics[scale = 0.45, trim = 10mm 5mm 0mm 0mm]{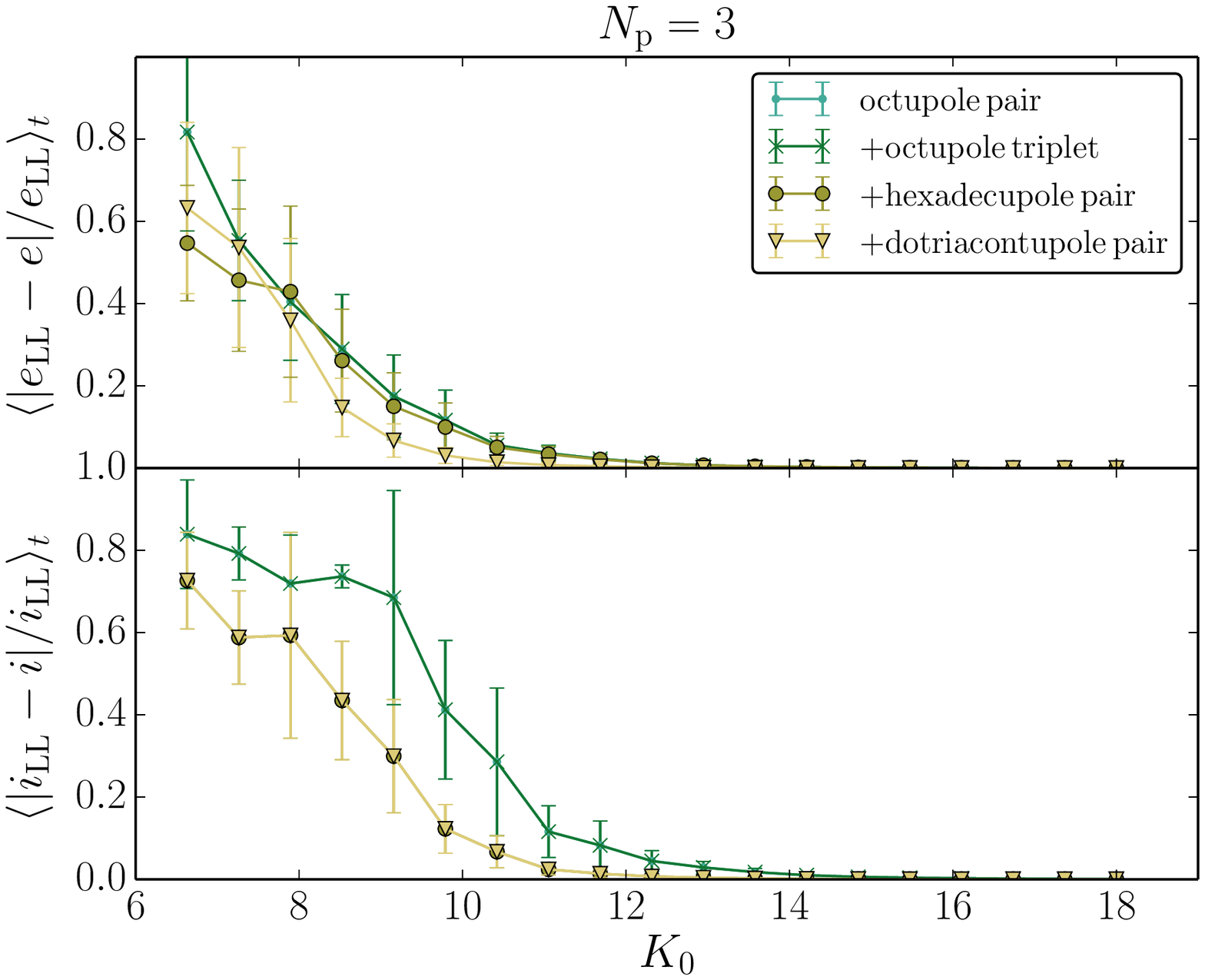}
\caption{\small Quantitative measures of agreement between \textsc{\codename} (with various terms, indicated in the legends) and LL theory, assuming $N_\mathrm{p}=3$. In the top two panels, we consider the fractional differences between the methods, of the amplitudes of the eccentricity oscillations (upper panel), and of the peak frequencies of the power spectra of the eccentricity oscillations (lower panel). Here, the solid lines represent the mean values of these measures of all planets; error bars show the standard deviations. In the middle two panels, we consider similar differences, now applied to the oscillations of the inclinations. Lastly, in the bottom two panels, we consider the time-averaged relative differences between the eccentricities (upper panel) and the inclinations (bottom panel). }
\label{fig:LL_sequence_e_paper_Np_3r}
\end{figure}

The initial inclinations are $i_i = 1^\circ$, where the $i_i$ are measured with respect to an arbitrary inertial reference frame. The longitudes of the ascending node, $\Omega_i$, are chosen randomly from a uniform distribution. Note that this implies that the orbits of the planets are (slightly) mutually inclined. The arguments of pericentre $\omega_i$ are also sampled randomly from a uniform distribution. Because the latter mainly affect the phases of the secular oscillations in the orbital elements, only a single realization of $\omega_i$ is taken for each sequence in $K_0$. For $\Omega_i$, five different values are sampled for each $K_0$, and results are averaged over the integrations with different $\Omega_i$. 

For the fixed parameters, we integrated a sequence of systems with different $K_{ij}=K_0$, $\Omega_i$ and $N_\mathrm{p}$. In \F\,\ref{fig:LL_example_all_planets_paper_Np_4e_example}, we show an example for $K_0=12$ and $N_\mathrm{p}=4$; the lines show the eccentricities (top panel) and the inclinations (bottom panel) of the planets as a function of time. Solid lines are according to \textsc{\codename} with the inclusion of all terms up and including octupole order, plus the hexadecupole and dotriacontupole binary pair terms. Dashed lines are according to second-order LL theory, computed using the equations given in \citet{1999ssd..book.....M}. Note that the latter are valid for any semimajor axis ratio and hence $K_0$, provided that the secular approximation is valid (i.e. mean motion resonances are not important). However, because an expansion is made in the eccentricities and inclinations, the latter variables should remain small. For the value of $K_0=12$ in \F\,\ref{fig:LL_example_all_planets_paper_Np_4e_example}, \textsc{\codename} is in good agreement with LL theory. 

Such agreement is contingent on the inclusion of sufficient high orders in \textsc{\codename}. To illustrate this, we show in \F\,\ref{fig:LL_example_all_orders_paper_Np_4e_example} the same setup as in \F\,\ref{fig:LL_example_all_planets_paper_Np_4e_example}, but now with the solid lines showing the elements of the innermost planet according to \textsc{\codename} with the inclusion of various terms. We consider the following combinations:
\begin{itemize}
\item quadrupole-order terms and octupole-order binary pair terms (i.e. without the octupole-order binary triplet terms);
\item the above, plus the octupole-order binary triplet terms;
\item the above, plus the hexadecupole-order binary pair terms;
\item the above, plus the dotriacontupole-order binary pair terms.
\end{itemize}

Clearly, the agreement with LL theory (dashed lines) becomes increasingly better when higher order terms are included. This is because of the expansion in terms of separation ratios, which reduces to ratios of semimajor axes after orbit averaging (note that the eccentricities and inclinations can, in principle, be arbitrarily large as long as orbit averaging is still appropriate and the condition of hierarchy is still satisfied). In the example, the semimajor axis ratio $a_i/a_j \approx 0.31$ is large compared to zero. Therefore, high orders are required for accurate results. Note that the octupole-order triplet terms do not have any noticeable effect. This was estimated before in \S\,\ref{sect:der:impl:higher}.

Generally, increasing $K_0$ implies smaller semimajor axis ratios, which improves the agreement of \textsc{\codename} with LL theory. To study this more quantitatively, and to determine how small $K_0$ can be chosen for still acceptable agreement, we show in \F\,\ref{fig:LL_sequence_e_paper_Np_3r} comparisons between \textsc{\codename} and LL theory, assuming $N_\mathrm{p}=3$. On the horizontal axes, $K_0$ is plotted, whereas the vertical axes show various quantitative measures for this agreement. In the top two panels of \F\,\ref{fig:LL_sequence_e_paper_Np_3r}, we consider the fractional differences between the methods, of the amplitudes of the eccentricity oscillations (upper panel) and of the peak frequencies of the power spectra of the eccentricity oscillations (lower panel). Here, the solid lines represent the mean values  of these measures of all planets; errorbars show the standard deviations. In the middle two panels of \F\,\ref{fig:LL_sequence_e_paper_Np_3r}, we consider similar differences, now applied to oscillations of the inclinations. Lastly, in the bottom two panels of \F\,\ref{fig:LL_sequence_e_paper_Np_3r}, we consider the time-averaged relative differences between the eccentricities (upper panel) and the inclinations (bottom panel). 

In all cases, there is no discernable difference between the inclusion of the octupole-order binary pair terms and the inclusion of all octupole-order terms (i.e. including the binary triplet terms); in \F\,\ref{fig:LL_sequence_e_paper_Np_3r}, the corresponding curves always overlap. When considering the {\it amplitude} of the eccentricity oscillations, the octupole-order terms, surprisingly, give better agreement with LL theory compared to the case of including the next higher order terms, the hexadecupole-order terms. However, the amplitude of the inclination oscillations tends to be better with the inclusion of the hexadecupole-order terms. Also, when considering the time-averaged measure (bottom two panels of \F\,\ref{fig:LL_sequence_e_paper_Np_3r}), including the hexadecupole-order terms also performs better compared to the octupole-order terms only. Note that for the inclination oscillations, there are little to no differences between the hexadecupole- and dotriacontupole-order terms. In contrast, the agreement with respect to eccentricity is typically much better when including dotriacontupole-order terms. 

\begin{figure}
\center
\includegraphics[scale = 0.45, trim = 10mm 0mm 0mm 0mm]{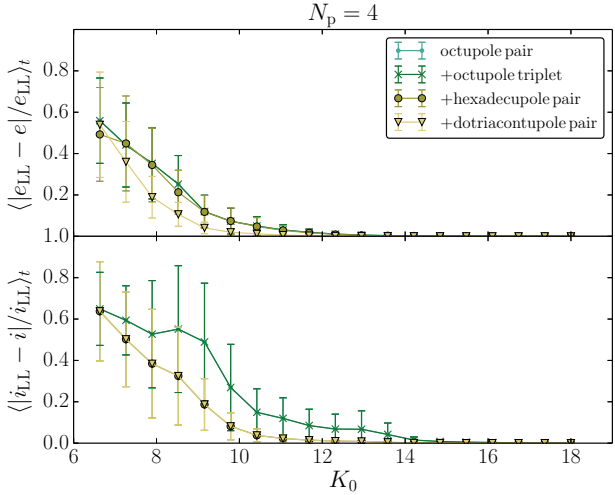}
\caption{\small Quantitative measures of agreement between \textsc{\codename} and LL theory as in \F\,\ref{fig:LL_sequence_e_paper_Np_3r}, now with four planets. Here, we only show the time-averaged relative differences between the eccentricities (upper panel) and the inclinations (bottom panel). }
\label{fig:LL_sequence_f_paper_Np_4r}
\end{figure}

\begin{figure}
\center
\includegraphics[scale = 0.45, trim = 10mm 0mm 0mm 0mm]{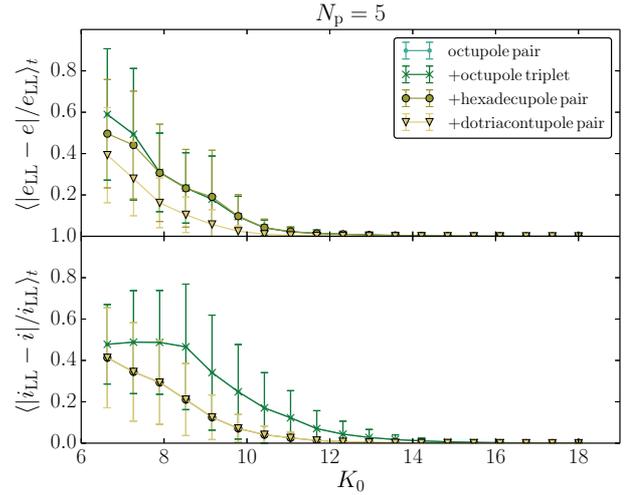}
\caption{\small Quantitative measures of agreement between \textsc{\codename} and LL theory as in \F\,\ref{fig:LL_sequence_f_paper_Np_4r}, now with five planets. }
\label{fig:LL_sequence_f_paper_Np_5r}
\end{figure}

In \F s\,\ref{fig:LL_sequence_f_paper_Np_4r} and \ref{fig:LL_sequence_f_paper_Np_5r}, we show similar figures (only considering the time-averaged measure) for $N_\mathrm{p}=4$ and 5, respectively. The trends described above for $N_\mathrm{p}=3$ also apply to larger planet numbers. As the number of planets is increased, the dependence of the measure of agreement as a function of $K_0$ becomes smoother. 

Generally, $\sim 2 \%$ discrepancy with respect to LL theory (in terms of the time-averaged measure) is reached with dotriacontupole-order terms if $K_0 \gtrsim 10$. When including only the octupole-order terms, the discrepancy is between $\sim 10$ and $\sim 20\%$, depending on the number of planets. In particular, the agreement with respect to the eccentricity becomes much better with the inclusion of the dotriacontupole-order terms. This demonstrates the need to include high-order terms in multiplanet systems.

\begin{figure}
\center
\includegraphics[scale = 0.42, trim = 0mm 15mm 0mm 0mm]{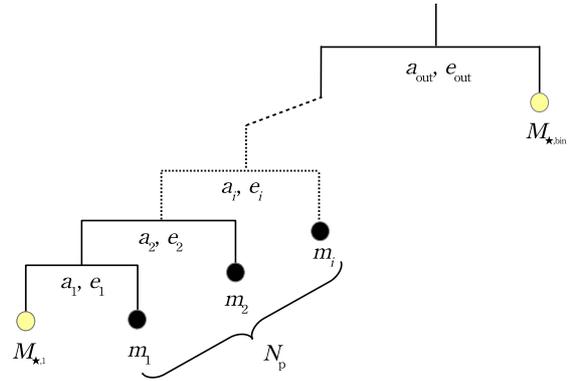}
\caption{\small A mobile diagram of a hierarchical multiplanet system in a stellar binary (cf. \S\,\ref{sect:multiplanet:binary}). The $N_\mathrm{p}$ planets orbit a star with mass $M_{\star,1}$. The centre of mass of this subsystem is orbited by a binary companion star with mass $M_{\star,\mathrm{bin}}$ in an orbit with semimajor axis and eccentricity $a_\mathrm{bin}$ and $e_\mathrm{bin}$, respectively. In \S\,\ref{sect:multiplanet:binary:comp}, we set $N_\mathrm{p} = 2$ (to match the setup of \citealt{2008ApJ...683.1063T}); in \S\,\ref{sect:multiplanet:binary:const_spacing}, we set $N_\mathrm{p} = 4$. }
\label{fig:mobile_planetary_binary}
\end{figure}

\begin{figure}
\center
\includegraphics[scale = 0.45, trim = 10mm 0mm 0mm 0mm]{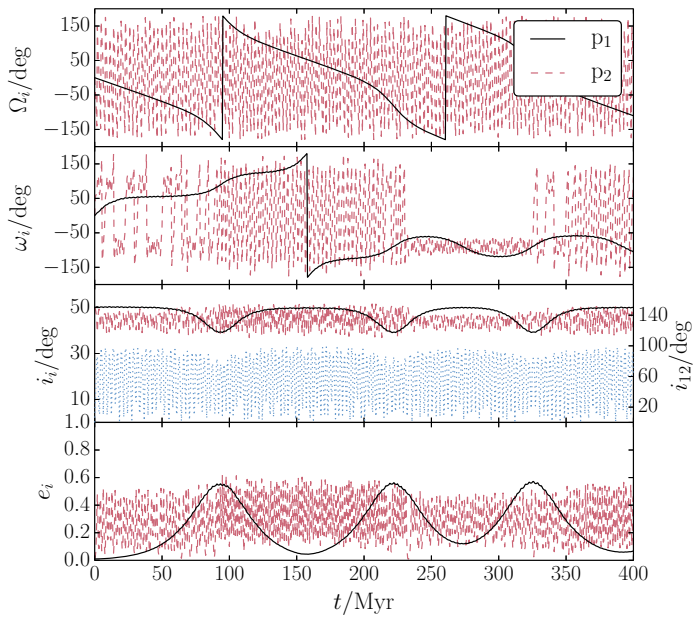}
\caption{\small Integration of a two-planet system with a stellar binary companion (cf. \F\,\ref{fig:mobile_planetary_binary}). The system parameters are adopted from Fig. 5 of \citet{2008ApJ...683.1063T}. The planetary parameters are $m_1 = 1\, M_\mathrm{J}$, $a_1 = 2 \, \mathrm{AU}$, $e_1 = 0.01$, $i_1 = 50^\circ$, $\omega_1 = 0^\circ$, $\Omega_1 = 0^\circ$ (inner planet; solid black lines), and $m_2 = 0.032\,M_\mathrm{J}$, $a_2 = 31.6 \, \mathrm{AU}$, $e_2 = 0.01$, $i_2 = 50^\circ$, $\omega_2 = 0^\circ$, $\Omega_2 = 0^\circ$ (outer planet; red dashed lines). The central star has mass $M_{\star,1} = 1 \, \mathrm{M}_\odot$, and the binary companion, with mass $M_{\star,\mathrm{bin}} = 1 \, \mathrm{M}_\odot$, is in an orbit with $a_\mathrm{out} = 750 \, \mathrm{AU}$, $e_\mathrm{out} = 0.2$, $i_\mathrm{out} = 0^\circ$, $\omega_\mathrm{b}=0^\circ$ and $\Omega_\mathrm{out} = 0^\circ$, i.e. the initial mutual inclination between the planetary orbits and the binary orbit is $50^\circ$. In the third panel, the blue dotted line shows the mutual inclination $i_{12}$ between the two planets; note the different scale on the vertical axis, indicated to the right of the panel. }
\label{fig:Comparison_Takeda08_takeda_Fig5}
\end{figure}

\begin{figure}
\center
\includegraphics[scale = 0.45, trim = 10mm 0mm 0mm 0mm]{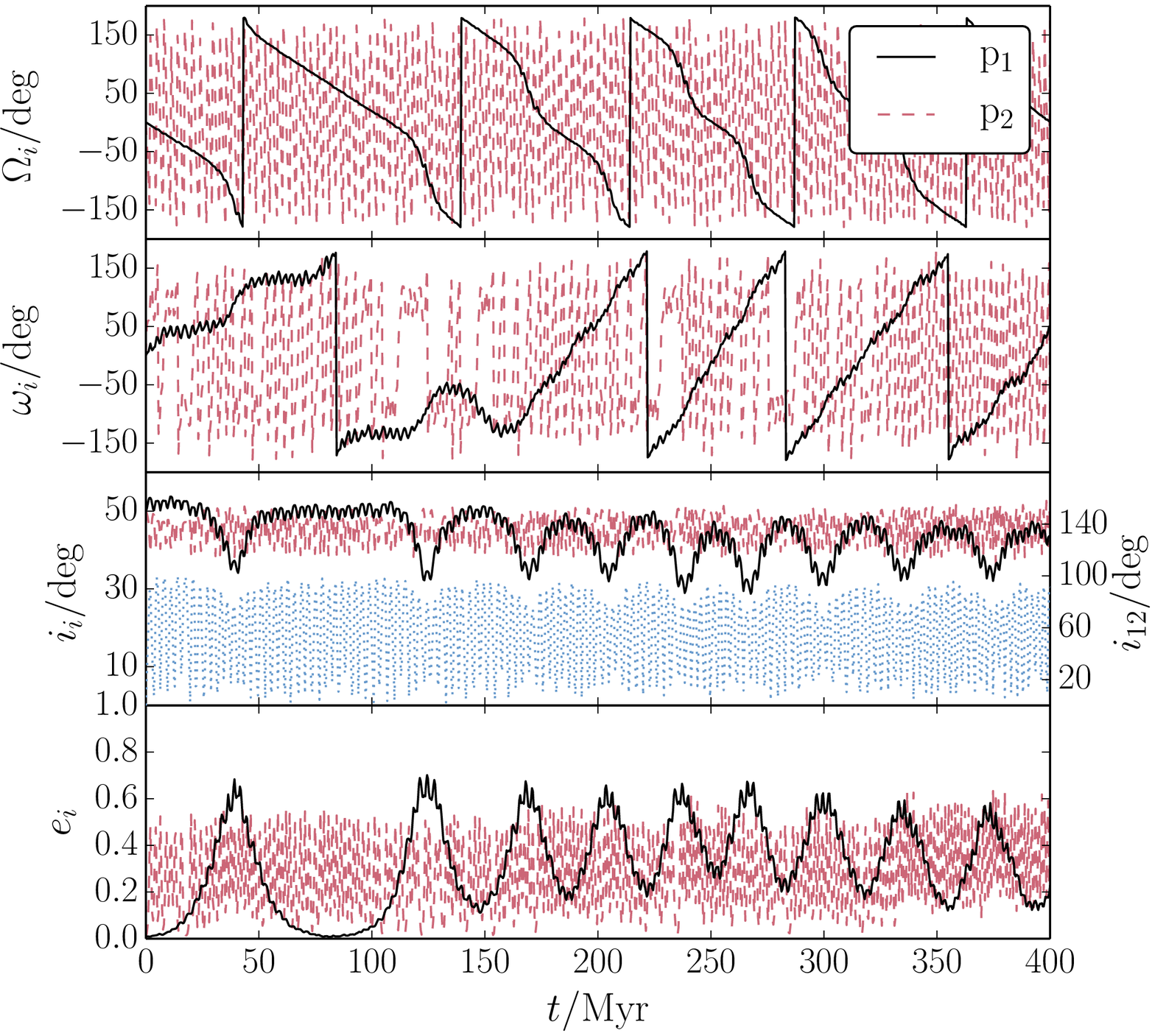}
\caption{\small As in \F\,\ref{fig:Comparison_Takeda08_takeda_Fig5}, now with $m_2 = 0.32\,M_\mathrm{J}$ (parameters adopted from Fig. 6 of \citealt{2008ApJ...683.1063T}). }
\label{fig:Comparison_Takeda08_takeda_Fig6}
\end{figure}

\subsection{Binary-star systems}
\label{sect:multiplanet:binary}
As mentioned in \S\,\ref{sect:introduction}, it is possible that a star hosting a multiplanet system is itself part of a multiple stellar system. Here, we consider the case of an S-type multiplanet system in which the host star has a more distant stellar binary companion well beyond the orbits of the planets. A general representation of the system (for $N_\mathrm{p}$ planets) is given in \F\,\ref{fig:mobile_planetary_binary}. In \S\,\ref{sect:multiplanet:triple}, we also consider the planet-hosting star to be the tertiary star in a stellar triple system. In the latter case, the stellar `binary companion' to the planet-hosting star is itself a stellar binary (cf. \F\,\ref{fig:mobile_planetary_triple}).

\subsubsection{Two-planet system in a stellar binary -- comparison with direct $N$-body integrations}
\label{sect:multiplanet:binary:comp}
First, we compare results of \textsc{\codename} to a number of direct $N$-body integrations of a system previously studied by \citet{2008ApJ...683.1063T}. The latter authors considered two-planet systems, and showed that two distinct dynamical classes exist with respect to the response of the multiplanet system to the secular torque of the binary companion. In one class, the orbits of the planets evolve `rigidly', i.e. closely maintaining their mutual inclinations and nodal angles despite an overall change of the orbital plane of the multiplanet system. In the second class, the orbits are coupled to the secular torque of the binary companion, thereby inducing potentially large mutual inclinations between the planets, and large excitations of the eccentricities. 

In \F s \ref{fig:Comparison_Takeda08_takeda_Fig5} and \ref{fig:Comparison_Takeda08_takeda_Fig6}, we show two integrations similar to figs 5 and 6 of \citet{2008ApJ...683.1063T}, respectively (refer to the captions for detailed parameters). In these figures, the planets are either decoupled from each other and respond to the secular torque of the binary companion (\F\,\ref{fig:Comparison_Takeda08_takeda_Fig5}, with $m_2=0.032\,M_\mathrm{J}$) or are weakly coupled (\F\,\ref{fig:Comparison_Takeda08_takeda_Fig6}, $m_2=0.32\,M_\mathrm{J}$). 

\begin{figure}
\center
\includegraphics[scale = 0.45, trim = 10mm 0mm 0mm 0mm]{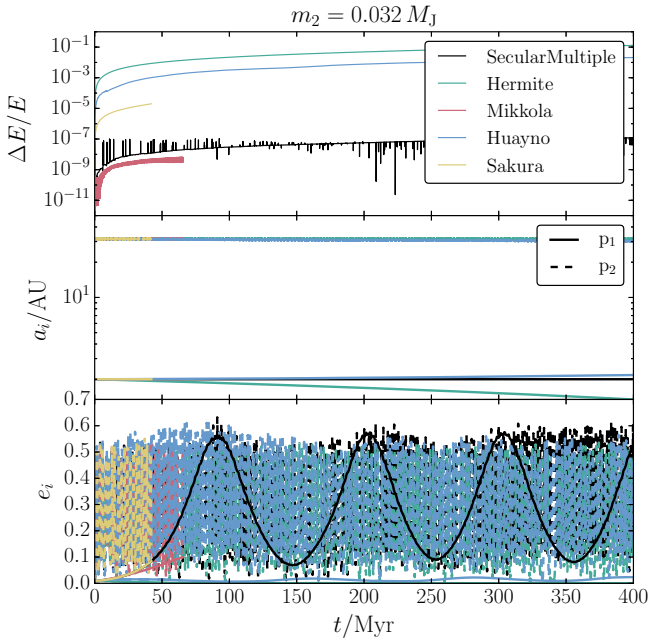}
\caption{\small Comparison between \textsc{\codename} and various direct $N$-body integrations using different codes (refer to the labels in the top panel) for the system from \F\,\ref{fig:Comparison_Takeda08_takeda_Fig5} ($m_2 = 0.032\,M_\mathrm{J}$). Top panel: the energy errors; middle panel: the semimajor axes of the planets (by construction, these remain exactly constant in \textsc{\codename}). Bottom panel: the eccentricities of the planets. }
\label{fig:paper_Takeda_Fig5_nbody_comparisons}
\end{figure}

\begin{figure}
\center
\includegraphics[scale = 0.45, trim = 10mm 0mm 0mm 0mm]{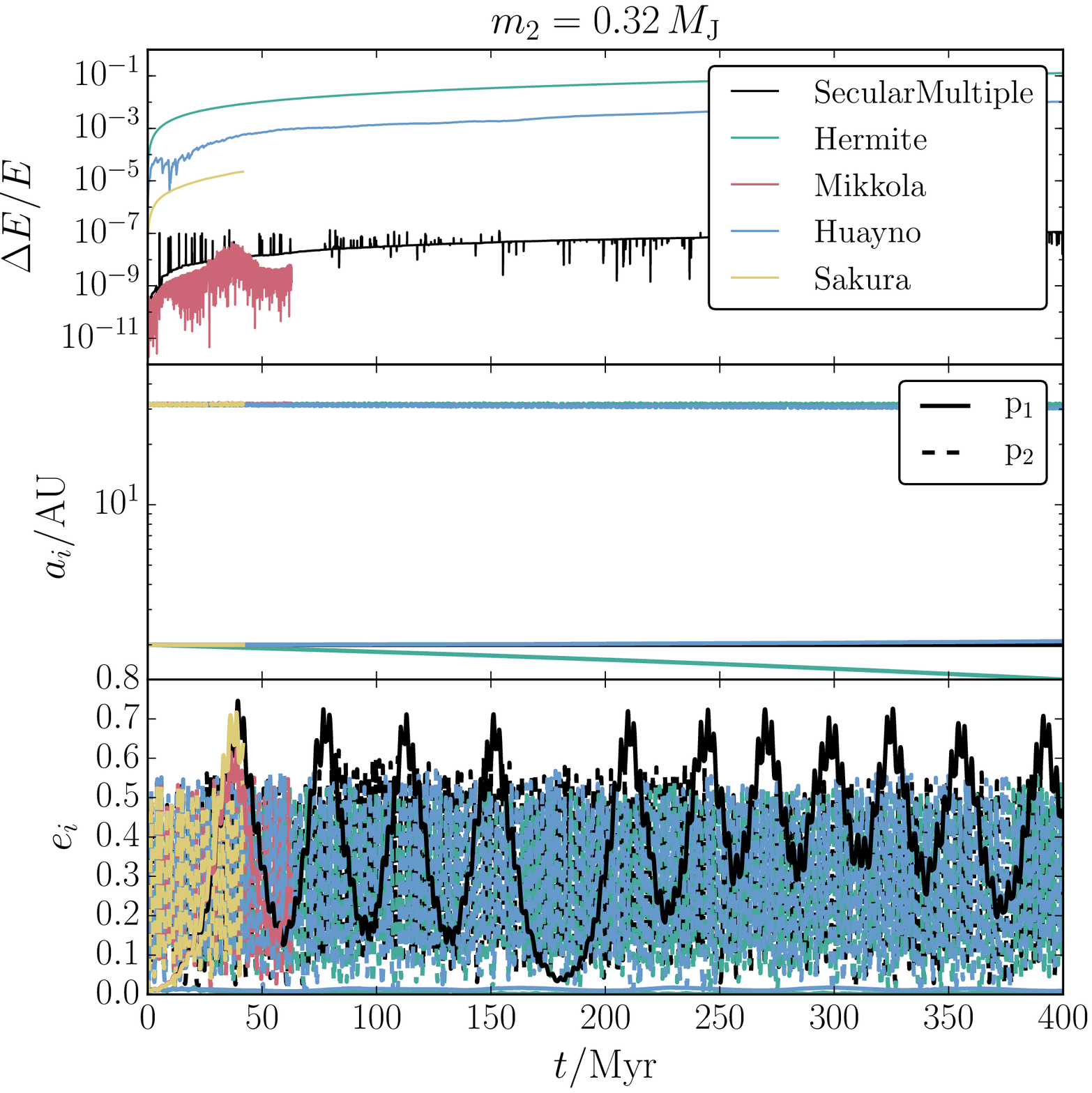}
\caption{\small As in \F\,\ref{fig:paper_Takeda_Fig5_nbody_comparisons}, now with $m_2 = 0.32\,M_\mathrm{J}$. }
\label{fig:paper_Takeda_Fig6_nbody_comparisons}
\end{figure}

For verification purposes\footnote{We were unable to obtain the data from \citet{2008ApJ...683.1063T}.}, we carried out a number of direct $N$-body integrations of these systems with various codes within \textsc{AMUSE}: \textsc{Hermite} \citep{1995ApJ...443L..93H}, \textsc{Mikkola} \citep{2008AJ....135.2398M}, \textsc{Huayno} \citep{2012NewA...17..711P} and \textsc{Sakura} \citep{2014MNRAS.440..719G}. In \F s\,\ref{fig:paper_Takeda_Fig5_nbody_comparisons} and \ref{fig:paper_Takeda_Fig6_nbody_comparisons}, we show, for $m_2=0.032\,M_\mathrm{J}$ and $m_2=0.32\,M_\mathrm{J}$ respectively, the energy errors, the semimajor axes and the eccentricities of the planets for the integrations with these codes. We also include integrations with \textsc{\codename} (black lines). We limited the run time of the $N$-body integrations to approximately three weeks, and only \textsc{Hermite} and \textsc{Huayno} were able to complete the 400 Myr integration within that time. The other codes, \textsc{Mikkola} and \textsc{Sakura}, computed $\approx 42\, \mathrm{Myr}$ of the evolution during our run time. Note that the 400 Myr integration with \textsc{\codename} takes a few minutes. 

For \textsc{Hermite} and \textsc{Huayno}, the energy errors increased to $\mathcal{O}(10^{-1})$ and $\mathcal{O}(10^{-2})$, respectively, and the semimajor axes of the planets started to deviate significantly from their original values (cf. the middle panels of \F s\,\ref{fig:paper_Takeda_Fig5_nbody_comparisons} and \ref{fig:paper_Takeda_Fig6_nbody_comparisons}). Given the large energy errors and the differences in behaviour in the innermost planet semimajor axis evolution between \textsc{Hermite} and \textsc{Huayno} (according to the former, $a_1$ decreases over several 100 Myr, whereas according to the latter, $a_1$ increases over the same time-scale), this is likely due to computational errors, and not due to the true dynamical evolution. 

\begin{figure}
\center
\includegraphics[scale = 0.45, trim = 10mm 0mm 0mm 0mm]{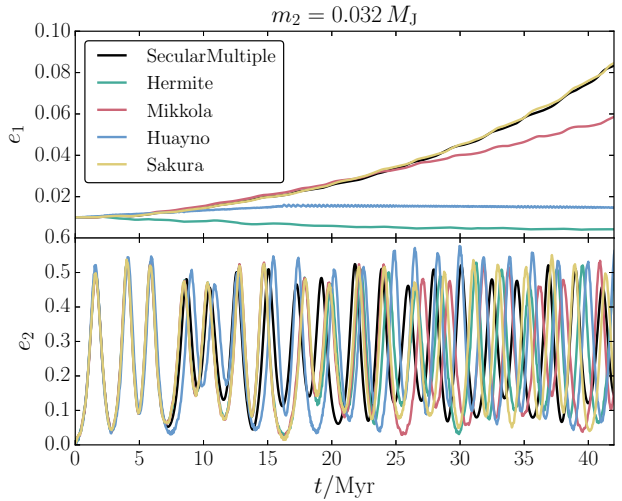}
\caption{\small A zoomed-in version of the bottom panel of \F\,\ref{fig:paper_Takeda_Fig5_nbody_comparisons} (with $m_2 = 0.032\,M_\mathrm{J}$), showing the eccentricities of planets 1 and 2 in the top and bottom panels, respectively. }
\label{fig:paper_Takeda_Fig5_nbody_comparisons_zoom}
\end{figure}

\begin{figure}
\center
\includegraphics[scale = 0.45, trim = 10mm 0mm 0mm 0mm]{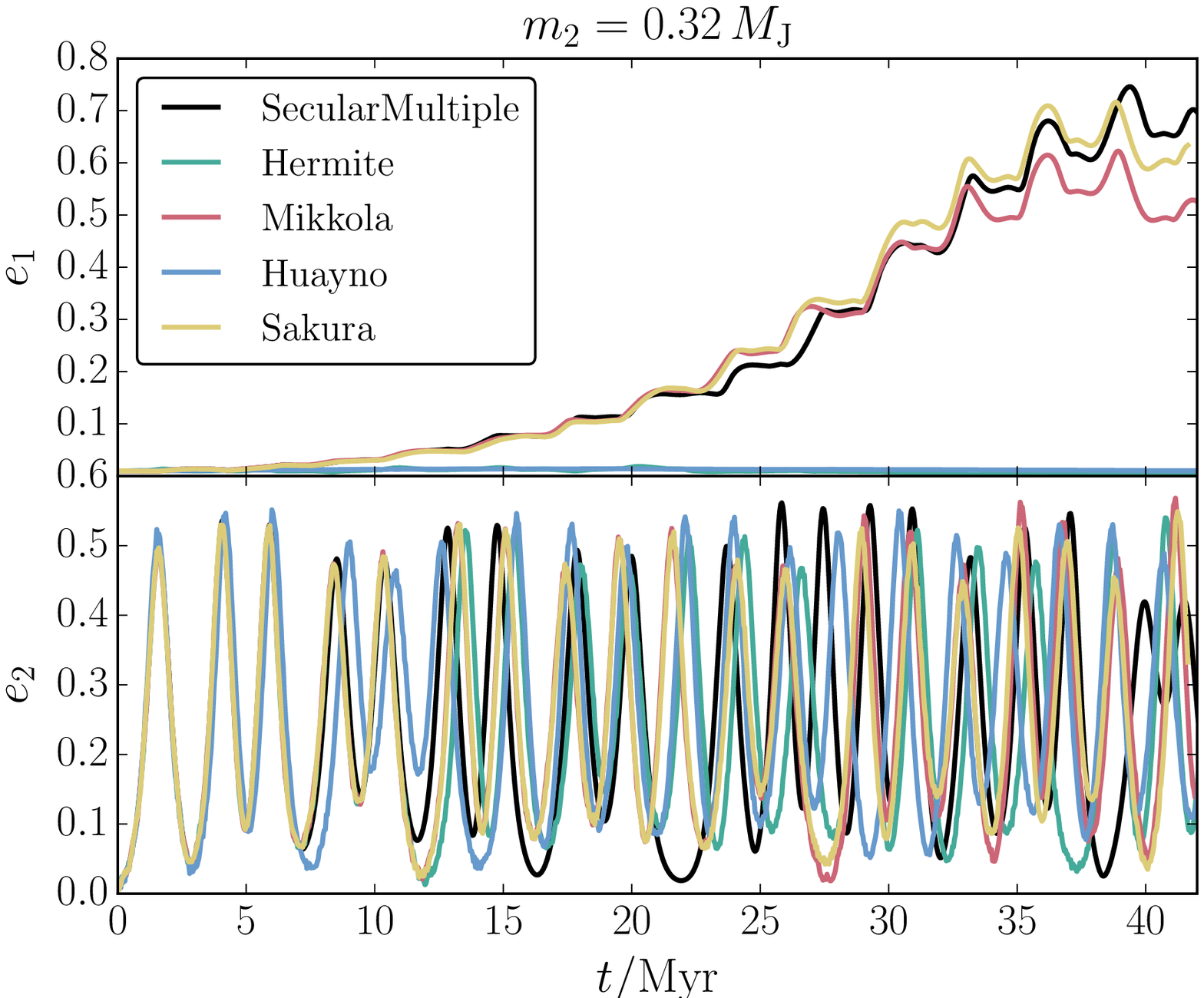}
\caption{\small As in \F\,\ref{fig:paper_Takeda_Fig5_nbody_comparisons_zoom}, now with $m_2 = 0.32\,M_\mathrm{J}$. }
\label{fig:paper_Takeda_Fig6_nbody_comparisons_zoom}
\end{figure}

The large energy errors in the case of \textsc{Hermite} and \textsc{Huayno} also likely explain why the eccentricity of the innermost planet is not excited to large values of $\approx 0.6$ and $\approx 0.7$ for $m_2=0.032\,M_\mathrm{J}$ and $m_2=0.32\,M_\mathrm{J}$, respectively, as computed by \textsc{\codename}, and according to the integrations of \citet{2008ApJ...683.1063T} (cf. Figures 5 and 6 of the latter paper). In \F s\,\ref{fig:paper_Takeda_Fig5_nbody_comparisons_zoom} and \ref{fig:paper_Takeda_Fig6_nbody_comparisons_zoom}, we show zoomed-in versions of the eccentricities shown in the bottom panels of \F s\,\ref{fig:paper_Takeda_Fig5_nbody_comparisons} and \ref{fig:paper_Takeda_Fig6_nbody_comparisons}, respectively. For both values of $m_2$ and the first $\approx 7 \, \mathrm{Myr}$, the evolution of $e_2$ matches well between the various $N$-body codes and \textsc{\codename}, and starts to deviate between the various methods after this time. Note that these deviations occur even for the two most accurate $N$-body integrations with \textsc{Mikkola} and \textsc{Sakura}. For $m_2 = 0.32\,M_\mathrm{J}$, the eccentricity evolution of the innermost planet ($e_1$) agrees between \textsc{Mikkola}, \textsc{Sakura} and \textsc{\codename} within a margin of $\sim 0.1$. Note that the differences between the secular and direct $N$-body codes are similar to the differences between the two direct codes \textsc{Mikkola} and \textsc{Sakura}. For $m_2 = 0.032\,M_\mathrm{J}$, there is a relatively larger discrepancy in $e_1$ between \textsc{\codename} and \textsc{Mikkola} of $\approx 0.02$ at $\approx 40\,\mathrm{Myr}$. However, \textsc{\codename} and \textsc{Sakura} agree very well. In contrast to \textsc{Mikkola}, \textsc{Sakura} and \textsc{\codename}, \textsc{Hermite} and \textsc{Huayno} give a very different evolution of $e_1$, and this is likely due to the large energy errors in the integrations with the latter codes.

We conclude that there is good agreement of \textsc{\codename} with the most accurate direct $N$-body integrations. We emphasize that \textsc{Hermite} and \textsc{Huayno}, even though slower compared to \textsc{\codename} by a factor of $\sim 10^3$, fail to accurately produce the eccentricity evolution of the innermost planet. To get accurate results on secular time-scales with direct $N$-body methods, it is crucial that energy errors remain small during the evolution (based on \F\,\ref{fig:paper_Takeda_Fig5_nbody_comparisons}, no larger than $\sim 10^{-4}$). Note that this minimum energy error is smaller than the value of $10^{-1}$ suggested by \citet{2015ComAC...2....2B}; however, this value applies in a statistical sense to an ensemble of systems, and not to individual systems. This requires the use of special integration techniques such as algorithm chain regularization (as in \textsc{Mikkola}) or Keplerian-based Hamiltonian splitting (as in \textsc{Sakura}). Alternatively, for the hierarchical systems considered here, \textsc{\codename} agrees well with the most accurate direct $N$-body codes, and it is faster by several orders of magnitude. 

In \S\,\ref{sect:multiplanet:binary:const_spacing} below, we take advantage of the speed of \textsc{\codename}, and study more generally the nature of the eccentricity oscillations as a function of the binary semimajor axis.

\subsubsection{Four-planet system with constant spacing in a stellar binary}
\label{sect:multiplanet:binary:const_spacing}
Here, we consider four-planet systems with fixed spacing and small initial eccentricities and inclinations as in \S\,\ref{sect:multiplanet:single}, now including a binary companion star with mass $M_{\star,\mathrm{bin}} = 1 \, \mathrm{M}_\odot$. For the multiplanet system, we set $M_{\star,1} = 1 \, \mathrm{M}_\odot$, $m_i = 1 \, M_\mathrm{J}$, $K_0=10$, $a_1 = 0.2 \, \mathrm{AU}$, $e_i = 0.01$ and $i_i = 1^\circ$. A range of semimajor axes for the binary companion orbit $a_\mathrm{out}$\footnote{We use the subscript `out' rather than e.g. `bin', in anticipation of the case of a triple star system discussed in \S\,\ref{sect:multiplanet:triple}.} is assumed: between 50 and 150 AU with $e_\mathrm{out} = 0.1$, and between 100 and 300 AU with $e_\mathrm{out} = 0.8$. The lower limits on $a_\mathrm{out}$ are chosen to ensure short-term dynamical stability according to the formulae of \citet{1999AJ....117..621H}. The initial inclination of the binary orbit is set to $80^\circ$ with $\Omega_\mathrm{out}=\omega_\mathrm{out} = 0^\circ$. For the two values of $e_\mathrm{out}$, we integrated 500 systems with 100 different values of $a_\mathrm{out}$, and for each value of $a_\mathrm{out}$, five realizations of $\Omega_i$ for the planetary system assuming that the $\Omega_i$ are randomly distributed. Because the planetary inclinations are small, the initial mutual inclinations of the planets with respect to the binary companion are very similar to each other, and are $\approx 80^\circ$. We integrate each system for a duration of $1 \, \mathrm{Myr}$, which is comparable to the LK time-scale of the binary with respect to the outermost planet. 

\begin{figure}
\center
\includegraphics[scale = 0.45, trim = 10mm 0mm 0mm 0mm]{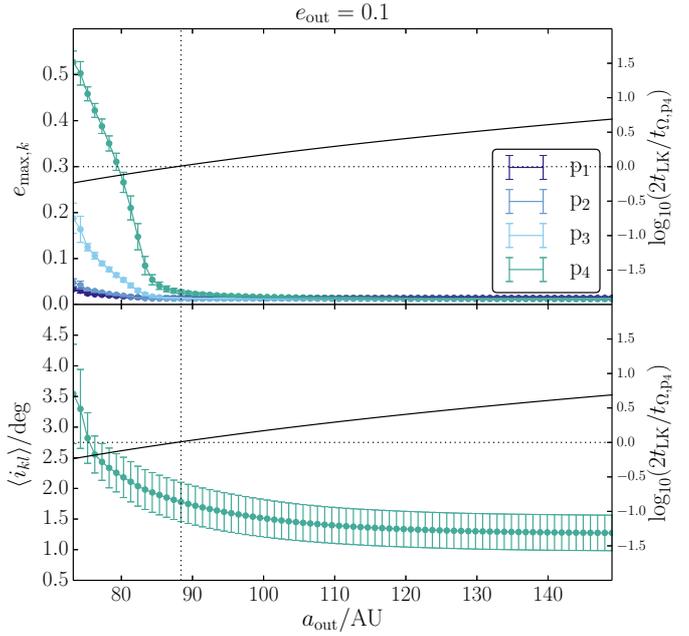}
\caption{\small For a four-planet system with a binary companion and eccentricity $e_\mathrm{out} = 0.1$ (cf. \F\,\ref{fig:mobile_planetary_binary}), the maximum values of the eccentricities of the planets as a function of $a_\mathrm{out}$, assuming $e_\mathrm{out} = 0.1$. Bullets show the average values over $\Omega_i$, whereas the error bars indicate the standard deviations. In the bottom panel, we show the mean values of all combinations of mutual inclinations between the planets, averaged over time. In each panel, the right axes show the logarithm of the ratio $2t_\mathrm{LK}/t_{\Omega,4}$, where $t_{\Omega,\mathrm{p}_4} \equiv 2\pi/f_4$, of the nodal precession time-scale associated with LK oscillations induced by the binary companion, compared to the nodal precession time-scale associated with the mutual planetary torques. }
\label{fig:binary_sequence_paper_binary_seq01_Np_4}
\end{figure}

\begin{figure}
\center
\includegraphics[scale = 0.45, trim = 10mm 0mm 0mm 0mm]{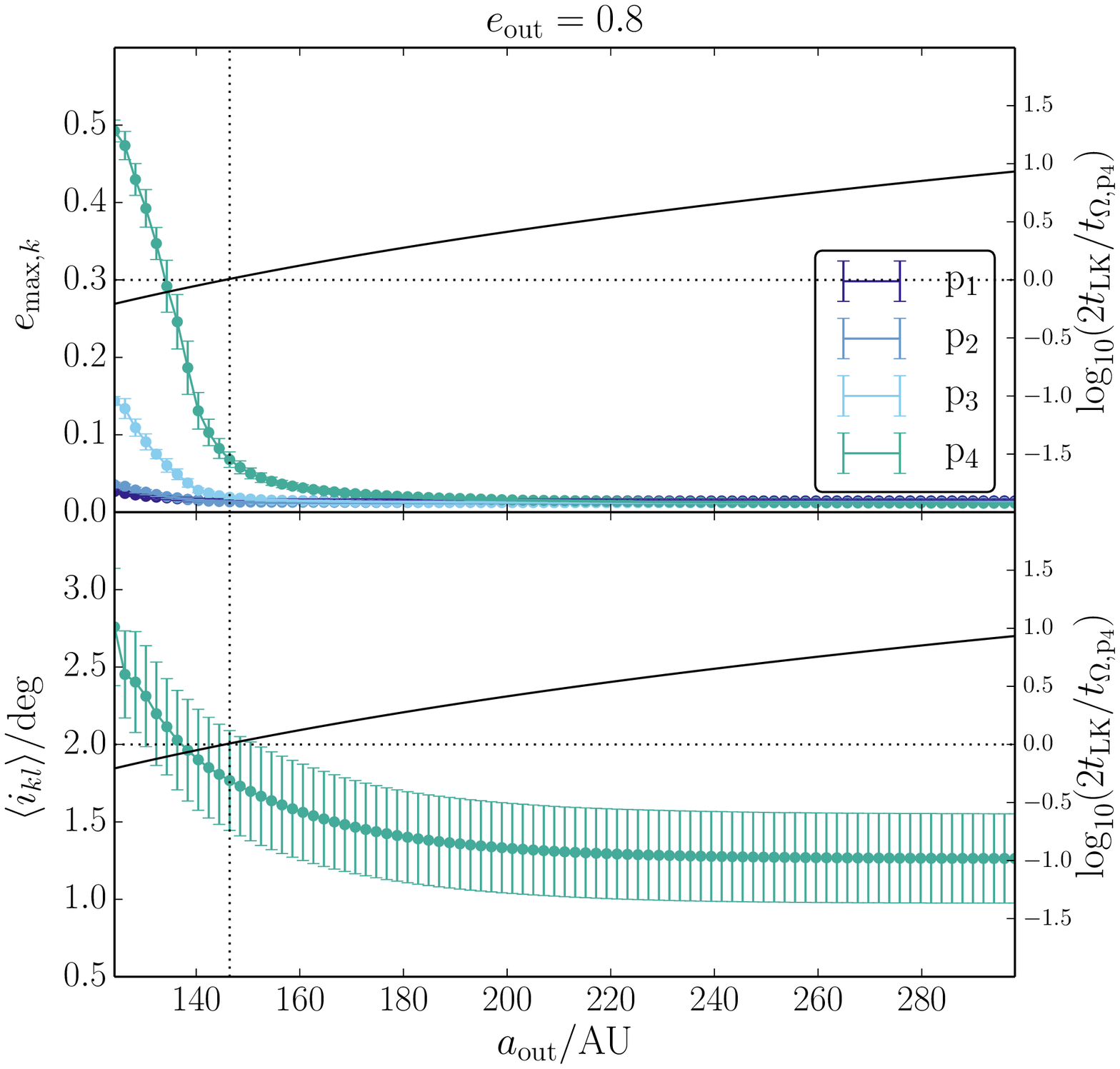}
\caption{\small Similar to \F\,\ref{fig:binary_sequence_paper_binary_seq01_Np_4}, now with $e_\mathrm{out} = 0.8$. }
\label{fig:binary_sequence_paper_binary_seq02_Np_4}
\end{figure}

In the top panel of \F\,\ref{fig:binary_sequence_paper_binary_seq01_Np_4}, we show the resulting maximum values of the eccentricities of the planets as a function of $a_\mathrm{out}$, assuming $e_\mathrm{out} = 0.1$. Bullets show the average values over $\Omega_i$, whereas the error bars indicate the standard deviations. In the bottom panel of  \F\,\ref{fig:binary_sequence_paper_binary_seq01_Np_4}, we show the mean values of all combinations of the mutual inclinations between the planets, averaged over time. 

\begin{figure}
\center
\includegraphics[scale = 0.45, trim = 10mm 0mm 0mm 0mm]{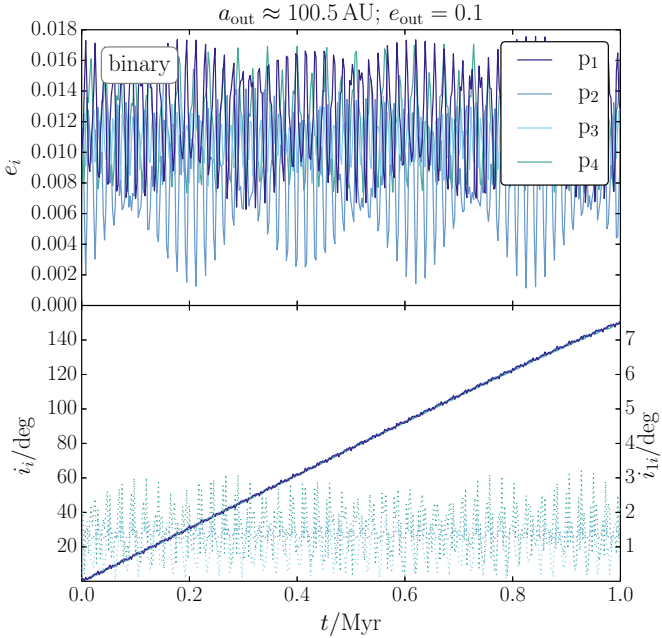}
\caption{\small The eccentricities and inclinations of a four-planet system in the presence of a binary companion, in the regime in which planet-planet torques dominate compared to the torque of the binary companion. The assumed values of $a_\mathrm{out}$ and $e_\mathrm{out}$ are indicated above the top panel. In the bottom panel, the absolute inclinations $i_i$ are shown with solid lines; the inclinations relative to planet 1, $i_{1i}$ (right axis), are shown with dotted lines. }
\label{fig:binary_single_paper_binary_seq01_Np_4_index_250}
\end{figure}

For large $a_\mathrm{out}$, the maximum eccentricities are small and so are the mutual inclinations. In this limit, the secular torque of the binary companion is weak compared to the mutual torques exerted by the planets. Although the absolute inclinations and longitudes of the ascending nodes change, the planets remain closely fixed to the mutual plane which is precessing because of torque of the binary companion. In \F\,\ref{fig:binary_single_paper_binary_seq01_Np_4_index_250}, we show an example of the evolution of the eccentricities and the inclinations in this regime, with $a_\mathrm{out} \approx 100.5 \, \mathrm{AU}$ and $e_\mathrm{out} = 0.1$. The {\it absolute} inclinations change by $\approx 150^\circ$ on a time-scale of $1\,\mathrm{Myr}$ (cf. the solid lines in the bottom panel of \F\,\ref{fig:binary_single_paper_binary_seq01_Np_4_index_250}), whereas the inclinations with respect to the innermost planet do not reach values above $\approx 3^\circ$ (cf. the dotted lines in the bottom panel of \F\,\ref{fig:binary_single_paper_binary_seq01_Np_4_index_250}).

\begin{figure}
\center
\includegraphics[scale = 0.45, trim = 10mm 0mm 0mm 0mm]{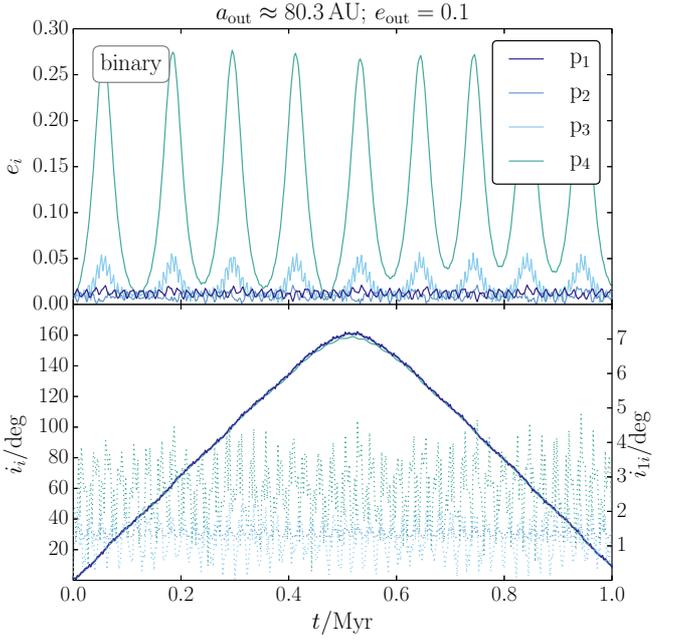}
\caption{\small Similar to \F\,\ref{fig:binary_single_paper_binary_seq01_Np_4_index_150}, now with a smaller binary semimajor axis. In this case, the eccentricities
of the planets are excited by the secular torque of the binary companion. } 
\label{fig:binary_single_paper_binary_seq01_Np_4_index_150}
\end{figure}

As $a_\mathrm{out}$ is decreased, the torque of the binary companion increases. For sufficiently small $a_\mathrm{out}$, the outermost planet is decoupled from the inner planets resulting in larger mutual inclinations; the outermost planet reaches high eccentricity. For $a_\mathrm{out}\lesssim 75\, \mathrm{AU}$, $e_4$ is high enough for planet 4 to cross its orbit with planet 3; consequently, we stopped the integration. These systems are not included in \F\,\ref{fig:binary_sequence_paper_binary_seq01_Np_4}. The maximum eccentricities of planets 1 through 3, in particular planet 3, are also affected when decreasing $a_\mathrm{out}$. In our systems, however, they do not reach high enough values for orbit crossings to occur. In \F\,\ref{fig:binary_single_paper_binary_seq01_Np_4_index_150}, we show an example of the time evolution with $a_\mathrm{out} \approx 80.3 \, \mathrm{AU}$ and $e_\mathrm{out} = 0.1$. The outermost two planets are markedly affected by LK cycles induced by the binary companion. For planet 3, secular oscillations due to planet-planet torques, which have shorter time-scales compared to the LK oscillations, are still clearly present. Note that even though the eccentricity of the outermost planet is strongly increased by the binary companion, the mutual inclination with respect to planet 1 (cf. the green dotted line in the bottom panel of \F\,\ref{fig:binary_single_paper_binary_seq01_Np_4_index_150}) remains small, i.e. $\lesssim 5^\circ$.

To estimate the value of $a_\mathrm{out}$ that separates the two regimes, we use the arguments given by \citet{2008ApJ...683.1063T}. We estimate the LK time-scale $t_\mathrm{LK}$ (cf. equation~\ref{eq:P_LK}) of the binary companion with respect to the outermost orbit by
\begin{align}
t_\mathrm{LK} \approx \frac{1}{3\pi} \left ( \frac{P_\mathrm{orb,out}^2}{P_\mathrm{orb,4}} \right ) \frac{M_{\star,1} + M_{\star,\mathrm{out}}}{M_{\star,\mathrm{out}}} \left ( 1 - e_\mathrm{out}^2 \right )^{3/2}.
\end{align}
We compare $2t_\mathrm{LK}$, the approximate time-scale for nodal precession due to the secular torque of the binary companion, to $t_{\Omega,\mathrm{p}_4} \equiv 2\pi/f_4$, an estimate of the time-scale associated with nodal oscillation of the fourth planet according to second-order LL theory \citep{1999ssd..book.....M}. 

In \F\,\ref{fig:binary_sequence_paper_binary_seq01_Np_4}, we show with the black solid lines the ratio $2t_\mathrm{LK}/t_{\Omega,\mathrm{p}_4}$ as a function of $a_\mathrm{out}$. The vertical dotted lines indicate the value of $a_\mathrm{out}$ for which the ratio is $\approx 1$. The latter lines indeed approximate the transition between the LK-dominated and planet-planet-dominated regimes. 

In \F\,\ref{fig:binary_sequence_paper_binary_seq02_Np_4}, we show a similar figure of the maximum eccentricities and averaged inclinations as a function of $a_\mathrm{out}$, now with $e_\mathrm{out} = 0.8$. The behaviour of the maximum eccentricities and the averaged mutual inclinations as a function of $a_\mathrm{out}$ is similar as for the case $e_\mathrm{out} = 0.1$. As expected, the transition occurs at larger $a_\mathrm{out}$, i.e. $a_\mathrm{out} \sim 150 \, \mathrm{AU}$ compared to $\sim 90 \, \mathrm{AU}$ if $e_\mathrm{out} = 0.1$. The method of comparing $2t_\mathrm{LK}$ to $t_{\Omega,\mathrm{p}_4} \equiv 2\pi/f_4$ again gives an estimate for the transition value of $a_\mathrm{out}$.

\begin{figure}
\center
\includegraphics[scale = 0.42, trim = 0mm 15mm 0mm 0mm]{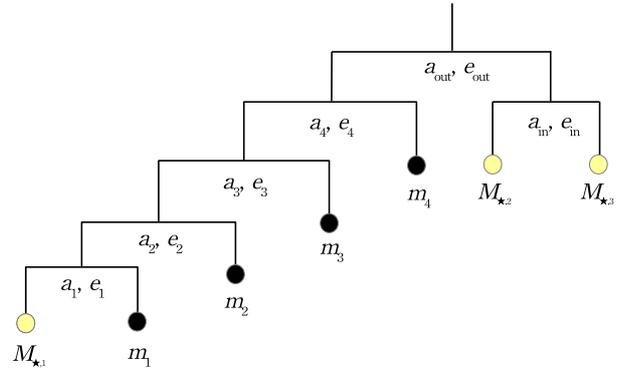}
\caption{\small A mobile diagram of a hierarchical four-planet system in a stellar triple (cf. \S\,\ref{sect:multiplanet:triple}). The planets are in S-type orbits around the tertiary star ($M_{\star,1}$) of a hierarchical triple system with inner and outer semimajor axes $a_\mathrm{in}$ and $a_\mathrm{out}$, respectively. }
\label{fig:mobile_planetary_triple}
\end{figure}

\subsection{Triple-star systems}
\label{sect:multiplanet:triple}
To our knowledge, the secular dynamics of multiplanet systems orbited by a stellar binary have not been explored. Here, we carry out a number of integrations similar to those of \S\,\ref{sect:multiplanet:binary:const_spacing}, but now with the binary companion {\it star} replaced by a stellar {\it binary} with semimajor axis $a_\mathrm{in}$. The hierarchy of the system is shown in a mobile diagram in \F\,\ref{fig:mobile_planetary_triple}. The system can be viewed as a stellar hierarchical triple system, in which the tertiary star has four S-type planets (effectively making it a heptuple system). 

In order to easily compare with the case of a stellar binary, we assume that the sum of the inner binary stellar masses is equal to the mass of the companion star, $M_{\star,\mathrm{bin}} = 1 \, \mathrm{M}_\odot$, that was assumed in \S\,\ref{sect:multiplanet:binary:const_spacing}, i.e. $M_{\star,2} + M_{\star,3} = M_{\star,\mathrm{bin}}$. Here, we set $M_{\star,2} = 0.6 \, \mathrm{M}_\odot$ and $M_{\star,3} = 0.4 \, \mathrm{M}_\odot$. The initial (absolute) inclination of the inner orbit is set to $i_\mathrm{in} = 0^\circ$; for the outer orbit, we set $i_\mathrm{out} = 80^\circ$. Furthermore, $\Omega_\mathrm{in} = \Omega_\mathrm{out} = \omega_\mathrm{in} = \omega_\mathrm{out} = 0^\circ$. Consequently, the mutual inclinations between the multiplanet system and the outer orbit are $\approx 80^\circ$ as in \S\,\ref{sect:multiplanet:binary:const_spacing}, and the inner and outer binaries are inclined by $80^\circ$. The initial eccentricities of the stellar orbits are $e_\mathrm{in} = e_\mathrm{out} = 0.1$. We integrated a grid of systems with various values of $a_\mathrm{in}$ and $a_\mathrm{out}$. As in \S s\,\ref{sect:multiplanet:single} and \ref{sect:multiplanet:binary}, for each combination of parameters, five integrations were carried out with all parameters fixed, except for the $\Omega_i$ of the planets, which were sampled randomly.

\begin{figure}
\center
\includegraphics[scale = 0.45, trim = 10mm 0mm 0mm 0mm]{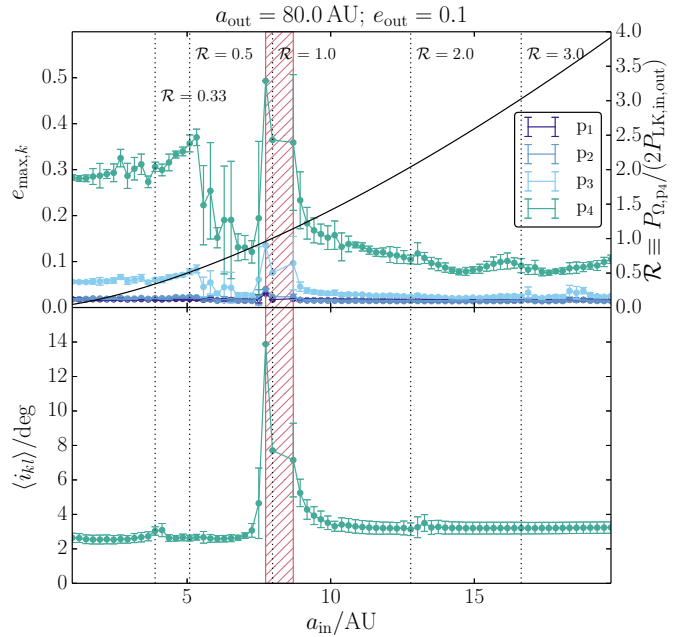}
\caption{\small For a four-planet system orbited by a stellar binary (cf. \F\,\ref{fig:mobile_planetary_triple}), the maximum eccentricities (top panel) and time-averaged mutual inclinations (bottom panel) of the planets as a function of $a_\mathrm{in}$, with fixed $a_\mathrm{out} = 80\, \mathrm{AU}$ and $e_\mathrm{out} = 0.1$. Each bullet corresponds to five different realizations with respect to $\Omega_i$ of the planets; the bullets (error bars) indicate the mean values (standard deviations). In the top panel, the black solid line shows the ratio $\mathcal{R}$ (right axis) of the nodal precession time-scale associated with the outermost planet due to planet-planet torques, to the nodal time-scale associated with inner-outer binary torques. Black vertical dotted lines show values of $a_\mathrm{in}$ corresponding to various integer ratio values of $\mathcal{R}$, indicated in the top panel. The red hatched region shows the values of $a_\mathrm{in}$ for which the orbit of the fourth planet is sufficiently eccentric to cross with that of the third planet.  }
\label{fig:triple_sequence_paper_triple_seq03_Np_4}
\end{figure}

In \F\,\ref{fig:triple_sequence_paper_triple_seq03_Np_4}, we show the maximum eccentricities (top panel) and the time-averaged mutual inclinations (bottom panel) of the planets as a function of $a_\mathrm{in}$, with fixed $a_\mathrm{out} = 80\, \mathrm{AU}$ and $e_\mathrm{out} = 0.1$. Note that in the equivalent system with the inner binary replaced by a point mass, the maximum eccentricities are approximately $0.02$, $0.02$, $0.05$ and $0.27$ for planets 1 through 4, respectively (see e.g. \F\,\ref{fig:binary_single_paper_binary_seq01_Np_4_index_150}). These values are indeed attained for small values of $a_\mathrm{in}$, i.e. $a_\mathrm{in} \lesssim 3 \, \mathrm{AU}$, in which case the inner binary is effectively a point mass. However, for $a_\mathrm{in} \gtrsim 3 \, \mathrm{AU}$, the maximum eccentricities are strongly affected.

\begin{figure}
\center
\includegraphics[scale = 0.45, trim = 10mm 0mm 0mm 0mm]{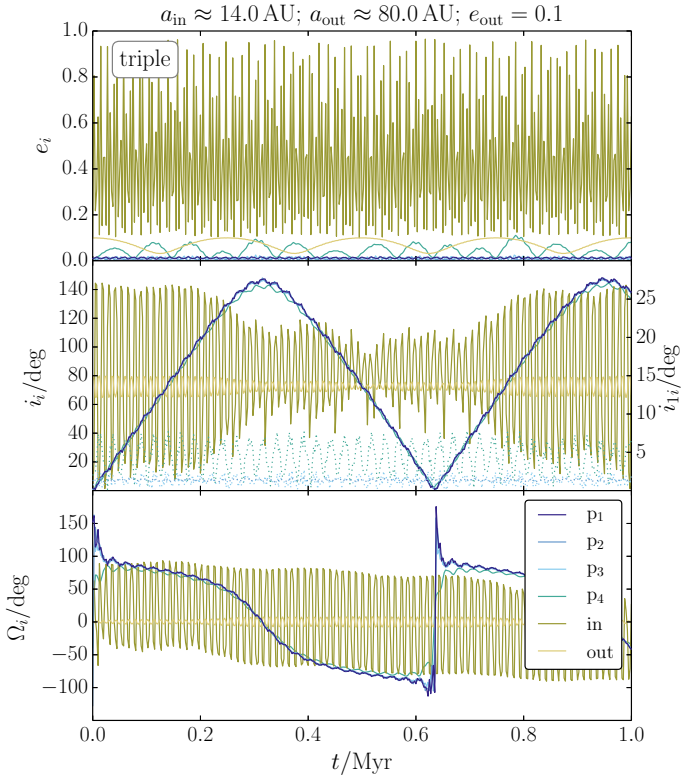}
\caption{\small Example evolution of a system from \F\,\ref{fig:triple_sequence_paper_triple_seq03_Np_4}, with $a_\mathrm{in} \approx 14.0 \, \mathrm{AU}$, $a_\mathrm{out}\approx 80.0 \, \mathrm{AU}$ and $e_\mathrm{out}=0.1$. Refer to  \F\,\ref{fig:binary_single_paper_binary_seq01_Np_4_index_150} for the evolution of the equivalent system in the case of a stellar binary. The top, middle and bottom panels show the eccentricities, inclinations and longitudes of the ascending nodes, respectively, of the four planets and the inner and outer binaries. In the middle panel, the absolute inclinations $i_i$ are shown with solid lines; for the planets, the inclinations relative to planet 1, $i_{1i}$ (right axis), are shown with dotted lines. }
\label{fig:triple_single_paper_triple_seq03_Np_4_index_260}
\end{figure}

\begin{figure}
\center
\includegraphics[scale = 0.45, trim = 10mm 0mm 0mm 0mm]{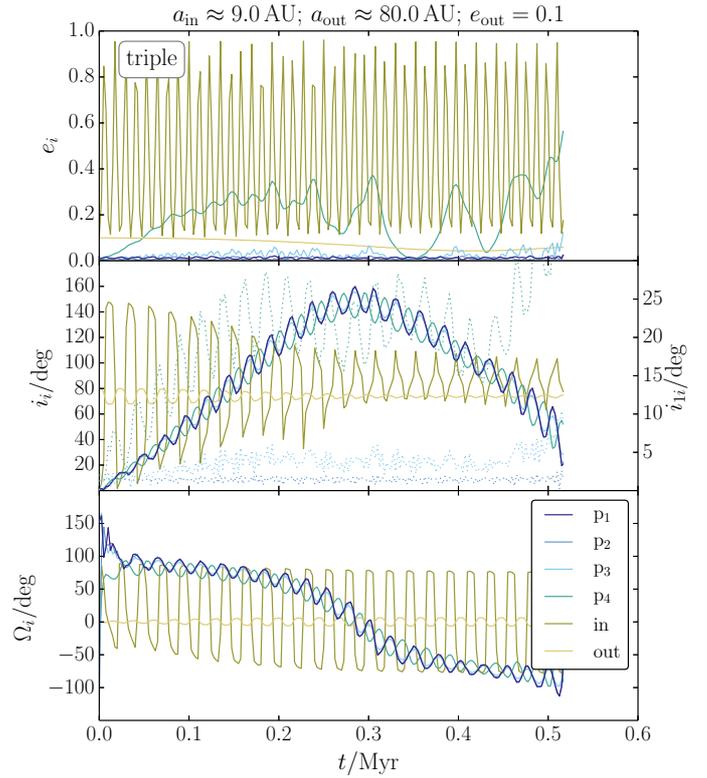}
\caption{\small Similar to \F\,\ref{fig:triple_single_paper_triple_seq03_Np_4_index_160}, now with $a_\mathrm{in} \approx 9.0 \, \mathrm{AU}$. The integration was stopped at $t\approx 0.5 \, \mathrm{Myr}$, when $e_3$ and $e_4$ became high enough for the third and fourth orbits to cross.}
\label{fig:triple_single_paper_triple_seq03_Np_4_index_160}
\end{figure}

For large $a_\mathrm{in}$, $a_\mathrm{in}\gtrsim 10 \, \mathrm{AU}$, the maximum eccentricities are {\it smaller} compared to the case of a stellar binary. This can be understood by considering the evolution of the nodal angle $\Omega_\mathrm{out}$ of the outer binary. An example of the evolution of the eccentricities, inclinations and nodal angles assuming $a_\mathrm{in} \approx 14.0 \, \mathrm{AU}$, $a_\mathrm{out} \approx 80.0 \, \mathrm{AU}$ and $e_\mathrm{out} = 0.1$ is shown in \F\,\ref{fig:triple_single_paper_triple_seq03_Np_4_index_260}. In the case of a stellar binary, $\Omega_\mathrm{out}$ does not oscillate because the planets are not massive enough to cause precession of the angular momentum vector of the outer binary, $\ve{h}_\mathrm{out}$. However, in the case of a stellar triple with sufficiently large $a_\mathrm{in}$, the inner binary causes a precession of $\ve{h}_\mathrm{out}$ on a time-scale $2 t_\mathrm{LK,in,out}$. In the case shown in \F\,\ref{fig:triple_single_paper_triple_seq03_Np_4_index_260}, the time-scale for oscillations of $\Omega_\mathrm{out}$ is short compared to the time-scale of LK oscillations of the outer orbit on the outermost planet (cf. \F\,\ref{fig:binary_single_paper_binary_seq01_Np_4_index_150}). Consequently, the rapid nodal precession of $\Omega_\mathrm{out}$ reduces the excitation of planetary eccentricities by the secular torque of the outer binary. 

However, the opposite effect can occur if the nodal precession time-scale of the inner and outer binaries, $2 t_\mathrm{LK,in,out}$, is similar to the time-scale $t_{\Omega,\mathrm{p}_4}$ for precession of the outermost planet nodal angle due to the planet-planet torques. In this case, the (nearly) synchronous nodal precession frequencies of the outer binary and the outermost planets cause an {\it additional} excitation of the eccentricity oscillations of the innermost planets, compared to the case of a stellar binary. 

An example is shown in \F\,\ref{fig:triple_single_paper_triple_seq03_Np_4_index_160}, in which $a_\mathrm{in} \approx 9.0 \, \mathrm{AU}$, $a_\mathrm{out} \approx 80.0 \, \mathrm{AU}$ and $e_\mathrm{out} = 0.1$ are assumed. The eccentricity of the outermost planet, planet 4, is excited to large values; in particular, it evolves in an intricate manner on a time-scale of approximately twice the time-scale of the eccentricity oscillations of the inner binary. The latter indeed occur on time-scales of half the nodal oscillation time-scales, cf. the bottom panel of \F\,\ref{fig:triple_single_paper_triple_seq03_Np_4_index_160}. Also, the inclination, of the outermost planet in particular, is strongly excited, reaching values as high as $\approx 30^\circ$ with respect to planet 1 (cf. the green dotted line in the middle panel of \F\,\ref{fig:triple_single_paper_triple_seq03_Np_4_index_160}). 

In this regime of excitation of the planetary eccentricities coupled to the LK oscillations of the inner and outer stellar binaries, high values of the planetary eccentricities can be reached. For example, in \F\,\ref{fig:triple_sequence_paper_triple_seq03_Np_4}, for $7 \lesssim a_\mathrm{in}/\mathrm{AU} \lesssim 8$, $e_4$ becomes large enough for orbit crossing with planet 3 (indicated with the red hatched region). In the right axis in the top panel of \F\,\ref{fig:triple_sequence_paper_triple_seq03_Np_4}, we show with the solid black line the ratio $\mathcal{R}$ of the nodal precession time-scale associated with the outermost planet due to planet-planet torques, to the nodal time-scale associated with the inner-outer binary torques. Note that the latter cause oscillations in $\Omega_\mathrm{in}$ and $\Omega_\mathrm{out}$ on the same time-scale (see e.g. \F\,\ref{fig:triple_single_paper_triple_seq03_Np_4_index_160}). Vertical dashed lines indicate values of $a_\mathrm{in}$ for which $\mathcal{R}$ is equal to various integer ratios. The strongest excitations, likely leading to a destabilization of the planetary system, occur when $\mathcal{R} \sim 1$. Weaker excitations occur near other integer ratios. 

\begin{figure}
\center
\includegraphics[scale = 0.45, trim = 10mm 0mm 0mm 0mm]{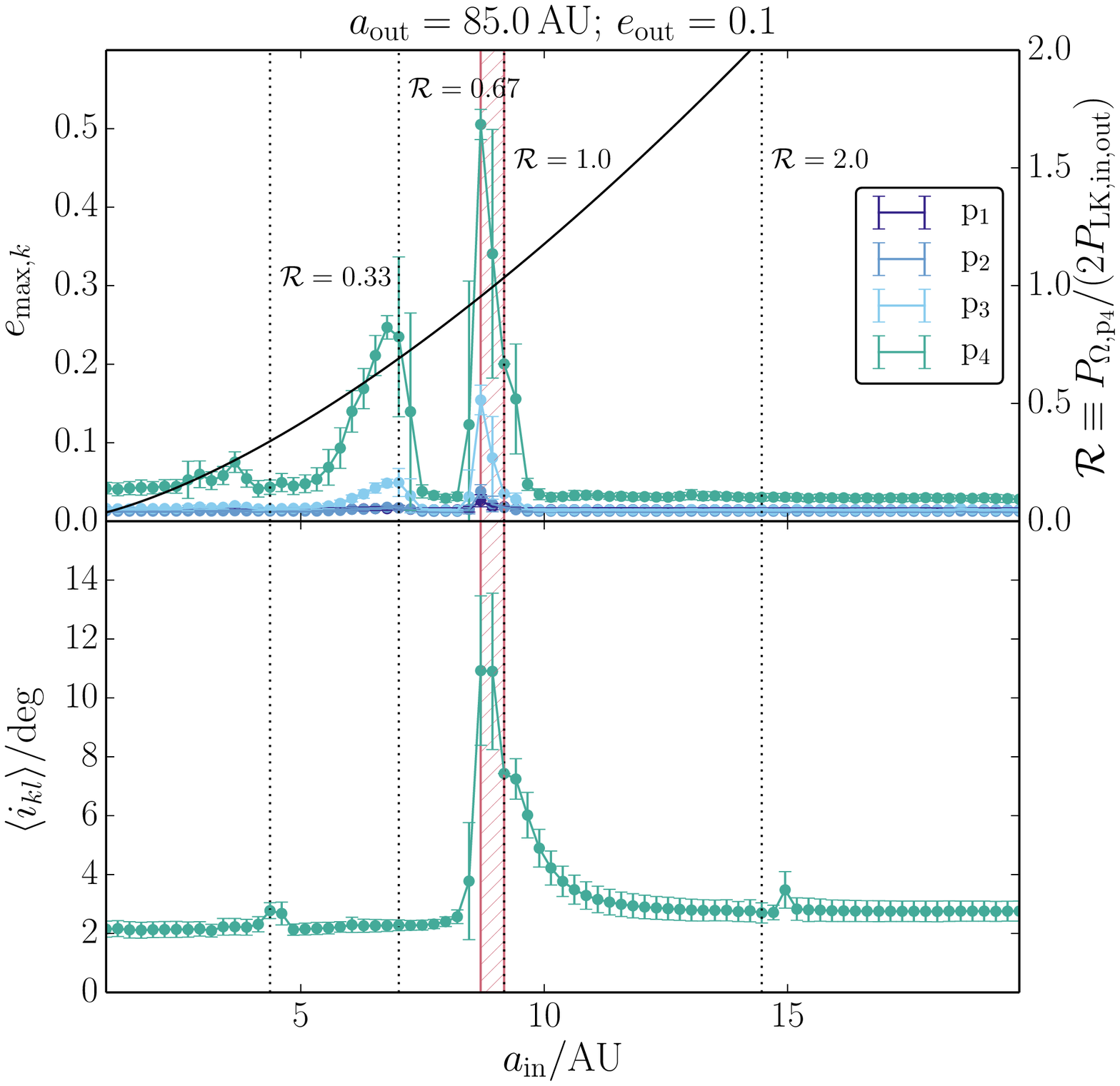}
\caption{\small Similar to \F\,\ref{fig:triple_sequence_paper_triple_seq03_Np_4}, now assuming $a_\mathrm{out} = 85\, \mathrm{AU}$ and $e_\mathrm{out} = 0.1$. }
\label{fig:triple_sequence_paper_triple_seq05_Np_4}
\end{figure}

\begin{figure}
\center
\includegraphics[scale = 0.45, trim = 10mm 0mm 0mm 0mm]{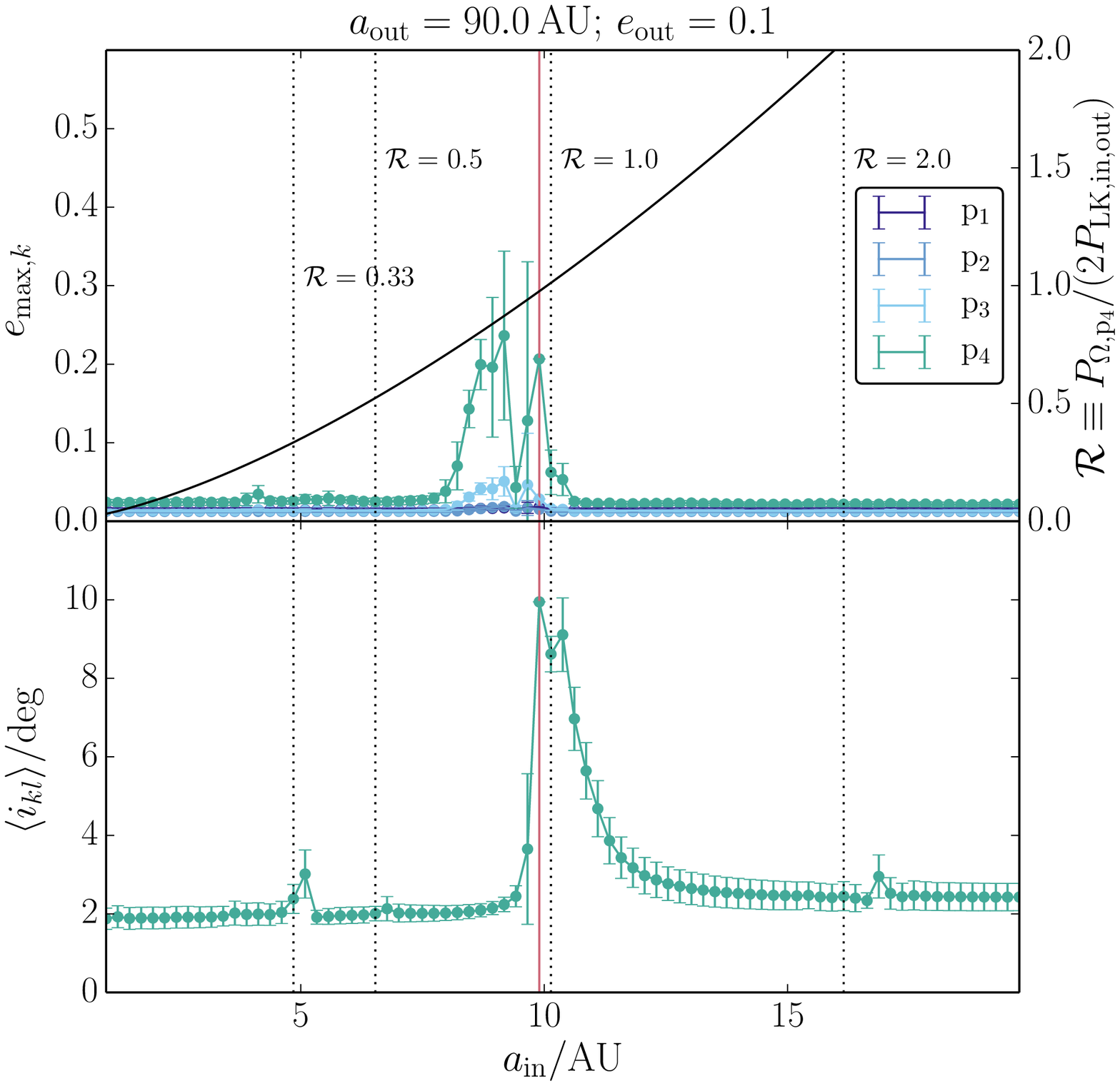}
\caption{\small Similar to \F\,\ref{fig:triple_sequence_paper_triple_seq03_Np_4}, now assuming $a_\mathrm{out} = 90\, \mathrm{AU}$ and $e_\mathrm{out} = 0.1$. }
\label{fig:triple_sequence_paper_triple_seq04_Np_4}
\end{figure}

In \F s\,\ref{fig:triple_sequence_paper_triple_seq05_Np_4} and \ref{fig:triple_sequence_paper_triple_seq04_Np_4}, we show two figures similar to \F\,\ref{fig:triple_sequence_paper_triple_seq03_Np_4} with the maximum eccentricities and time-averaged mutual inclinations as a function of $a_\mathrm{in}$, for $a_\mathrm{out} \approx 85.0 \, \mathrm{AU}$ and $a_\mathrm{out} \approx 90.0 \, \mathrm{AU}$, respectively. For $a_\mathrm{out} \approx 85.0 \, \mathrm{AU}$, there is a mild excitation of $e_4$ in the case of a stellar binary, with $e_\mathrm{max,4} \approx 0.04$ (cf. \F\,\ref{fig:binary_sequence_paper_binary_seq01_Np_4}). However, in the case of a stellar triple, large values of $e_4$ can be reached, even leading to orbit crossings, if $\mathcal{R}\sim 1$. Interestingly, there is a high and wide peak in $e_{\mathrm{max},4}$ near $\mathcal{R}=2/3$; this is not the case for $a_\mathrm{out} \approx 80.0 \, \mathrm{AU}$ (cf. the top panel of \F\,\ref{fig:triple_sequence_paper_triple_seq03_Np_4}). For $a_\mathrm{out} \approx 90.0 \, \mathrm{AU}$ (cf. \F\,\ref{fig:triple_sequence_paper_triple_seq04_Np_4}), there is no strong excitation of $e_4$ in the case of a stellar binary (cf. \F\,\ref{fig:binary_sequence_paper_binary_seq01_Np_4}). In the case of a stellar triple, excitation is still possible if $\mathcal{R}\sim 1$.

Generally, the locations and relative heights of the peaks with respect to $\mathcal{R}$ change with different parameters of the stellar binary orbit. Further investigation into this behaviour is left for future work. Nevertheless, the results found here already suggest that compared to the case of a single stellar companion, the `binary nature' of the companion in the case of a stellar triple can result in both dynamical {\it protection} and {\it destabilization} of the planets against the stellar torques, depending on the parameters. Given that destabilization only occurs for specific values of the ratio $\mathcal{R}$ whereas protection occurs in the other cases, we expect that the typical effect is that of protection.

\section{Secular constraints in observed systems}
\label{sect:observed}

\begin{figure}
\center
\includegraphics[scale = 0.42, trim = 0mm 15mm 0mm 10mm]{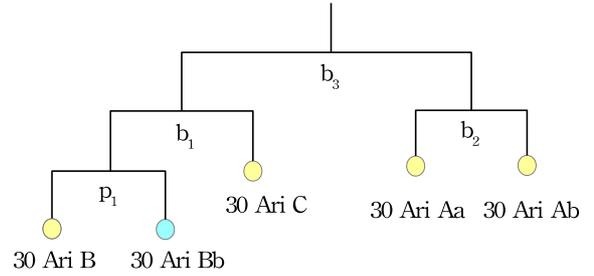}
\caption{\small 30 Ari represented in a mobile diagram. The assumed orbital parameters are listed in Table \ref{table:30_Ari}.}
\label{fig:mobile_30_Ari.eps}
\end{figure}

\subsection{30 Arietis}
\label{sect:observed:30_ari}
30 Arietis (commonly abbreviated as 30 Ari) is a `2+2' quadruple star system consisting of F-type main-sequence stars, with an age of $\approx 0.9 \, \mathrm{Gyr}$ \citep{2009A&A...507.1659G}. The subsystem 30 Ari A is a spectroscopic binary with a period of $1.109526 \, \mathrm{d}$ and an eccentricity of $0.062$ \citep{1974PASP...86..455M}. This subsystem forms a visual binary, with a period of $\approx 34000$ yr, with a subsystem containing 30 Ari B, an F6V star  \citep{2006A&A...450..681T}. The latter star harbours 30 Ari Bb, a massive planet ($ m \sin i \approx 9.88 \, M_\mathrm{J}$) orbiting 30 Ari B in $\approx 335\, \mathrm{d}$ with an eccentricity of $0.289$ \citep{2009A&A...507.1659G}. Recently, \citet{2015ApJ...799....4R} found an $\approx 0.5 \, \mathrm{M}_\odot$ companion star to 30 Ari B, 30 Ari C, in an orbit with a projected separation of $\approx 22.3 \, \mathrm{AU}$ \citep{2015AJ....149..118R}. In \F\,\ref{fig:mobile_30_Ari.eps}, we show, in a mobile diagram, the hierarchy of the system, which is effectively a hierarchical quintuple system. To date, 30 Ari is the second confirmed stellar quadruple system known to host at least one exoplanet; the first is Ph1b \citep{2013ApJ...768..127S}.

Apart from the masses of the components and (most of) the semimajor axes and eccentricities, the orbital properties of 30 Ari, in particular the relative inclinations, are unknown. Here, we investigate the secular dynamical evolution of the system, focusing on the orbit of the planet 30 Ari Bb (\S\,\ref{sect:observed:30_ari:single}). Using secular stability arguments, we constrain the relative inclinations. Furthermore, we investigate the possibility of additional planets around 30 Ari B (\S\,\ref{sect:observed:30_ari:multiple}). 

\begin{table}
\begin{threeparttable}
\begin{tabular}{llll}
Parameter & Value(s) & Parameter & Value(s) \\
\toprule
$m_\mathrm{B}/\mathrm{M}_\odot$ & 1.13 \tnote{a} & $P_\mathrm{b_2}/\mathrm{d}$ & 1.109526 \tnote{d} \\
$m_\mathrm{Bb}/M_\mathrm{J} $ & 9.88 \tnote{a} & $e_\mathrm{b_2}$ & 0.062 \tnote{d} \\
$m_\mathrm{C}/\mathrm{M}_\odot$ & 0.5 \tnote{b} & $P_\mathrm{b_3}/\mathrm{yr}$ & 34,000 \tnote{e} \\
$m_\mathrm{Aa}/\mathrm{M}_\odot$ & 1.31 \tnote{a} & $e_\mathrm{b_3}$ & $0.01-0.9$ \tnote{c} \\
$m_\mathrm{Ab}/\mathrm{M}_\odot$ & 0.14 \tnote{a} & $P_\mathrm{p_1}/\mathrm{d}$ & 335.1 \tnote{a} \\
$a_\mathrm{b_1}/\mathrm{AU}$ & 22.3 \tnote{b} & $e_\mathrm{p_1}$ & 0.289 \tnote{a} \\
$e_\mathrm{b_1}$ & $0.01-0.80$ \tnote{c} \\
\bottomrule
\end{tabular}
\begin{tablenotes}
            \item[a] \citet{2009A&A...507.1659G}
            \item[b] \citet{2015AJ....149..118R}
            \item[c] Rayleigh distribution $\mathrm{d} N/\mathrm{d} e_i \propto e_i \exp(-\beta e_i^2)$; $ \langle e_i^2  \rangle^{1/2} = \beta^{-1/2} = 0.3$ \citep{2010ApJS..190....1R}
            \item[d] \citet{1974PASP...86..455M}            
            \item[e] \citet{2006A&A...450..681T}            
\end{tablenotes}
\caption{ Assumed parameters for 30 Ari. Refer to \F\,\ref{fig:mobile_30_Ari.eps} for the definitions of the orbits. }
\label{table:30_Ari}
\end{threeparttable}
\end{table}

Our assumed parameters are listed in Table \ref{table:30_Ari}. For 30 Ari C (i.e. orbit $\mathrm{b_1}$), only the projected separation of $22.3\,\mathrm{AU}$ is known. Here, we adopt a semimajor axis of $a_\mathrm{b_1} = 22.3 \, \mathrm{AU}$, and consider a distribution of different eccentricities (see below). 

First, we consider the importance of orbit $\mathrm{b_2}$ with regard to the secular evolution of orbit $\mathrm{b_1}$ and its constituents. As shown in \S\,\ref{sect:der:impl}, this can be estimated by considering the ratio $\mathcal{R}^{(\mathrm{bb})}_{\mathrm{b_1;b_2;b_3}}$ of the LK time-scale for the $(\mathrm{b_1,b_3})$ pair, compared to that for the $(\mathrm{b_2,b_3})$ pair. Setting the coefficients $C$ in equation~(\ref{eq:R_def_bb}) equal to unity for simplicity, this ratio is given by
\begin{align}
\nonumber \mathcal{R}^{(\mathrm{bb})}_{\mathrm{b_1;b_2;b_3}} &\sim \left ( \frac{a_\mathrm{b_2}}{a_\mathrm{b_1}} \right )^{3/2} \left ( \frac{M_\mathrm{b_1}}{M_\mathrm{b_2}} \right )^{3/2} \\
&\approx \left ( \frac{2.4 \times 10^{-2} \, \mathrm{AU}}{22.3 \, \mathrm{AU}} \right )^{3/2} \left ( \frac{1.63 \, \mathrm{M}_\odot}{1.45 \, \mathrm{M}_\odot} \right )^{3/2} \approx 4.1 \times 10^{-5},
\end{align}
which is $\ll 1$. This shows that binary $\mathrm{b_2}$, i.e. the 30 Ari A spectroscopic binary, is effectively a point mass from a secular dynamical point of view. Nevertheless, in the numerical integrations below, we do not make this approximation, i.e. we resolve all secular interactions.

\subsubsection{Single planet}
\label{sect:observed:30_ari:single}

We carried out an ensemble of integrations of 5000 systems in which the unknown orbital parameters were sampled from assumed distributions. In particular, $e_\mathrm{b_1}$ and $e_\mathrm{b_3}$ were sampled from a Rayleigh distribution  $\mathrm{d} N/\mathrm{d} e_i \propto e_i \exp(-\beta e_i^2)$ with $ \langle e_i^2  \rangle^{1/2} = \beta^{-1/2} = 0.3$ \citep{2010ApJS..190....1R} in the range $0.01<e_\mathrm{b_1}<0.8$ and $0.01<e_\mathrm{b_3}<0.9$. The upper limit on $e_\mathrm{b_1}$ is motivated by the requirement of (short-term) dynamical stability of the planet with respect to orbit $\mathrm{b_1}$. Using the formulae of \citet{1999AJ....117..621H}, we find that the maximum value of $e_\mathrm{b_1}$ such that the planet in an S-type orbit around 30 Ari B is stable against the perturbation of 30 Ari C, is $\approx 0.80$. The upper limit on $e_\mathrm{b_3}$ is motivated by the requirement of (short-term) dynamical stability of orbit $\mathrm{b_1}$ with respect to orbit $\mathrm{b_3}$. Using the stability criterion on \citet{2001MNRAS.321..398M}, we find that the largest value of $e_\mathrm{b_3}$ for dynamical stability is $\approx 0.90$. The inclinations of all orbits were sampled from a distribution that is linear in $\cos(i_i)$. The arguments of pericentre $\omega_i$ and the longitudes of the ascending nodes $\Omega_i$ were sampled assuming random distributions. 

We integrated the systems for a duration of $t_\mathrm{end} = 1 \, \mathrm{Gyr}$. We also included precession due to relativity in all orbits (cf. Appendix\,\ref{app:der:1PN}) and precession due to tidal and rotational bulges in the planet (cf. \S\,\ref{sect:der:alg}). For the planet, we assumed an apsidal motion constant of $k_\mathrm{AM} = 0.52$ and a spin frequency of $1 \, \mathrm{d^{-1}}$, with the spin vector initially aligned with binary $\mathrm{p_1}$. Tidal dissipation in the planet was not included in detail; instead, we assumed that strong tidal dissipation would occur as soon as the pericentre of the orbit of the planet reached a value smaller than $3\, \mathrm{R}_\odot$. In this case, we considered the system to be unstable, in the sense that the semimajor axis of the planet would decrease to a smaller value compared to the observed value. Furthermore, we considered a realization to be unstable if the eccentricities of any the orbits would imply orbit crossing. 

\begin{figure}
\center
\includegraphics[scale = 0.45, trim = 10mm 0mm 0mm 0mm]{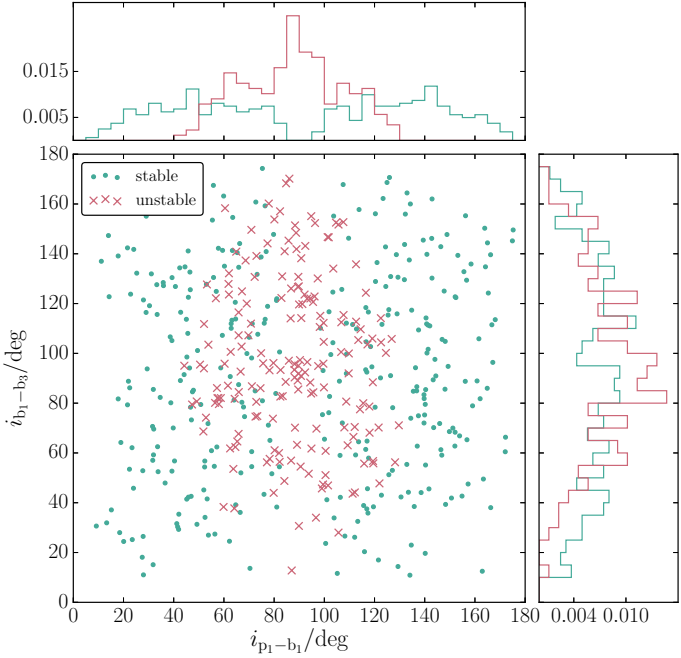}
\caption{\small Outcomes of integrations of Monte Carlo realizations of 30 Ari in the $(i_\mathrm{p_1-b_1},i_\mathrm{b_1-b_3})$-plane. Systems that are stable (unstable) are shown with green bullets (red crosses). The insets show distributions of these quantities for these two cases with green and red lines, respectively. }
\label{fig:MC_Np1_stable_vs_unstable_INCLs_paper_Np1_grid03_L1}
\end{figure}

In \F\,\ref{fig:MC_Np1_stable_vs_unstable_INCLs_paper_Np1_grid03_L1}, we show in the $(i_\mathrm{p_1-b_1},i_\mathrm{b_1-b_3})$-plane the systems that remained stable (green bullets) and those that became unstable (red crosses). Distributions of these quantities for the two cases are shown in the top and right insets, respectively. There is a strong dependence of instability on $i_\mathrm{p_1-b_1}$, with instability occurring between $\approx 40^\circ$ and $140^\circ$. For $85^\circ \lesssim i_\mathrm{p_1-b_1} \lesssim 95^\circ$, all systems are unstable, regardless of $i_\mathrm{b_1-b_3}$, $e_\mathrm{b_1}$ or $e_\mathrm{b_3}$. 

\begin{figure}
\center
\includegraphics[scale = 0.45, trim = 10mm 0mm 0mm 0mm]{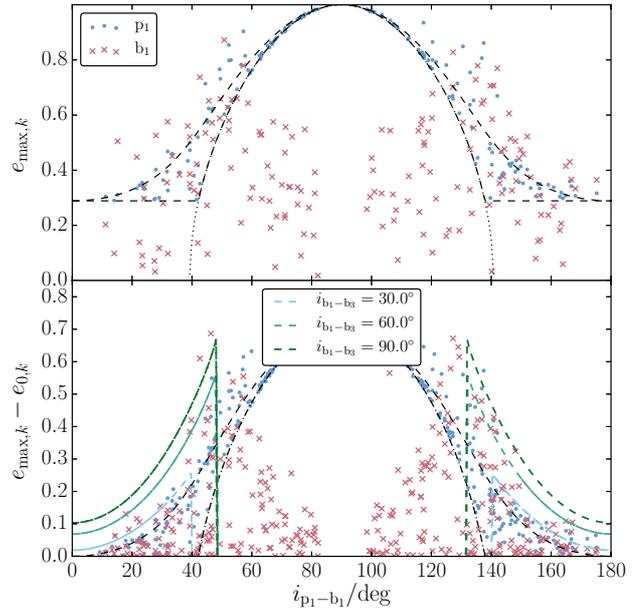}
\caption{\small For the integrations of Monte Carlo realizations of 30 Ari, the maximum eccentricities reached in orbits $\mathrm{p_1}$ (blue bullets) and $\mathrm{b_1}$ (red crosses) as a function of $i_\mathrm{p_1-b_1}$. In the bottom panel, the maximum eccentricities are normalized to the initial values; note that in the case of $e_\mathrm{b_1}$ the initial eccentricities vary per system. In the top panel, the two black dashed lines show an analytic calculation of the maximum eccentricity for two limiting values of $\omega_\mathrm{p_1}$ \citep{2007CeMDA..98...67K}, $\omega_\mathrm{p_1} = 0^\circ$ (upper dashed line) and $\omega_\mathrm{p_1} = 90^\circ$ (lower dashed line). The black dotted line shows the well known relation $e_\mathrm{max,p_1} = \sqrt{1-(5/3) \cos^2(i_\mathrm{p_1-b_1})}$, in which $\omega_\mathrm{p_1} = 90^\circ$ and an initial eccentricity of zero are assumed. In the bottom panel, coloured dashed lines show semi-analytic calculations of the maximum eccentricity in orbit $\mathrm{b_1}$ for various values of $i_\mathrm{b_1-b_3}$, using a method similar to that of \citet{2015MNRAS.449.4221H}. Note that for $50^\circ \lesssim i_\mathrm{p_1-b_1} \lesssim 130^\circ$, this method does not yield a maximum eccentricity (i.e. no solutions can be found). }
\label{fig:MC_Np1_stable_e_max_paper_Np1_grid03_L1}
\end{figure}

\begin{figure}
\center
\includegraphics[scale = 0.45, trim = 10mm 0mm 0mm 0mm]{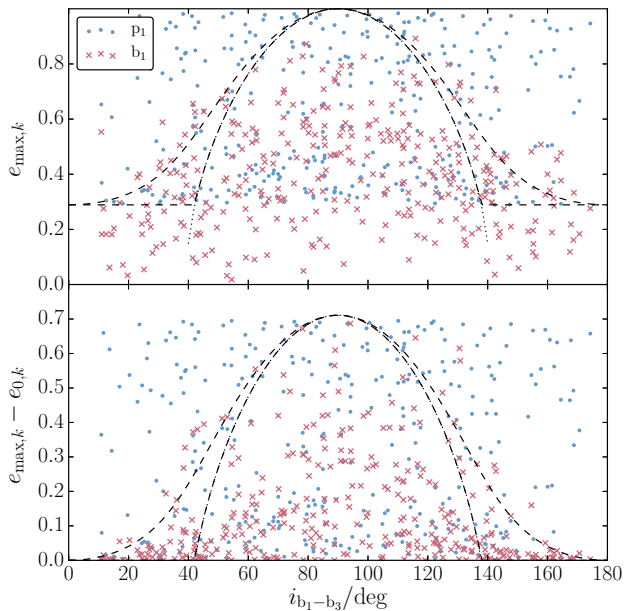}
\caption{\small Similar to \F\,\ref{fig:MC_Np1_stable_e_max_paper_Np1_grid03_L1}, now showing the maximum eccentricities as a function of $i_\mathrm{b_1-b_3}$. In the bottom panel, the maximum eccentricities are normalized to the initial values. The black lines are the same as in \F\,\ref{fig:MC_Np1_stable_e_max_paper_Np1_grid03_L1}. }
\label{fig:MC_Np1_stable_e_max_b1_b3_paper_Np1_grid03_L1}
\end{figure}

This strong dependence is the result of LK cycles induced in orbit $\mathrm{p_1}$ by orbit $\mathrm{b_1}$. To describe this more quantitatively, we show in \F\,\ref{fig:MC_Np1_stable_e_max_paper_Np1_grid03_L1} the maximum eccentricities reached in orbits $\mathrm{p_1}$ (blue bullets) and $\mathrm{b_1}$ (red crosses) as a function of $i_\mathrm{p_1-b_1}$. In the bottom panel, the maximum eccentricities are normalized to the initial values; note that the initial value of $e_\mathrm{p_1}$ is $0.289$ for all systems, whereas the initial eccentricities $e_\mathrm{b_1}$ vary per system (cf. Table~\ref{table:30_Ari}).

The relative inclination $i_\mathrm{p_1-b_1}$ should be high enough for eccentricity excitation in orbit $\mathrm{p_1}$, as expected for LK cycles. The dependence of $e_\mathrm{max,p_1}$ on $i_\mathrm{p_1-b_1}$ is symmetric with respect to $i_\mathrm{p_1-b_1} = 90^\circ$. When $i_\mathrm{p_1-b_1}$ approaches $90 ^\circ$, $e_\mathrm{max,p_1}$ approaches unity. Consequently, there are no stable systems with $85^\circ \lesssim i_\mathrm{p_1-b_1} \lesssim 95^\circ$. Depending on the initial argument of pericentre of orbit $\mathrm{p_1}$, there is a sharp cutoff ($\omega_\mathrm{p_1} = 90^\circ$) or a smooth transition ($\omega_\mathrm{p_1} = 0^\circ$) from low to high inclinations. This is illustrated by the two black dashed lines in \F\,\ref{fig:MC_Np1_stable_e_max_paper_Np1_grid03_L1}, which show an analytic calculation of the maximum eccentricity for these two limiting values of $\omega_\mathrm{p_1}$ for hierarchical triples, computed using the analytic solutions of \citet{2007CeMDA..98...67K}. Note that the latter authors assumed the quadrupole-order limit, and that the `outer' semimajor axis (in this case, $a_\mathrm{b_1}$) is much larger than the `inner' semimajor axis (in this case, $a_\mathrm{p_1}$). Nevertheless, the analytic solutions generally give good estimates for the largest and smallest values of the numerically calculated $e_\mathrm{max,p_1}$ as a function of $i_\mathrm{p_1-b_1}$.

Exceptions of the latter occur near $i_\mathrm{p_1-b_1} \approx 60^\circ$ and $i_\mathrm{p_1-b_1}\approx 120^\circ$ (approximately symmetrically with respect to $i_\mathrm{p_1-b_1} = 90^\circ$), where $e_\mathrm{max,p_1}$ can be higher than the top black analytic line in \F\,\ref{fig:MC_Np1_stable_e_max_paper_Np1_grid03_L1}. This can be understood by noting that $e_\mathrm{b_1}$ can be highly excited for $40^\circ \lesssim i_\mathrm{p_1-b_1} \lesssim 60^\circ$ (see below). In this case, the high eccentricities of orbit $\mathrm{b_1}$ lead to strong perturbations of $\mathrm{p_1}$. 

In \F s\,\ref{fig:MC_Np1_stable_e_max_paper_Np1_grid03_L1} and \ref{fig:MC_Np1_stable_e_max_b1_b3_paper_Np1_grid03_L1}, the dependence of $e_\mathrm{max,b_1}$ as a function of $i_\mathrm{p_1-b_1}$ and $i_\mathrm{b_1-b_3}$, respectively, is shown with the red crosses. There is a clear dependence of $e_\mathrm{max,b_1}$ on $i_\mathrm{p_1-b_1}$, with excitations of $e_\mathrm{b_1}$ reaching $\approx 0.8$ if $i_\mathrm{p_1-b_1}$ is around $50^\circ$ or $130^\circ$; note that $e_\mathrm{b_1} = 0.8$ is close to the limit of dynamical stability. This dependence can be understood by noting that orbit $\mathrm{p_1}$ can induce precession on orbit $\mathrm{b_1}$, reducing the amplitude of eccentricity oscillations in $\mathrm{b_1}$ (this effect was discussed previously in \citealt{2015MNRAS.449.4221H}). The torque of orbit $\mathrm{p_1}$ on $\mathrm{b_1}$ is weakest when $\mathrm{p_1}$ and $\mathrm{b_1}$ are inclined by $\sim 50$ or $130^\circ$. Therefore, $e_\mathrm{max,b_1}$ shows a maximum near these values. 

To understand this more quantitatively, we show in the bottom panel of \F\,\ref{fig:MC_Np1_stable_e_max_paper_Np1_grid03_L1} with the dashed lines semi-analytic calculations of the maximum eccentricity in orbit $\mathrm{b_1}$ as a function of $i_\mathrm{p_1-b_1}$ for various values of $i_\mathrm{b_1-b_3}$, using a method similar to that of \citet{2015MNRAS.449.4221H}. Note that for $50^\circ \lesssim i_\mathrm{p_1-b_1} \lesssim 130^\circ$, this method does not yield a maximum eccentricity (i.e. no solutions can be found). The semi-analytic curves approximately capture the low-inclination dependence of $e_\mathrm{max,b_1}$ on $i_\mathrm{p_1-b_1}$. The larger $i_\mathrm{b_1-b_3}$, the larger $e_\mathrm{max,b_1}$. The dependence of $e_\mathrm{max,b_1}$ on $i_\mathrm{p_1-b_1}$ explains why in \F\,\ref{fig:MC_Np1_stable_e_max_b1_b3_paper_Np1_grid03_L1}, where $i_\mathrm{p_1-b_1}$ varies per point, $e_\mathrm{max,b_1}$ is typically below the black dashed analytic curves, which assume only a perturbation by $\mathrm{b_3}$, and no additional precession due to orbit $\mathrm{p_1}$. 

These results show that it is unlikely that orbits $\mathrm{p_1}$ and $\mathrm{b_1}$ are highly inclined (i.e., a mutual inclination close to $90^\circ$). If this were the case, then the planet would become highly eccentric and tidally interact with, or collide with 30 Ari B. Furthermore, more moderate inclinations of $40^\circ \lesssim i_\mathrm{p_1-b_1} \lesssim 60^\circ$ and $120^\circ \lesssim i_\mathrm{p_1-b_1} \lesssim 140^\circ$ are also not likely because orbit $\mathrm{b_1}$ could be excited to high eccentricity, up to $\approx 0.8$. This would trigger a dynamical instability in the planetary orbit, possibly resulting in the ejection of the planet from the system, or a collision with one of the stars. Therefore, low inclinations are more likely, i.e. $0^\circ \lesssim i_\mathrm{p_1-b_1} \lesssim 40^\circ$ or $140^\circ \lesssim i_\mathrm{p_1-b_1} \lesssim 180^\circ$ (based on these results, we cannot make constraints with respect to prograde or retrograde orbits). Because of the degeneracy of $e_\mathrm{max,b_1}$ with respect to $i_\mathrm{p_1-b_1}$ and $i_\mathrm{b_1-b_3}$, no strong constraints can be put on $i_\mathrm{b_1-b_3}$.

\subsubsection{Constraints on additional planets}
\label{sect:observed:30_ari:multiple}

As a further application, we consider one plausible realization of 30 Ari, and insert an additional planet in orbit around 30 Ari B, outside of the orbit of the observed planet 30 Ari Bb and within the orbit of 30 Ari C. Effectively, the system is then a hierarchical sextuple system. The orbit of the additional planet (`$\mathrm{p_2}$') is assumed to be initially coplanar with orbit $\mathrm{p_1}$, and its eccentricity is $e_\mathrm{p_2} = 0.01$. Furthermore, we adopt a mutual inclination of $\mathrm{p_1}$ and $\mathrm{p_2}$ with respect to orbit $\mathrm{b_1}$ of $i_\mathrm{p_1-b_1} = i_\mathrm{p_2-b_1} = 40^\circ$. The inclination between orbits $\mathrm{b_1}$ and $\mathrm{b_3}$ is also assumed to be $i_\mathrm{b_1-b_3} = 40^\circ$. We assume $e_\mathrm{b_1} = 0.01$ and $e_\mathrm{b_3} = 0.21$.

As free parameters, we consider the mass $m_\mathrm{p_2}$ and semimajor axis $a_\mathrm{p_2}$. Using the formulae of \citet{1999AJ....117..621H}, the lower limit on $a_\mathrm{p_2}$ based on dynamical stability with respect to 30 Ari B and 30 Ari Bb (i.e. $\mathrm{p_1}$) is $a_\mathrm{p_2}\gtrsim 2.9 \, \mathrm{AU}$. Similar arguments with respect to 30 Ari C require $a_\mathrm{p_2}\lesssim 7.6 \, \mathrm{AU}$ (assuming $e_\mathrm{b_1} = 0.01$). Based on this, we take a linear grid in $a_\mathrm{p_2}$ with $3.5 \leq a_\mathrm{p_2}/\mathrm{AU} \leq 7.5$. The masses are taken from a linear grid in $\log_{10}(m_\mathrm{p_2}/M_\mathrm{J})$, with $-3 \leq \log_{10} (m_\mathrm{p_2}/M_\mathrm{J}) \leq 1$. 

\begin{figure}
\center
\includegraphics[scale = 0.45, trim = 10mm 0mm 0mm 0mm]{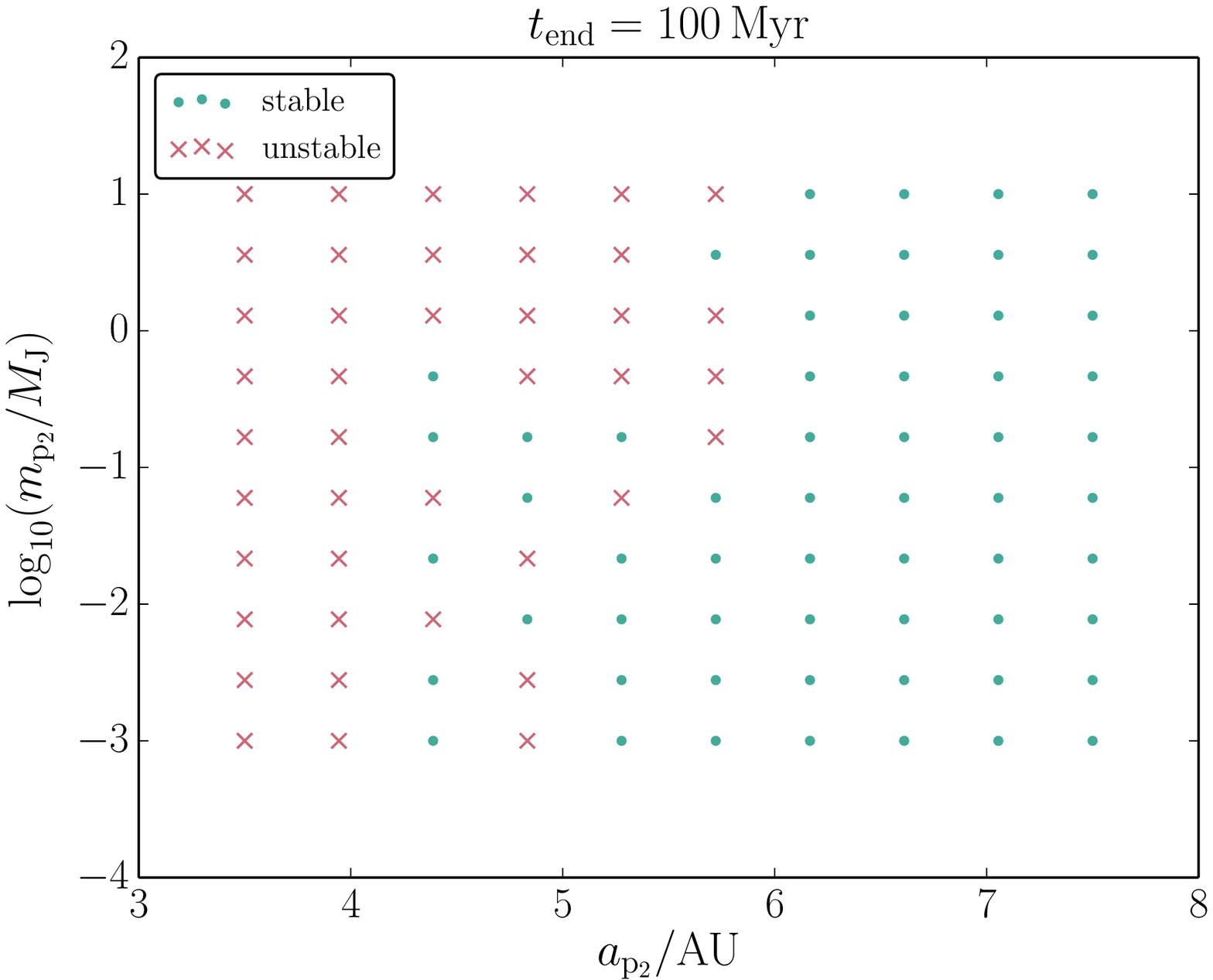}
\includegraphics[scale = 0.45, trim = 10mm 0mm 0mm 0mm]{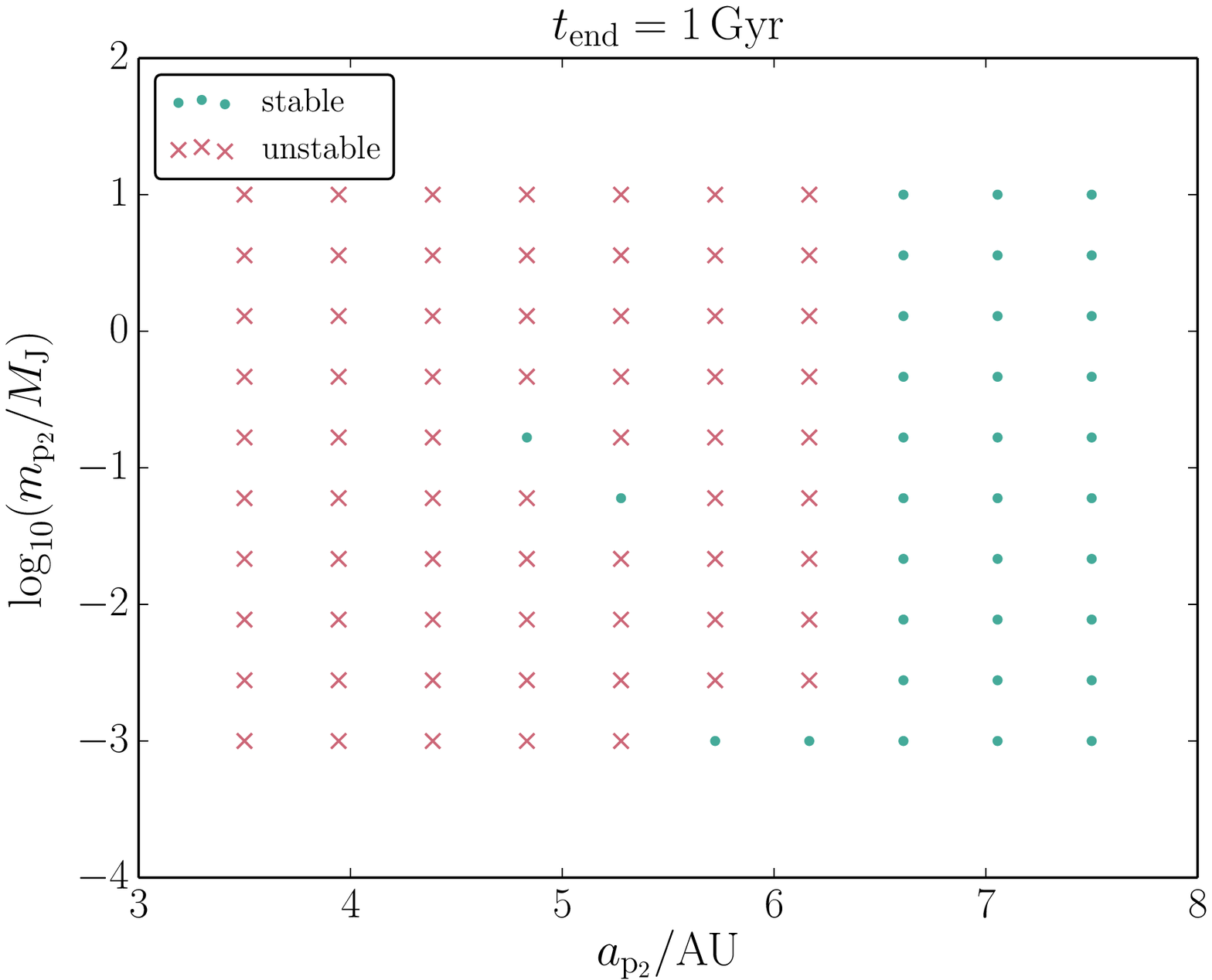}
\caption{\small Stable (green bullets) and unstable (red crosses) systems in the $(a_\mathrm{p_2},m_\mathrm{p_2}/M_\mathrm{J})$-plane when taking into account an additional planet $\mathrm{p_2}$ orbiting beyond 30 Ari Bb, and inside of the orbit of 30 Ari C. The integration time is either 100 Myr (top panel) or 1 Gyr (bottom panel). }
\label{fig:MC_Np2_stable_vs_unstable_paper_Np2_grid01}
\end{figure}

In \F\,\ref{fig:MC_Np2_stable_vs_unstable_paper_Np2_grid01}, we show in the $(a_\mathrm{p_2},m_\mathrm{p_2}/M_\mathrm{J})$-plane the systems that remain stable (green points), and that become unstable (red crosses). Here, unstable systems are due to collisions or tidal interactions of $\mathrm{p_1}$ with the star 30 Ari B ($\approx 0.06$ of the systems), or orbit crossings of the additional planet $\mathrm{p_2}$ with the first planet ($\approx 0.60$ of the systems). There is a noticeable dependence of stability on time; to illustrate this, we include two integration times in \F\,\ref{fig:MC_Np2_stable_vs_unstable_paper_Np2_grid01}: 100 Myr (top panel) and 1 Gyr (bottom panel). As can be expected, more systems become unstable as time progresses. Regardless of age, the additional planet should be placed sufficiently far from the innermost planet. Whereas short-term dynamical stability would allow $3.5 \lesssim a_\mathrm{p_2}/\mathrm{AU} \lesssim 6$, this is no longer the case when taking into account secular evolution. For $t_\mathrm{end} = 1 \,\mathrm{Gyr}$, the minimum semimajor axis, $\approx 6 \, \mathrm{AU}$, is a factor of $\sim 2$ larger compared to the situation when only short-term dynamical stability is taken into account ($2.9 \, \mathrm{AU}$). 

Generally, there are no strong constraints on the mass of the additional planet. This can be understood by noting that the excitation of the eccentricity of $\mathrm{p_2}$ by $\mathrm{b_1}$ is independent of $m_\mathrm{p_2}$ as long as $m_\mathrm{p_2}\ll m_\mathrm{C}$.

\begin{figure}
\center
\includegraphics[scale = 0.45, trim = 10mm 0mm 0mm 0mm]{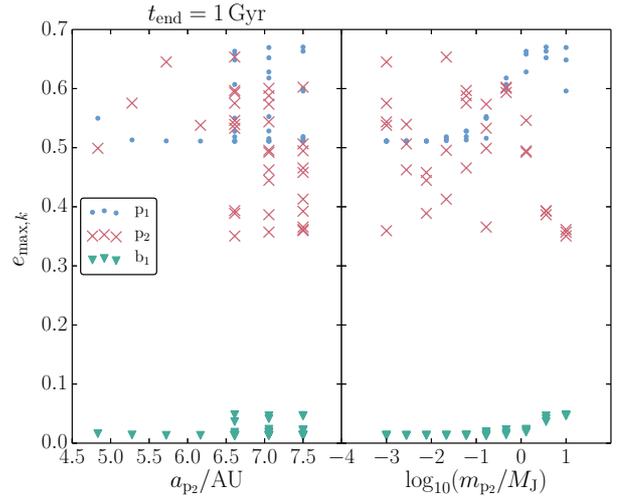}
\caption{\small The maximum eccentricities in orbits $\mathrm{p_1}$ (blue bullets), $\mathrm{p_2}$ (red crosses) and $\mathrm{b_1}$ (green triangles), as a function of the parameters $a_\mathrm{p_2}$ (left-hand panel) and $m_\mathrm{p_2}/M_\mathrm{J}$ (right-hand panel). The integration time is $t_\mathrm{end} = 1 \, \mathrm{Gyr}$. }
\label{fig:MC_Np2_e_max_paper_Np2_grid01_L1}
\end{figure}

Furthermore, in \F\,\ref{fig:MC_Np2_e_max_paper_Np2_grid01_L1}, we show the maximum eccentricities in orbits $\mathrm{p_1}$, $\mathrm{p_2}$ and $\mathrm{b_1}$ as a function of the parameters $a_\mathrm{p_2}$ (left-hand panel) and $m_\mathrm{p_2}/M_\mathrm{J}$ (right-hand panel), with $t_\mathrm{end} = 1 \, \mathrm{Gyr}$. For sufficiently massive planets and sufficiently large semimajor axes, the additional planet can increase the maximum eccentricities of the orbit of $\mathrm{p_1}$, and, to a minor extent, the orbit of $\mathrm{b_1}$. 

To conclude, assuming a low mutual inclination between orbits $\mathrm{p_1}$ and $\mathrm{b_1}$ and $e_\mathrm{b_1}=0.01$ and $e_\mathrm{b_3} = 0.21$, we find a region of secular dynamical stability of an additional planet around 30 Ari B, outside of and coplanar with the orbit of 30 Ari Bb. The semimajor axis of the orbit of this additional planet should be confined to a narrow range, $6 \lesssim a_\mathrm{p_2}/\mathrm{AU} \lesssim 7.6$, whereas there are no strong constraints on mass of the planet (cf. the bottom panel of \F\,\ref{fig:MC_Np2_stable_vs_unstable_paper_Np2_grid01}). Furthermore, the eccentricity of the orbit is likely high, possibly as high as $\sim 0.6$ (cf. \F\,\ref{fig:MC_Np2_e_max_paper_Np2_grid01_L1}).

\subsection{Mizar and Alcor}
\label{sect:observed:mizar_alcor}
Mizar and Alcor is perhaps one of the most well-known nearby hierarchical stellar multiple systems. To date, it is known to be a sextuple system. Mizar and Alcor form a visual stellar binary, which is famous for being used as a vision test \citep{1899sntm.book.....A}. Mizar itself is also a visual binary, known from as early as 1617, when Benedetto Castelli reported resolving it in a letter to Galileo Galilei \citep{2004S&T...108a..74O,2005JHA....36..251S}. Both components of Mizar, Mizar A and B, are spectroscopic binaries. Mizar A is the first spectroscopic binary known, found by Antonia Maury and reported by \citet{1890Obs....13...80P}; the spectroscopic binary Mizar B was discovered independently by \citet{1908AN....177..171F} and \citet{1908AN....177....7L}. Recently, Alcor was found to be a binary as well \citep{2010AJ....139..919M,2010ApJ...709..733Z}. It has been disputed whether or not Mizar is gravitationally bound to Alcor. In the recent works of \citet{2010AJ....139..919M,2010ApJ...709..733Z}, it was found that this is likely the case. To our knowledge, the secular dynamics of the system have not been explored.

\begin{figure}
\center
\includegraphics[scale = 0.45, trim = 10mm 0mm 0mm 0mm]{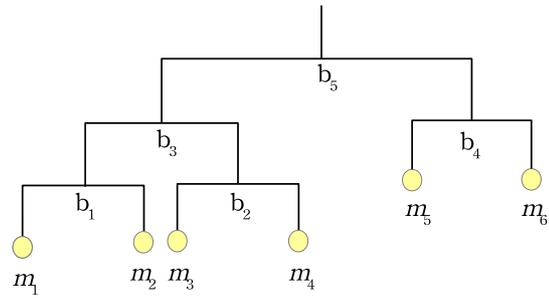}
\caption{\small A mobile diagram of Mizar (the subsystem with binaries $\mathrm{b_1}$, $\mathrm{b_2}$ and $\mathrm{b_3}$) and Alcor (the subsystem with binary $\mathrm{b_4}$). }
\label{fig:mobile_Mizar_Alcor}
\end{figure}

In \F\,\ref{fig:mobile_Mizar_Alcor}, we show the hierarchy of the system in a mobile diagram. Mizar forms a `2+2' quadruple system, orbited by the binary Alcor. Based on the studies mentioned above, we assume masses $m_i/\mathrm{M}_\odot = \{2.43, 2.50, 1.6, 1.6, 1.8, 0.3\}$, semimajor axes $a_i/\mathrm{AU} = \{0.25, 0.9,  5.9 \times 10^2, 2.8 \times10^2, 7.4 \times 10^4\}$ and eccentricities $e_i = \{0.53, 0.46, 0.3, 0.3, 0.3\}$.  For the unknown eccentricities of $\mathrm{b_3}$, $\mathrm{b_4}$ and $\mathrm{b_5}$, we assume a value of 0.3, a typical value for long-period binaries \citep{2010ApJS..190....1R}. 

Below, we evaluate the characteristic quantities $\mathcal{R}^{\mathrm{(bb)}}$ (cf. equation~\ref{eq:R_def_bb}) for the relevant binary combinations, setting the quantities $C$ to unity. For the `outermost' binary-binary combination $(\mathrm{b_3,b_4,b_5})$, we find
\begin{align}
\mathcal{R}^{\mathrm{(bb)}}_{345} \sim \left ( \frac{a_\mathrm{b_4}}{a_\mathrm{b_3}} \right )^{3/2} \left ( \frac{m_1+m_2+m_3+m_4}{m_5+m_6} \right )^{3/2} \approx 8 \times 10^{-2},
\end{align}
indicating that the binary nature of $\mathrm{b_4}$ can typically be neglected. Moreover, even for the pair $(\mathrm{b_3},\mathrm{b_5})$, the LK time-scale (cf. equation~\ref{eq:P_LK} with $C=1$) is $P_\mathrm{LK;35} \sim 33 \, \mathrm{Gyr}$, which is much longer than the age of the system, $\approx 0.5 \, \mathrm{Gyr}$. Therefore, LK cycles in the outermost binaries $\mathrm{b_3}$, $\mathrm{b_4}$ and $\mathrm{b_5}$ are not important during the lifetime of the system. 

For the `innermost' combination $(\mathrm{b_1,b_2,b_3})$, we find
\begin{align}
\mathcal{R}^{\mathrm{(bb)}}_{123} \sim \left ( \frac{a_\mathrm{b_2}}{a_\mathrm{b_1}} \right )^{3/2} \left ( \frac{m_1+m_2}{m_3+m_4} \right )^{3/2} \approx 1 \times 10^{1},
\end{align}
showing that LK cycles are typically more important in the $(\mathrm{b_2,b_3})$ pair compared to the $(\mathrm{b_1,b_3})$ pair. Nevertheless, resonant behaviour might be expected in some cases. Using equation~(\ref{eq:P_LK}), the individual LK time-scales are
\begin{align}
P_\mathrm{LK;13} \sim 980 \, \mathrm{Myr}; \quad \quad P_\mathrm{LK;23} \sim 74 \, \mathrm{Myr},
\end{align}
showing that during the lifetime of the system, of the order of 10 eccentricity oscillations could have occurred in binary $\mathrm{b_2}$.

However, the above does not take into account relativistic precession. In fact, the 1PN time-scales in both binaries $\mathrm{b_1}$ and $\mathrm{b_2}$ are much shorter compared to their respective LK time-scales. Using equation~(\ref{eq:t_1PN}), we find
\begin{align}
t_\mathrm{1PN;b_1} \approx 0.1 \, \mathrm{Myr}; \quad \quad t_\mathrm{1PN;b_2} \approx 4 \, \mathrm{Myr},
\end{align}
showing that LK cycles are likely (nearly) completely quenched. 

\begin{figure}
\center
\includegraphics[scale = 0.45, trim = 10mm 0mm 0mm 0mm]{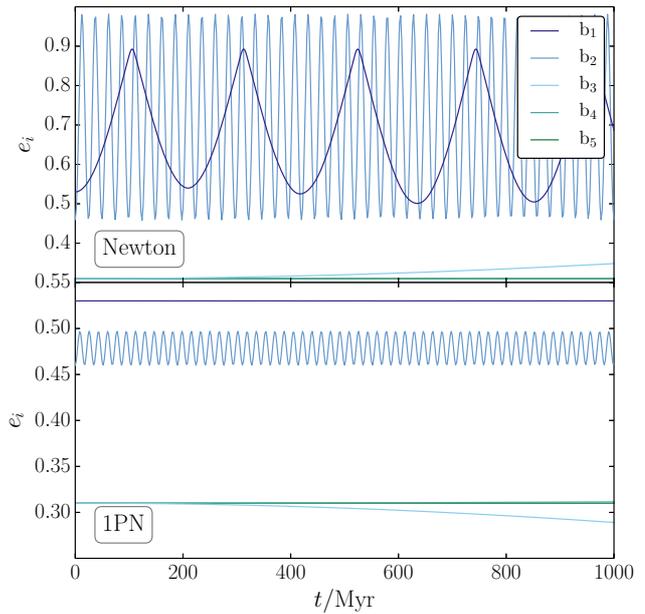}
\caption{\small Example eccentricity evolution for Mizar and Alcor. In the top panel, we included only Newtonian terms, whereas in the bottom panel, we also included the pairwise 1PN terms. The initial mutual inclinations are $i_\mathrm{13} = 80^\circ$ and $i_\mathrm{23} = 85^\circ$. }
\label{fig:Mizar_Alcor_single_test01_index_0}
\end{figure}

We show an example in \F\,\ref{fig:Mizar_Alcor_single_test01_index_0}, where in the top panel, we show the eccentricities assuming only Newtonian terms, and in the bottom panel the pairwise 1PN terms are also included. Even though the initial mutual inclinations $i_\mathrm{13} = 80^\circ$ and $i_\mathrm{23} = 85^\circ$ are high, with the inclusion of the 1PN terms, oscillations are severely quenched. We conclude that secular evolution is not important in the Mizar and Alcor system.

\section{Discussion}
\label{sect:discussion}

\subsection{The validity of orbit averaging}
\label{sect:discussion:orbit_averaging}
In our derivations, we orbit averaged the Hamiltonian after the expansion in ratios of separation ratios over {\it all} orbits (cf. \S\,\ref{sect:der:oa}). As shown recently by \citet{2016MNRAS.458.3060L}, in the case of hierarchical triple systems with moderately large octupole parameters ($ \epsilon_\mathrm{oct} \gtrsim 10^{-3}$) and large mass ratios of the tertiary body with respect to the inner binary ($m_3/(m_1+m_2)\gtrsim 10^2$), the {\it double} averaging technique can result in errors compared to the true secular evolution (for example, as measured by the fraction in an ensemble of systems in which orbital flips occur). Physically, this discrepancy arises from oscillations of the inner orbital angular momentum vector on time-scales of the outer orbital period \citep{2012ApJ...757...27A,2012arXiv1211.4584K,2014MNRAS.439.1079A}, which invalidate the assumption of a fixed inner orbit during the orbit of the tertiary. These effects very likely also apply, to some cases, to the more general hierarchical $N$-body systems considered in this paper. A similar technique as used by \citet{2016MNRAS.458.3060L}, i.e. to apply a `correction' in the process of averaging over the outer body, might also be applied to hierarchical $N$-body systems. However, such an analysis is beyond the scope of this work. 

\subsection{Short-term instabilities}
\label{sect:discussion:range}
The assumptions on which the \textsc{\codename} algorithm is based (cf. \S\,\ref{sect:introduction}) should be considered not only for the initial conditions of the systems, but also for the subsequent secular evolution. In particular, when eccentricities become high, there is the possibility that dynamical instabilities are triggered on short-term scales, i.e. of the order of the orbital periods. These dynamical instabilities are not taken into account in the secular equations of motion. For example, in multiplanet systems, dynamical stabilities are triggered when the Hill or Laplace boundaries are approached for sufficiently eccentric orbits. 

For this reason, a number of stopping conditions are implemented in \textsc{\codename} that use various stability criteria to evaluate the importance of short-term dynamical instabilities. These criteria are approximate, however. In order to investigate more accurately whether the system is short-term dynamically stable, \textsc{\codename} allows for the user to easily switch to direct $N$-body integration. A complication of the latter is that the orbital phases are inherently not modelled in \textsc{\codename} but evidently, they are required for direct $N$-body integration. In practice, it is therefore necessary to carry out an ensemble of $N$-body integrations with e.g. randomly sampled orbital phases. 

Typically, when a dynamical instability occurs, this happens on a time-scale that is short compared to secular time-scales. The resulting configuration is likely again hierarchical. Therefore, although \textsc{\codename} cannot be used to model the short-term phase of dynamical instability itself, once a new stable configuration has been attained, it can again be used to model the subsequent secular evolution. In future work, we intend to use this approach to study the secular stability of multiplanet systems, both in single- and multistar configurations.

\section{Conclusions}
\label{sect:conclusions}

We have presented a new method for studying the secular dynamics of hierarchical $N$-body systems composed of nested binaries with an arbitrary number of bodies and an arbitrary hierarchical configuration. This algorithm is suitable for studying the secular evolution of hierarchical systems with complex hierarchies, making long-term integrations of such systems feasible. We have applied the method to various multiplanet and multistar systems. Our main conclusions are as follows.

\medskip \noindent 1. We have derived the expanded and orbit-averaged Hamiltonian for the system (a complete self-contained derivation is given in Appendix \ref{app:der}). Our main assumptions are that the system is hierarchical, i.e. the ratios $x_i$ of the separation vectors are small, and that orbit averaging is applicable (i.e. the time-scales for angular momentum changes are much longer than orbital time-scales). The results from the expansion are summarized below. 
	\begin{itemize}
	\item To first order in $x_i$ (`dipole order'), all terms vanish identically for {\it any} hierarchical configuration. 
	\item To order $x_i^2$ (`quadrupole order'), the Hamiltonian consists of terms which, individually, depend on binary {\it pairs}. These pairwise terms are mathematically equivalent to the quadrupole-order terms of the hierarchical three-body Hamiltonian. 
	\item To order $x_i^3$ (`octupole order'), the Hamiltonian consists of two types of terms: terms which, individually, depend on binary {\it pairs}, and terms which, individually, depend on binary {\it triplets}. The pairwise terms are mathematically equivalent to the octupole-order terms of the hierarchical three-body Hamiltonian. The triplet terms are mathematically equivalent to the term depending on three binaries at the octupole order in the hierarchical four-body Hamiltonian (`3+1' configuration; this term was first derived by \citealt{2015MNRAS.449.4221H}).
	\item Generally, to order $x_i^n$, there are $n-1$ types of terms (i.e. depending on binary pairs, triplets, etc.). The individual terms depend on at most $n$ binaries. Whether all these terms appear is contingent on the system: if the system does not contain a sufficiently complex hierarchy, then not all types of terms may appear.
	\end{itemize}
In summary, the general hierarchical $N$-body Hamiltonian can be constructed from the Hamiltonians of smaller subsystems. The higher the order of the expansion, the larger the number of binaries within these subsystems.

\medskip \noindent 2. We have explicitly derived the expanded Hamiltonian up and including octupole-order, and including the pairwise hexadecupole- and dotriacontupole-order terms (cf. Appendix \ref{app:der:av:HO}). We orbit-averaged these terms, and also gave a general expression for the orbit-averaged binary pair term to order $n$ in terms of derivatives of polynomial functions (cf. Appendix \ref{app:der:av:gen_pair}). Note that for hierarchical triple systems, the Hamiltonian consists exclusively of these pairwise terms. Also, in the case of systems with more complex hierarchies, the pairwise terms are expected to be typically dominant. 

\medskip \noindent 3. We presented a new algorithm within the \textsc{AMUSE} framework, \textsc{\codename}, to numerically solve the resulting equations of motion. The new algorithm allows for fast long-term integration of hierarchical systems, opening up the possibility for studying the long-term dynamics of complex hierarchical systems without having to resort to costly $N$-body integrations. Tidal friction and precession due to relativity and tidal bulges are also included in an approximate approach, treating each binary as an isolated system. 

\medskip \noindent 4. As a first demonstration, we showed that \textsc{\codename} can be used for the secular dynamics of multiplanet systems in both single-star and multiple-star systems. For the case of a single-star system (cf. \S\,\ref{sect:multiplanet:single}), we compared our results to Laplace-Lagrange (LL) theory in the regime of low eccentricities and inclinations, in which LL theory is essentially exact (barring the effects of mean motion resonances). We found that \textsc{\codename} produces results that differ by only $\sim 2 \%$ with LL theory, provided that the spacing between the planets is $K_0 \gtrsim 10$ (equivalently, a semimajor axis ratio $\lesssim 0.4$), and that sufficiently high-order pairwise binary terms, as high as the dotriacontupole-order (fifth order in $x_i$), are included. In the systems that we considered, the octupole-order binary triplet terms do not affect the secular dynamics. 
More generally, for any order $n$, we estimated that the non-pairwise binary terms are not important for nested systems with a central massive body, and with the other bodies of comparable mass. 

\medskip \noindent 5. Whereas LL theory only applies to planetary systems with low eccentricities and inclinations and with a central massive object, our method can also be used in the case of arbitrary eccentricities (provided that the $x_i$ remain sufficiently small), inclinations and hierarchies. In particular, we showed that \textsc{\codename} can be used to efficiently study multiplanet systems in binary and triple-star systems (cf. \S s\,\ref{sect:multiplanet:binary} and \ref{sect:multiplanet:triple}, respectively). In the case of triple-star systems in which the tertiary star has multiple S-type planets, we showed that, compared to the case of a binary-star system, the `binary nature' of the companion in the case of a stellar triple can result in both dynamical {\it protection} and {\it destabilization} of the planets against the stellar torques, depending on the configuration.

\medskip \noindent 6. We applied our method to the observed stellar quadruple system 30 Arietis (30 Ari), which harbours a massive planet, 30 Ari Bb. The mutual inclinations in this system are unknown. Using secular stability arguments, we showed that the orbit of the planet is likely not highly inclined with respect to 30 Ari C. Furthermore, assuming the latter, we found that there is a narrow region in semimajor axis space ($6 \lesssim a_\mathrm{p_2}/\mathrm{AU} \lesssim 7.6$) which allows for the presence of an additional planet beyond the orbit of 30 Ari Bb, and inside of the orbit of 30 Ari C. The eccentricity of the orbit of the additional planet would likely be high, possibly as high as $\sim 0.6$. 

We will make \textsc{\codename} publicly available within \textsc{AMUSE}\footnote{\href{http://www.amusecode.org}{amusecode.org}} by 2016.

\section*{Acknowledgements}
We thank the referee, Will Farr, for providing very useful comments, and for motivating us to further generalize our results. This work was supported by the Netherlands Research Council NWO (grants \#639.073.803 [VICI],  \#614.061.608 [AMUSE] and \#612.071.305 [LGM]) and the Netherlands Research School for Astronomy (NOVA).

\bibliographystyle{mnras}
\bibliography{literature}

\appendix
%\newpage

\onecolumn

\section{Derivation of the expanded and orbit-averaged hierarchical multiple Hamiltonian}
\label{app:der}
In this appendix, we derive the Hamiltonian $H$ for a hierarchical multiple system consisting of nested binaries with an arbitrary number of bodies and hierarchy, expanded in ratios of the binary separations (cf. Appendices\,\ref{app:der:struc}, \ref{app:der:pot}, \ref{app:der:kin} and \ref{app:der:ham}). Subsequently, we orbit average the Hamiltonian (cf. Appendix\,\ref{app:der:av}). Lastly, we give an ad hoc expression for the 1PN Hamiltonian (cf. Appendix\,\ref{app:der:1PN}).

\subsection{Description of the system structure}
\label{app:der:struc}
Consider a system of $N$ point masses $m_i \in \{m_1\cdots m_N\}$ with position vectors $\ve{R}_i \in \{\ve{R}_1 \cdots \ve{R}_N\}$, arranged in $N-1$ binary orbits (see \F\,\ref{fig:example_quintuple} for an example quintuple system). The relative position vectors between binaries and/or bodies are denoted by $\ve{r}_i$, where generally $i\leq N-1$; let $\mathrm{B} = \{i:i\leq N-1\}$ denote the corresponding set of indices. The hierarchy of the system is specified using the mass ratio matrix $\bf A$ according to
\begin{align}
\label{eq:R_to_r}
\ve{r}_i = \sum_{k=1}^N A_{ik} \ve{R}_k.
\end{align}
By definition, the last row in $\bf A$, i.e. $A_{Nk}$, corresponds to the centre of mass $\ve{r}_N$, and is given by $A_{Nk} = m_k/M$, where
\begin{align}
M \equiv \sum_{i=1}^N m_i
\end{align}
is the total mass of the system. For convenience in a number of examples below, we also introduce the following notation for the partial sum over the masses,
\begin{align}
\label{eq:M_ij_def}
M_{i,j} \equiv \sum_{n=i}^j m_n.
\end{align}

\begin{table}
\def\arraystretch{2.0}
\begin{tabular}{c|cccc|cccccc}
\toprule
\multicolumn{10}{c}{$i\rightarrow j$} \\
\cmidrule(lr){2-11}
\multirow{2}{*}{$k$} & \multicolumn{4}{c}{1} & \multicolumn{3}{c}{2} & \multicolumn{2}{c}{3} & \multicolumn{1}{c}{4} \\
\cmidrule(lr){2-5} \cmidrule(lr){6-8} \cmidrule(lr){9-10} \cmidrule(lr){11-11} 
& 2 & 3 & 4 & 5 & 3 & 4 & 5 & 4 & 5 & 5 \\
\midrule
1 & 1 & $\frac{m_2}{M_{1,2}}$ & $\frac{m_2}{M_{1,2}}$ & $\frac{m_2}{M_{1,2}}$ & $-\frac{m_1}{M_{1,2}}$ & $-\frac{m_1}{M_{1,2}}$ & $-\frac{m_1}{M_{1,2}}$ & 0 & 0 & 0 \\
2 & 0 & 1 & $\frac{m_3}{M_{1,3}}$ & $\frac{m_3}{M_{1,3}}$ & 1 & $\frac{m_3}{M_{1,3}}$ & $\frac{m_3}{M_{1,3}}$ & $-\frac{M_{1,2}}{M_{1,3}}$ & $-\frac{M_{1,2}}{M_{1,3}}$ & 0 \\
3 & 0 & 0 & $-\frac{m_5}{M_{4,5}}$ & $\frac{m_4}{M_{4,5}}$ & 0 & $-\frac{m_5}{M_{4,5}}$ & $\frac{m_4}{M_{4,5}}$ & $-\frac{m_5}{M_{4,5}}$ & $\frac{m_4}{M_{4,5}}$ & 1 \\
4 & 0 & 0 & 1 & 1 & 0 & 1 & 1 & 1 & 1 & 0 \\
\bottomrule
\end{tabular}
\caption{ The values of $B_{ijk}$ (cf. equation~\ref{eq:B_def}) for the quintuple system depicted in \F\,\ref{fig:example_quintuple}. Columns correspond to pairs of bodies $(i,j)$; rows correspond to binaries $k$. The partial mass sums $M_{i,j}$ are defined in equation~(\ref{eq:M_ij_def}). Note that $B_{ijk}=-B_{jik}$ and $B_{iik}=0$ (cf. equation~\ref{eq:B_def}).}
\label{table:example_quintuple_B}
\end{table}

\begin{table}
\def\arraystretch{2.0}
\begin{tabular}{c|cccc|cccccc}
\toprule
\multicolumn{10}{c}{$k\rightarrow l$} \\
\cmidrule(lr){2-11}
\multirow{2}{*}{$i$} & \multicolumn{4}{c}{1} & \multicolumn{3}{c}{2} & \multicolumn{2}{c}{3} & \multicolumn{1}{c}{4} \\
\cmidrule(lr){2-5} \cmidrule(lr){6-8} \cmidrule(lr){9-10} \cmidrule(lr){11-11} 
& 1 & 2 & 3 & 4 & 2 & 3 & 4 & 3 & 4 & 4 \\
\midrule
1 & $\frac{m_2^2}{M_{1,2}^2}$ & $\frac{m_2 m_3}{M_{1,2} M_{1,3}}$ & 0 & $\frac{m_2 M_{4,5}}{M_{1,2} M}$ & $\frac{m_3^2}{M{1,3}^2}$ & 0 & $\frac{m_3 M_{4,5}}{M_{1,3} M}$ & 0 & 0 & $\frac{M_{4,5}^2}{M^2}$ \\
2 & $\frac{m_1^2}{M_{1,2}^2}$ & $- \frac{m_1 m_3}{M_{1,2} M_{1,3}}$ & 0 & $-\frac{m_1M_{4,5}}{M_{1,2}M}$ & $\frac{m_3^2}{M_{1,3}^2}$ & 0 & $\frac{m_3 M_{4,5}}{M_{1,3} M}$ & 0 & 0 & $\frac{M_{4,5}^2}{M^2}$ \\
3 & 0 & 0 & 0 & 0 & $\frac{M_{1,2}^2}{M_{1,3}^2}$ & 0 & $-\frac{M_{1,2} M_{4,5}}{M_{1,3} M}$ & 0 & 0 & $\frac{M_{4,5}^2}{M^2}$ \\
4 & 0 & 0 & 0 & 0 & 0 & 0 & 0 & $\frac{m_5^2}{M_{4,5}^2}$ & $-\frac{M_{1,3}m_5}{M_{4,5}M}$ & $\frac{M_{1,3}^2}{M^2}$ \\
5 & 0 & 0 & 0 & 0 & 0 & 0 & 0 & $\frac{m_4^2}{M_{4,5}^2}$ & $\frac{M_{1,3}m_4}{M_{4,5}M}$ & $\frac{M_{1,3}^2}{M^2}$ \\
\bottomrule
\end{tabular}
\caption{ The values of $C_{ikl}$ (cf. equation~\ref{eq:C_def}) for the quintuple system depicted in \F\,\ref{fig:example_quintuple}. Columns correspond to pairs of binaries $(k,l)$; rows correspond to bodies $i$. The partial mass sums $M_{i,j}$ are defined in equation~(\ref{eq:M_ij_def}). Note that $C_{ikl}=C_{ilk}$ (cf. equation~\ref{eq:C_def}). }
\label{table:example_quintuple_C}
\end{table}

For example, a hierarchical triple system (the simplest hierarchical multiple system) is represented by
\begin{align}
\renewcommand\arraystretch{2}
\bf A = \begin{pmatrix}
1 & -1 & 0 \\ \frac{m_1}{M_{1,2}} & \frac{m_2}{M_{1,2}} & -1 \\
\frac{m_1}{M} & \frac{m_2}{M} & \frac{m_3}{M}
\end{pmatrix},
\end{align}
which implies $\ve{r}_1 = \ve{R}_1 - \ve{R}_2$ and $\ve{r}_2 = (m_1 \ve{R}_1 + m_2 \ve{R}_2)/(m_1+m_2) - \ve{R}_3$. With this definition of $\bf A$, $\ve{r}_1$ corresponds to the relative separation vector of the inner binary between $\ve{R}_1$ and $\ve{R}_2$, whereas $\ve{r}_2$ corresponds to the relative separation vector in the outer binary between the centre of mass of the inner binary and $\ve{R}_3$. Another example, which is also used for further illustration below, is a hierarchical quintuple consisting of a triple-binary pair,
\begin{align}
\renewcommand\arraystretch{2}
\bf A = \begin{pmatrix}
1 & -1 & 0 & 0 & 0 \\
\frac{m_1}{M_{1,2}} & \frac{m_2}{M_{1,2}} & -1 & 0 & 0 \\
0 & 0 & 0 & 1 & -1 \\
\frac{m_1}{M_{1,3}} & \frac{m_2}{M_{1,3}} & \frac{m_3}{M_{1,3}} & -\frac{m_4}{M_{4,5}} & -\frac{m_5}{M_{4,5}} \\
\frac{m_1}{M} & \frac{m_2}{M} & \frac{m_3}{M} & \frac{m_4}{M} & \frac{m_5}{M}
\end{pmatrix}.
\end{align}
A schematic representation of this system in a mobile diagram is given in \F\,\ref{fig:example_quintuple}. Note that the order of the rows in $\bf{A}$ sets the labels on the binary separation vectors $\ve{r}_i$, and the signs determine the directions of the $\ve{r}_i$.

The inverse relation between (absolute) position vectors and relative binary separation vectors is given by the inverse matrix of $\bf{A}$, i.e.
\begin{align}
\label{eq:r_to_R}
\ve{R}_i = \sum_{k=1}^N A_{ik}^{-1} \ve{r}_k.
\end{align}

The potential energy (cf. Appendix\,\ref{app:der:pot}) is expressed in terms of distances between pairs of bodies, $||\ve{R}_i - \ve{R}_j||$. Using equation~(\ref{eq:r_to_R}), the difference vectors can be written as
\begin{align}
\label{eq:Ri_minus_Rj1}
\ve{R}_i - \ve{R}_j = \sum_{k=1}^N \left ( A_{ik}^{-1} - A_{jk}^{-1} \right ) \ve{r}_k \equiv \sum_{k=1}^N B_{ijk} \ve{r}_k,
\end{align}
where we defined the quantity
\begin{align}
\label{eq:B_def}
B_{ijk} \equiv A_{ik}^{-1} - A_{jk}^{-1}.
\end{align}
Another quantity derived from $\bf{A}$, used for the kinetic energy (cf. Appendix\,\ref{app:der:kin}), is
\begin{align}
\label{eq:C_def}
C_{ikl} \equiv A_{ik}^{-1} A_{il}^{-1}.
\end{align}
Before proceeding with the expansion of the potential energy in Appendix\,\ref{app:der:pot}, we first introduce a number of useful definitions, and discuss some general properties of the three-index quantities $B_{ijk}$ (Appendix\,\ref{app:der:struc:B}) and $C_{ikl}$ (Appendix\,\ref{app:der:struc:C}).

\paragraph*{Children, siblings and descendants.} We refer to the two members of a binary $k$ as the `children' of that binary, i.e. child 1 and child 2, denoted by `$k.\mathrm{C1}$' and `$k.\mathrm{C2}$'. The children can be either bodies or binaries themselves. Each child has a sibling, i.e. the binary or body with the same parent binary. For a binary pair $(k,l)$, where $l$ is a direct or indirect child of $k$, we use `$k.\mathrm{CS}(l)$' to denote the sibling in $k$ of the child in $k$ that is connected to $l$. 

We define a descendant of a binary $k$ as a body or binary that is connected to $k$ via one of the children of $k$, not necessarily directly. For example, in \F\,\ref{fig:example_quintuple}, bodies 1 and 2 are descendants of binary 4, whereas they are not descendants of binary 3. The set of all descendant {\it bodies} of a binary $k$ is denoted with `$\{k.\mathrm{C}\}$', and the body descendant sets within the children of $k$ are denoted with `$\{k.\mathrm{C}n\}$' for child $n$. These notations will be used frequently below. 

\paragraph*{Masses.}
The mass of binary $k$, $M_k$, is defined as the sum of the masses of {\it all bodies} contained within the hierarchy of binary $k$. The masses of the children of binary $k$ are denoted with $M_{k.\mathrm{C1}}$ and $M_{k.\mathrm{C2}}$, which, evidently, satisfy
\begin{align}
M_{k.\mathrm{C1}} + M_{k.\mathrm{C2}} = M_k.
\end{align}
Furthermore,
\begin{align}
\sum_{i\in \{k.\mathrm{C}n \} } m_i = M_{\mathrm{C}n}.
\end{align}

\paragraph*{Levels.} We define the `level' of a binary $k$ as the number of binaries that needs to be traversed to reach $k$, starting from the `absolute top' binary of the system, i.e. the binary for which all other binaries are children. For example, in \F\,\ref{fig:example_quintuple}, the top binary is binary 4 with level 0. We denote the level of binary $k$ with `$k.\mathrm{L}$'.

\paragraph*{The `sign quantity' $\alpha(i,j;k)$.} For any pair of bodies or binaries $(i,j)$, there is a unique path, i.e. a set of binaries, connecting $i$ and $j$. Let this path be denoted by $\mathrm{B_p}=\mathrm{B_p}(i,j)$. For a given path $\mathrm{B_p}$ and a given binary $k$, we define the quantity $\alpha = \alpha(i,j;k)$ for which
\begin{align}
\alpha(i,j;k) = \pm1.
\end{align}
The positive (negative) sign applies when in binary $k$, the path $\mathrm{B_p}$ is opposed to (directed along) the direction of $\ve{r}_k$. In other words, if $\ve{r}_k$ is defined such that it points from `child 2' to `child 1' in binary $k$, then the sign is positive if the path in $\mathrm{B_p}$ leads from child 1 to child 2 (directed against $\ve{r}_k$), and negative if path in $\mathrm{B_p}$ leads from child 2 to child 1 (directed along $\ve{r}_k$). From this, it also immediately follows that $\alpha(i,j;k) = -\alpha(j,i;k)$. Also, for $j \notin \{k.\mathrm{C1} \}$ and $j \notin \{k.\mathrm{C2} \}$, $\alpha(k.\mathrm{C1},j;k) = - \alpha(k.\mathrm{C2},j;k)$.

For example, in \F\,\ref{fig:example_quintuple}, $\alpha(1,2;1) = 1$, $\alpha(2,1;1) = -1$, $\alpha(1,5;4) = 1$ and $\alpha(5,1;4) = -1$. 

\subsubsection{General properties of $B_{ijk}$}
\label{app:der:struc:B}
By definition (cf. equation~\ref{eq:Ri_minus_Rj1}), the first two indices in $B_{ijk}$ refer to bodies, whereas the third index refers to a binary. Given pair $(i,j)$ with path $\mathrm{B_p}=\mathrm{B_p}(i,j)$ and given binary $k$, the following properties of $B_{ijk}$ apply.
\begin{itemize}
\item $B_{ijk} = -B_{jik}$ (this follows immediately from equation~\ref{eq:B_def}).
\item $B_{ijk}=0$ if $k\notin \mathrm{B_p}(i,j)$ or $k=N$ (in the latter case, $k$ corresponds to the centre of mass).
\item $B_{ijk}= \alpha(i,j;m)$ if binary $k\in \mathrm{B_p}$ and both children of $k$ are part of $\mathrm{B_p}$. In other words, $k$ is the lowest level, or `top' binary in $\mathrm{B_p}$. Below, we refer to this binary with the `special' index $m=m(i,j)$. There is a unique binary $m$ for every path $\mathrm{B_p}$ between bodies $i$ and $j$. For example, in \F\,\ref{fig:example_quintuple}, for $\mathrm{B_p}(i,j)=\mathrm{B_p}(1,4)$, the top binary is $m(1,4)=4$.
\item Otherwise, $B_{ijk}= \alpha(i,j;k) \, M_{k.{\mathrm{C}(3-n)}}/M_k$ if $i$ and $j$ are connected to each other through child $n$ of $k$ (here, $n$ is either 1 or 2). 
\end{itemize}
To illustrate these properties, we give in Table \ref{table:example_quintuple_B} the values of $B_{ijk}$ for our example quintuple system depicted in \F\,\ref{fig:example_quintuple}, computed directly from the definition equation~(\ref{eq:B_def}).

\subsubsection{General properties of $C_{ikl}$}
\label{app:der:struc:C}
In $C_{ikl}$, the first index refers to a body, whereas the second and third indices refer to binaries. From the definition in equation~(\ref{eq:C_def}), it immediately follows that $C_{ikl} = C_{ilk}$. For a pair of binaries $\{k,l\}$, let $p$ denote the binary with the highest level, and $q$ the binary with the lowest level (i.e. binary $p$ is `below' $q$). The following properties apply.
\begin{itemize}
\item $C_{ikl} = 0$ if $k\neq l$ and binaries $k$ and $l$ have the same level.
\item $C_{ikl} = 0$ if $i$ is not a descendant of both binaries $k$ and $l$.
\item $C_{ikl} = M_{k.\mathrm{C}(3-n)}^2 / M_k^2$ if $k=l$ and $i$ is connected to $k$ through child $n$ of $k$.
\item $C_{ikl} = \alpha(i,q;p) \, \alpha(i,q;q) \, M_{p.\mathrm{C}(n-3)} M_{q.\mathrm{CS}(p)} / (M_p M_q)$ if $i$ is a descendant of $p$, and $i$ is connected to $p$ through child $n$ of $p$. The product of two sign quantities implies that $C_{ikl}$ is positive if the path from body $i$ to binary $q$ is either opposed to, or directed along {\it both} directions of $\ve{r}_p$ and $\ve{r}_q$. The sign of $C_{ikl}$ is negative if the path in from body $i$ to binary $q$ is opposed to the direction of $\ve{r}_p$ and directed along $\ve{r}_q$, or if the path in from body $i$ to binary $q$ is directed along to the direction of $\ve{r}_p$ and opposed to $\ve{r}_q$.
\end{itemize}
The values of $C_{ikl}$ for our example quintuple system depicted in \F\,\ref{fig:example_quintuple}, computed from equation~(\ref{eq:C_def}),  are given for illustration in Table \ref{table:example_quintuple_C}.

\subsection{Expansion of the potential energy}
\label{app:der:pot}
The Newtonian potential energy $V$ is given by
\begin{align}
\label{eq:V1}
V = -G \sum_{i<j} \frac{m_i m_j}{||\ve{R}_i - \ve{R}_j||}.
\end{align}
Our approach is to write $V$ in terms of the $\ve{r}_i$, and to expand it in terms of the (assumed to be) small separation ratios $x_i$ (defined more precisely below). 

Using equation~(\ref{eq:Ri_minus_Rj1}), the norm of the separation vector between two pairs of bodies can be written as
\begin{align}
\label{eq:Ri_minus_Rj2}
\nonumber ||\ve{R}_i - \ve{R}_j|| &= \left [ \left ( \ve{R}_i - \ve{R}_j \right ) \cdot \left ( \ve{R}_i - \ve{R}_j \right ) \right ]^{1/2} = \left [ \sum_{k=1}^N \sum_{l=1}^N B_{ijk} B_{ijl} \left(  \ve{r}_k\cdot \ve{r}_l \right )\right ]^{1/2} = \left [ \sum_{k\in \mathrm{B}} \sum_{l\in \mathrm{B}} B_{ijk} B_{ijl} \left( \ve{r}_k\cdot \ve{r}_l \right ) \right ]^{1/2} \\
\nonumber &= \left [ \sum_{k\in \Bm} \sum_{l\in \Bm} B_{ijk} B_{ijl} \left (\ve{r}_k\cdot \ve{r}_l\right ) + \sum_{l\in \Bm} B_{ijm} B_{ijl} \left (\ve{r}_m\cdot \ve{r}_l\right) + \sum_{k\in \Bm}  B_{ijk} B_{ijm} \left(\ve{r}_k\cdot \ve{r}_m\right) + B_{ijm} B_{ijm} \left( \ve{r}_m\cdot \ve{r}_m \right )\right ]^{1/2} \\
&= r_m \left [ \sum_{k\in \Bm} \sum_{l\in \Bm} B_{ijk} B_{ijl} \left(  \unit{r}_k\cdot \unit{r}_l \right ) \left ( \frac{r_k}{r_m} \right ) \left ( \frac{r_l}{r_m} \right ) + 2\, \alpha(i,j;m) \sum_{k\in \Bm} B_{ijk} \left ( \unit{r}_k\cdot \unit{r}_m \right ) \left ( \frac{r_k}{r_m} \right ) + 1 \right ]^{1/2}.
\end{align}
Here, we defined $\Bm \equiv \mathrm{B} \setminus \{m\}$ as the set of binaries excluding binary $m$, where $m=m(i,j)$ is the lowest level, or `top' binary in $\mathrm{B_p}(i,j)$, the path of binaries between bodies $i$ and $j$. Evidently, $\unit{r}_m\cdot \unit{r}_m=1$, and as mentioned in Appendix\,\ref{app:der:struc:B}, $B_{ijm} = \alpha(i,j;m)$, hence $B_{ijm}^2 = (\pm 1)^2=1$.

By our assumption of hierarchy, $r_m$ is the largest of the separations $r_k$ in the path $\mathrm{B_p}(i,j)$, i.e. if $k\in \mathrm{B_p}(i,j)$, then $r_k\ll r_m$ if $k\neq m$. Furthermore, $B_{ijk}$ and $\alpha(i,j;m)$ always satisfy $|B_{ijk}|\leq 1$ and $|\alpha(i,j;m)| = 1$. Therefore, for (sufficiently) hierarchical systems, it is appropriate to expand the potential in terms of $x_k\equiv r_k/r_m \ll 1$. The less hierarchical the system, the more terms of higher orders need to be included.

The expression for $||\ve{R}_i - \ve{R}_j||^{-1}$ is a function of a set of small (compared to zero) variables $\{x_n\}$, where $n \in \Bm$. The multivariate function that needs to be expanded is
\begin{align}
f(\{x_n\}) &= \frac{1}{r_m}  \left [ \sum_{k\in \Bm} \sum_{l\in \Bm} B_{ijk} B_{ijl} \left(\unit{r}_k\cdot \unit{r}_l\right ) x_k x_l + 2 \, \alpha(i,j;m) \sum_{k\in \Bm} B_{ijk} \left( \unit{r}_k\cdot \unit{r}_m\right) x_k + 1 \right ]^{-1/2}.
\end{align}
To $\mathcal{O} \left (x^5 \right)$, the general Taylor expansion of $f(\{x_n\})$ for all $x_n$ near 0 reads
\begin{align}
\nonumber f(\{x_n\}) &= f(\{x_n=0\}) + \sum_{p \in \Bm} \left. \frac{\partial f}{\partial x_p} \right |_{\{x_n\}=0} \, x_p + \frac{1}{2} \sum_{p \in \Bm} \sum_{q \in \Bm} \left. \frac{\partial^2 f}{\partial x_p \partial x_q} \right |_{\{x_n\}=0} \, x_p x_q + \frac{1}{2\cdot 3} \sum_{p \in \Bm} \sum_{q \in \Bm} \sum_{u \in \Bm} \left. \frac{\partial^3 f}{\partial x_p \partial x_q \partial x_u} \right |_{\{x_n\}=0} \, x_p x_q x_u \\
&\quad+ \frac{1}{2\cdot 3 \cdot 4} \sum_{p \in \Bm} \sum_{q \in \Bm} \sum_{u \in \Bm} \sum_{v \in \Bm} \left. \frac{\partial^4 f}{\partial x_p \partial x_q \partial x_u \partial x_v} \right |_{\{x_n\}=0} \, x_p x_q x_u x_v \\
\nonumber &\quad+ \frac{1}{2\cdot 3 \cdot 4\cdot 5} \sum_{p \in \Bm} \sum_{q \in \Bm} \sum_{u \in \Bm} \sum_{v \in \Bm} \sum_{w \in \Bm} \left. \frac{\partial^5 f}{\partial x_p \partial x_q \partial x_u \partial x_v \partial x_w} \right |_{\{x_n\}=0} \, x_p x_q x_u x_v x_w + \mathcal{O} \left (x^6 \right ). 
\end{align}
Computing the required partial derivatives of $f$, we find
\begin{align}
\label{eq:V2}
\frac{1}{||\ve{R}_i-\ve{R}_j||} &= \frac{1}{r_m} \left \{ S_0 + S_1 + S_2 + S_3 + S_4 + S_5 + \mathcal{O} \left(x^6\right) \right\},
\end{align}
where $S_n$ denote terms of order $x^n$. Up and including fifth order, they are given by
\begin{align}
\label{eq:S_0_full}
S_0 &= 1; \\
\label{eq:S_1_full}
S_1 &= - \alpha(i,j;m) \sum_{p \in \Bm} B_{ijp} \beta_1(\ve{r}_p,\ve{r}_m); \\
\label{eq:S_2_full}
S_2 &= \frac{1}{2} \sum_{p \in \Bm} \sum_{q \in \Bm} B_{ijp} B_{ijq} \beta_2(\ve{r}_p,\ve{r}_q,\ve{r}_m); \\
\label{eq:S_3_full}
S_3 &= - \frac{1}{2} \alpha(i,j;m) \sum_{p \in \Bm} \sum_{q \in \Bm} \sum_{u \in \Bm} B_{ijp} B_{ijq} B_{iju} \beta_3 (\ve{r}_p,\ve{r}_q,\ve{r}_u,\ve{r}_m) \\
\label{eq:S_4_full}
S_4 &= \frac{1}{8} \sum_{p \in \Bm} \sum_{q \in \Bm} \sum_{u \in \Bm} \sum_{v \in \Bm} B_{ijp} B_{ijq} B_{iju} B_{ijv} \beta_4 (\ve{r}_p,\ve{r}_q,\ve{r}_u,\ve{r}_v,\ve{r}_m) \\
\label{eq:S_5_full}
S_5 &= -\frac{1}{8} \alpha(i,j;m) \sum_{p \in \Bm} \sum_{q \in \Bm} \sum_{u \in \Bm} \sum_{v \in \Bm} \sum_{w \in \Bm} B_{ijp} B_{ijq} B_{iju} B_{ijv} B_{ijw} \beta_5 (\ve{r}_p,\ve{r}_q,\ve{r}_u,\ve{r}_v, \ve{r}_w, \ve{r}_m). 
\end{align}
Here, the functions $\beta_n$ are given by
\begin{align}
\label{eq:beta_1_def}
&\beta_1 (\ve{r}_p,\ve{r}_m) \equiv \frac{r_p}{r_m} \, \left ( \unit{r}_p \cdot \unit{r}_m \right ); \\
\label{eq:beta_2_def}
&\beta_2 (\ve{r}_p,\ve{r}_q,\ve{r}_m) \equiv \frac{r_p}{r_m} \frac{r_q}{r_m}
	\left [ 3 \left ( \unit{r}_p \cdot \unit{r}_m \right ) \left ( \unit{r}_q \cdot \unit{r}_m \right ) - \left ( \unit{r}_p \cdot \unit{r}_q \right ) \right ]; \\
\label{eq:beta_3_def}
&\beta_3 (\ve{r}_p,\ve{r}_q,\ve{r}_u,\ve{r}_m) \equiv \frac{r_p}{r_m} \frac{r_q}{r_m} \frac{r_u}{r_m} 
	\left [ 5 \left ( \unit{r}_p \cdot \unit{r}_m \right ) \left ( \unit{r}_q \cdot \unit{r}_m \right ) \left ( \unit{r}_u \cdot \unit{r}_m \right ) 
	- \left ( \unit{r}_p \cdot \unit{r}_m \right ) \left ( \unit{r}_q \cdot \unit{r}_u \right ) 
	- \left ( \unit{r}_q \cdot \unit{r}_m \right ) \left ( \unit{r}_p \cdot \unit{r}_u \right ) - \left ( \unit{r}_u \cdot \unit{r}_m \right ) \left ( \unit{r}_p \cdot \unit{r}_q 			\right ) \right ]; \\
\label{eq:beta_4_def}
&\nonumber \beta_4 (\ve{r}_p,\ve{r}_q,\ve{r}_u,\ve{r}_v,\ve{r}_m) \equiv \frac{r_p}{r_m} \frac{r_q}{r_m} \frac{r_u}{r_m} \frac{r_v}{r_m} 
	\left [ 35 \left ( \unit{r}_p \cdot \unit{r}_m \right ) \left ( \unit{r}_q \cdot \unit{r}_m \right ) \left ( \unit{r}_u \cdot \unit{r}_m \right ) \left ( \unit{r}_v \cdot \unit{r}_m \right ) \right. \\
	\nonumber &\quad \left. - 5 \left ( \unit{r}_p \cdot \unit{r}_m \right ) \left ( \unit{r}_q \cdot \unit{r}_m \right ) \left ( \unit{r}_u \cdot \unit{r}_v \right ) 
	- 5 \left ( \unit{r}_p \cdot \unit{r}_m \right ) \left ( \unit{r}_u \cdot \unit{r}_m \right ) \left ( \unit{r}_q \cdot \unit{r}_v \right ) 
	- 5 \left ( \unit{r}_q \cdot \unit{r}_m \right ) \left ( \unit{r}_u \cdot \unit{r}_m \right ) \left ( \unit{r}_p \cdot \unit{r}_v \right ) \right. \\
	\nonumber &\quad \left. - 5 \left ( \unit{r}_p \cdot \unit{r}_m \right ) \left ( \unit{r}_v \cdot \unit{r}_m \right ) \left ( \unit{r}_q \cdot \unit{r}_u \right ) - 5 \left ( \unit{r}_q \cdot \unit{r}_m \right ) \left ( \unit{r}_v \cdot \unit{r}_m \right ) \left ( \unit{r}_p \cdot \unit{r}_u \right ) - 5 \left ( \unit{r}_u \cdot \unit{r}_m \right ) \left ( \unit{r}_v \cdot \unit{r}_m \right ) \left ( \unit{r}_p \cdot \unit{r}_q \right ) \right. \\
	&\quad \left. + \left ( \unit{r}_p \cdot \unit{r}_v \right ) \left ( \unit{r}_q \cdot \unit{r}_u \right ) + \left ( \unit{r}_p \cdot \unit{r}_u \right ) \left ( \unit{r}_q \cdot 	\unit{r}_v \right ) + \left ( \unit{r}_p \cdot \unit{r}_q \right ) \left ( \unit{r}_u \cdot \unit{r}_v \right ) \right ];
\end{align}
\begin{align}
&\nonumber \beta_5 (\ve{r}_p,\ve{r}_q,\ve{r}_u,\ve{r}_v, \ve{r}_w, \ve{r}_m) \equiv \frac{r_p}{r_m} \frac{r_q}{r_m} \frac{r_u}{r_m} \frac{r_v}{r_m} \frac{r_w}{r_m} \left [ 63 \left ( \unit{r}_p \cdot \unit{r}_m \right ) \left ( \unit{r}_q \cdot \unit{r}_m \right ) \left ( \unit{r}_u \cdot \unit{r}_m \right ) \left ( \unit{r}_v \cdot \unit{r}_m \right ) \left ( \unit{r}_w \cdot \unit{r}_m \right ) \right. \\
\nonumber &\quad \left. - 7 \left ( \unit{r}_u \cdot \unit{r}_m \right ) \left ( \unit{r}_v \cdot \unit{r}_m \right ) \left ( \unit{r}_w \cdot \unit{r}_m \right ) \left ( \unit{r}_p \cdot \unit{r}_q \right ) - 7 \left ( \unit{r}_q \cdot \unit{r}_m \right ) \left ( \unit{r}_v \cdot \unit{r}_m \right ) \left ( \unit{r}_w \cdot \unit{r}_m \right ) \left ( \unit{r}_p \cdot \unit{r}_u \right ) - 7 \left ( \unit{r}_q \cdot \unit{r}_m \right ) \left ( \unit{r}_u \cdot \unit{r}_m \right ) \left ( \unit{r}_w \cdot \unit{r}_m \right ) \left ( \unit{r}_p \cdot \unit{r}_v \right )  \right. \\
\nonumber &\quad \left. - 7 \left ( \unit{r}_q \cdot \unit{r}_m \right ) \left ( \unit{r}_u \cdot \unit{r}_m \right ) \left ( \unit{r}_v \cdot \unit{r}_m \right ) \left ( \unit{r}_p \cdot \unit{r}_w \right ) - 7 \left ( \unit{r}_p \cdot \unit{r}_m \right ) \left ( \unit{r}_v \cdot \unit{r}_m \right ) \left ( \unit{r}_w \cdot \unit{r}_m \right ) \left ( \unit{r}_q \cdot \unit{r}_u \right ) - 7 \left ( \unit{r}_p \cdot \unit{r}_m \right ) \left ( \unit{r}_u \cdot \unit{r}_m \right ) \left ( \unit{r}_w \cdot \unit{r}_m \right ) \left ( \unit{r}_q \cdot \unit{r}_v \right ) \right. \\
\nonumber &\quad \left. - 7 \left ( \unit{r}_p \cdot \unit{r}_m \right ) \left ( \unit{r}_u \cdot \unit{r}_m \right ) \left ( \unit{r}_v \cdot \unit{r}_m \right ) \left ( \unit{r}_q \cdot \unit{r}_w \right ) - 7 \left ( \unit{r}_p \cdot \unit{r}_m \right ) \left ( \unit{r}_q \cdot \unit{r}_m \right ) \left ( \unit{r}_w \cdot \unit{r}_m \right ) \left ( \unit{r}_u \cdot \unit{r}_v \right ) - 7 \left ( \unit{r}_p \cdot \unit{r}_m \right ) \left ( \unit{r}_q \cdot \unit{r}_m \right ) \left ( \unit{r}_v \cdot \unit{r}_m \right ) \left ( \unit{r}_u \cdot \unit{r}_w \right ) \right. \\
\nonumber &\quad \left. - 7 \left ( \unit{r}_p \cdot \unit{r}_m \right ) \left ( \unit{r}_q \cdot \unit{r}_m \right ) \left ( \unit{r}_u \cdot \unit{r}_m \right ) \left ( \unit{r}_v \cdot \unit{r}_w \right ) \right. \\
\nonumber &\quad \left. + \left ( \unit{r}_w \cdot \unit{r}_m \right ) \left ( \unit{r}_p \cdot \unit{r}_q \right ) \left ( \unit{r}_u \cdot \unit{r}_v \right ) 
		+ \left ( \unit{r}_w \cdot \unit{r}_m \right ) \left ( \unit{r}_p \cdot \unit{r}_u \right ) \left ( \unit{r}_q \cdot \unit{r}_v \right ) 
		+ \left ( \unit{r}_w \cdot \unit{r}_m \right ) \left ( \unit{r}_p \cdot \unit{r}_v \right ) \left ( \unit{r}_q \cdot \unit{r}_u \right ) \right. \\
\nonumber &\quad \left. + \left ( \unit{r}_v \cdot \unit{r}_m \right ) \left ( \unit{r}_p \cdot \unit{r}_q \right ) \left ( \unit{r}_u \cdot \unit{r}_w \right ) 
		+ \left ( \unit{r}_v \cdot \unit{r}_m \right ) \left ( \unit{r}_p \cdot \unit{r}_u \right ) \left ( \unit{r}_q \cdot \unit{r}_w \right ) 
		+ \left ( \unit{r}_v \cdot \unit{r}_m \right ) \left ( \unit{r}_p \cdot \unit{r}_w \right ) \left ( \unit{r}_q \cdot \unit{r}_u \right ) \right. \\
\nonumber &\quad \left. + \left ( \unit{r}_u \cdot \unit{r}_m \right ) \left ( \unit{r}_p \cdot \unit{r}_q \right ) \left ( \unit{r}_v \cdot \unit{r}_w \right ) 
		+ \left ( \unit{r}_u \cdot \unit{r}_m \right ) \left ( \unit{r}_p \cdot \unit{r}_v \right ) \left ( \unit{r}_q \cdot \unit{r}_w \right ) 
		+ \left ( \unit{r}_u \cdot \unit{r}_m \right ) \left ( \unit{r}_p \cdot \unit{r}_w \right ) \left ( \unit{r}_q \cdot \unit{r}_v \right ) \right. \\
\nonumber &\quad \left. + \left ( \unit{r}_q \cdot \unit{r}_m \right ) \left ( \unit{r}_p \cdot \unit{r}_u \right ) \left ( \unit{r}_v \cdot \unit{r}_w \right ) 
		+ \left ( \unit{r}_q \cdot \unit{r}_m \right ) \left ( \unit{r}_p \cdot \unit{r}_v \right ) \left ( \unit{r}_u \cdot \unit{r}_w \right ) 
		+ \left ( \unit{r}_q \cdot \unit{r}_m \right ) \left ( \unit{r}_p \cdot \unit{r}_w \right ) \left ( \unit{r}_u \cdot \unit{r}_v \right ) \right. \\
 &\quad \left. + \left ( \unit{r}_p \cdot \unit{r}_m \right ) \left ( \unit{r}_q \cdot \unit{r}_u \right ) \left ( \unit{r}_v \cdot \unit{r}_w \right ) 
 		+ \left ( \unit{r}_p \cdot \unit{r}_m \right ) \left ( \unit{r}_q \cdot \unit{r}_v \right ) \left ( \unit{r}_u \cdot \unit{r}_w \right ) 
		+ \left ( \unit{r}_p \cdot \unit{r}_m \right ) \left ( \unit{r}_q \cdot \unit{r}_w \right ) \left ( \unit{r}_u \cdot \unit{r}_v \right ) \right ].
\end{align}
Generally, we formally write for order $n$,
\begin{align}
\label{eq:S_n}
S_n = (-1)^n \alpha(i,j;m)^n \sum_{p_1,... p_n \in \Bm} \underbrace{B_{ijp_1} \, ... \, B_{ijp_n}}_{n\times} c_n \beta_n ( \underbrace{ \ve{r}_{p_1}, ... \,, \ve{r}_{p_n}}_{n \times}, \ve{r}_m ),
\end{align}
where $c_n$ are constants. For pairwise terms, i.e. $p_1=...=p_n=p$, we formally write $c_n \beta_n$ as
\begin{align}
\label{eq:beta_n}
c_n \beta_n ( \underbrace{ \ve{r}_{p_1}, ...\,, \ve{r}_{p_n}}_{n \times}, \ve{r}_m) = c_n \beta_n ( \underbrace{ \ve{r}_p, ...\,, \ve{r}_p}_{n \times}, \ve{r}_m) = \sum_{j=0}^n \frac{ \left ( \ve{r}_p \cdot \ve{r}_m \right )^j r_p^{n-j} }{r_m^{n+j}} \mathcal{A}_j^{(n)},
\end{align}
where $\mathcal{A}_j^{(n)}$ are integer ratio coefficients. They are the same as the coefficients appearing in the Legendre polynomials. The latter can be
obtained by e.g. Rodrigues's formula, i.e.
\begin{align}
\label{eq:A_j}
\sum_{j=0}^n \mathcal{A}_j^{(n)} x^j = \frac{1}{2^n n!} \frac{\mathrm{d}^n }{\mathrm{d} x^n} \left [ \left (x^2 -1 \right )^n \right ].
\end{align}

Note that the terms $S_i$ in which $\unit{r}_m$ appears an odd number of times (i.e. if $n$ is odd), contain the factor $\alpha(i,j;m)$. This is to be expected: when changing the definition of $\unit{r}_m$ such that it now points in the opposite direction, i.e. $\unit{r}_m' = - \unit{r}_m$, $\alpha$ also changes sign: $\alpha(i,j;m)' = -\alpha(i,j;m)$. If $\unit{r}_m$ appears an odd number of times, the resulting even number of minus signs cancel such that the expression of the expansion in terms of $\unit{r}_m'$ is unaltered. The same property applies to the other separation vectors ($\unit{r}_p$, $\unit{r}_q$, etc.), which always appear only one time in each individual term (e.g., the term $(\unit{r}_p \cdot \unit{r}_m )^2$ never appears). When reversing the direction of these vectors, the resulting minus sign is absorbed by the mass ratio quantities ($B_{ijp}$, $B_{ijq}$, etc.), which each contain a factor $\alpha$ that also changes sign ($\alpha(i,j,p)$, $\alpha(i,j,q)$, etc.). Physically, this means that the expansion of $||\ve{R}_i-\ve{R}_j||^{-1}$ does not depend on the choice of direction of the relative vectors, as it should be.

\subsection{Derivation of the kinetic energy term}
\label{app:der:kin}
The Newtonian kinetic energy of the system is given by 
\begin{align}
\label{eq:T1}
T = \frac{1}{2} \sum_{i=1}^N m_i \left ( \dot{\ve{R}}_i \cdot \dot{\ve{R}}_i \right ),
\end{align}
where the dots denote derivates with respect to time. Using equation~(\ref{eq:r_to_R}) and assuming constant masses, this can be written in terms of time derivatives of the binary separation vectors as
\begin{align}
\label{eq:T2}
\nonumber T &= \frac{1}{2} \sum_{i=1}^N \sum_{k=1}^{N} \sum_{l=1}^{N} m_i A_{ik}^{-1} A_{il}^{-1} \left( \dot{\ve{r}}_k \cdot \dot{\ve{r}}_l \right )
= \frac{1}{2} \sum_{i=1}^N \sum_{k=1}^{N-1} \sum_{l=1}^{N-1} m_i A_{ik}^{-1} A_{il}^{-1} \left( \dot{\ve{r}}_k \cdot \dot{\ve{r}}_l \right ) =  \frac{1}{2} \sum_{i=1}^N \sum_{k\in\mathrm{B}} \sum_{l\in\mathrm{B}} m_i A_{ik}^{-1} A_{il}^{-1} \left( \dot{\ve{r}}_k \cdot \dot{\ve{r}}_l \right ) \\
&\equiv \frac{1}{2} \sum_{i=1}^N \sum_{k\in\mathrm{B}} \sum_{l\in\mathrm{B}} m_i C_{ikl} \left( \dot{\ve{r}}_k \cdot \dot{\ve{r}}_l \right),
\end{align}
where we used that the centre of mass, $\ve{r}_N$, satisfies $\dot{\ve{r}}_N=\boldsymbol{0}$, and where we defined $C_{ikl} \equiv A_{ik}^{-1} A_{il}^{-1}$ (cf. equation~\ref{eq:C_def}). Equation~(\ref{eq:T2}) can be simplified, as shown below.

The order of the summation over $i$ and $k$ and $l$ can be reversed because $\mathrm{B}$ and $i$ are independent (in contrast to $\Bm$ and $i$), i.e.
\begin{align}
\label{eq:T3}
T=\frac{1}{2} \sum_{k\in\mathrm{B}} \sum_{l\in\mathrm{B}} \left (\dot{\ve{r}}_k \cdot \dot{\ve{r}}_l\right ) \sum_{i=1}^N m_i C_{ikl}.
\end{align}
We split the double summations over binaries into several double summations over binaries, making a distinction between the relative binary levels, and into a single summation corresponding to $k=l$ in equation~(\ref{eq:T3}),
\begin{align}
\label{eq:T4}
T &= \frac{1}{2} \sum_{\substack{ p,q\in\mathrm{B} \\ p \neq q \\p.\mathrm{L}>q.\mathrm{L}} } \left (\dot{\ve{r}}_p \cdot \dot{\ve{r}}_q\right ) \sum_{i =1}^N m_i C_{ipq} 
	+ \frac{1}{2} \sum_{\substack{ p,q\in\mathrm{B} \\ p\neq q \\q.\mathrm{L}>p.\mathrm{L}} } \left (\dot{\ve{r}}_p \cdot \dot{\ve{r}}_q\right ) \sum_{i =1}^N m_i C_{ipq} 
	+ \frac{1}{2} \sum_{\substack{ p,q\in\mathrm{B}\\ p\neq q \\ p.\mathrm{L}=q.\mathrm{L}} } \left (\dot{\ve{r}}_p \cdot \dot{\ve{r}}_q\right ) \sum_{i =1}^N m_i C_{ipq} + \frac{1}{2} \sum_{ p\in\mathrm{B}} \left (\dot{\ve{r}}_p \cdot \dot{\ve{r}}_p\right ) \sum_{i =1}^N m_i C_{ipp}.
\end{align}
When $p\neq q$ and the binary levels are the same ($p.\mathrm{L}=q.\mathrm{L}$), then $C_{ipq} = 0$ (cf. Appendix\,\ref{app:der:struc:C}). Also, $C_{ipq}=0$ if $i$ is not a descendant of the highest level binary in $\{p,q\}$. Therefore, the summation over bodies $i$ from $i=1$ to $N$ can be rewritten in terms of a summation over bodies only in the highest level binary, i.e.
\begin{align}
\label{eq:T5}
\nonumber T=&\frac{1}{2} \sum_{\substack{ p,q\in\mathrm{B} \\ p\neq q \\ p.\mathrm{L}>q.\mathrm{L}} } \left (\dot{\ve{r}}_p \cdot \dot{\ve{r}}_q\right ) \left ( \sum_{i \in \{p.\mathrm{C1}\}} + \sum_{i \in \{p.\mathrm{C2}\}} \right ) m_i C_{ipq} 
	+ \frac{1}{2} \sum_{\substack{ p,q\in\mathrm{B} \\ p\neq q \\ q.\mathrm{L}>p.\mathrm{L}} } \left (\dot{\ve{r}}_p \cdot \dot{\ve{r}}_q\right ) \left ( \sum_{i \in \{q.\mathrm{C1}\}} + \sum_{i \in \{q.\mathrm{C2}\}} \right ) m_i C_{ipq} \\
&\quad + \frac{1}{2} \sum_{ p\in\mathrm{B}} \left (\dot{\ve{r}}_p \cdot \dot{\ve{r}}_p\right ) \left ( \sum_{i \in \{p.\mathrm{C1}\}} + \sum_{i \in \{p.\mathrm{C2}\}} \right ) m_i C_{ipp}.
\end{align}
We recall that $\{p.\mathrm{C}n\}$ denotes the set of all descendant bodies of child $n$ of binary $p$. Substituting the explicit expressions for $C_{ipq}$ and $C_{ipp}$ (cf. Appendix\,\ref{app:der:struc:C}), we find
\begin{align}
\label{eq:T5}
\nonumber T&=\frac{1}{2} \sum_{\substack{ p,q\in\mathrm{B} \\ p \neq q \\ \\p.\mathrm{L}>q.\mathrm{L}} } \left (\dot{\ve{r}}_p \cdot \dot{\ve{r}}_q\right ) \left ( \sum_{i \in \{p.\mathrm{C1}\}} m_i \frac{ \alpha(p.\mathrm{C1},q;p) \alpha(p.\mathrm{C1},q;q) M_{p.\mathrm{C2}} M_{q.\mathrm{CS}(p)}}{M_p M_q} + \sum_{i \in \{p.\mathrm{C2}\}} m_i \frac{  \alpha(p.\mathrm{C2},q;p) \alpha(p.\mathrm{C2},q;q) M_{p.\mathrm{C1}} M_{q.\mathrm{CS}(p)}}{M_p M_q} \right )  \\
\nonumber &\quad + \frac{1}{2} \sum_{\substack{ p,q\in\mathrm{B} \\ p \neq q \\q.\mathrm{L}>p.\mathrm{L}} } \left (\dot{\ve{r}}_p \cdot \dot{\ve{r}}_q\right ) \left ( \sum_{i \in \{q.\mathrm{C1}\}} m_i \frac{ \alpha(q.\mathrm{C1},p;q) \alpha(q.\mathrm{C1},p;p) M_{q.\mathrm{C2}} M_{p.\mathrm{CS}(q)}}{M_p M_q} + \sum_{i \in \{q.\mathrm{C2}\}} m_i \frac{ \alpha(q.\mathrm{C2},p;q) \alpha(q.\mathrm{C2},p;p) M_{q.\mathrm{C1}} M_{p.\mathrm{CS}(p)}}{M_p M_q} \right ) \\
\nonumber &\quad + \frac{1}{2} \sum_{p\in\mathrm{B}} \left (\dot{\ve{r}}_p \cdot \dot{\ve{r}}_p\right ) \left ( \sum_{i \in \{p.\mathrm{C1}\}} m_i \frac{ M_{p.\mathrm{C2}}^2}{M_p^2} + \sum_{i \in \{p.\mathrm{C2}\}} m_i \frac{ M_{p.\mathrm{C1}}^2}{M_p^2} \right ) 
\end{align}
\begin{align}
\nonumber &= \frac{1}{2} \sum_{\substack{ p,q\in\mathrm{B} \\ p \neq q \\ \\p.\mathrm{L}>q.\mathrm{L}} } \left (\dot{\ve{r}}_p \cdot \dot{\ve{r}}_q\right ) \left ( M_{p.\mathrm{C1}} \frac{ \alpha(p.\mathrm{C1},q;p) \alpha(p.\mathrm{C1},q;q) M_{p.\mathrm{C2}} M_{q.\mathrm{CS}(p)}}{M_p M_q} + M_{p.\mathrm{C}2} \frac{ -\alpha(p.\mathrm{C1},q;p) \alpha(p.\mathrm{C1},q;q) M_{p.\mathrm{C1}} M_{q.\mathrm{CS}(p)}}{M_p M_q} \right )  \\
\nonumber &\quad + \frac{1}{2} \sum_{\substack{ p,q\in\mathrm{B} \\ p \neq q \\q.\mathrm{L}>p.\mathrm{L}} } \left (\dot{\ve{r}}_p \cdot \dot{\ve{r}}_q\right ) \left ( M_{q.\mathrm{C1}} \frac{ \alpha(q.\mathrm{C1},p;q) \alpha(q.\mathrm{C1},p;p) M_{q.\mathrm{C2}} M_{p.\mathrm{CS}(q)}}{M_p M_q} + M_{q.\mathrm{C2}} \frac{ -\alpha(q.\mathrm{C1},p;q) \alpha(q.\mathrm{C1},p;p) M_{q.\mathrm{C1}} M_{p.\mathrm{CS}(p)}}{M_p M_q} \right ) \\
\nonumber &\quad +  \frac{1}{2} \sum_{p\in\mathrm{B}} \left (\dot{\ve{r}}_p \cdot \dot{\ve{r}}_p\right ) \left ( \frac{  M_{p.\mathrm{C1}} M_{p.\mathrm{C2}}^2}{M_p^2} + \frac{ M_{p.\mathrm{C2}} M_{p.\mathrm{C1}}^2}{M_p^2} \right ) \\
&= \frac{1}{2} \sum_{p\in\mathrm{B}} \frac{  M_{p.\mathrm{C1}} M_{p.\mathrm{C2}}}{M_p} \left (\dot{\ve{r}}_p \cdot \dot{\ve{r}}_p\right ).
\end{align}
Here, we used that $\alpha(p.\mathrm{C1},q;q) = \alpha(p.\mathrm{C2},q;q)$ and $\alpha(p.\mathrm{C1},q;p) = -\alpha(p.\mathrm{C2},q;p)$ in the summation with $p.\mathrm{L}>q.\mathrm{L}$. Similarly, in the summation with $p.\mathrm{L}<q.\mathrm{L}$, $\alpha(q.\mathrm{C1},p;p) = \alpha(q.\mathrm{C2},p;p)$ and $\alpha(q.\mathrm{C1},p;q) = -\alpha(q.\mathrm{C2},p;q)$. The expression in equation~(\ref{eq:T5}) shows that the total kinetic energy is just the sum of the kinetic energies of all the individual binaries; all terms depending on velocities of pairs of binaries cancel.

\subsection{Rewriting summations in the Hamiltonian}
\label{app:der:ham}
The Hamiltonian $H$ is given by the sum of the kinetic and potential energies, $H=T+V$. Equations~(\ref{eq:V1}), (\ref{eq:V2}) and (\ref{eq:T5}) give the following expression for the Hamiltonian accurate to fifth order in ratios of the binary separations,
\begin{align}
\label{eq:H1}
\nonumber H &= \frac{1}{2} \sum_{p\in\mathrm{B}} \frac{  M_{p.\mathrm{C1}} M_{p.\mathrm{C2}}}{M_p} \left (\dot{\ve{r}}_p \cdot \dot{\ve{r}}_p\right ) -  \sum_{i<j} \frac{Gm_im_j}{r_{m(i,j)}} \left \{ S_0 +  S_1 + S_2 + S_3 + S_4 + S_5 + \mathcal{O} \left(x^6\right) \right\} \\
&\equiv  \frac{1}{2} \sum_{p\in\mathrm{B}} \frac{  M_{p.\mathrm{C1}} M_{p.\mathrm{C2}}}{M_p} \left (\dot{\ve{r}}_p \cdot \dot{\ve{r}}_p\right ) + S_0' + S_1'+S_2'+S_3'+S_4'+S_5' + \mathcal{O} \left(x^6\right),
\end{align}
where the primed quantities $S_n'$ are defined as
\begin{align}
S_n' \equiv -\sum_{i<j} \frac{Gm_im_j}{r_{m}} S_n.
\label{eq:S_n_prime}
\end{align}

The terms $S_n'$ in equation~(\ref{eq:H1}) can be simplified substantially. A key ingredient in the simplifications is to sum over binaries and their children, rather than explicitly over all pairs of bodies in the first summation of equation~(\ref{eq:H1}), and to apply the properties of the mass ratio quantities $B_{ijk}$. Below we carry out these simplifications order by order.

\subsubsection{Monopole-order term}
\label{app:der:ham:S0}
The `monopole-order' term $S_0'$ can be simplified by summing over all binaries and, subsequently, the children within those binaries, rather than summing explicitly over all pairs of bodies, i.e.
\begin{align}
\label{eq:Sm1}
\sum_{i<j} \frac{Gm_im_j}{r_{m}} = \sum_{p\in \mathrm{B}} \sum_{i \in \{p.\mathrm{C1}\}} \sum_{j \in \{p.\mathrm{C2}\}} \frac{Gm_i m_j}{r_m} = \sum_{p\in \mathrm{B}} \sum_{i \in \{p.\mathrm{C1}\}} \sum_{j \in \{p.\mathrm{C2}\}} \frac{Gm_i m_j}{r_p}.
\end{align}
The last step in equation~(\ref{eq:Sm1}) follows from the constancy of the separation $r_m$ for fixed $p$ and varying $i$ and $j$, and which is given by $r_p$. Therefore,
\begin{align}
\label{eq:Sm2}
-S_0' \equiv \sum_{i<j} \frac{Gm_im_j}{r_{m}} = \sum_{p\in \mathrm{B}} \frac{G}{r_p} \sum_{i \in \{p.\mathrm{C1}\}} m_i \sum_{j \in \{p.\mathrm{C2}\}} m_j = \sum_{p\in \mathrm{B}} \frac{G}{r_p} \sum_{i \in \{p.\mathrm{C1}\}} m_i M_{p.\mathrm{C2}} = \sum_{p\in \mathrm{B}} \frac{G}{r_p} M_{p.\mathrm{C2}} \sum_{i \in \{p.\mathrm{C1}\}} m_i = \sum_{p\in \mathrm{B}} \frac{GM_{p.\mathrm{C1}} M_{p.\mathrm{C2}}}{r_p}.
\end{align}

\begin{figure}
\center
\includegraphics[scale = 0.47, trim = 0mm 15mm 0mm 10mm]{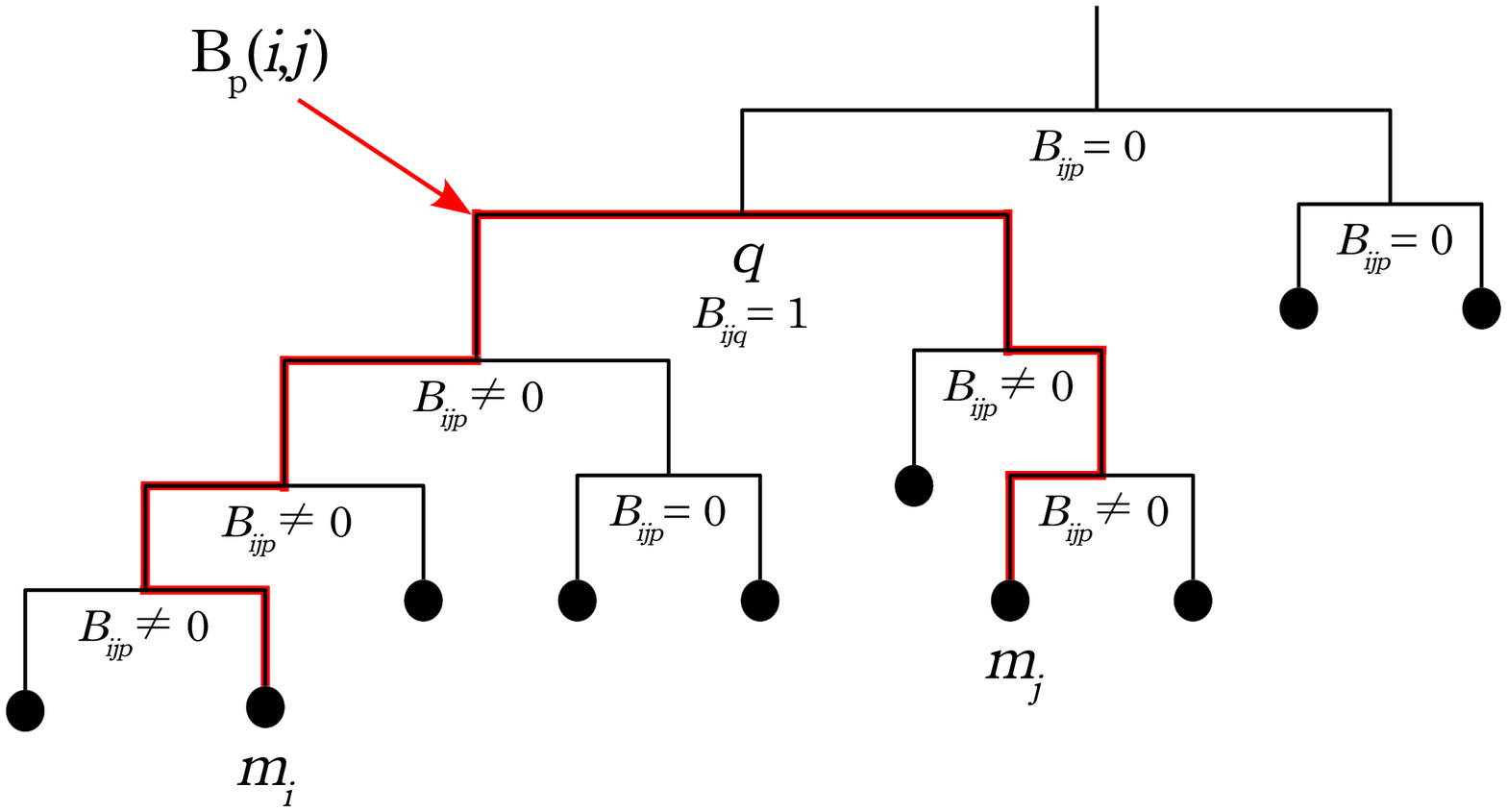}
\caption{\small A mobile diagram with a particular path $\mathrm{B_p}(i,j)$, and a `top' level binary $q$ (cf. Appendix\,\ref{app:der:ham:S1}). For each binary, we indicate whether or not $B_{ijp}$ is zero. }
\label{fig:mobile_app_bijp}
\end{figure}

\subsubsection{Dipole-order term}
\label{app:der:ham:S1}
Here, we show that the `dipole-order' term $S_1'$ (cf. equation~\ref{eq:S_1_full}) vanishes identically for any system. Using equation~(\ref{eq:Sm1}) to rewrite the summations of pairs of bodies, $S_1'$ reads
\begin{align}
\nonumber S_1' &\equiv \sum_{i<j} \frac{Gm_im_j}{r_m} \alpha(i,j;m) \sum_{p \in \Bm} B_{ijp} \beta_1(\ve{r}_p,\ve{r}_m) = \sum_{q\in \mathrm{B}} \sum_{i \in \{q.\mathrm{C1} \}} \sum_{j \in \{q.\mathrm{C2} \}} \frac{Gm_im_j}{r_q} \alpha(i,j;q) \sum_{p \in \Bm} B_{ijp}  \beta_1(\ve{r}_p,\ve{r}_q) \\
&=\sum_{q\in \mathrm{B}} \sum_{i \in \{q.\mathrm{C1} \}} \sum_{j \in \{q.\mathrm{C2} \}} \sum_{p \in \Bm} B_{ijp} \frac{Gm_im_j}{r_q} \alpha(i,j;q)  \beta_1(\ve{r}_p,\ve{r}_q),
\end{align}
where in the last step we used that $m_i$, $m_j$ and $r_q$ are independent of $p$. We recall that $\Bm \equiv \mathrm{B} \setminus \{m\} = \mathrm{B} \setminus \{q\}$ depends on $i$ and $j$. Therefore, $p$ runs over all binaries except $q$. Whenever $p$ is not part of the descendants of $q$, it is also not  within the path $\mathrm{B_p}(i,j)$ because $i$ and $j$ run over the children of $q$. Therefore, $B_{ijp}=0$ in that case (cf. Appendix\,\ref{app:der:struc:B}, see also \F\,\ref{fig:mobile_app_bijp}). From this it follows that the summation of $p$ over $\Bm$ can be rewritten as a summation of $p$ over all binaries $\mathrm{B}$, where $p$ is a descendant of $q$, i.e.
\begin{align}
S_1' = \sum_{q\in \mathrm{B}} \sum_{i \in \{q.\mathrm{C1} \}} \sum_{j \in \{q.\mathrm{C2} \}} \sum_{\substack{ p \in \mathrm{B} \\ p\in \{q.\mathrm{C} \}}} B_{ijp} \frac{Gm_im_j}{r_q} \alpha(i,j;q) \beta_1(\ve{r}_p,\ve{r}_q).
\end{align}
The summation over $p$ is now independent of $i$ and $j$, and can therefore be placed before the summations over $i$ and $j$. Also placing other quantities independent of $i$ and $j$ before the summations over $i$ and $j$, we find
\begin{align}
\nonumber S_1' &= \sum_{q\in \mathrm{B}}  \sum_{\substack{ p \in \mathrm{B} \\ p\in \{q.\mathrm{C} \}}} \frac{G}{r_q} \beta_1(\ve{r}_p,\ve{r}_q) \sum_{i \in \{q.\mathrm{C1} \}} \sum_{j \in \{q.\mathrm{C2} \}} \alpha(i,j;q) B_{ijp} m_i m_j \\
\nonumber &= \sum_{q\in \mathrm{B}}  \left ( \sum_{\substack{ p \in \mathrm{B} \\ p\in \{q.\mathrm{C1} \}}} + \sum_{\substack{ p \in \mathrm{B} \\ p\in \{q.\mathrm{C2} \}}} \right ) \frac{G}{r_q} \beta_1(\ve{r}_p,\ve{r}_q) \sum_{i \in \{q.\mathrm{C1} \}} \sum_{j \in \{q.\mathrm{C2} \}} \alpha(i,j;q) B_{ijp} m_i m_j .
\end{align}
Here, we explicitly wrote the summations of $p$ over the children of $q$ into summations over child 1 and 2 of $q$. If $p$ sums over $\{q.\mathrm{C1} \}$, then $B_{ijp}=0$ unless $i\in \{q.\mathrm{C1}\}$ is a descendant of $p$ (cf. Appendix\,\ref{app:der:struc:B}). Conversely, if $p$ sums over $\{q.\mathrm{C2} \}$, then $B_{ijp}=0$ unless $j\in \{q.\mathrm{C2}\}$ is a descendant of $p$. Therefore, either summations of $i$ and $j$ over the children of $q$ can be rewritten as summations over the descendants of binary $p$, i.e.
\begin{align}
\nonumber S_1' &= \sum_{q\in \mathrm{B}}  \sum_{\substack{ p \in \mathrm{B} \\ p\in \{q.\mathrm{C1} \}}}  \frac{G}{r_q} \beta_1(\ve{r}_p,\ve{r}_q) \sum_{i \in \{p.\mathrm{C} \}} \sum_{j \in \{q.\mathrm{C2} \}} \alpha(i,j;q) B_{ijp} m_i m_j  +  \sum_{q\in \mathrm{B}}  \sum_{\substack{ p \in \mathrm{B} \\ p\in \{q.\mathrm{C2} \}}}  \frac{G}{r_q} \beta_1(\ve{r}_p,\ve{r}_q) \sum_{i \in \{q.\mathrm{C1} \}} \sum_{j \in \{p.\mathrm{C} \}} \alpha(i,j;q) B_{ijp} m_i m_j \\
\nonumber &= \sum_{q\in \mathrm{B}}  \sum_{\substack{ p \in \mathrm{B} \\ p\in \{q.\mathrm{C1} \}}}  \frac{G}{r_q} \beta_1(\ve{r}_p,\ve{r}_q) \sum_{j \in \{q.\mathrm{C2} \}} m_j \sum_{i \in \{p.\mathrm{C} \}} \alpha(i,j;q) B_{ijp} m_i + \sum_{q\in \mathrm{B}}  \sum_{\substack{ p \in \mathrm{B} \\ p\in \{q.\mathrm{C2} \}}} \frac{G}{r_q} \beta_1(\ve{r}_p,\ve{r}_q) \sum_{i \in \{q.\mathrm{C1} \}} m_i \sum_{j \in \{p.\mathrm{C} \}} \alpha(i,j;q) B_{ijp} m_j \\
\nonumber &= \sum_{q\in \mathrm{B}}  \sum_{\substack{ p \in \mathrm{B} \\ p\in \{q.\mathrm{C1} \}}}  \frac{G}{r_q} \beta_1(\ve{r}_p,\ve{r}_q) \sum_{j \in \{q.\mathrm{C2} \}} m_j  \, \alpha(p,j;q) \sum_{i \in \{p.\mathrm{C} \}} B_{ijp} m_i + \sum_{q\in \mathrm{B}}  \sum_{\substack{ p \in \mathrm{B} \\ p\in \{q.\mathrm{C2} \}}}  \frac{G}{r_q} \beta_1(\ve{r}_p,\ve{r}_q) \sum_{i \in \{q.\mathrm{C1} \}} m_i \, \alpha(i,p;q) \sum_{j \in \{p.\mathrm{C} \}} B_{ijp} m_j.
\end{align}
In the last line, we used that in the terms where the outermost summation runs over $i$, $i$ runs over the descendants of $p$, therefore $\alpha(i,j;q)$ is equivalent to $\alpha(p,j;q)$; the same applies when the outermost summation runs over $j$, in which case $\alpha(i,j;q)$ is equivalent to $\alpha(i,p;q)$. Focusing on these outermost summations over children of $p$, we split the outermost summation over the children of $p$; using the properties of $B_{ijp}$ (cf. Appendix\,\ref{app:der:struc:B}), we find
\begin{align}
\label{eq:sum_vanish}
\nonumber \sum_{i \in \{p.\mathrm{C} \}} B_{ijp} m_i &= \sum_{i \in \{p.\mathrm{C1} \}} B_{ijp} m_i + \sum_{i \in \{p.\mathrm{C2} \}} B_{ijp} m_i = \sum_{i \in \{p.\mathrm{C1} \}} m_i \frac{\alpha(p.\mathrm{C1},q.\mathrm{C2};p) M_{p.\mathrm{C2}}}{M_p} + \sum_{i \in \{p.\mathrm{C2} \}} m_i \frac{\alpha(p.\mathrm{C2},q.\mathrm{C2};p) M_{p.\mathrm{C1}}}{M_p} \\
\nonumber &= \frac{\alpha(p.\mathrm{C1},q.\mathrm{C2};p) M_{p.\mathrm{C2}}}{M_p}  \sum_{i \in \{p.\mathrm{C1} \}} m_i + \frac{-\alpha(p.\mathrm{C1},q.\mathrm{C2};p) M_{p.\mathrm{C1}}}{M_p}  \sum_{i \in \{p.\mathrm{C2} \}} m_i \\
&= \frac{\alpha(p.\mathrm{C1},q.\mathrm{C2};p) M_{p.\mathrm{C2}}}{M_p} M_{p.\mathrm{C1}} - \frac{\alpha(p.\mathrm{C1},q.\mathrm{C2};p) M_{p.\mathrm{C1}}}{M_p}  M_{p.\mathrm{C2}} = 0,
\end{align}
where we used that $\alpha(p.\mathrm{C1},q.\mathrm{C2};p) = -\alpha(p.\mathrm{C2},q.\mathrm{C2};p)$. By the same arguments,
\begin{align}
\sum_{j \in \{p.\mathrm{C} \}} B_{ijp} m_j = 0.
\end{align}
This shows that $S_1'=0$.

\subsubsection{Quadrupole-order term}
\label{app:der:ham:S2}
In the summations of the `quadrupole-order' term $S_2'$ (cf. equation~\ref{eq:S_2_full}), the individual terms contain, in principle, three different binaries. We show that only terms with binary pairs remain, significantly simplifying the expression. Using equation~(\ref{eq:Sm1}) to rewrite the summations of pairs of bodies, and using similar arguments as in Appendix\,\ref{app:der:ham:S1}, we find
\begin{align}
\nonumber -S_2' &= \sum_{i<j} \frac{Gm_im_j}{r_m} \frac{1}{2} \sum_{p \in \Bm} \sum_{q \in \Bm} B_{ijp} B_{ijq} \beta_2(\ve{r}_p,\ve{r}_q,\ve{r}_m) = \sum_{k \in \mathrm{B}} \sum_{i \in \{k.\mathrm{C1} \}} \sum_{j \in \{k.\mathrm{C2} \}} \frac{Gm_im_j}{r_k} \frac{1}{2} \sum_{p \in \Bm} \sum_{q \in \Bm} B_{ijp} B_{ijq} \beta_2(\ve{r}_p,\ve{r}_q,\ve{r}_k) \\
\nonumber &= \frac{1}{2} \sum_{k \in \mathrm{B}} \sum_{i \in \{k.\mathrm{C1} \}} \sum_{j \in \{k.\mathrm{C2} \}} \sum_{p \in \Bm} \sum_{q \in \Bm} \frac{Gm_im_j}{r_k} B_{ijp} B_{ijq} \beta_2(\ve{r}_p,\ve{r}_q,\ve{r}_k) \\
\nonumber &= \frac{1}{2} \sum_{k \in \mathrm{B}} \sum_{i \in \{k.\mathrm{C1} \}} \sum_{j \in \{k.\mathrm{C2} \}} \sum_{\substack{p \in \mathrm{B} \\ p \in \{k.\mathrm{C} \} }} \sum_{\substack{q \in \mathrm{B} \\ q \in \{k.\mathrm{C} \} }} \frac{Gm_im_j}{r_k} B_{ijp} B_{ijq} \beta_2(\ve{r}_p,\ve{r}_q,\ve{r}_k) \\
\nonumber &= \frac{1}{2} \sum_{k \in \mathrm{B}}  \sum_{\substack{p \in \mathrm{B} \\ p \in \{k.\mathrm{C} \} }} \sum_{\substack{q \in \mathrm{B} \\ q \in \{k.\mathrm{C} \} }} \frac{G}{r_k}  \beta_2(\ve{r}_p,\ve{r}_q,\ve{r}_k) \sum_{i \in \{k.\mathrm{C1} \}} \sum_{j \in \{k.\mathrm{C2} \}} m_i m_j  B_{ijp} B_{ijq} \\
\nonumber &= \underbrace{ \frac{1}{2} \sum_{k \in \mathrm{B}}  \sum_{\substack{p \in \mathrm{B} \\ p \in \{k.\mathrm{C1} \} }} \sum_{\substack{q \in \mathrm{B} \\ q \in \{k.\mathrm{C} \} }} \frac{G}{r_k}  \beta_2(\ve{r}_p,\ve{r}_q,\ve{r}_k) \sum_{j \in \{k.\mathrm{C2} \}} m_j \sum_{i \in \{p.\mathrm{C} \}} m_i  B_{ijp} B_{ijq}}_{\equiv R_1} \\
&\quad + \underbrace{ \frac{1}{2} \sum_{k \in \mathrm{B}}  \sum_{\substack{p \in \mathrm{B} \\ p \in \{k.\mathrm{C2} \} }} \sum_{\substack{q \in \mathrm{B} \\ q \in \{k.\mathrm{C} \} }} \frac{G}{r_k}  \beta_2(\ve{r}_p,\ve{r}_q,\ve{r}_k) \sum_{i \in \{k.\mathrm{C1} \}} m_i \sum_{j \in \{p.\mathrm{C} \}} m_j  B_{ijp} B_{ijq}}_{\equiv R_2}.
\end{align}
In the last line, we split the summation of $p \in \{k.\mathrm{C}\}$ into separate summations over child $n$ of $k$, and we denote the corresponding terms with $R_n$. Alternatively, because the expression is fully symmetric with respect to $p$ and $q$ (note that $\beta_2(\ve{r}_p,\ve{r}_q,\ve{r}_k) = \beta_2(\ve{r}_q,\ve{r}_p,\ve{r}_k)$), the split can also be applied to the summation of $q \in \{k.\mathrm{C}\}$.

The following arguments apply similarly to $R_1$ and $R_2$; here, we specify to $R_1$. $B_{ijq}=0$ if $q \notin \mathrm{B_p}(i,j)$. Because $i \in \{p.\mathrm{C}\}$, this implies that for nonzero terms, 
\begin{enumerate}
\item $q=p$ or
\item $q\neq p$ with
\begin{enumerate}
\item $p\in \{q.\mathrm{C}\}$,
\item $q\in \{p.\mathrm{C} \}$ or
\item $q \in \{k.\mathrm{C2} \}$.
\end{enumerate}
\end{enumerate}
We address these cases below individually, starting with the case $q\neq p$ first.

\paragraph*{(ii) (a) $q\neq p$ and $p\in \{q.\mathrm{C}\}$.} In this case, $B_{ijq}$ in the summation of $i \in \{p.\mathrm{C}\}$ is constant, and is given by
\begin{align}
B_{ijq} = \alpha(p,j;q) \frac{M_{q.\mathrm{CS}(p)}}{M_q}.
\end{align}
Therefore, $B_{ijq}$ can be taken outside of the summation of $i \in \{p.\mathrm{C}\}$, i.e.
\begin{align}
\sum_{i \in \{p.\mathrm{C} \}} m_i  B_{ijp} B_{ijq} = B_{ijq} \sum_{i \in \{p.\mathrm{C} \}} m_i B_{ijp} = 0,
\end{align}
where the last property was shown explicitly in equation~(\ref{eq:sum_vanish}). In other words, the terms associated with case (ii) (a) cancel. 

\paragraph*{(ii) (b) $q\neq p$ and $q\in \{p.\mathrm{C}\}$.}
If $q\in \{p.\mathrm{C}\}$, then $B_{ijq} \neq 0$ only if $i \in \{q.\mathrm{C}\}$. Furthermore, $B_{ijp}$ is constant, giving
\begin{align}
\sum_{i \in \{p.\mathrm{C} \}} m_i  B_{ijp} B_{ijq} = B_{ijp} \sum_{i \in \{q.\mathrm{C} \}} m_i B_{ijq} = 0,
\end{align}
where we again used equation~(\ref{eq:sum_vanish}). Therefore, the terms associated with case (ii) (b) cancel as well.

\paragraph*{(ii) (c) $q\neq p$ and $q \in \{k.\mathrm{C2} \}$.}
Note that in the case of $R_1$, $j \in \{k.\mathrm{C2}\}$. If $q \in \{k.\mathrm{C2} \}$, then $B_{ijq} \neq 0$ only if $j \in \{q.\mathrm{C}\}$. Furthermore, $B_{ijp}$ is constant when summing over $j$, whereas $B_{ijq}$ is constant when summing over $i \in \{p.\mathrm{C}\}$. Therefore,
\begin{align}
 \sum_{j \in \{k.\mathrm{C2} \}} m_j \sum_{i \in \{p.\mathrm{C} \}} m_i  B_{ijp} B_{ijq} = \left ( \sum_{i \in \{p.\mathrm{C} \}} m_i B_{ijp} \right ) \left ( \sum_{j \in \{q.\mathrm{C} \}} m_j B_{ijq} \right ) = 0,
\end{align}
showing that terms associated with case (ii) (c) are zero. 

\paragraph*{(i) $q=p$.}
In the case $q=p$, the associated terms do not cancel. The double summation over $p$ and $q$ reduces to a single summation and $B_{ijq} = B_{ijp}$, hence
\begin{align}
R_1 = \frac{1}{2} \sum_{k \in \mathrm{B}}  \sum_{\substack{p \in \mathrm{B} \\ p \in \{k.\mathrm{C1} \} }} \frac{G}{r_k}  \beta_2(\ve{r}_p,\ve{r}_p,\ve{r}_k)  \sum_{j \in \{k.\mathrm{C2} \}} m_j \sum_{i \in \{p.\mathrm{C} \}} m_i  B_{ijp}^2.
\end{align}
The outermost summation of $i \in \{p.\mathrm{C} \}$ does not cancel, and evaluates to
\begin{align}
\label{eq:sum_sq}
\nonumber \sum_{i \in \{p.\mathrm{C} \}} m_i  B_{ijp}^2 &= \left ( \sum_{i \in \{p.\mathrm{C1} \}} + \sum_{i \in \{p.\mathrm{C2} \}} \right ) m_i  B_{ijp}^2 = \frac{[\alpha(p.\mathrm{C1},k.\mathrm{C2};p)]^2 M^2_{p.\mathrm{C2}}}{M_p^2}  \sum_{i \in \{p.\mathrm{C1} \}} m_i + \frac{[\alpha(p.\mathrm{C2},k.\mathrm{C2};p)]^2 M^2_{p.\mathrm{C1}}}{M_p^2}  \sum_{i \in \{p.\mathrm{C2} \}} m_i \\
&= \frac{M_{p.\mathrm{C2}}^2 M_{p.\mathrm{C1}} + M_{p.\mathrm{C1}}^2 M_{p.\mathrm{C2}}}{M_p^2} = \frac{M_{p.\mathrm{C1}} M_{p.\mathrm{C2}}}{M_p}.
\end{align}
The latter expression is independent of $j$ and can therefore be taken in front of the summation of $j \in \{k.\mathrm{C2}\}$. 
Therefore,
\begin{align}
\nonumber R_1 &= \frac{1}{2} \sum_{k \in \mathrm{B}}  \sum_{\substack{p \in \mathrm{B} \\ p \in \{k.\mathrm{C1} \} }}  \frac{M_{p.\mathrm{C1}} M_{p.\mathrm{C2}}}{M_p} \frac{G}{r_k}  \beta_2(\ve{r}_p,\ve{r}_p,\ve{r}_k)  \sum_{j \in \{k.\mathrm{C2} \}} m_j = \frac{1}{2} \sum_{k \in \mathrm{B}}  \sum_{\substack{p \in \mathrm{B} \\ p \in \{k.\mathrm{C1} \} }}  \frac{M_{p.\mathrm{C1}} M_{p.\mathrm{C2}}}{M_p} \frac{GM_{k.\mathrm{C2}}}{r_k}  \beta_2(\ve{r}_p,\ve{r}_p,\ve{r}_k) \\
&= \frac{1}{2} \sum_{k \in \mathrm{B}}  \sum_{\substack{p \in \mathrm{B} \\ p \in \{k.\mathrm{C1} \} }}  \frac{M_{p.\mathrm{C1}} M_{p.\mathrm{C2}}}{M_p} \frac{GM_{k.\mathrm{CS}(p)}}{r_k} \beta_2(\ve{r}_p,\ve{r}_p,\ve{r}_k),
\end{align}
where, as before, we used the notation $M_{k.\mathrm{CS}(p)}$ to denote the mass of the child of $k$ that is the sibling of the child of $k$ that is connected to $p$ (cf. Appendix\,\ref{app:der:struc}). 

The same arguments apply to $R_2$, in which case $p \in \{k.\mathrm{C2}\}$. The result is
\begin{align}
R_2 = \frac{1}{2} \sum_{k \in \mathrm{B}}  \sum_{\substack{p \in \mathrm{B} \\ p \in \{k.\mathrm{C2} \} }}  \frac{M_{p.\mathrm{C1}} M_{p.\mathrm{C2}}}{M_p} \frac{GM_{k.\mathrm{CS}(p)}}{r_k} \beta_2(\ve{r}_p,\ve{r}_p,\ve{r}_k).
\end{align}

Adding $R_1$ and $R_2$, writing the separate summations of $p\in \{k.\mathrm{C1}\}$ and $p\in \{k.\mathrm{C2}\}$ again as a single summation of $p\in \{k.\mathrm{C}\}$, and substituting the explicit expression for $\beta_2$ (cf. equation~\ref{eq:beta_2_def}), we find the final simplified expression for the quadrupole-order term,
\begin{align}
\label{eq:S_2_der}
-S_2' = R_1+R_2 = \frac{1}{2} \sum_{k \in \mathrm{B}}  \sum_{\substack{p \in \mathrm{B} \\ p \in \{k.\mathrm{C} \} }}  \frac{M_{p.\mathrm{C1}} M_{p.\mathrm{C2}}}{M_p} \frac{GM_{k.\mathrm{CS}(p)}}{r_k} \left ( \frac{r_p}{r_k} \right )^2 \left [ 3 \left ( \unit{r}_p \cdot \unit{r}_k \right )^2 - 1 \right].
\end{align}

Therefore, the quadrupole-order term can be written as a double sum over binaries, where one of the binaries is a descendant of the other. As a consequence of the topology of the system and the resulting properties of the quantity $B_{ijk}$, all the terms depending on three different binaries cancel. Only terms with $p=q$ survive, and the resulting quadrupole-order terms individually only depend on {\it two} binary separation vectors. 

The mathematical form of each of these terms is identical to the quadrupole-order term in the hierarchical three-body system. In particular, the `angular part' $(1/2)[3 (\unit{r}_p \cdot \unit{r}_k )^2 - 1]$ in equation~(\ref{eq:S_2_der}) is the same as the second Legendre polynomial $P_2(x) = (1/2)(3x^2-1)$ with $x = \cos(i_{pk})$ the cosine of the angle between the `inner' and `outer' separation vectors $\ve{r}_p$ and $\ve{r}_k$, respectively, that appears in the hierarchical three-body Hamiltonian \citep{1968AJ.....73..190H}. In other words, at quadrupole order, the Hamiltonian of the hierarchical $N$-body system can be constructed from the three-body Hamiltonian.

\subsubsection{Octupole-order term}
\label{app:der:ham:S3}
For the `octupole-order' term $S_3'$ (cf. equation~\ref{eq:S_3_full}), the individual terms contain, in principle, four different binaries. After rewriting, we find that terms with binary pairs and triplets remain; terms depending on quadlets cancel. Following similar steps as in Appendix\,\ref{app:der:ham:S2},
\begin{align}
\nonumber S_3' &= \sum_{i<j} \frac{Gm_im_j}{r_m} \frac{1}{2} \alpha(i,j;m) \sum_{p \in \Bm} \sum_{q \in \Bm}  \sum_{u \in \Bm} B_{ijp} B_{ijq} B_{iju} \beta_3(\ve{r}_p,\ve{r}_q,\ve{r}_u,\ve{r}_m) \\
\nonumber &= \sum_{k \in \mathrm{B}} \sum_{i \in \{k.\mathrm{C1} \}} \sum_{j \in \{k.\mathrm{C2} \}} \frac{Gm_im_j}{r_k}  \frac{1}{2} \alpha(i,j;k) \sum_{\substack{p \in \mathrm{B} \\ p\in\{k.\mathrm{C} \}}} \sum_{\substack{q \in \mathrm{B} \\ q\in\{k.\mathrm{C} \}}}  \sum_{\substack{u \in \mathrm{B} \\ u\in\{k.\mathrm{C} \}}} B_{ijp} B_{ijq} B_{iju} \beta_3(\ve{r}_p,\ve{r}_q,\ve{r}_u,\ve{r}_k) \\
\nonumber &= \frac{1}{2} \sum_{k \in \mathrm{B}} \sum_{\substack{p \in \mathrm{B} \\ p\in\{k.\mathrm{C} \}}} \sum_{\substack{q \in \mathrm{B} \\ q\in\{k.\mathrm{C} \}}}  \sum_{\substack{u \in \mathrm{B} \\ u\in\{k.\mathrm{C} \}}} \frac{G}{r_k} \beta_3(\ve{r}_p,\ve{r}_q,\ve{r}_u,\ve{r}_k) \sum_{i \in \{k.\mathrm{C1} \}} \sum_{j \in \{k.\mathrm{C2} \}} \alpha(i,j;k) \, m_i m_j B_{ijp} B_{ijq} B_{iju}. 
\end{align}
As before, we split the summation of $p\in \{k.\mathrm{C}\}$ into two summations, $R_1$ and $R_2$, over the children of $k$ ($q$ and $u$ being equally valid choices). In the case of $R_1$,
\begin{align}
\nonumber R_1 &= \frac{1}{2} \sum_{k \in \mathrm{B}} \sum_{\substack{p \in \mathrm{B} \\ p\in\{k.\mathrm{C1} \}}} \sum_{\substack{q \in \mathrm{B} \\ q\in\{k.\mathrm{C} \}}}  \sum_{\substack{u \in \mathrm{B} \\ u\in\{k.\mathrm{C} \}}} \frac{G}{r_k} \beta_3(\ve{r}_p,\ve{r}_q,\ve{r}_u,\ve{r}_k) \sum_{i \in \{p.\mathrm{C} \}} \sum_{j \in \{k.\mathrm{C2} \}} \alpha(i,j;k) \, m_i m_j B_{ijp} B_{ijq} B_{iju} \\
&= \frac{1}{2} \sum_{k \in \mathrm{B}} \sum_{\substack{p \in \mathrm{B} \\ p\in\{k.\mathrm{C1} \}}} \sum_{\substack{q \in \mathrm{B} \\ q\in\{k.\mathrm{C} \}}}  \sum_{\substack{u \in \mathrm{B} \\ u\in\{k.\mathrm{C} \}}} \frac{G}{r_k} \beta_3(\ve{r}_p,\ve{r}_q,\ve{r}_u,\ve{r}_k) \sum_{j \in \{k.\mathrm{C2} \}} m_j \alpha(p,j;k)\sum_{i \in \{p.\mathrm{C} \}} m_i B_{ijp} B_{ijq} B_{iju},
\end{align}
where we used that $\alpha(i,j;k) = \alpha(p,j;k)$ is constant in the summation of $i \in \{p.\mathrm{C} \}$. The triple summations of $p$, $q$ and $u$ can be divided into the following five main cases, with several subcases. The subcases are those for which the corresponding $B_{ijp}$, $B_{ijq}$, etc. are nonzero. Specifying to $R_1$ (i.e. $p \in \{k.\mathrm{C1} \}$), the cases are as follows.
\begin{enumerate}
\item $u=q=p$;
\item $q=p$ \& $u\neq q$ with
	\begin{enumerate}
	\item $p \in \{u.\mathrm{C} \}$,
	\item $u \in \{p.\mathrm{C} \}$ or
	\item $u \in \{k.\mathrm{C2} \}$;
	\end{enumerate}
\item $u=p$ \& $q \neq u$ with
	\begin{enumerate}
	\item $p \in \{q.\mathrm{C} \}$,
	\item $q \in \{p.\mathrm{C} \}$ or
	\item $q \in \{k.\mathrm{C2} \}$;
	\end{enumerate}
\item $u=q$ \& $p\neq u$ with
	\begin{enumerate}
	\item $p \in \{u.\mathrm{C} \}$,
	\item $u \in \{p.\mathrm{C} \}$ or
	\item $u \in \{k.\mathrm{C2} \}$;
	\end{enumerate}
\item $u \neq q$ \& $q \neq p$ with
	\begin{enumerate}
	\item $p \in \{q.\mathrm{C}\}$ \& $p \in \{u.\mathrm{C} \}$,
	\item $p \in \{q.\mathrm{C}\}$ \& $u \in \{p.\mathrm{C} \}$,	
	\item $p \in \{q.\mathrm{C}\}$ \& $u \in \{k.\mathrm{C2} \}$,		
	\item $q \in \{p.\mathrm{C}\}$ \& $p \in \{u.\mathrm{C} \}$,	
	\item $q \in \{p.\mathrm{C}\}$ \& $u \in \{p.\mathrm{C} \}$,	
	\item $q \in \{p.\mathrm{C}\}$ \& $u \in \{k.\mathrm{C2} \}$,	
	\item $q \in \{k.\mathrm{C2}\}$ \& $p \in \{u.\mathrm{C} \}$,	
	\item $q \in \{k.\mathrm{C2}\}$ \& $u \in \{p.\mathrm{C} \}$,	
	\item $q \in \{k.\mathrm{C2}\}$ \& $u \in \{k.\mathrm{C2} \}$,	
	\item $u \in \{p.\mathrm{C}\}$ \& $q \in \{p.\mathrm{C} \}$ or
	\item $u \in \{p.\mathrm{C}\}$ \& $q \in \{k.\mathrm{C2} \}$.
	\end{enumerate}
\end{enumerate}

\paragraph*{(i) $u=q=p$.} With $u=q=p$, the corresponding outermost summation in $R_1$ becomes
\begin{align}
\label{eq:sum_3}
\nonumber \sum_{i \in \{p.\mathrm{C} \}} m_i  B_{ijp}^3 &= \left ( \sum_{i \in \{p.\mathrm{C1} \}} + \sum_{i \in \{p.\mathrm{C2} \}} \right ) m_i  B_{ijp}^3 = \frac{[\alpha(p.\mathrm{C1},k.\mathrm{C2};p)]^3 M^3_{p.\mathrm{C2}}}{M_p^3}  \sum_{i \in \{p.\mathrm{C1} \}} m_i + \frac{[\alpha(p.\mathrm{C2},k.\mathrm{C2};p)]^3 M^3_{p.\mathrm{C1}}}{M_p^3}  \sum_{i \in \{p.\mathrm{C2} \}} m_i \\
\nonumber &= \frac{ [\alpha(p.\mathrm{C1},k.\mathrm{C2};p)] M_{p.\mathrm{C2}}^3 M_{p.\mathrm{C1}} + [\alpha(p.\mathrm{C2},k.\mathrm{C2};p)] M_{p.\mathrm{C1}}^3 M_{p.\mathrm{C2}}}{M_p^3} = \alpha(p.\mathrm{C1},k.\mathrm{C2};p) M_{p.\mathrm{C1}} M_{p.\mathrm{C2}} \frac{ M_{p.\mathrm{C2}}^2 - M_{p.\mathrm{C1}}^2}{M_p^3} \\
&= \frac{M_{p.\mathrm{C1}} M_{p.\mathrm{C2}} }{M_p} \alpha(p.\mathrm{C1},k.\mathrm{C2};p) \frac{ M_{p.\mathrm{C2}} - M_{p.\mathrm{C1}}}{M_p}.
\end{align}
The contribution to $R_1$ due to case (i) is, therefore,
\begin{align}
\nonumber R_{1\mathrm{(i)}} &= \frac{1}{2} \sum_{k \in \mathrm{B}} \sum_{\substack{p \in \mathrm{B} \\ p\in\{k.\mathrm{C1} \}}} \frac{G}{r_k} \beta_3(\ve{r}_p,\ve{r}_p,\ve{r}_p,\ve{r}_k) \sum_{j \in \{k.\mathrm{C2} \}} m_j \alpha(p,j;k) \frac{M_{p.\mathrm{C1}} M_{p.\mathrm{C2}} }{M_p} \alpha(p.\mathrm{C1},k.\mathrm{C2};p) \frac{ M_{p.\mathrm{C2}} - M_{p.\mathrm{C1}}}{M_p} \\
\nonumber &= \frac{1}{2} \sum_{k \in \mathrm{B}} \sum_{\substack{p \in \mathrm{B} \\ p\in\{k.\mathrm{C1} \}}} \frac{M_{p.\mathrm{C1}} M_{p.\mathrm{C2}} }{M_p} \alpha(p.\mathrm{C1},k.\mathrm{C2};p)  \frac{ M_{p.\mathrm{C2}} - M_{p.\mathrm{C1}}}{M_p} \alpha(p,k.\mathrm{C2};k) \frac{G M_{k.\mathrm{C2}}}{r_k}  \beta_3(\ve{r}_p,\ve{r}_p,\ve{r}_p,\ve{r}_k) \\
&= \frac{1}{2} \sum_{k \in \mathrm{B}} \sum_{\substack{p \in \mathrm{B} \\ p\in\{k.\mathrm{C1} \}}} \frac{M_{p.\mathrm{C1}} M_{p.\mathrm{C2}} }{M_p} \alpha(p.\mathrm{C1},k.\mathrm{CS}(p);p)  \frac{ M_{p.\mathrm{C2}} - M_{p.\mathrm{C1}}}{M_p} \alpha(p,k.\mathrm{CS}(p);k) \frac{G M_{k.\mathrm{CS}(p)}}{r_k} \beta_3(\ve{r}_p,\ve{r}_p,\ve{r}_p,\ve{r}_k).
\end{align}
A similar result holds for $R_{2\mathrm{(i)}}$, i.e.
\begin{align}
R_{2\mathrm{(i)}} &= \frac{1}{2} \sum_{k \in \mathrm{B}} \sum_{\substack{p \in \mathrm{B} \\ p\in\{k.\mathrm{C2} \}}} \frac{M_{p.\mathrm{C1}} M_{p.\mathrm{C2}} }{M_p} \alpha(p.\mathrm{C1},k.\mathrm{CS}(p);p)  \frac{ M_{p.\mathrm{C2}} - M_{p.\mathrm{C1}}}{M_p} \alpha(p,k.\mathrm{CS}(p);k) \frac{G M_{k.\mathrm{CS}(p)}}{r_k} \beta_3(\ve{r}_p,\ve{r}_p,\ve{r}_p,\ve{r}_k),
\end{align}
and, therefore,
\begin{align}
\nonumber R_\mathrm{(i)} &\equiv R_{1\mathrm{(i)}} + R_{2\mathrm{(i)}} = \frac{1}{2} \sum_{k \in \mathrm{B}} \sum_{\substack{p \in \mathrm{B} \\ p\in\{k.\mathrm{C} \}}} \frac{M_{p.\mathrm{C1}} M_{p.\mathrm{C2}} }{M_p} \alpha(p.\mathrm{C1},k.\mathrm{CS}(p);p)  \frac{ M_{p.\mathrm{C2}} - M_{p.\mathrm{C1}}}{M_p} \alpha(p,k.\mathrm{CS}(p);k) \frac{G M_{k.\mathrm{CS}(p)}}{r_k} \left ( \frac{r_p}{r_k} \right )^3 \\
&\quad \times \left [5 \left ( \unit{r}_p \cdot \unit{r}_k \right )^3 - 3  \left ( \unit{r}_p \cdot \unit{r}_k \right ) \right ],
\end{align}
where we substituted the explicit expression for $\beta_3$ using equation~(\ref{eq:beta_3_def}). 

Note the presence of the `sign quantities' $\alpha$ in $R_{1\mathrm{(i)}}$ and $R_{2\mathrm{(i)}}$. They appear because reversing the definition of the direction of $\ve{r}_p$ and $\ve{r}_k$, i.e. $\ve{r}_p' = -\ve{r}_p$ and $\ve{r}_k' = -\ve{r}_k$, should not change the new expressions in terms of the primed vectors. The minus signs arising from the redefinition are absorbed into the new quantities $\alpha'$, which also change sign. A similar argument applies to the mass difference term $M_{p.\mathrm{C2}} - M_{p.\mathrm{C1}}$ (see also Appendix\,\ref{app:der:av:HO}).

\paragraph*{(ii) (a) $q=p$ \& $u\neq q$ \& $p \in \{u.\mathrm{C} \}$.} While summing over $i \in \{p.\mathrm{C} \}$, $B_{iju}$ is constant, giving
\begin{align}
\sum_{i \in \{p.\mathrm{C} \}} m_i  B_{ijp}^2 B_{iju} &= \frac{\alpha(p,k.\mathrm{C2};u) M_{u.\mathrm{CS}(p)}}{M_u} \sum_{i \in \{p.\mathrm{C} \}} m_i  B_{ijp}^2 = \frac{\alpha(p,k.\mathrm{C2};u) M_{u.\mathrm{CS}(p)}}{M_u} \frac{M_{p.\mathrm{C1}} M_{p.\mathrm{C2}}}{M_p},
\end{align}
where we used equation~(\ref{eq:sum_sq}).

\paragraph*{(ii) (b) $q=p$ \& $u\neq q$ \& $u \in \{p.\mathrm{C} \}$.} While summing over $i \in \{p.\mathrm{C} \}$, $B_{ijp}$ is constant, and $B_{iju} \neq 0$ only if $i \in \{u.\mathrm{C} \}$. Therefore,
\begin{align}
\sum_{i \in \{p.\mathrm{C} \}} m_i  B_{ijp}^2 B_{iju} = B_{ijp}^2 \sum_{i \in \{ u.\mathrm{C} \}} m_i B_{iju} = 0,
\end{align}
where the last step follows from equation~(\ref{eq:sum_vanish}).

\paragraph*{(ii) (c) $q=p$ \& $u\neq q$ \& $u \in \{k.\mathrm{C2} \}$.} While summing over $i \in \{p.\mathrm{C} \}$, $B_{iju}$ is constant, whereas $B_{ijp}$ is constant while summing over $j \in \{k.\mathrm{C2} \}$. Furthermore, $B_{iju}\neq 0$ only if $j \in \{u.\mathrm{C} \}$. This implies
\begin{align}
\sum_{j \in \{k.\mathrm{C2} \}} m_j \sum_{i \in \{p.\mathrm{C} \}} m_i  B_{ijp}^2 B_{iju} = \left ( \sum_{j \in \{u.\mathrm{C} \}} m_j B_{iju} \right ) \left (  \sum_{i \in \{p.\mathrm{C} \}} m_i  B_{ijp}^2 \right ) = 0.
\end{align}
Therefore, for $R_1$ the total contribution due to case (ii) is
\begin{align}
\nonumber R_{1\mathrm{(ii)}} &= \frac{1}{2} \sum_{k \in \mathrm{B}} \sum_{\substack{p \in \mathrm{B} \\ p\in\{k.\mathrm{C1} \}}} \sum_{\substack{u \in \mathrm{B} \\ u\in\{k.\mathrm{C} \} \\ p \in \{u.\mathrm{C} \}}} \frac{G}{r_k} \beta_3(\ve{r}_p,\ve{r}_p,\ve{r}_u,\ve{r}_k) \sum_{j \in \{k.\mathrm{C2} \}} m_j \alpha(p,j;k) \frac{\alpha(p,k.\mathrm{C2};u) M_{u.\mathrm{CS}(p)}}{M_u} \frac{M_{p.\mathrm{C1}} M_{p.\mathrm{C2}}}{M_p} \\
\nonumber &= \frac{1}{2} \sum_{k \in \mathrm{B}} \sum_{\substack{p \in \mathrm{B} \\ p\in\{k.\mathrm{C1} \}}} \sum_{\substack{u \in \mathrm{B} \\ u\in\{k.\mathrm{C} \} \\ p \in \{u.\mathrm{C} \} }}  \frac{M_{p.\mathrm{C1}} M_{p.\mathrm{C2}}}{M_p} \alpha(p,k.\mathrm{C2};k) \frac{G M_{k.\mathrm{C2}}}{r_k} \frac{\alpha(p,k.\mathrm{C2};u) M_{u.\mathrm{CS}(p)}}{M_u} \beta_3(\ve{r}_p,\ve{r}_p,\ve{r}_u,\ve{r}_k) \\
\nonumber &= \frac{1}{2} \sum_{k \in \mathrm{B}} \sum_{\substack{p \in \mathrm{B} \\ p\in\{k.\mathrm{C1} \}}} \sum_{\substack{u \in \mathrm{B} \\ u\in\{k.\mathrm{C} \} \\ p \in \{u.\mathrm{C} \} }}  \frac{M_{p.\mathrm{C1}} M_{p.\mathrm{C2}}}{M_p} \alpha(p,k.\mathrm{CS}(p);k) \frac{G M_{k.\mathrm{CS}(p)}}{r_k} \frac{\alpha(p,k.\mathrm{CS}(p);u) M_{u.\mathrm{CS}(p)}}{M_u} \left ( \frac{r_p}{r_k} \right )^2 \left ( \frac{r_u}{r_k} \right ) \\
&\quad \times \left [ 5 \left ( \unit{r}_p \cdot \unit{r}_k \right )^2 \left ( \unit{r}_u \cdot \unit{r}_k \right ) - 2 \left ( \unit{r}_p \cdot \unit{r}_k \right ) \left ( \unit{r}_p \cdot \unit{r}_u \right ) - \left ( \unit{r}_u \cdot \unit{r}_k \right ) \right ].
\end{align}
A similar result applies to $R_{2(\mathrm{ii})}$. This gives
\begin{align}
\nonumber R_\mathrm{(ii)} &\equiv R_{1\mathrm{(ii)}} + R_{2\mathrm{(ii)}} = \frac{1}{2} \sum_{k \in \mathrm{B}} \sum_{\substack{p \in \mathrm{B} \\ p\in\{k.\mathrm{C} \}}} \sum_{\substack{u \in \mathrm{B} \\ u\in\{k.\mathrm{C} \} \\ p \in \{u.\mathrm{C} \} }}  \frac{M_{p.\mathrm{C1}} M_{p.\mathrm{C2}}}{M_p} \alpha(p,k.\mathrm{CS}(p);k) \frac{G M_{k.\mathrm{CS}(p)}}{r_k} \frac{\alpha(p,k.\mathrm{CS}(p);u) M_{u.\mathrm{CS}(p)}}{M_u} \left ( \frac{r_p}{r_k} \right )^2 \left ( \frac{r_u}{r_k} \right ) \\
&\quad \times \left [ 5 \left ( \unit{r}_p \cdot \unit{r}_k \right )^2 \left ( \unit{r}_u \cdot \unit{r}_k \right ) - 2 \left ( \unit{r}_p \cdot \unit{r}_k \right ) \left ( \unit{r}_p \cdot \unit{r}_u \right ) - \left ( \unit{r}_u \cdot \unit{r}_k \right ) \right ].
\end{align}
Note that the sign quantities $\alpha(p,k.\mathrm{CS}(p);k)$ and $\alpha(p,k.\mathrm{CS}(p);u)$ ensure invariance with respect to the definition of the directions of $\ve{r}_k$ and $\ve{r}_u$, respectively.

\paragraph*{(iii) (a,b,c) $u=p$ \& $q\neq u$.} This case is completely analogous to case (ii), now with the roles of $q$ and $u$ reversed. Note that $\beta_3$ satisfies $\beta_3(\ve{r}_p,\ve{r}_p,\ve{r}_u,\ve{r}_k) = \beta_3(\ve{r}_p,\ve{r}_u,\ve{r}_p,\ve{r}_k)$. Therefore, $\nonumber R_{\mathrm{(ii)}} = \nonumber R_{\mathrm{(iii)}}$. 

The case (iv) also turns out to satisfy $R_{\mathrm{(iv)}} = R_{\mathrm{(ii)}} = \nonumber R_{\mathrm{(iii)}}$, although the equality of $R_{\mathrm{(iv)}}$ to $R_{\mathrm{(ii)}}$ and $R_{\mathrm{(iii)}}$ is not immediately clear. For this reason, we treat this case in detail. 

\paragraph*{(iv) (a) $u=q$ \& $p\neq u$ \& $p \in \{u.\mathrm{C} \}$.} While summing over $i \in \{p.\mathrm{C} \}$, $B_{iju}$ is constant, giving
\begin{align}
\sum_{i \in \{p.\mathrm{C} \}} m_i  B_{ijp} B_{iju}^2 &= B_{iju}^2 \sum_{i \in \{p.\mathrm{C} \}} m_i  B_{ijp} = 0.
\end{align}

\paragraph*{(iv) (b) $u=q$ \& $p\neq u$ \& $u \in \{p.\mathrm{C} \}$.} While summing over $i \in \{p.\mathrm{C} \}$, $B_{iju} \neq 0$ only if $i \in \{u.\mathrm{C} \}$, and $B_{ijp}$ is constant. Therefore,
\begin{align}
\sum_{i \in \{p.\mathrm{C} \}} m_i  B_{ijp} B_{iju}^2 = B_{ijp} \sum_{i \in \{ u.\mathrm{C} \}} m_i B_{iju}^2 = \alpha(u,k.\mathrm{C2};p) \frac{M_{p.\mathrm{CS}(u)}}{M_p} \frac{M_{u.\mathrm{C1}}M_{u.\mathrm{C2}}}{M_u}.
\end{align}
In this case, $\alpha(i,j;k) = \alpha(u,k.\mathrm{C2};k)$.

\paragraph*{(iv) (c) $u=q$ \& $p\neq u$ \& $u \in \{k.\mathrm{C2} \}$.} While summing over $i \in \{p.\mathrm{C} \}$, $B_{iju}$ is constant, whereas $B_{ijp}$ is constant while summing over $j \in \{k.\mathrm{C2} \}$. Furthermore, $B_{iju}\neq 0$ only if $j \in \{u.\mathrm{C} \}$. This implies
\begin{align}
\sum_{j \in \{k.\mathrm{C2} \}} m_j \sum_{i \in \{p.\mathrm{C} \}} m_i  B_{ijp} B_{iju}^2 = \left ( \sum_{j \in \{u.\mathrm{C} \}} m_j B_{iju}^2 \right ) \left (  \sum_{i \in \{p.\mathrm{C} \}} m_i  B_{ijp} \right ) = 0.
\end{align}

Specifying to $R_1$, the total contribution due to case (iv) is
\begin{align}
\nonumber R_{1\mathrm{(iv)}} &= \frac{1}{2} \sum_{k \in \mathrm{B}} \sum_{\substack{p \in \mathrm{B} \\ p\in\{k.\mathrm{C1} \}}} \sum_{\substack{u \in \mathrm{B} \\ u\in\{k.\mathrm{C} \} \\ u \in \{p.\mathrm{C} \}}} \frac{G}{r_k} \beta_3(\ve{r}_p,\ve{r}_u,\ve{r}_u,\ve{r}_k) \sum_{j \in \{k.\mathrm{C2} \}} m_j \alpha(u,k.\mathrm{C2};k) \frac{\alpha(u,k.\mathrm{C2};p) M_{p.\mathrm{CS}(u)}}{M_p} \frac{M_{u.\mathrm{C1}} M_{u.\mathrm{C2}}}{M_u} \\
\nonumber &= \frac{1}{2} \sum_{k \in \mathrm{B}} \sum_{\substack{p \in \mathrm{B} \\ p\in\{k.\mathrm{C1} \} }} \sum_{\substack{u \in \mathrm{B} \\ u\in\{k.\mathrm{C} \} \\ u \in \{p.\mathrm{C} \}}}  \frac{M_{u.\mathrm{C1}} M_{u.\mathrm{C2}}}{M_u} \alpha(u,k.\mathrm{C2};k) \frac{G M_{k.\mathrm{C2}}}{r_k} \frac{\alpha(u,k.\mathrm{C2};p) M_{p.\mathrm{CS}(u)}}{M_p} \beta_3(\ve{r}_p,\ve{r}_u,\ve{r}_u,\ve{r}_k) \\
&= \frac{1}{2} \sum_{k \in \mathrm{B}} \sum_{\substack{p \in \mathrm{B} \\ p\in\{k.\mathrm{C1} \} }} \sum_{\substack{u \in \mathrm{B} \\ u\in\{k.\mathrm{C} \} \\ u \in \{p.\mathrm{C} \}}}  \frac{M_{u.\mathrm{C1}} M_{u.\mathrm{C2}}}{M_u} \alpha(u,k.\mathrm{CS}(p);k) \frac{G M_{k.\mathrm{CS}(p)}}{r_k} \frac{\alpha(u,k.\mathrm{CS}(p);p) M_{p.\mathrm{CS}(u)}}{M_p} \beta_3(\ve{r}_p,\ve{r}_u,\ve{r}_u,\ve{r}_k).
\end{align}
This gives
\begin{align}
\label{eq:R_oct_4} R_{\mathrm{(iv)}} &\equiv R_{1\mathrm{(iv)}} + R_{2\mathrm{(iv)}} = \frac{1}{2} \sum_{k \in \mathrm{B}} \sum_{\substack{p \in \mathrm{B} \\ p\in\{k.\mathrm{C} \} }} \sum_{\substack{u \in \mathrm{B} \\ u\in\{k.\mathrm{C} \} \\ u \in \{p.\mathrm{C} \}}}  \frac{M_{u.\mathrm{C1}} M_{u.\mathrm{C2}}}{M_u} \alpha(u,k.\mathrm{CS}(p);k) \frac{G M_{k.\mathrm{CS}(p)}}{r_k} \frac{\alpha(u,k.\mathrm{CS}(p);p) M_{p.\mathrm{CS}(u)}}{M_p} \beta_3(\ve{r}_p,\ve{r}_u,\ve{r}_u,\ve{r}_k).
\end{align}
In equation~(\ref{eq:R_oct_4}), $p,u \in \{k.\mathrm{C} \}$ and $u \in \{p.\mathrm{C} \}$. Therefore, 
\begin{align}
\nonumber \alpha(u,k.\mathrm{CS}(p);k) &= \alpha(u, k.\mathrm{CS}(u); k); \\
\nonumber M_{k.\mathrm{CS}(p)} &= M_{k.\mathrm{CS}(u)}; \\
\alpha(u,k.\mathrm{CS}(p);p) &= \alpha(u,k.\mathrm{CS}(u); p).
\end{align}
Inserting these relations into equation~(\ref{eq:R_oct_4}) and, subsequently, interchanging $p$ and $u$ (i.e. only changing notation), we find
\begin{align}
R_{\mathrm{(iv)}} &= \frac{1}{2} \sum_{k \in \mathrm{B}} \sum_{\substack{p \in \mathrm{B} \\ p\in\{k.\mathrm{C} \} }} \sum_{\substack{u \in \mathrm{B} \\ u\in\{k.\mathrm{C} \} \\ u \in \{p.\mathrm{C} \}}}  \frac{M_{u.\mathrm{C1}} M_{u.\mathrm{C2}}}{M_u} \alpha(u,k.\mathrm{CS}(u);k) \frac{G M_{k.\mathrm{CS}(u)}}{r_k} \frac{\alpha(u,k.\mathrm{CS}(u);p) M_{p.\mathrm{CS}(u)}}{M_p} \beta_3(\ve{r}_p,\ve{r}_u,\ve{r}_u,\ve{r}_k) \\
\nonumber &= \frac{1}{2} \sum_{k \in \mathrm{B}} \sum_{\substack{u \in \mathrm{B} \\ u\in\{k.\mathrm{C} \} }} \sum_{\substack{p \in \mathrm{B} \\ p\in\{k.\mathrm{C} \} \\ p \in \{u.\mathrm{C} \}}}  \frac{M_{p.\mathrm{C1}} M_{p.\mathrm{C2}}}{M_p} \alpha(p,k.\mathrm{CS}(p);k) \frac{G M_{k.\mathrm{CS}(p)}}{r_k} \frac{\alpha(p,k.\mathrm{CS}(p);u) M_{u.\mathrm{CS}(p)}}{M_u} \beta_3(\ve{r}_u,\ve{r}_p,\ve{r}_p,\ve{r}_k) \\
&= \frac{1}{2} \sum_{k \in \mathrm{B}} \sum_{\substack{p \in \mathrm{B} \\ p\in\{k.\mathrm{C} \}}} \sum_{\substack{u \in \mathrm{B} \\ u\in\{k.\mathrm{C} \} \\ p \in \{u.\mathrm{C} \} }}  \frac{M_{p.\mathrm{C1}} M_{p.\mathrm{C2}}}{M_p} \alpha(p,k.\mathrm{CS}(p);k) \frac{G M_{k.\mathrm{CS}(p)}}{r_k} \frac{\alpha(p,k.\mathrm{CS}(p);u) M_{u.\mathrm{CS}(p)}}{M_u} \left ( \frac{r_p}{r_k} \right )^2 \left ( \frac{r_u}{r_k} \right ) \\
&\quad \times \left [ 5 \left ( \unit{r}_p \cdot \unit{r}_k \right )^2 \left ( \unit{r}_u \cdot \unit{r}_k \right ) - 2 \left ( \unit{r}_p \cdot \unit{r}_k \right ) \left ( \unit{r}_p \cdot \unit{r}_u \right ) - \left ( \unit{r}_u \cdot \unit{r}_k \right ) \right ] = R_\mathrm{(ii)} = R_\mathrm{(iii)}.
\end{align}
Note that $\beta_3(\ve{r}_u,\ve{r}_p,\ve{r}_p,\ve{r}_k) = \beta_3(\ve{r}_p,\ve{r}_p,\ve{r}_u,\ve{r}_k)$. 

To conclude, the nonzero terms from cases (ii), (iii) and (iv) are equal, and so adding them gives $R_\mathrm{(ii)} + R_\mathrm{(iii)} + R_\mathrm{(iv)} = 3R_\mathrm{(ii)}$.

\paragraph*{(v) $u \neq q$ \& $q \neq p$.} In all subcases listed above, two of the quantities in $B_{ijp} B_{ijq} B_{iju}$ are always constant when summing over a particular binary. Consequently, all terms include the term of the form
\begin{align}
\sum_{i \in \{w.\mathrm{C}\}} m_i B_{ijw} = 0,
\end{align}
where $w$ is any of $p$, $q$ and $u$. Therefore, all terms associated with $u \neq q$ and $q \neq p$ are zero, i.e. $R_{\mathrm{(v)}} = 0$.

We conclude that the octupole-order term can be written as
\begin{align}
\nonumber S_3' &= R_\mathrm{(i)} + R_\mathrm{(ii)} + R_\mathrm{(iii)} + R_\mathrm{(iv)} + R_\mathrm{(v)} \\
\nonumber &= \frac{1}{2} \sum_{k \in \mathrm{B}} \sum_{\substack{p \in \mathrm{B} \\ p\in\{k.\mathrm{C} \}}} \frac{M_{p.\mathrm{C1}} M_{p.\mathrm{C2}} }{M_p} \alpha(p.\mathrm{C1},k.\mathrm{CS}(p);p)  \frac{ M_{p.\mathrm{C2}} - M_{p.\mathrm{C1}}}{M_p} \alpha(p,k.\mathrm{CS}(p);k) \frac{G M_{k.\mathrm{CS}(p)}}{r_k} \left ( \frac{r_p}{r_k} \right )^3  \left [5 \left ( \unit{r}_p \cdot \unit{r}_k \right )^3 - 3  \left ( \unit{r}_p \cdot \unit{r}_k \right ) \right ] \\
\nonumber & \quad +  \frac{3}{2} \sum_{k \in \mathrm{B}} \sum_{\substack{p \in \mathrm{B} \\ p\in\{k.\mathrm{C} \}}} \sum_{\substack{u \in \mathrm{B} \\ u\in\{k.\mathrm{C} \} \\ p \in \{u.\mathrm{C} \} }}  \frac{M_{p.\mathrm{C1}} M_{p.\mathrm{C2}}}{M_p} \alpha(p,k.\mathrm{CS}(p);k) \frac{G M_{k.\mathrm{CS}(p)}}{r_k} \frac{\alpha(p,k.\mathrm{CS}(p);u) M_{u.\mathrm{CS}(p)}}{M_u} \left ( \frac{r_p}{r_k} \right )^2 \left ( \frac{r_u}{r_k} \right ) \\
&\quad \quad \times \left [ 5 \left ( \unit{r}_p \cdot \unit{r}_k \right )^2 \left ( \unit{r}_u \cdot \unit{r}_k \right ) - 2 \left ( \unit{r}_p \cdot \unit{r}_k \right ) \left ( \unit{r}_p \cdot \unit{r}_u \right ) - \left ( \unit{r}_u \cdot \unit{r}_k \right ) \right ].
\label{eq:S3_comp_man}
\end{align}
The first major term consists of binary pair terms, and is mathematically equivalent to the octupole-order term in the hierarchical three-body problem. The second major term consists of binary triplet terms, and is mathematically equivalent to the octupole-order `cross term' in the hierarchical four-body problem (`3+1' configuration), which was derived previously by \citet{2015MNRAS.449.4221H} (cf. Eq. A7d of the latter paper).

\subsubsection{Hexadecupole-order term (pairwise)}
\label{app:der:ham:S4}
Extrapolating the above results, at the `hexadecupole' order, i.e. $\mathcal{O} \left (x^4 \right )$, terms in the Hamiltonian after rewriting contain terms depending on binary pairs, triplets and quadlets. For all these combinations, the binaries need to be connected for the terms to be nonvanishing. Furthermore, terms depending on five binaries, i.e. quintlet terms, vanish because of the property equation~(\ref{eq:sum_vanish}). 

Using equation~(\ref{eq:Sm1}), $S_4'$ (cf. equation~\ref{eq:S_4_full}) is given by
\begin{align}
\label{eq:hex1}
\nonumber -S_4' &= \sum_{i<j} \frac{Gm_im_j}{r_m} \frac{1}{8} \sum_{p \in \Bm} \sum_{q \in \Bm}  \sum_{u \in \Bm} \sum_{v \in \Bm} B_{ijp} B_{ijq} B_{iju} B_{ijv} \beta_4(\ve{r}_p,\ve{r}_q,\ve{r}_u,\ve{r}_v,\ve{r}_m) \\
\nonumber &= \sum_{k \in \mathrm{B}} \sum_{i \in \{k.\mathrm{C1} \}} \sum_{j \in \{k.\mathrm{C2} \}} \frac{Gm_im_j}{r_k}  \frac{1}{8} \sum_{\substack{p \in \mathrm{B} \\ p\in\{k.\mathrm{C} \}}} \sum_{\substack{q \in \mathrm{B} \\ q\in\{k.\mathrm{C} \}}}  \sum_{\substack{u \in \mathrm{B} \\ u\in\{k.\mathrm{C} \}}}  \sum_{\substack{v \in \mathrm{B} \\ v\in\{k.\mathrm{C} \}}} B_{ijp} B_{ijq} B_{iju} B_{ijv} \beta_4(\ve{r}_p,\ve{r}_q,\ve{r}_u,\ve{r}_v,\ve{r}_k) \\
\nonumber &= \frac{1}{8} \sum_{k \in \mathrm{B}} \sum_{\substack{p \in \mathrm{B} \\ p\in\{k.\mathrm{C} \}}} \sum_{\substack{q \in \mathrm{B} \\ q\in\{k.\mathrm{C} \}}}  \sum_{\substack{u \in \mathrm{B} \\ u\in\{k.\mathrm{C} \}}} \sum_{\substack{v \in \mathrm{B} \\ v\in\{k.\mathrm{C} \}}} \frac{G}{r_k} \beta_4(\ve{r}_p,\ve{r}_q,\ve{r}_u,\ve{r}_v,\ve{r}_k) \sum_{i \in \{k.\mathrm{C1} \}} \sum_{j \in \{k.\mathrm{C2} \}} \, m_i m_j B_{ijp} B_{ijq} B_{iju} B_{ijv} \\
&= -\left ( S'_{4;2} + S'_{4;3} + S'_{4;4} \right ),
\end{align}
where in the last line, we introduced the notation $S'_{n;l}$ to denote terms of order $n$ that contain summations with individual terms depending on $l$ binaries; generally, $l\leq n$. In this case, these terms are the binary pair terms ($S'_{4;2}$), binary triplet terms ($S'_{4;3}$) and binary quadlet terms ($S'_{4;4}$). 

Rather than deriving all terms at hexadecupole order explicitly, we here restrict to the binary pair terms, which we expect to be typically most important. We refer to Appendix\,\ref{app:der:ham:gen} below for a formal generalization which includes all terms at any order. 

Setting $p=q=u=v$ in equation~(\ref{eq:hex1}), we find
\begin{align}
\nonumber -S'_{4;2} &= \frac{1}{8} \sum_{k \in \mathrm{B}} \sum_{\substack{p \in \mathrm{B} \\ p\in\{k.\mathrm{C} \}}} \frac{G}{r_k} \beta_4(\ve{r}_p,\ve{r}_p,\ve{r}_p,\ve{r}_p,\ve{r}_k) \sum_{i \in \{k.\mathrm{C1} \}} \sum_{j \in \{k.\mathrm{C2} \}} \, m_i m_j B_{ijp}^4 \\
\nonumber &= \frac{1}{8} \sum_{k \in \mathrm{B}} \sum_{\substack{p \in \mathrm{B} \\ p\in\{k.\mathrm{C1} \}}} \frac{G}{r_k} \beta_4(\ve{r}_p,\ve{r}_p,\ve{r}_p,\ve{r}_p,\ve{r}_k) \sum_{j \in \{k.\mathrm{C2} \}} m_j \sum_{i \in \{p.\mathrm{C} \}} \, m_i B_{ijp}^4 \\
&\quad + \frac{1}{8} \sum_{k \in \mathrm{B}} \sum_{\substack{p \in \mathrm{B} \\ p\in\{k.\mathrm{C2} \}}} \frac{G}{r_k} \beta_4(\ve{r}_p,\ve{r}_p,\ve{r}_p,\ve{r}_p,\ve{r}_k) \sum_{i \in \{k.\mathrm{C1} \}} m_i \sum_{j \in \{p.\mathrm{C} \}} \, m_j B_{ijp}^4.
\end{align}
The outermost summation over bodies, in the case of $i \in \{p.\mathrm{C}\}$, is given by
\begin{align}
\nonumber \sum_{i \in \{p.\mathrm{C} \}} \, m_i B_{ijp}^4 &= \left ( \sum_{i\in \{p.\mathrm{C1}\}} m_i + \sum_{i\in \{p.\mathrm{C2}\}} m_i \right ) B_{ijp}^4 = \frac{[\alpha(p.\mathrm{C1},k.\mathrm{C2};p)]^4 M^4_{p.\mathrm{C2}}}{M_p^4}  \sum_{i \in \{p.\mathrm{C1} \}} m_i + \frac{[\alpha(p.\mathrm{C2},k.\mathrm{C2};p)]^4 M^4_{p.\mathrm{C1}}}{M_p^4}  \sum_{i \in \{p.\mathrm{C2} \}} m_i \\
&= \frac{M_{p.\mathrm{C2}}^4 M_{p.\mathrm{C1}} + M_{p.\mathrm{C1}}^4 M_{p.\mathrm{C2}}}{M_p^4} = M_{p.\mathrm{C1}} M_{p.\mathrm{C2}} \frac{ M_{p.\mathrm{C1}}^3 + M_{p.\mathrm{C2}}^3}{M_p^4} = \frac{M_{p.\mathrm{C1}} M_{p.\mathrm{C2}} }{M_p} \frac{M_{p.\mathrm{C1}}^2 - M_{p.\mathrm{C1}} M_{p.\mathrm{C2}} + M_{p.\mathrm{C2}}^2}{M_p^2}.
\end{align}
The same result applies to the summation of $j \in \{p.\mathrm{C}\}$. We conclude that $S'_{4;2}$ is given by
\begin{align}
\nonumber -S'_{4;2} &= \frac{1}{8} \sum_{k \in \mathrm{B}} \sum_{\substack{p \in \mathrm{B} \\ p\in\{k.\mathrm{C} \}}} \frac{M_{p.\mathrm{C1}} M_{p.\mathrm{C2}} }{M_p} \frac{M_{p.\mathrm{C1}}^2 - M_{p.\mathrm{C1}} M_{p.\mathrm{C2}} + M_{p.\mathrm{C2}}^2}{M_p^2} \frac{GM_{k.\mathrm{CS}(p)}}{r_k} \beta_4(\ve{r}_p,\ve{r}_p,\ve{r}_p,\ve{r}_p,\ve{r}_k) \\
&= \frac{1}{8} \sum_{k \in \mathrm{B}} \sum_{\substack{p \in \mathrm{B} \\ p\in\{k.\mathrm{C} \}}} \frac{M_{p.\mathrm{C1}} M_{p.\mathrm{C2}} }{M_p} \frac{M_{p.\mathrm{C1}}^2 - M_{p.\mathrm{C1}} M_{p.\mathrm{C2}} + M_{p.\mathrm{C2}}^2}{M_p^2} \frac{GM_{k.\mathrm{CS}(p)}}{r_k} \left ( \frac{r_p}{r_k} \right )^4 \left [ 35 \left ( \unit{r}_p \cdot \unit{r}_k \right )^4- 30 \left ( \unit{r}_p \cdot \unit{r}_k \right )^2 + 3 \right ].
\end{align}
Note that $M_{p.\mathrm{C1}}^2 - M_{p.\mathrm{C1}} M_{p.\mathrm{C2}} + M_{p.\mathrm{C2}}^2>0$ for any combination of non-zero and positive $M_{p.\mathrm{C1}}$ and $M_{p.\mathrm{C2}}$. Therefore, unlike the pairwise octupole-order terms (cf. Appendix\,\ref{app:der:ham:S3}), the pairwise hexadecupole-order terms never exactly vanish. 

\subsubsection{Dotriacontupole-order term (pairwise)}
\label{app:der:ham:S5}
As for the hexadecupole-order term, we only explicitly derive the binary pair terms for terms of order $x^5$, i.e. the `dotriacontupole'-order terms. For formal expressions for all binary interactions, we refer to Appendix\,\ref{app:der:ham:gen}.

Using equations~(\ref{eq:S_5_full}) and (\ref{eq:Sm1}),
\begin{align}
\nonumber S_5' &= \sum_{i<j} \frac{Gm_im_j}{r_m} \frac{1}{8} \alpha(i,j;m) \sum_{p \in \Bm} \sum_{q \in \Bm}  \sum_{u \in \Bm} \sum_{v \in \Bm} \sum_{w \in \Bm} B_{ijp} \sum_{w \in \Bm} B_{ijp} B_{ijq} B_{iju} B_{ijv} B_{ijw} \beta_5(\ve{r}_p,\ve{r}_q,\ve{r}_u,\ve{r}_v,\ve{r}_w,\ve{r}_m) \\
\nonumber &= \sum_{k \in \mathrm{B}} \sum_{i \in \{k.\mathrm{C1} \}} \sum_{j \in \{k.\mathrm{C2} \}} \frac{Gm_im_j}{r_k}  \frac{1}{8} \alpha(i,j;k) \sum_{\substack{p \in \mathrm{B} \\ p\in\{k.\mathrm{C} \}}} \sum_{\substack{q \in \mathrm{B} \\ q\in\{k.\mathrm{C} \}}}  \sum_{\substack{u \in \mathrm{B} \\ u\in\{k.\mathrm{C} \}}} \sum_{\substack{v \in \mathrm{B} \\ v\in\{k.\mathrm{C} \}}} \sum_{\substack{w \in \mathrm{B} \\ w\in\{k.\mathrm{C} \}}} B_{ijp} B_{ijq} B_{iju} B_{ijv} B_{ijw} \beta_5(\ve{r}_p,\ve{r}_q,\ve{r}_u,\ve{r}_v,\ve{r}_w,\ve{r}_k) \\
\nonumber &= \frac{1}{8} \sum_{k \in \mathrm{B}} \sum_{\substack{p \in \mathrm{B} \\ p\in\{k.\mathrm{C} \}}} \sum_{\substack{q \in \mathrm{B} \\ q\in\{k.\mathrm{C} \}}}  \sum_{\substack{u \in \mathrm{B} \\ u\in\{k.\mathrm{C} \}}} \sum_{\substack{v \in \mathrm{B} \\ v\in\{k.\mathrm{C} \}}} \sum_{\substack{w \in \mathrm{B} \\ w\in\{k.\mathrm{C} \}}} \frac{G}{r_k} \beta_5(\ve{r}_p,\ve{r}_q,\ve{r}_u,\ve{r}_v,\ve{r}_w,\ve{r}_k) \sum_{i \in \{k.\mathrm{C1} \}} \sum_{j \in \{k.\mathrm{C2} \}} \alpha(i,j;k) \, m_i m_j B_{ijp} B_{ijq} B_{iju} B_{ijv} B_{ijw} \\
&= S'_{5;2} + S'_{5;3} + S'_{5;4} + S'_{5;5},
\end{align}
with
\begin{align}
\nonumber S'_{5;2} &= \frac{1}{8} \sum_{k \in \mathrm{B}} \sum_{\substack{p \in \mathrm{B} \\ p\in\{k.\mathrm{C} \}}} \frac{G}{r_k} \beta_5(\ve{r}_p,\ve{r}_p,\ve{r}_p,\ve{r}_p,\ve{r}_p,\ve{r}_k) \sum_{i \in \{k.\mathrm{C1} \}} \sum_{j \in \{k.\mathrm{C2} \}} \, \alpha(i,j;k) \, m_i m_j B_{ijp}^5 \\
&= \frac{1}{8} \sum_{k \in \mathrm{B}} \sum_{\substack{p \in \mathrm{B} \\ p\in\{k.\mathrm{C1} \}}} \frac{G}{r_k} \beta_5(\ve{r}_p,\ve{r}_p,\ve{r}_p,\ve{r}_p,\ve{r}_p,\ve{r}_k) \, \alpha(p,k.\mathrm{C2};k) \sum_{j \in \{k.\mathrm{C2} \}} m_j \sum_{i \in \{p.\mathrm{C} \}} m_i B_{ijp}^5 \\
&\quad +\frac{1}{8} \sum_{k \in \mathrm{B}} \sum_{\substack{p \in \mathrm{B} \\ p\in\{k.\mathrm{C2} \}}} \frac{G}{r_k} \beta_5(\ve{r}_p,\ve{r}_p,\ve{r}_p,\ve{r}_p,\ve{r}_p,\ve{r}_k) \, \alpha(p,k.\mathrm{C1};k) \sum_{i \in \{k.\mathrm{C1} \}} m_i \sum_{j \in \{p.\mathrm{C} \}} m_j B_{ijp}^5.
\end{align}

The outermost summation over bodies, in the case of $i \in \{p.\mathrm{C}\}$, is given by
\begin{align}
\nonumber \sum_{i \in \{p.\mathrm{C} \}} \, m_i B_{ijp}^5 &= \left ( \sum_{i\in \{p.\mathrm{C1}\}} m_i + \sum_{i\in \{p.\mathrm{C2}\}} m_i \right ) B_{ijp}^5 = \frac{[\alpha(p.\mathrm{C1},k.\mathrm{C2};p)]^5 M^5_{p.\mathrm{C2}}}{M_p^5}  \sum_{i \in \{p.\mathrm{C1} \}} m_i + \frac{[\alpha(p.\mathrm{C2},k.\mathrm{C2};p)]^5 M^5_{p.\mathrm{C1}}}{M_p^5}  \sum_{i \in \{p.\mathrm{C2} \}} m_i \\
\nonumber &= \frac{ [\alpha(p.\mathrm{C1},k.\mathrm{C2};p)] M_{p.\mathrm{C2}}^5 M_{p.\mathrm{C1}} + [\alpha(p.\mathrm{C2},k.\mathrm{C2};p)] M_{p.\mathrm{C1}}^5 M_{p.\mathrm{C2}}}{M_p^5} = \alpha(p.\mathrm{C1},k.\mathrm{C2};p) M_{p.\mathrm{C1}} M_{p.\mathrm{C2}} \frac{ M_{p.\mathrm{C2}}^4 - M_{p.\mathrm{C1}}^4}{M_p^5} \\
&= \alpha(p.\mathrm{C1},k.\mathrm{C2};p) \frac{M_{p.\mathrm{C1}} M_{p.\mathrm{C2}}}{M_p} \frac{M_{p.\mathrm{C2}} - M_{p.\mathrm{C1}}}{M_p} \frac{M_{p.\mathrm{C1}}^2 + M_{p.\mathrm{C2}}^2}{M_p^2}.
\end{align}
The same result holds for $j \in \{p.\mathrm{C}\}$. Therefore,
\begin{align}
\nonumber S'_{5;2} &= \frac{1}{8} \sum_{k \in \mathrm{B}} \sum_{\substack{p \in \mathrm{B} \\ p\in\{k.\mathrm{C} \}}} \frac{M_{p.\mathrm{C1}} M_{p.\mathrm{C2}} }{M_p} \, \alpha(p.\mathrm{C1},p.\mathrm{CS}(p);p) \frac{M_{p.\mathrm{C2}} - M_{p.\mathrm{C1}}}{M_p} \frac{M_{p.\mathrm{C1}}^2 + M_{p.\mathrm{C2}}^2}{M_p^2} \, \alpha(p,k.\mathrm{CS}(p);k) \frac{GM_{k.\mathrm{CS}(p)}}{r_k} \beta_5(\ve{r}_p,\ve{r}_p,\ve{r}_p,\ve{r}_p,\ve{r}_p,\ve{r}_k) \\
\nonumber &= \frac{1}{8} \sum_{k \in \mathrm{B}} \sum_{\substack{p \in \mathrm{B} \\ p\in\{k.\mathrm{C} \}}} \frac{M_{p.\mathrm{C1}} M_{p.\mathrm{C2}} }{M_p} \, \alpha(p.\mathrm{C1},p.\mathrm{CS}(p);p) \frac{M_{p.\mathrm{C2}} - M_{p.\mathrm{C1}}}{M_p} \frac{M_{p.\mathrm{C1}}^2 + M_{p.\mathrm{C2}}^2}{M_p^2} \, \alpha(p,k.\mathrm{CS}(p);k) \frac{GM_{k.\mathrm{CS}(p)}}{r_k} \left ( \frac{r_p}{r_k} \right )^5 \\
&\quad \times \left [ 63 \left ( \unit{r}_p \cdot \unit{r}_k \right )^5 - 70 \left ( \unit{r}_p \cdot \unit{r}_k \right )^3 + 15 \left ( \unit{r}_p \cdot \unit{r}_k \right ) \right ].
\end{align}

\subsubsection{Generalisation}
\label{app:der:ham:gen}
In the above appendices, we have shown that $H$ can be written in terms of summations of only binaries, and we have derived explicit expressions for all terms up and including octupole order, and including the binary pair terms at hexadecupole and dotriacontupole order. Here, using similar arguments as before, we generalize these results and derive a formal expression for $S_n'$ at {\it arbitrary} order $n$ including {\it all} binary interactions (i.e. pairwise terms, triplet terms, etc.).

Equations~(\ref{eq:S_n}) and (\ref{eq:S_n_prime}) combined read
\begin{align}
S_n' = (-1)^{n+1} \sum_{i<j} \frac{Gm_im_j}{r_{m(i,j)}} \alpha(i,j;m)^n \sum_{u_1,...,u_n \in \Bm} c_n \beta_n(\ve{r}_{u_1},...,\ve{r}_{u_n},\ve{r}_m) \, B_{iju_1} ... B_{iju_n},
\end{align}
where we recall that $m=m(i,j)$ denotes the lowest level binary in $\mathrm{B_p}(i,j)$. Using equation~(\ref{eq:Sm1}), we rewrite the summation over pairs of bodies to summations over binaries and children within those binaries, giving
\begin{align}
S_n' = (-1)^{n+1} \sum_{k \in \mathrm{B}} \sum_{i \in \{k.\mathrm{C1}\}} \sum_{j \in \{k.\mathrm{C2} \}} \frac{Gm_im_j}{r_k} \alpha(i,j;k)^n \sum_{u_1,...,u_n \in \Bm} c_n \beta_n(\ve{r}_{u_1},...,\ve{r}_{u_n},\ve{r}_k) \, B_{iju_1} ... B_{iju_n}.
\end{align}
By definition, none of the binaries $u_l$ in the summations $u_1,...,u_n \in \Bm$  are the same as binary $k$. Furthermore, because $i \in \{k.\mathrm{C1}\}$ and $j \in \{k.\mathrm{C2}\}$, if $u_l$ is not part of the children of binary $k$ (i.e. if $u_l \not\in \{k.\mathrm{C}\}$), then $B_{iju_l} = 0$. Therefore, the summations $u_1,...,u_n \in \Bm$ can be written as summations $u_1,...,u_n \in \mathrm{B}$ with $u_l \in \{k.\mathrm{C}\}$, i.e.
\begin{align}
\nonumber S_n' &= (-1)^{n+1} \sum_{k \in \mathrm{B}} \sum_{i \in \{k.\mathrm{C1}\}} \sum_{j \in \{k.\mathrm{C2} \}} \frac{Gm_im_j}{r_k} \alpha(i,j;k)^n \sum_{\substack{u_1,...,u_n \in \mathrm{B} \\ u_1,...,u_n \in \{k.\mathrm{C}\}}} c_n \beta_n(\ve{r}_{u_1},...,\ve{r}_{u_n},\ve{r}_k) \, B_{iju_1} ... B_{iju_n} \\
&= (-1)^{n+1} \sum_{k \in \mathrm{B}}  \frac{G}{r_k}  \sum_{\substack{u_1,...,u_n \in \mathrm{B} \\ u_1,...,u_n \in \{k.\mathrm{C}\}}} c_n \beta_n(\ve{r}_{u_1},...,\ve{r}_{u_n},\ve{r}_k) 
    \sum_{i \in \{k.\mathrm{C1}\}} \sum_{j \in \{k.\mathrm{C2} \}} \alpha(i,j;k)^n m_i m_j\, B_{iju_1} ... B_{iju_n},
\label{eq:S_n_gen1}
\end{align}
where in the second line, we used that $u_1,...,u_n$ are (no longer) dependent on $i$ and $j$. 

In equation~(\ref{eq:S_n_gen1}), each binary $u_l$ in the summation $u_1,...,u_n \in \mathrm{B}$ is associated with a factor $B_{iju_l}$. A given binary $u_l$ may occur zero to multiple times, up to $n$ times. For example, the pairwise binary terms correspond to $u_1=...=u_n$, such that there are two unique binaries: $k$ and the equal $u$'s. For binary $u_l$, let the number of recurrences be denoted by $d_{u_l}$; generally, $0\leq d_{u_l}\leq n$. Consequently, the power of $B_{iju_l}$ is $d_{u_l}$, i.e. the mass ratio factor in $S_n'$ associated with binary $u_l$ is $B_{iju_l}^{d_{u_l}}$. 

In the summations $u_1,...,u_n \in \mathrm{B}$, the binaries $u_1,...,u_n$ occur in combinations with various divisions over the two children of binary $k$. For a particular combination, let the binaries in child 1 (2) of binary $k$ be denoted by $p_l$ ($q_l$). Furthermore, for child 1 (2) of $k$, let $p$ ($q$) denote the binary with the highest level within that child (see below for the case when the highest level within a child of binary $k$ is shared amongst two or more binaries). The number of recurrences associated with binaries $p$ and $q$ are $d_p$ and $d_q$, respectively. Note that if $d_p=0$ ($d_q=0$), then this implies that there are {\it no} binaries in child 1 (2) of binary $k$.

Furthermore, let $n_1$ $(n_2$) denote the combined exponent of all mass ratio quantities $B_{iju_l}$ associated with child 1 (2) of binary $k$. The allowed values of $n_1$ and $n_2$ are $0\leq n_1\leq n$ and $0\leq n_2\leq n$, respectively, and because the combined power of all mass ratio factors is $n$, $n_1+n_2=n$. For a given nonzero exponent $d_p$ ($d_q$) of the mass ratio factor associated with binary $p$ ($q$), there are $n_1-d_p$ ($n_2-d_q$) binaries in child 1 (2) distinct from $p$ ($q$), such that the combined mass ratio exponent in child 1 (2) is $n_1$ ($n_2$). Note that these $n_1-d_p$ ($n_2-d_q$) binaries are not necessarily distinct; each is associated with a recurrence number $d_{p_l}$ ($d_{q_l}$). The maximum recurrence number for binary $p$ ($q$) is $n_1$ ($n_2$).

Subsequently, we rewrite the summations $u_1,...,u_n \in \mathrm{B}$ as separate summations over the binaries $p$, $p_1,...,p_{n_1-d_p}$ and $q$, $q$, $q_1,...,q_{n_2-d_q}$, for all possible combinations of $n_1$, $n_2$, $d_p$ and $d_q$. Note that there should only be summations over $p$ and $p_1,...,p_{n_1-d_p}$ ($q$ and $q_1,...,q_{n_2-d_q}$) if $d_p>0$ ($d_q>0$). Also, if both $d_p$ and $d_q$ are zero, then there are no binaries in either child of binary $k$ with the highest level. In that case, the corresponding terms vanish (see below).

When rewriting the summations $u_1,...,u_n \in \mathrm{B}$ to summations over both children of binary $k$, it should be taken into account that for given $n_1$, $n_2$, $d_p$, $d_{p_1},...,d_{p_{n_1-d_p}}$, $d_q$ and $d_{q_1},...,d_{q_{n_2-d_q}}$, there are multiple possible permutations of the binaries $p,p_1,...,p_{n_1-d_p},q,...,q_{n_2-d_q}$ from the set $u_1,...,u_n \in \mathrm{B}$. The number of permutations is determined by $n$ and the number of recurrences of each of the binaries. This is equivalent to placing numbers in $n$ slots, where there are $d_p$ identical numbers, $d_{p_1}$ other identical numbers, $d_{p_2}$ other identical numbers, etc. This gives a number of permutations
\begin{align}
\mathcal{P}^{n_1,n_2}_{d_p,d_{p_1},...,d_{p_{n_1-d_p}},d_q,d_{q_1},...,d_{q_{n_2-d_q}}} \equiv \mathcal{P}^{n_1,n_2}_{d_p;d_q} = \frac{(n_1+n_2)!}{(d_p)!(d_{p_1})!...(d_{p_{n_1-d_p}})!(d_q)!(d_{q_1})!...(d_{q_{n_2-d_q}})!},
\label{eq:perm}
\end{align}
where $\mathcal{P}^{n_1,n_2}_{d_p;d_q}$ is short-hand notation, and $n_2=n-n_1$. Note that the quantity $\beta_n(\ve{r}_{u_1},...,\ve{r}_{u_n},\ve{r}_k)$ is invariant under each permutation because it does not depend on the order of any of the first $n$ arguments (cf. Appendix\,\ref{app:der:pot}). Evidently, the same applies to the scalar multiplication $B_{iju_1}...B_{iju_n}$ in equation~(\ref{eq:S_n_gen1}).

We thus arrive at the following rewriting of summations,
\begin{align}
\nonumber &\sum_{\substack{u_1,...,u_n \in \mathrm{B} \\ u_1,...,u_n \in \{k.\mathrm{C}\}}} c_n \beta_n(\ve{r}_{u_1},...,\ve{r}_{u_n},\ve{r}_k) 
    \sum_{i \in \{k.\mathrm{C1}\}} \sum_{j \in \{k.\mathrm{C2} \}} \alpha(i,j;k)^n m_i m_j\, B_{iju_1} ... B_{iju_n} \\
\nonumber &\quad = \sum_{n_1=0}^n \sum_{d_p=0}^{n_1} \sum_{d_q=0}^{n_2=n-n_1} \sum_{\substack{\mathrm{if}\,d_p>0: \\ p,p_1,...,p_{n_1-d_p} \in \mathrm{B} \\ p,p_l \in \{k.\mathrm{C1}\} \\ p \in \{p_l.\mathrm{C}\}}} \sum_{\substack{\mathrm{if}\,d_q>0: \\ q,q_1,...,q_{n_2-d_q} \in \mathrm{B} \\ q,q_l \in \{k.\mathrm{C2}\} \\ q \in \{q_l.\mathrm{C}\}}} \mathcal{P}^{n_1,n_2}_{d_p;d_q} c_n \beta_n(\underbrace{\ve{r}_p,...,\ve{r}_p}_{d_p \times}, \ve{r}_{p_1},...,\ve{r}_{p_{n_1-d_p}},\underbrace{\ve{r}_q,...,\ve{r}_q}_{d_q\times},\ve{r}_{q_1},...\ve{r}_{q_{n_2-d_q}},\ve{r}_k) \\
    &\quad \quad \times f(n_1,n_2,d_p,d_q) \sum_{i \in \{k.\mathrm{C1}\}} \sum_{j \in \{k.\mathrm{C2} \}} \alpha(i,j;k)^n m_i m_j\, \left ( B_{ijp}^{d_p} B_{ijp_1}^{d_{p_1}} ... B_{ijp_{n_1-d_p}}^{d_{n_1-d_p}} \right ) \left ( B_{ijq}^{d_q} B_{ijq_1}^{d_{q_1}} ... B_{ijq_{n_2-d_q}}^{d_{n_2-d_q}} \right ).
\label{eq:sum_split1}
\end{align}
Note that if $d_p=0$ ($d_q=0$), then also the associated recurrences $d_{p_1},...,d_{n_1-d_p}$ ($d_{q_1},...,d_{n_2-d_q}$) are zero, and the corresponding product $B_{ijp_1}^{d_{p_1}} ... B_{ijp_{n_1-d_p}}^{d_{n_1-d_p}}$ ($B_{ijq_1}^{d_{q_1}} ... B_{ijq_{n_2-d_q}}^{d_{n_2-d_q}} $) is unity. The factor $f(n_1,n_2,d_p,d_q)$ takes into account that the terms corresponding to $d_p=d_q=0$ are zero (see below), and that terms with $d_p=0$ and $n_1-d_p>0$, and $d_q=0$ and $n_2-d_q>0$, should be discarded (by definition of the $p$ and $q$ binaries). Therefore,
\begin{align}
\label{eq:f_def}
f(n_1,n_2,d_p,d_q) = \left \{
\begin{array}{ll}
0, & d_p=0 \, \& \, d_q=0; \\
0, & d_p=0 \, \& \, n_1-d_p>0; \\
0, & d_q=0 \, \& \, n_2-d_q>0; \\
1, & \mathrm{otherwise}.
\end{array} \right.
\end{align}
Note that the use of $f(n_1,n_2,d_p,d_q)$ can be avoided by separating out the terms corresponding to $d_p=0$ and/or $d_q=0$. We nevertheless include it, in order to keep the notation relatively compact. 

When summing over body $i\in \{k.\mathrm{C1}\}$, $B_{ijp_l}=0$ for any $p_l$ within the binaries $p_1,...,p_{n_1-d_p}$, unless $i \in \{p.\mathrm{C}\}$ and $p \in \{p_l.\mathrm{C}\}$. Similarly, $B_{ijq_l}=0$ for any $q_l$ within the binaries $q_1,...,q_{n_2-d_q}$, unless $j \in \{q.\mathrm{C}\}$ and $q \in \{q_l.\mathrm{C}\}$. Furthermore, any of the mass ratio quantities $B_{ijp_1},...,B_{ijp_{n_1-d_p}}$, which are nonzero if $p \in \{p_l.\mathrm{C}\}$, are constant when summing over $i\in \{p.\mathrm{C}\}$ because $p$ is the highest-level binary in child 1 of binary $k$. Also, for $i\in \{p.\mathrm{C}\}$, the quantities $ B_{ijq},B_{ijq_1},...,B_{ijq_{n_2-d_q}}$ are constant because the `$q$'-binaries are, by definition, part of child 2 of binary $k$, whereas $p \in \{k.\mathrm{C1}\}$. Similarly, $B_{ijq_1},...,B_{ijq_{n_2-d_q}}$, $B_{ijp}$ and $B_{ijp_1},...,B_{ijp_{n_1-d_p}}$ are constant when summing over body $j\in \{q.\mathrm{C}\}$. Note that when summing $i \in \{p.\mathrm{C}\}$ and $j \in \{q.\mathrm{C}\}$, $\alpha(i,j;k) = \alpha(p,q;k)$. 

Therefore, the summations over bodies $i$ and $j$ in equation~(\ref{eq:sum_split1}) can be written as
\begin{align}
\nonumber &\sum_{i \in \{k.\mathrm{C1}\}} \sum_{j \in \{k.\mathrm{C2} \}} \alpha(i,j;k)^n m_i m_j\, \left ( B_{ijp}^{d_p} B_{ijp_1}^{d_{p_1}} ... B_{ijp_{n_1-d_p}}^{d_{n_1-d_p}} \right ) \left ( B_{ijq}^{d_q} B_{ijq_1}^{d_{q_1}} ... B_{ijq_{n_2-d_q}}^{d_{n_2-d_q}} \right ) \\
\nonumber &\quad = \sum_{i \in \{p.\mathrm{C}\}} \sum_{j \in \{q.\mathrm{C} \}} \alpha(p,q;k)^n m_i m_j\, \left ( B_{ijp}^{d_p} B_{ijp_1}^{d_{p_1}} ... B_{ijp_{n_1-d_p}}^{d_{n_1-d_p}} \right ) \left ( B_{ijq}^{d_q} B_{ijq_1}^{d_{q_1}} ... B_{ijq_{n_2-d_q}}^{d_{n_2-d_q}} \right ) \\
&\quad = \alpha(p,q;k)^n  \left ( B_{ijp_1}^{d_{p_1}} ... B_{ijp_{n_1-d_p}}^{d_{n_1-d_p}} \right ) \left ( B_{ijq_1}^{d_{q_1}} ... B_{ijq_{n_2-d_q}}^{d_{n_2-d_q}} \right ) \left ( \sum_{i \in \{p.\mathrm{C}\}} m_i B_{ijp}^{d_p} \right ) \left ( \sum_{j \in \{q.\mathrm{C} \}} m_j B_{ijq}^{d_q} \right ).
\label{eq:B_moved}
\end{align}
Note that the mass ratio factors outside of the summations of $i$ and $j$ are in fact independent of $i$ and $j$. One could e.g. write $B_{ijp_1}=B_{pqp_1}$.

Similarly to the previous appendices, the expressions in equation~(\ref{eq:B_moved}) with remaining summations over bodies can be simplified by summing separately over the two children of $p$ and $q$, and by substituting the explicit expression for the mass ratio quantities. For binary $p$,
\begin{align}
\label{eq:mBn}
\nonumber \sum_{i \in \{p.\mathrm{C} \}} \, m_i B_{ijp}^{d_p} &= \left ( \sum_{i\in \{p.\mathrm{C1}\}} + \sum_{i\in \{p.\mathrm{C2}\}} \right ) m_i B_{ijp}^{d_p} = \left ( \frac{\alpha(p.\mathrm{C1},k.\mathrm{CS}(p);p) M_{p.\mathrm{C2}}}{M_p} \right )^{d_p}  \sum_{i \in \{p.\mathrm{C1} \}} m_i + \left ( \frac{\alpha(p.\mathrm{C2},k.\mathrm{CS}(p);p) M_{p.\mathrm{C1}}}{M_p} \right )^{d_p}  \sum_{i \in \{p.\mathrm{C2} \}} m_i \\
\nonumber &= \frac{ [\alpha(p.\mathrm{C1},qk.\mathrm{CS}(p);p)]^{d_p} M_{p.\mathrm{C2}}^{d_p} M_{p.\mathrm{C1}} + [\alpha(p.\mathrm{C2},k.\mathrm{CS}(p);p)]^{d_p} M_{p.\mathrm{C1}}^{d_p} M_{p.\mathrm{C2}}}{M_p^{d_p}} \\
\nonumber&= \alpha(p.\mathrm{C1},k.\mathrm{CS}(p);p)^{d_p} \frac{M_{p.\mathrm{C1}} M_{p.\mathrm{C2}}}{M_p} \frac{ M_{p.\mathrm{C2}}^{{d_p}-1} + (-1)^{d_p} M_{p.\mathrm{C1}}^{{d_p}-1}}{M_p^{{d_p}-1}} \\
&\equiv \mathcal{M}_{p;k.\mathrm{CS}(p)}^{(d_p)},
\end{align}
where, for convenience, we introduced the short-hand notation $\mathcal{M}_{p;k.\mathrm{CS}(p)}^{(d_p)}$. Analogously,
\begin{align}
\sum_{j \in \{q.\mathrm{C} \}} \, m_j B_{ijq}^{d_q} = \mathcal{M}_{q;k.\mathrm{CS}(q)}^{(d_q)}.
\end{align}
Note that equation~(\ref{eq:mBn}) is essentially a generalized version of the quantity `$M_j$' that appears at order $j$ in the hierarchical three-body problem as given by equation (1) of \citet{1968AJ.....73..190H} -- in the latter paper, $M_j$ is multiplied by the tertiary mass `$m_2$'; in our case, this corresponds to $\mathcal{M}_{q;k.\mathrm{CS}(q)}^{(d_q)}$. Special values of $\mathcal{M}_{p;k.\mathrm{CS}(p)}^{(d_p)}$ of interest here are
\begin{align}
\mathcal{M}_{p,k.\mathrm{CS}(p)}^{(0)} = \alpha(p.\mathrm{C1},k.\mathrm{CS}(p);p)^0 \frac{M_{p.\mathrm{C1}} M_{p.\mathrm{C2}}}{M_p} \frac{M_{p.\mathrm{C2}}^{-1} + M_{p.\mathrm{C1}}^{-1}}{M_p^{-1}} = M_{p.\mathrm{C1}} + M_{p.\mathrm{C2}} = M_p,
\label{eq:mBn0}
\end{align}
and
\begin{align}
\mathcal{M}_{p,k.\mathrm{CS}(p)}^{(1)} = \alpha(p.\mathrm{C1},k.\mathrm{CS}(p);p) \frac{M_{p.\mathrm{C1}} M_{p.\mathrm{C2}}}{M_p} \frac{M_{p.\mathrm{C2}}^{0} - M_{p.\mathrm{C1}}^{0}}{M_p^{0}} = 0.
\label{eq:mBn1}
\end{align}

Substituting explicit expressions for the remaining mass ratio quantities $B_{ijp_1},...,B_{ijp_{n_1-d_p}}$ and $B_{ijq_1},...,B_{ijq_{n_2-d_q}}$, we find the following general expression for the potential energy,
\begin{align}
\nonumber S_n' &= (-1)^{n+1} \sum_{k \in \mathrm{B}} \frac{G}{r_k} \sum_{n_1=0}^n \sum_{d_p=0}^{n_1} \sum_{d_q=0}^{n_2=n-n_1} \sum_{\substack{\mathrm{if}\,d_p>0: \\ p,p_1,...,p_{n_1-d_p} \in \mathrm{B} \\ p,p_l \in \{k.\mathrm{C1}\} \\ p \in \{p_l.\mathrm{C}\}}} \sum_{\substack{\mathrm{if}\,d_q>0: \\ q,q_1,...,q_{n_2-d_q} \in \mathrm{B} \\ q,q_l \in \{k.\mathrm{C2}\} \\ q \in \{q_l.\mathrm{C}\}}} c_n 
\beta_n(\underbrace{\ve{r}_p,...,\ve{r}_p}_{d_p \times}, \ve{r}_{p_1},...,\ve{r}_{p_{n_1-d_p}},\underbrace{\ve{r}_q,...,\ve{r}_q}_{d_q\times},\ve{r}_{q_1},...\ve{r}_{q_{n_2-d_q}},\ve{r}_k) \\
\nonumber &\quad \times \alpha(p,q;k)^n f(n_1,n_2,d_p,d_q) \, \mathcal{P}^{n_1,n_2}_{d_p;d_q} \left \{ \prod_{l=1}^{n_1-d_p} \left [ \frac{ \alpha(p_l,k.\mathrm{CS}(p);k) M_{p_l.\mathrm{CS}(p)} }{M_{p_l}} \right ]^{d_{p_l}} \right \} \left \{ \prod_{l=1}^{n_2-d_q} \left [ \frac{ \alpha(q_{l},k.\mathrm{CS}(q);k) M_{q_{l}.\mathrm{CS}(q)} }{M_{q_{l}}} \right ]^{d_{q_{l}}} \right \} \\
&\quad \times \mathcal{M}_{p;k.\mathrm{CS}(p)}^{(d_p)} \mathcal{M}_{q;k.\mathrm{CS}(q)}^{(d_q)},
\label{eq:Sn_gen1}
\end{align}
where $f(n_1,n_2,d_p,d_q)$ was defined in equation~(\ref{eq:f_def}). Note that the products in curly brackets are to be evaluated to unity if the upper limit of the variable $l$, i.e. $n_1-d_p$ or $n_2-d_q$, is zero (this corresponds to no binaries $p_l$ or $q_l$ such that $p\in\{p_l.\mathrm{C}\}$ or $q \in \{q_l.\mathrm{C}\}$). 

Each individual term in equation~(\ref{eq:Sn_gen1}) depends on binary $k$ and, depending on the values of $d_p$ and $d_q$, binaries $p$ and at most $n_1-d_p$ distinct binaries $p_1,...,p_{n_1-d_p}$, and binaries $q$ and at most $n_2-d_q$ distinct binaries $q_1,...,q_{n_2-d_q}$. If $d_p\neq 0$ and $d_q \neq 0$, then the number of distinct binaries is at most $N_\mathrm{bin,dis} = 1 + (1 + n_1-d_p) + (1 + n_2-d_q) = 3 + n - d_p - d_q$. This number is maximized when $d_p$ and $d_q$ are smallest. Because the product $\mathcal{M}_{p;k.\mathrm{CS}(p)}^{(d_p)} \mathcal{M}_{q;k.\mathrm{CS}(q)}^{(d_q)}$ is zero whenever one or both of $d_p$ and $d_q$ are unity (cf. equation~\ref{eq:mBn1}), the lowest allowed values of $d_p>0$ and $d_q>0$ are $d_p=2$ and $d_q=2$, which corresponds to $N_\mathrm{bin,dis} = 3 + n - 4 = n - 1$. 

However, the case that either $d_p$ or $d_q$ are zero should also be considered. If e.g. $d_q=0$ and $d_p\neq 0$, then the occurring binaries are $k$, $p$ and at most $n_1-d_p$ distinct binaries $p_1,...,p_{n_1-d_p}$. The corresponding number of distinct binaries is $N_\mathrm{bin,dis} = 1 + (1 + n_1-d_p) = 2 + n_1 - d_p$. The maximum $N_\mathrm{bin,dis}$ occurs when $n_1=n$ and $d_p=2$ (again, terms corresponding to $d_p=1$ vanish due to equation~\ref{eq:mBn1}), giving $N_\mathrm{bin,dis} = 2 + n - 2 = n$. Evidently, the same result applies when choosing $d_p=0$ and $d_q\neq 0$.

We conclude that the individual terms in the Hamiltonian depend on at most $n$ different binaries. Because `singlets' (terms depending on just one binary) are not allowed (in addition to binary $k$, there will always be another binary in either $p$ or $q$ families), the maximum number of types of terms (i.e. binary pairs, triplets, quadlets, etc.) is $n-1$.

\paragraph*{More than one binary with the highest level in either child of binary $k$. } As mentioned above, it is possible that in either child of binary $k$, there is no unique binary with the highest level. In this case, let $v_1,...,v_s$ denote the $s$ binaries with the common highest level in either of the two children of binary $k$. When summing over $i\in \{k.\mathrm{C}\}$, $B_{ijv_l} = 0$ for binary $v_l$ unless $i\in \{v_l.\mathrm{C}\}$. However, when summing $i\in \{v_l.\mathrm{C}\}$, the other $B_{ijv_1},...,B_{ijv_s}$ (excluding $B_{ijv_l}$) are zero, because, by definition, the $v$-binaries are on the same level. Therefore, if there is more than one binary with the highest level in either child of binary $k$, then the corresponding terms vanish. This is taken into equation~(\ref{eq:Sn_gen1}) with the factor $f(n_1,n_2,d_p,d_q)$ (cf. equation~\ref{eq:f_def}).

\paragraph*{Pairwise binary terms.} Equation~(\ref{eq:Sn_gen1}) can be made more explicit for the pairwise binary terms. In that equation, the pairwise binary terms correspond to (1) $n_1=n$ ($n_2=0$), $d_p=n_1=n$ and $d_q=0$, and (2) $n_1=0$ ($n_2=n$), $d_q=n_2=n$ and $d_p=0$. In these cases, $n_1-d_p=0$ and $n_2-d_q=0$, and there are no summations of the binaries $p_1,...,p_{n_1-d_p}$ and $q_1,...,q_{n_2-d_q}$. Also, in case (1), there is no summation over the $q$-binaries; in case (2), there is no summation over the $p$-binaries. Furthermore, the number of permutations is $\mathcal{P}^{n,n_1}_{d_p;d_q} = n!/n!=1$, and the products in the curly brackets evaluate to unity. The resulting expression is

\begin{align}
\label{eq:S_n_pair_gen0}
\nonumber S'_{n;2} &= (-1)^{n+1} \sum_{k \in \mathrm{B}} \frac{G}{r_k} \left [ 
    \sum_{\substack{p \in \mathrm{B} \\ p \in \{ k.\mathrm{C1} \} }} c_n \beta_n ( \underbrace{\ve{r}_p, ...,\,  \ve{r}_p}_{n \times}, \ve{r}_k ) \mathcal{M}_{p;k.\mathrm{CS}(p)}^{(n)} \mathcal{M}_{k.\mathrm{CS}(p);p}^{(0)}
    + \sum_{\substack{q \in \mathrm{B} \\ q \in \{ k.\mathrm{C2} \} }} c_n \beta_n ( \underbrace{\ve{r}_q, ...,\,  \ve{r}_q}_{n \times}, \ve{r}_k ) \mathcal{M}_{q;k.\mathrm{CS}(q)}^{(n)} \mathcal{M}_{k.\mathrm{CS}(q);q}^{(0)} \right ] \\
&= \sum_{k \in \mathrm{B}} \frac{G}{r_k} \sum_{\substack{p \in \mathrm{B} \\ p \in \{ k.\mathrm{C} \} }} c_n \beta_n ( \underbrace{\ve{r}_p, ...,\,  \ve{r}_p}_{n \times}, \ve{r}_k ) \mathcal{M}_{p;k.\mathrm{CS}(p)}^{(n)} \mathcal{M}_{k.\mathrm{CS}(p);p}^{(0)}
\end{align}
where in the second line, we changed the variable $q$ in the second term to $p$, and combined the two separate summations over child 1 and child 2 of $k$ into one summation over all children. Note that $\mathcal{M}_{k.\mathrm{CS}(p);p}^{(0)}$ simply evaluates to $M_{k.\mathrm{CS}(p)}$ (cf. equation~\ref{eq:mBn0}). This gives
\begin{align}
\label{eq:S_n_pair_gen}
\nonumber S'_{n;2} &= (-1)^{n+1} \sum_{k \in \mathrm{B}} \sum_{\substack{p \in \mathrm{B} \\ p \in \{ k.\mathrm{C} \} }} \frac{M_{p.\mathrm{C1}} M_{p.\mathrm{C2}}}{M_p} \alpha(p.\mathrm{C1},k.\mathrm{C2};p)^n \frac{ M_{p.\mathrm{C2}}^{n-1} + (-1)^n M_{p.\mathrm{C1}}^{n-1}}{M_p^{n-1}} \frac{GM_{k.\mathrm{CS}(p)}}{r_k} c_n \beta_n ( \underbrace{\ve{r}_p, ...,\,  \ve{r}_p}_{n \times}, \ve{r}_k ) \\
&=  (-1)^{n+1} \sum_{k \in \mathrm{B}} \sum_{\substack{p \in \mathrm{B} \\ p \in \{ k.\mathrm{C} \} }} \frac{M_{p.\mathrm{C1}} M_{p.\mathrm{C2}}}{M_p} \alpha(p.\mathrm{C1},k.\mathrm{C2};p)^n \frac{ M_{p.\mathrm{C2}}^{n-1} + (-1)^n M_{p.\mathrm{C1}}^{n-1}}{M_p^{n-1}} \frac{GM_{k.\mathrm{CS}(p)}}{r_k} \sum_{m=0}^n \frac{ \left ( \ve{r}_p \cdot \ve{r}_k \right )^m r_p^{n-m} }{r_k^{n+m}} \mathcal{A}_m^{(n)}.
\end{align}
In the second line, we used equation~(\ref{eq:beta_n}) (recall that the $\mathcal{A}_m^{(n)}$ are integer ratio coefficients, which are the same coefficients appearing in the $n$th Legendre polynomials, cf. equations~\ref{eq:beta_n} and \ref{eq:A_j}).

\begin{table}
\begin{tabular}{ccccccl}
\toprule
$n_1$ & $n_2$ & $d_p$ & $d_q$ & $n_1-d_p$ & $n_2-d_q$ & Notes \\
\midrule
0 & 3 & 0 & 0 & 0 & 3 & Not included: $d_p=0$ and $d_q=0$ ($f=0$). \\
0 & 3 & 0 & 1 & 0 & 2 & Not included: $\mathcal{M}_{p;k.\mathrm{CS}(p)}^{(d_p)} \mathcal{M}_{q;k.\mathrm{CS}(q)}^{(d_q)} = 0$. \\
0 & 3 & 0 & 2 & 0 & 1 & Part of triplet terms (child 2 of $k$). \\
0 & 3 & 0 & 3 & 0 & 0 & Part of pairwise terms (child 2 of $k$). \\
1 & 2 & 0 & 0 & 1 & 2 & Not included: $d_p=0$ and $d_q=0$ ($f=0$). \\
1 & 2 & 0 & 1 & 1 & 1 & Not included: $d_p=0$ and $n_1-d_p>0$ ($f=0$). \\
1 & 2 & 0 & 2 & 1 & 0 & Not included: $d_p=0$ and $n_1-d_p>0$ ($f=0$). \\
1 & 2 & 1 & 0 & 0 & 2 & Not included: $d_q=0$ and $n_2-d_q>0$ ($f=0$). \\
1 & 2 & 1 & 1 & 0 & 1 & Not included: $\mathcal{M}_{p;k.\mathrm{CS}(p)}^{(d_p)} \mathcal{M}_{q;k.\mathrm{CS}(q)}^{(d_q)} = 0$. \\
1 & 2 & 1 & 2 & 0 & 0 & Not included: $\mathcal{M}_{p;k.\mathrm{CS}(p)}^{(d_p)} \mathcal{M}_{q;k.\mathrm{CS}(q)}^{(d_q)} = 0$. \\
2 & 1 & 0 & 0 & 2 & 1 & Not included: $d_p=0$ and $d_q=0$ ($f=0$). \\
2 & 1 & 0 & 1 & 2 & 0 & Not included: $d_p=0$ and $n_1-d_p>0$ ($f=0$). \\
2 & 1 & 1 & 0 & 1 & 1 & Not included: $\mathcal{M}_{p;k.\mathrm{CS}(p)}^{(d_p)} \mathcal{M}_{q;k.\mathrm{CS}(q)}^{(d_q)} = 0$. \\
2 & 1 & 1 & 1 & 1 & 0 & Not included: $\mathcal{M}_{p;k.\mathrm{CS}(p)}^{(d_p)} \mathcal{M}_{q;k.\mathrm{CS}(q)}^{(d_q)} = 0$. \\
2 & 1 & 2 & 0 & 0 & 1 & Not included: $d_q=0$ and $n_2-d_q>0$ ($f=0$). \\
2 & 1 & 2 & 1 & 0 & 0 & Not included: $\mathcal{M}_{p;k.\mathrm{CS}(p)}^{(d_p)} \mathcal{M}_{q;k.\mathrm{CS}(q)}^{(d_q)} = 0$. \\
3 & 0 & 0 & 0 & 3 & 0 & Not included: $d_p=0$ and $d_q=0$ ($f=0$). \\
3 & 0 & 1 & 0 & 2 & 0 & Not included: $\mathcal{M}_{p;k.\mathrm{CS}(p)}^{(d_p)} \mathcal{M}_{q;k.\mathrm{CS}(q)}^{(d_q)} = 0$. \\
3 & 0 & 2 & 0 & 1 & 0 & Part of triplet terms (child 1 of $k$). \\
3 & 0 & 3 & 0 & 0 & 0 & Part of pairwise terms (child 1 of $k$). \\
\bottomrule
\end{tabular}
\caption{ Combinations of $n_1$, $n_2$, $d_p$ and $d_q$ occurring in equation~(\ref{eq:Sn_gen1}) for $n=3$. For each combination, we note whether or not it is included in the summations. }
\label{table:S_gen_ex_n3}
\end{table}

\paragraph*{Demonstration of equation~(\ref{eq:Sn_gen1}).} As a demonstration, we apply equation~(\ref{eq:Sn_gen1}) to the case $n=3$ (octupole order), and show that the resulting expression is the same as the expression that was previously derived in Appendix\,\ref{app:der:ham:S3}. For $n=3$, the combinations of $n_1$, $n_2$, $d_p$ and $d_q$ are enumerated in Table~\ref{table:S_gen_ex_n3}. For each combination, we note whether or not it occurs in the summation, either because of the properties of the quantity $f(n_1,n_2,d_p,d_q)$ (cf. equation~\ref{eq:f_def}), or because of property equation~(\ref{eq:mBn1}). For the triplet terms, note that $\mathcal{P}^{n_1,n_2}_{d_p;d_q} = 3!/2! = 3$. This gives
\begin{align}
\nonumber S_3' &= \sum_{k\in \mathrm{B}} \sum_{\substack{q \in \mathrm{B} \\ q \in \{k.\mathrm{C2}\}}} \frac{G}{r_k} c_3 \beta_3(\ve{r}_q,\ve{r}_q,\ve{r}_q;\ve{r}_k) \times 1 \times \alpha(q,k.\mathrm{CS}(q);k) M_{k.\mathrm{CS}(q)} \frac{M_{q.\mathrm{C1}} M_{q.\mathrm{C2}}}{M_q} \alpha(q.\mathrm{C1},k.\mathrm{CS}(q);q) \frac{M_{q.\mathrm{C2}} - M_{q.\mathrm{C1}}}{M_q} \\
\nonumber &\quad + \sum_{k\in \mathrm{B}} \sum_{\substack{p \in \mathrm{B} \\ p \in \{k.\mathrm{C1}\}}} \frac{G}{r_k} c_3 \beta_3(\ve{r}_p,\ve{r}_p,\ve{r}_p;\ve{r}_k) \times 1 \times \alpha(p,k.\mathrm{CS}(p);k) M_{k.\mathrm{CS}(p)} \frac{M_{p.\mathrm{C1}} M_{p.\mathrm{C2}}}{M_p} \alpha(p.\mathrm{C1},k.\mathrm{CS}(p);p) \frac{M_{p.\mathrm{C2}} - M_{p.\mathrm{C1}}}{M_p} \\
\nonumber &\quad  + \sum_{k\in \mathrm{B}} \sum_{\substack{q,q_1 \in \mathrm{B} \\ q,q_1 \in \{k.\mathrm{C2}\} \\ q\in \{q_1.\mathrm{C}\} }} \frac{G}{r_k} c_3 \beta_3(\ve{r}_q,\ve{r}_q,\ve{r}_{q_1};\ve{r}_k) \times 3 \times \frac{\alpha(q,k.\mathrm{CS}(q);k) M_{q_1.\mathrm{CS}(q)}}{M_{q_1}} \alpha(q,k.\mathrm{CS}(q);k) M_{k.\mathrm{CS}(q)} \frac{M_{q.\mathrm{C1}} M_{q.\mathrm{C2}}}{M_q} \\
&\quad  + \sum_{k\in \mathrm{B}} \sum_{\substack{p,p_1 \in \mathrm{B} \\ p,p_1 \in \{k.\mathrm{C2}\} \\ p\in \{p_1.\mathrm{C}\} }} \frac{G}{r_k} c_3 \beta_3(\ve{r}_p,\ve{r}_p,\ve{r}_{p_1};\ve{r}_k) \times 3 \times \frac{\alpha(p,k.\mathrm{CS}(p);k) M_{p_1.\mathrm{CS}(p)}}{M_{p_1}} \alpha(p,k.\mathrm{CS}(p);k) M_{k.\mathrm{CS}(p)} \frac{M_{p.\mathrm{C1}} M_{p.\mathrm{C2}}}{M_p}.
\end{align}
Taking together the summations over $p$ and $q$ into a single summation of $p$ over both children of $k$ for the binary pair and triplet terms, relabelling $q_1 = u$ and $p_1=u$ and substituting the expressions for $c_3 \beta_3$, we recover equation~(\ref{eq:S3_comp_man}).

\subsubsection{Summary}
\label{app:der:ham:sum}
In the previous appendices, we explicitly carried out the rewriting procedure for all terms up and including octupole order, and including the binary pair terms at hexadecupole and dotriacontupole order. We also derived a general, more implicit expression valid for any order $n$ and including all binary interactions (cf. Appendix\,\ref{app:der:ham:gen}). For the purposes of giving a useful overview, we repeat the explicit expressions here. The simplified explicit expression reads
\begin{align}
\label{eq:H2}
H &= \frac{1}{2} \sum_{k\in\mathrm{B}} \left [ \frac{  M_{k.\mathrm{C1}} M_{k.\mathrm{C2}}}{M_k} \left (\dot{\ve{r}}_k \cdot \dot{\ve{r}}_k\right )
- \frac{GM_{k.\mathrm{C1}} M_{k.\mathrm{C2}}}{r_k} \right ] + S_2' + S_3' + S_4' + S_5' + ...
\end{align}
where
\begin{align}
\label{eq:S2_rewr}
S_2' &= -\frac{1}{2} \sum_{k \in \mathrm{B}}  \sum_{\substack{p \in \mathrm{B} \\ p \in \{k.\mathrm{C} \} }}  \frac{M_{p.\mathrm{C1}} M_{p.\mathrm{C2}}}{M_p} \frac{GM_{k.\mathrm{CS}(p)}}{r_k} \left ( \frac{r_p}{r_k} \right )^2 \left [ 3 \left ( \unit{r}_p \cdot \unit{r}_k \right )^2 - 1 \right];
\end{align}
\begin{align}
\label{eq:S3_rewr}
\nonumber S_3' &= \frac{1}{2} \sum_{k \in \mathrm{B}} \sum_{\substack{p \in \mathrm{B} \\ p\in\{k.\mathrm{C} \}}} \frac{M_{p.\mathrm{C1}} M_{p.\mathrm{C2}} }{M_p} \alpha(p.\mathrm{C1},k.\mathrm{CS}(p);p)  \frac{ M_{p.\mathrm{C2}} - M_{p.\mathrm{C1}}}{M_p} \alpha(p,k.\mathrm{CS}(p);k) \frac{G M_{k.\mathrm{CS}(p)}}{r_k} \left ( \frac{r_p}{r_k} \right )^3  \left [5 \left ( \unit{r}_p \cdot \unit{r}_k \right )^3 - 3  \left ( \unit{r}_p \cdot \unit{r}_k \right ) \right ] \\
&\quad +  \frac{3}{2} \sum_{k \in \mathrm{B}} \sum_{\substack{p \in \mathrm{B} \\ p\in\{k.\mathrm{C} \}}} \sum_{\substack{u \in \mathrm{B} \\ u\in\{k.\mathrm{C} \} \\ p \in \{u.\mathrm{C} \} }}  \frac{M_{p.\mathrm{C1}} M_{p.\mathrm{C2}}}{M_p} \alpha(p,k.\mathrm{CS}(p);k) \frac{G M_{k.\mathrm{CS}(p)}}{r_k} \frac{\alpha(p,k.\mathrm{CS}(p);u) M_{u.\mathrm{CS}(p)}}{M_u} \left ( \frac{r_p}{r_k} \right )^2 \left ( \frac{r_u}{r_k} \right ) \\
&\quad \quad \times \left [ 5 \left ( \unit{r}_p \cdot \unit{r}_k \right )^2 \left ( \unit{r}_u \cdot \unit{r}_k \right ) - 2 \left ( \unit{r}_p \cdot \unit{r}_k \right ) \left ( \unit{r}_p \cdot \unit{r}_u \right ) - \left ( \unit{r}_u \cdot \unit{r}_k \right ) \right ];
\end{align}
\begin{align}
\label{eq:S4_rewr}
S_4' &= -\frac{1}{8} \sum_{k \in \mathrm{B}} \sum_{\substack{p \in \mathrm{B} \\ p\in\{k.\mathrm{C} \}}} \frac{M_{p.\mathrm{C1}} M_{p.\mathrm{C2}} }{M_p} \frac{M_{p.\mathrm{C1}}^2 - M_{p.\mathrm{C1}} M_{p.\mathrm{C2}} + M_{p.\mathrm{C2}}^2}{M_p^2} \frac{GM_{k.\mathrm{CS}(p)}}{r_k} \left ( \frac{r_p}{r_k} \right )^4 \left [ 35 \left ( \unit{r}_p \cdot \unit{r}_k \right )^4- 30 \left ( \unit{r}_p \cdot \unit{r}_k \right )^2 + 3 \right ]+ S'_{4;3} + S'_{4;4};
\end{align}
\begin{align}
\label{eq:S5_rewr}
\nonumber S_5' &= \frac{1}{8} \sum_{k \in \mathrm{B}} \sum_{\substack{p \in \mathrm{B} \\ p\in\{k.\mathrm{C} \}}} \frac{M_{p.\mathrm{C1}} M_{p.\mathrm{C2}} }{M_p} \, \alpha(p.\mathrm{C1},p.\mathrm{CS}(p);p) \frac{M_{p.\mathrm{C2}} - M_{p.\mathrm{C1}}}{M_p} \frac{M_{p.\mathrm{C1}}^2 + M_{p.\mathrm{C2}}^2}{M_p^2} \, \alpha(p,k.\mathrm{CS}(p);k) \frac{GM_{k.\mathrm{CS}(p)}}{r_k} \left ( \frac{r_p}{r_k} \right )^5 \\
&\quad \times \left [ 63 \left ( \unit{r}_p \cdot \unit{r}_k \right )^5 - 70 \left ( \unit{r}_p \cdot \unit{r}_k \right )^3 + 15 \left ( \unit{r}_p \cdot \unit{r}_k \right ) \right ] + S'_{5;3} + S'_{5;4} + S'_{5;5}.
\end{align}
Here, the dots in equation~(\ref{eq:H2}) denote higher order terms. Note that in equation~(\ref{eq:H2}), no approximations are made apart from the expansions in the binary separation ratios. Below, in Appendix\,\ref{app:der:av}, we orbit average equation~(\ref{eq:H2}).

\subsection{Orbit averaging}
\label{app:der:av}
We average the expanded Hamiltonian over the $N-1$ binary orbits, assuming that in each orbit, the motion on suborbital time-scales can be well approximated by a bound Kepler orbit. We define the orbit-averaged Hamiltonian as
\begin{align}
\overline{H} \equiv \frac{1}{(2\pi)^{N-1}} \underbrace{\int_0^{2\pi} \cdots \int_0^{2\pi}}_{N-1} H \prod_{k\in\mathrm{B}} \mathrm{d} l_k,
\label{eq:app:avdef}
\end{align}
where $l_k$ is the mean anomaly of orbit $k$. We express the angular momenta and orientations of each binary in terms of the triad of perpendicular orbital state vectors $(\boldsymbol{j}_k,\boldsymbol{e}_k,\boldsymbol{q}_k)$, where $\boldsymbol{q}_k \equiv \boldsymbol{j}_k \times \boldsymbol{e}_k$. Here, $\boldsymbol{j}_k$ is the specific angular momentum vector with magnitude $j_k=\sqrt{1-e_k^2 }$, where $e_k$, $0\leq e_k<1$, is the orbital eccentricity. The specific orbital angular momentum vector $\boldsymbol{j}_k$ is related to the orbital angular momentum vector $\boldsymbol{h}_k \equiv (M_{k.\mathrm{C1}} M_{k.\mathrm{C2}}/M_k)\, \boldsymbol{r}_k \times \dot{\boldsymbol{r}}_k$ via $\boldsymbol{h}_k = \Lambda_k \, \boldsymbol{j}_k$, where $\Lambda_k$ is the angular momentum of a circular orbit,
\begin{align}
\label{eq:Lambda}
\Lambda_k = M_{k.\mathrm{C1}} M_{k.\mathrm{C2}} \, \sqrt{ \frac{G a_k}{M_k}},
\end{align}
with $a_k$ the semimajor axis. Furthermore, $\boldsymbol{e}_k$ is the eccentricity or Laplace-Runge-Lenz vector that is aligned with the major axis and that has magnitude $e_k$. 

In terms of $\boldsymbol{e}_k$ and $\boldsymbol{q}_k$ and the orbital phase (we use either the true anomaly $f_k$ or the eccentric anomaly $\varepsilon_k$ as integration variables), the dot product of $\boldsymbol{r}_k$ with some other (constant) vector $\boldsymbol{u}$ is given by two equivalent expressions,
\begin{align}
\label{eq:app:r_k_dot_r_u}
\nonumber \boldsymbol{r}_k \cdot \boldsymbol{u} &= \frac{a_k\left(1-e_k^2\right)}{1+e_k \cos(f_k)} \left [ \cos(f_k) \, \left( \unit{e}_k \cdot \boldsymbol{u} \right ) + \sin(f_k) \, \left ( \unit{q}_k \cdot \boldsymbol{u} \right ) \right ] \\
&= a_k \left [ \cos(\varepsilon_k) - e_k \right ] \, \left ( \unit{e}_k \cdot \boldsymbol{u} \right ) + a_k \sqrt{1-e_k^2} \sin(\varepsilon_k) \, \left ( \unit{q}_k \cdot \boldsymbol{u} \right ).
\end{align}
The simplified Hamiltonian, equation~(\ref{eq:H2}), is expressed in terms of these dot products. Furthermore, the separations are given by the well-known expressions
\begin{align}
\label{eq:app:r_p}
r_k &= \frac{a_k\left(1-e_k^2\right)}{1+e_k \cos(f_k)} = a_k \left [ 1 - e_k \cos(\varepsilon_k) \right ].
\end{align}

In equation~(\ref{eq:app:avdef}), it is convenient to use the true anomaly $f_k$ as the integration variable whenever $r_k$ appears with a high negative power, and, otherwise, the eccentric anomaly $\varepsilon_k$. The relevant Jacobians are given by
\begin{align}
\label{eq:app:av_jac}
\mathrm{d} l_k = \frac{1}{\sqrt{1-e_k^2}} \left ( \frac{r_k}{a_k} \right )^2 \, \mathrm{d} f_k = \left ( \frac{r_k}{a_k} \right ) \, \mathrm{d} \varepsilon_k.
\end{align}
In order to express the orbit-averaged quantities in terms of dot products of $\boldsymbol{e}_k$ and $\boldsymbol{j}_k$ with those of other orbits (i.e. to remove any reference to the auxiliary vector $\boldsymbol{q}_k$), we repeatedly use the following vector identity for the scalar product of two scalar triple products,
\begin{align}
\label{eq:vec_id}
\left ( \unit{q}_k \cdot \boldsymbol{u} \right ) \left ( \unit{q}_k \cdot \boldsymbol{w} \right ) &= \left [ \left ( \unit{j}_k \times \unit{e}_k \right ) \cdot \boldsymbol{u} \right ] \left [ \left ( \unit{j}_k \times \unit{e}_k \right ) \cdot \boldsymbol{w} \right ] = \boldsymbol{u} \cdot \boldsymbol{w} - \left ( \unit{e}_k \cdot \boldsymbol{u} \right ) \left ( \unit{e}_k \cdot \boldsymbol{w} \right ) - \left ( \unit{j}_k \cdot \boldsymbol{u} \right ) \left ( \unit{j}_k \cdot \boldsymbol{w} \right ),
\end{align}
where $\boldsymbol{u}$ and $\boldsymbol{w}$ are two arbitrary vectors. 

For example, the orbit average of the term $(\boldsymbol{r}_p \cdot \boldsymbol{r}_k)^2/r_k^5$ that appears in the quadrupole-order term of the Hamiltonian (cf. equation~\ref{eq:H2}) is given by
\begin{subequations}
\begin{align}
\left \langle \frac{ \left (\boldsymbol{r}_p \cdot \boldsymbol{r}_k \right )^2}{r_k^5} \right \rangle_p &= \frac{a_p^2}{2 r_k^5} \left [ j_p^2 r_k^2 - \left ( \ve{j}_p \cdot \ve{r}_k \right )^2 + 5 \left ( \ve{e}_p \cdot \ve{r}_k \right )^2 \right ]; \\
\left \langle \left \langle \frac{ \left (\boldsymbol{r}_p \cdot \boldsymbol{r}_k \right )^2}{r_k^5} \right \rangle_p \right \rangle_k &= \frac{a_p^2}{4a_k^3 j_k^5} \left [ \left ( 1 + 4 e_p^2\right ) j_k^2 + \left ( \ve{j}_p \cdot \ve{j}_k \right )^2 - 5 \left ( \ve{e}_p \cdot \ve{j}_k \right )^2 \right ].
\end{align}
\end{subequations}
Note that this expression is well defined in the limit $e_p\rightarrow 0$, in which case $\ve{e}_p \rightarrow \ve{0}$.

Formally, the averaging of the Hamiltonian over the binary orbits as described by equation~(\ref{eq:app:avdef}) is not a canonical transformation. However, applying the Von Zeipel transformation technique to the unaveraged Hamiltonian \citep{1959AJ.....64..378B}, a canonical transformation can be found that eliminates the short-period terms $l_k$ from the Hamiltonian (cf. appendix A2 of \citealt{2013MNRAS.431.2155N}; the derivation presented there is straightforwardly extended to $N-1$, rather than two short-period variables). This transformation leads to a transformed Hamiltonian that is equivalent to equation~(\ref{eq:app:avdef}). The transformed coordinates $\boldsymbol{e}_k^*$ and $\boldsymbol{q}_k^*$ differ from the original ones, $\boldsymbol{e}_k$ and $\boldsymbol{q}_k$. However, as noted by \citet{2013MNRAS.431.2155N}, the untransformed and the transformed coordinates differ by order $x^2$. Therefore, this difference can usually be neglected. This is borne out by our tests with other methods (cf. \S\,\ref{sect:multiplanet}).

\subsubsection{Binding energies}
\label{app:der:av:BE}
If Kepler orbits are assumed, the first two summation terms in equation~(\ref{eq:H2}) immediately reduce to the sum of the binary binding energies, without having to explicitly average over the orbits. This directly follows from the canonical expressions for the Kepler binary separation (equation~\ref{eq:app:r_p}) and the squared velocity,
\begin{align}
\dot{\ve{r}}_k\cdot \dot{\ve{r}}_k = \frac{G M_k}{a_k\left(1-e_k^2\right)} \left [1 + 2 e_k \cos(f_k) + e_k^2 \right].
\end{align}
Substituting these expressions into the term in equation~(\ref{eq:H2}) with the single summation of $k \in \mathrm{B}$, we find
\begin{align}
\label{eq:bin_energy_simple}
\nonumber & \sum_{k\in\mathrm{B}} \left [  \frac{1}{2}  \frac{  M_{k.\mathrm{C1}} M_{k.\mathrm{C2}}}{M_k} \left (\dot{\ve{r}}_k \cdot \dot{\ve{r}}_k\right )
- \frac{GM_{k.\mathrm{C1}} M_{k.\mathrm{C2}}}{r_k} \right ] \\
&= \sum_{k\in\mathrm{B} }M_{k.\mathrm{C1}} M_{k.\mathrm{C2}} \left \{ \frac{1}{2} \frac{ GM_k}{M_k a_k \left(1-e_k^2\right)} \left [1+2e_k \cos(f_k) + e_k^2 \right ] - \frac{G}{a_k\left(1-e_k^2\right)} \left [1+e_k \cos(f_k) \right ] \right \} = \sum_{k\in \mathrm{B}} \frac{G M_{k.\mathrm{C1}} M_{k.\mathrm{C2}}}{-2a_k}.
\end{align}
Each of the individual terms in the latter expression is the Keplerian binary binding energy $E_k<0$. Because they only depend on the semimajor axes (and not on $\ve{e}_k$ and $\ve{j}_k$), equation~(\ref{eq:bin_energy_simple}) does not give rise to secular changes of the orbits (cf. equations~\ref{eq:EOM}).

\subsubsection{Higher order terms}
\label{app:der:av:HO}
Secular changes arise from the higher-order orbit-averaged terms $\overline{S}_n'$ with $n\geq 2$, which depend on $\ve{e}_k$ and $\ve{j}_k$. These terms combined form the disturbing function of the system. Given their mathematical equivalence to the corresponding terms in the hierarchical three-body problem, orbit-averaged expressions for the binary pair terms in the Hamiltonian up and including octupole order can be directly taken from previous studies (e.g. \citealt{2014ApJ...789..110B}, where the same vector formalism was adopted). The orbit averaged octupole-order term depending on three binaries and the hexadecupole-order term depending on two binaries were derived previously by \citet{2015MNRAS.449.4221H}. In the latter work, these terms were expressed in terms of unit vectors $\hat{\ve{e}}_k$, which are ill-defined in the limit $e_k \rightarrow 0$. Here, we have rederived the orbit average of all terms up and including dotriacontupole order as given explicitly in equation~(\ref{eq:H2}), where for the hexadecupole and dotriacontupole-order terms, we only included the binary pair terms. We have expressed the result in terms of the non-unit vectors $\ve{e}_k$, which are well defined in the limit $e_k \rightarrow 0$, i.e., trivially, $\ve{e}_k \rightarrow \ve{0}$. We have verified the expressions of \citet{2015MNRAS.449.4221H} numerically for cases with non-zero eccentricities. 

After averaging over the orbits, the Hamiltonian is no longer a function of the phases of the orbits $l_k$. Consequently, the sign quantities $\alpha$ must be set to unity. A subtlety arises with the binary pair octupole- and dotriacontupole-order terms (or, generally odd order $n$), which contain the factor
\begin{align}
\label{eq:mass_diff} \alpha(p.\mathrm{C1},k.\mathrm{CS}(p);p) \left ( M_{p.\mathrm{C2}} - M_{p.\mathrm{C1}} \right ) = \alpha(p.\mathrm{C2},k.\mathrm{CS}(p);p) \left ( M_{p.\mathrm{C1}} - M_{p.\mathrm{C2}} \right ).
\end{align}
After averaging, interchanging the children of $p$ should not change the expression. Therefore, the expression equation~(\ref{eq:mass_diff}) should be replaced by the absolute mass difference, i.e.
\begin{align}
\alpha(p.\mathrm{C1},k.\mathrm{CS}(p);p) \left ( M_{p.\mathrm{C2}} - M_{p.\mathrm{C1}} \right ) \rightarrow \left | M_{p.\mathrm{C1}} - M_{p.\mathrm{C2}} \right |.
\end{align}

The explicit orbit-averaged Hamiltonian, to dotriacontupole order, reads
\begin{align}
\nonumber \overline{H} &= \sum_{k\in \mathrm{B}} \left ( \frac{G M_{k.\mathrm{C1}} M_{k.\mathrm{C2}}}{-2a_k} \right ) + \overline{S}_2' + \overline{S}_3' + \overline{S}_4' + \overline{S}_5' + ...,
\end{align}
where
\begin{align}
\label{eq:S2_av}
\overline{S}_2' &=  \sum_{k\in \mathrm{B}} \sum_{ \substack{ p \in \mathrm{B} \\ p \in \{k.\mathrm{C} \}}} \frac{M_{p.\mathrm{C1}} M_{p.\mathrm{C2}}}{M_p}
\frac{G M_{k.\mathrm{CS}(p)}}{a_k} \left ( \frac{a_p}{a_k} \right )^2 \frac{1}{8 j_k^5} \left[ \left( 1-6 e_p^2\right )j_k^2 + 15 \left ( \ve{e}_p \cdot \ve{j}_k \right )^2 - 3 \left(\ve{j}_p\cdot \ve{j}_k\right)^2 \right ];
\end{align}
\begin{align}
\label{eq:S3_av}
\nonumber \overline{S}_3' &= \sum_{k\in \mathrm{B}} \sum_{ \substack{ p \in \mathrm{B} \\ p \in \{k.\mathrm{C} \}}} \frac{M_{p.\mathrm{C1}} M_{p.\mathrm{C2}}}{M_p} \frac{\left | M_{p.\mathrm{C1}} - M_{p.\mathrm{C2}} \right | }{M_p} \frac{G M_{k.\mathrm{CS}(p)}}{a_k} \left ( \frac{a_p}{a_k} \right )^3 \frac{15}{64 j_k^7} \left \{ 10 \, (\ve{e}_p \cdot \ve{j}_k) (\ve{e}_k \cdot \ve{j}_p) (\ve{j}_p \cdot \ve{j}_k) \right. \\
\nonumber&\quad \quad \left. - (\ve{e}_p \cdot \ve{e}_k) \left [ \left(1-8 e_p^2 \right) j_k^2 + 35 (\ve{e}_p \cdot \ve{j}_k)^2 - 5 (\ve{j}_p \cdot \ve{j}_k)^2 \right ] \right \} \\
\nonumber&\quad + \sum_{k \in \mathrm{B}} \sum_{\substack{p \in \mathrm{B} \\ p\in\{k.\mathrm{C} \}}} \sum_{\substack{u \in \mathrm{B} \\ u\in\{k.\mathrm{C} \} \\ p \in \{u.\mathrm{C} \} }} \frac{M_{p.\mathrm{C1}} M_{p.\mathrm{C2}}}{M_p} \frac{M_{u.\mathrm{CS}(p)}}{M_u} \frac{G M_{k.\mathrm{CS}(p)}}{a_k} \left ( \frac{a_p}{a_k} \right )^2 \left ( \frac{a_u}{a_k} \right ) \frac{9}{32 j_k^7} \left \{ 10 (\ve{e}_p \cdot \ve{e}_u) (\ve{e}_p \cdot \ve{e}_k) j_k^2 \right. \\
\nonumber &\quad \quad \left. - 50 (\ve{e}_p \cdot \ve{e}_k) (\ve{e}_p \cdot \ve{j}_k) (\ve{e}_u \cdot \ve{j}_k) - 2 (\ve{e}_k \cdot \ve{j}_p) (\ve{e}_u \cdot \ve{j}_p) j_k^2 + 10 (\ve{e}_u \cdot \ve{j}_k) (\ve{e}_k \cdot \ve{j}_p) (\ve{j}_p \cdot \ve{j}_k) \right. \\
&\quad \quad \left. - (\ve{e}_u \cdot \ve{e}_k) \left [ \left(1-6e_p^2\right)j_k^2 + 25 (\ve{e}_p \cdot \ve{j}_k)^2 - 5 (\ve{j}_p \cdot \ve{j}_k)^2 \right ] \right \};
\end{align}
\begin{align}
\label{eq:S4_av}
\nonumber \overline{S}_4' &= \sum_{k\in \mathrm{B}} \sum_{ \substack{ p \in \mathrm{B} \\ p \in \{k.\mathrm{C} \}}}\frac{M_{p.\mathrm{C1}} M_{p.\mathrm{C2}} }{M_p} \frac{M_{p.\mathrm{C1}}^2 - M_{p.\mathrm{C1}} M_{p.\mathrm{C2}} + M_{p.\mathrm{C2}}^2}{M_p^2} \frac{G M_{k.\mathrm{CS}(p)}}{a_k} \left ( \frac{a_p}{a_k} \right )^4 \frac{3}{1024 j_k^{11}} \left \{ 3 j_k^4 \left [ -6 + e_k^2 + 40 e_p^2 \left(1+8e_k^2 \right) \right. \right. \\
\nonumber &\quad \quad \left. \left. - 20 e_p^4 \left (8 + 15 e_k^2 \right ) \right ] + 420 \left ( \ve{e}_p \cdot \ve{e}_k \right )^2 j_k^2 \left [ - \left(2 + e_p^2 \right ) j_k^2 + 21 \left ( \ve{e}_p \cdot \ve{j}_k \right )^2 \right ] - 5880 j_k^2 \left(\ve{e}_p \cdot \ve{e}_k\right) \left(\ve{e}_k \cdot \ve{j}_p\right) \left(\ve{e}_p \cdot \ve{j}_k\right) \left(\ve{j}_p \cdot \ve{j}_k\right) \right. \\
\nonumber &\quad \left. + 5 \left [ 28 j_k^2 \left(1-10e_p^2 \right)\left(4+3e_k^2\right) \left(\ve{e}_p \cdot \ve{j}_k\right)^2 - 6 j_k^2 \left (8 + 6e_k^2 + e_p^2 \left(6 + 29 e_k^2 \right ) \right ) \left (\ve{j}_p \cdot \ve{j}_k \right )^2 - 12 j_k^2 \left ( \ve{e}_k \cdot \ve{j}_p \right )^2 \left (\left (1 + 13 e_p^2 \right ) j_k^2 - 7 \left ( \ve{j}_p \cdot \ve{j}_k \right )^2 \right ) \right. \right. \\
\nonumber &\quad \left. \left. + 98 j_k^2 \left ( -2 -3 e_k^2 + 4e_p^2 \left (5 + 3 e_k^2 \right ) \right ) \left ( \ve{e}_p \cdot \ve{j}_k \right )^2 - 441 \left (2 + e_k^2 \right ) \left ( \ve{e}_p \cdot \ve{j}_k \right )^4 + 42 \left (2 + e_k^2 \right) \left ( j_p^2 j_k^2 + 7 \left (\ve{e}_p \cdot \ve{j}_k \right )^2 \right ) \left ( \ve{j}_p \cdot \ve{j}_k \right )^2 \right. \right. \\
&\quad \left. \left. - 21 \left (2 + e_k^2 \right) \left ( \ve{j}_p \cdot \ve{j}_k \right )^4 \right ]\right \} + \overline{S}'_{4;3} + \overline{S}'_{4;4};
\end{align}
\begin{align}
\label{eq:S5_av}
\nonumber \overline{S}_5' &= \sum_{k\in \mathrm{B}} \sum_{ \substack{ p \in \mathrm{B} \\ p \in \{k.\mathrm{C} \}}}\frac{M_{p.\mathrm{C1}} M_{p.\mathrm{C2}} }{M_p} \frac{ \left | M_{p.\mathrm{C1}} - M_{p.\mathrm{C2}} \right |}{M_p} \frac{M_{p.\mathrm{C1}}^2  + M_{p.\mathrm{C2}}^2}{M_p^2} \frac{G M_{k.\mathrm{CS}(p)}}{a_k} \left ( \frac{a_p}{a_k} \right )^5 \frac{105}{4096 j_k^{13}} \left \{ 3024 \left ( \ve{e}_p \cdot \ve{e}_k \right)^2 \left ( \ve{e}_p \cdot \ve{j}_k \right) \left ( \ve{e}_k \cdot \ve{j}_p \right) \left ( \ve{j}_p \cdot \ve{j}_k \right) j_k^2 \right. \\
\nonumber &\quad \left. - 28 j_k^2 \left (\ve{e}_p \cdot \ve{e}_k \right )^3 \left [ \left ( -26 + 15 e_p^2 \right ) j_k^2 + 18 \left (\ve{j}_p \cdot \ve{j}_k \right )^2 + 99 \left ( \ve{e}_p \cdot \ve{j}_k \right )^2 \right ] - 28 \left ( \ve{e}_p \cdot \ve{j}_k \right) \left ( \ve{e}_k \cdot \ve{j}_p \right) \left ( \ve{j}_p \cdot \ve{j}_k \right) \left [ \left (1-4e_p^2 \right ) \left (8 + e_k^2 \right )j_k^2 \right. \right. \\
\nonumber &\quad \quad \left. \left. + 9 \left(8 + 3 e_k^2 \right) \left ( \ve{e}_p \cdot \ve{j}_k \right )^2 + 6 j_k^2 \left ( \ve{e}_k \cdot \ve{j}_p \right )^2 - 3 \left ( 8 + 3 e_k^2 \right ) \left ( \ve{j}_p \cdot \ve{j}_k \right )^2 \right ] - \left ( \ve{e}_p \cdot \ve{e}_k \right ) \left [ \left \{-8 + e_k^2 + e_p^2 \left (64 + 748 e_k^2 \right ) - 4 e_p^4 \left (80 + 179 e_k^2 \right) \right \} j_k^4 \right. \right. \\
\nonumber &\quad \quad \left. \left. -693 \left (8 + 3 e_k^2 \right ) \left ( \ve{e}_p \cdot \ve{j}_k \right )^4 + 42 \left ( \ve{e}_p \cdot \ve{j}_k \right )^2 \left \{ \left (-8 - 19e_k^2 + 6 e_p^2 \left (16 + 5 e_k^2 \right ) \right ) j_k^2 + 9 \left (8 + 3e_k^2 \right) \left ( \ve{j}_p \cdot \ve{j}_k \right )^2 \right \} \right. \right. \\
\nonumber &\quad \quad \left. \left. + 14 j_k^2 \left (8 + e_k^2 - 2 e_p^2 \left (16 + 29 e_k^2 \right ) \right ) \left ( \ve{j}_p \cdot \ve{j}_k \right )^2 - 21 \left (8 + 3 e_k^2 \right ) \left ( \ve{j}_p \cdot \ve{j}_k \right )^4 - 28 \left ( \ve{e}_k \cdot \ve{j}_p \right )^2 j_k^2 \left ( \left (1 + 23 e_p^2 \right ) j_k^2 - 9 \left ( \ve{j}_p \cdot \ve{j}_k \right )^2 \right ) \right ] \right \} \\
&+ \overline{S}'_{5;3} + \overline{S}'_{5;4} + \overline{S}'_{5;5}.
\end{align}

\subsubsection{General formal expressions for the orbit-averaged binary pair terms at order $n$}
\label{app:der:av:gen_pair}
For the binary pair terms, we show how the orbit-averaged integrals at any order can be evaluated using complex integration techniques. We note that \citet{2013MNRAS.435.2187M} derived the equivalent orbit-averaged terms for hierarchical three-body systems in terms of Hansen coefficients which are also valid for arbitrary order, but assuming coplanar orbits.

From equation~(\ref{eq:S_n_pair_gen}), the terms depending on binary separations in the unaveraged binary pair Hamiltonian at order $n$ are given by
\begin{align}
\label{eq:S_n_pair_gen_form}
\sum_{m=0}^n \frac{ \left ( \ve{r}_p \cdot \ve{r}_k \right )^m r_p^{n-m} }{r_k^{n+m+1}} \mathcal{A}_m^{(n)},
\end{align}
where binary $p \in \{k.\mathrm{C}\}$ (i.e. $r_p\ll r_k$), and $\mathcal{A}_m^{(n)}$ are coefficients that follow from the Taylor expansion of the potential energy and which are the same as the coefficients appearing in the Legendre polynomials (cf. Appendix\,\ref{app:der:pot}, in particular equations~\ref{eq:beta_n} and \ref{eq:A_j}). Generally, $\mathcal{A}_m^{(n)}=0$ if $n$ is even and $m$ is odd, or if $n$ is odd and $m$ is even. 

We first average over the `inner' orbit $\ve{r}_p$ with constant $\ve{r}_k$ using the eccentric anomaly $\varepsilon_p$, giving (cf. equations~\ref{eq:app:avdef}, \ref{eq:app:r_k_dot_r_u}, \ref{eq:app:r_p} and \ref{eq:app:av_jac})
\begin{align}
\left \langle \left ( \ve{r}_p \cdot \ve{r}_k \right )^m r_p^{n-m} \right \rangle_p = \frac{a_p^n}{2 \pi} \int_0^{2 \pi} \mathrm{d} \varepsilon_p \left [1 - e_p \cos(\varepsilon_p) \right ]^{n-m+1} \left [ \left \{\cos(\varepsilon_p)-e_p\right \} \left ( \unit{e}_p \cdot \ve{r}_k \right ) + \sqrt{1-e_p^2} \sin(\varepsilon_p) \left ( \unit{q}_p \cdot \ve{r}_k \right ) \right ]^m.
\end{align}
Subsequently, we transform to the complex variable $z=\exp(\complexi \varepsilon_p)$ where $\complexi = \sqrt{-1}$, giving
\begin{align}
\label{eq:int_gen1}
\nonumber \left \langle \left ( \ve{r}_p \cdot \ve{r}_k \right )^m r_p^{n-m} \right \rangle_p &= \frac{a_p^n}{2\pi} \oint_C \frac{\mathrm{d} z}{\complexi z} \left [ 1 - \frac{e_p}{2} \left (z+z^{-1} \right ) \right ]^{n-m+1} \left [ \frac{1}{2} \left(z + z^{-1} - 2e_p \right ) \left ( \unit{e}_p \cdot \ve{r}_k \right ) - \frac{1}{2} \complexi \sqrt{1-e_p^2} \left (z - z^{-1} \right ) \left ( \unit{q}_p \cdot \ve{r}_k \right ) \right ]^m \\
&= \frac{a_p^n}{2\pi \complexi} \left ( \frac{1}{2} \right )^{n+1} \oint_C \frac{\mathrm{d} z}{z^{n+2}} \left ( -e_p z^2 +2z - e_p  \right )^{n-m+1} \left ( \frac{\widetilde{C}_1}{e_p} z^2 + \widetilde{C}_3 z + \frac{\widetilde{C}_2}{e_p} \right )^m.
\end{align}
Here, the integration path $C$ is the unit circle $z=\exp(\complexi \varepsilon_p)$, $\varepsilon_p \in [0,2\pi)$ in the complex plane, and the constants $\widetilde{C}_1$, $\widetilde{C}_2$ and $\widetilde{C}_3$ are defined by
\begin{subequations}
\label{eq:tildeC_def}
\begin{align}
\widetilde{C}_1 &\equiv \left ( \ve{e}_p \cdot \ve{r}_k \right ) - \complexi \sqrt{1-e_p^2} \, e_p \left ( \unit{q}_p \cdot \ve{r}_k \right ) =  \left ( \ve{e}_p \cdot \ve{r}_k \right ) - \complexi \sqrt{1-e_p^2} (\pm1 ) \left [ e_p^2 r_k^2 - \left ( \ve{e}_p \cdot \ve{r}_k \right )^2 - \frac{e_p^2}{1-e_p^2} \left ( \ve{j}_p \cdot \ve{r}_k \right)^2 \right ]^{1/2}; \\
\widetilde{C}_2 &\equiv \left ( \ve{e}_p \cdot \ve{r}_k \right ) + \complexi \sqrt{1-e_p^2} \, e_p \left ( \unit{q}_p \cdot \ve{r}_k \right ) = \left ( \ve{e}_p \cdot \ve{r}_k \right ) + \complexi \sqrt{1-e_p^2} (\pm1 ) \left [ e_p^2 r_k^2 - \left ( \ve{e}_p \cdot \ve{r}_k \right )^2 - \frac{e_p^2}{1-e_p^2} \left ( \ve{j}_p \cdot \ve{r}_k \right)^2 \right ]^{1/2}; \\
\widetilde{C}_3 &\equiv -2 \left ( \ve{e}_p \cdot \ve{r}_k \right ),
\end{align}
\end{subequations}
where we used the vector identity equation~(\ref{eq:vec_id}) to eliminate explicit reference to $\unit{q}_p$.

Taking into account that $0\leq m \leq n$ and $n>0$, the integrand in equation~(\ref{eq:int_gen1}) is a meromorphic function of $z$ with a single pole at $z=0$. Using a standard Taylor series, the integrand can be written as
\begin{align}
\frac{1}{z^{n+2}} \left ( -e_p z^2 +2z - e_p  \right )^{n-m+1} \left ( \frac{\widetilde{C}_1}{e_p} z^2 + \widetilde{C}_3 z + \frac{\widetilde{C}_2}{e_p} \right )^m &= \frac{1}{z^{n+2}} \sum_{j\geq 0} a_j z^j = \sum_{j\geq 0} a_j \frac{1}{z^{n+2-j}},
\end{align}
where the coefficients $a_j$ are given by
\begin{align}
a_j = \frac{1}{j!} \frac{ \mathrm{d}^j}{\mathrm{d} z^j} \left [  \left ( -e_p z^2 +2z - e_p  \right )^{n-m+1} \left ( \frac{\widetilde{C}_1}{e_p} z^2 + \widetilde{C}_3 z + \frac{\widetilde{C}_2}{e_p} \right )^m \right ]_{z=0}.
\end{align}
Using Cauchy's theorem, 
\begin{align}
\oint_C \frac{\mathrm{d}z}{z^k} = \left \{ \begin{array}{cc}
0, & k\neq 1; \\
2 \pi \complexi, & k=1,
\end{array} \right.
\end{align}
it follows that the integral in equation~(\ref{eq:int_gen1}) is zero, unless $n+2-j=1$, or $j=n+1$. Therefore, equation~(\ref{eq:int_gen1}) can be written as
\begin{align}
\label{eq:int_gen2}
\left \langle \left ( \ve{r}_p \cdot \ve{r}_k \right )^m r_p^{n-m} \right \rangle_p &= a_p^n \left ( \frac{1}{2} \right )^{n+1} \frac{1}{(n+1)!}\frac{ \mathrm{d}^{n+1}}{\mathrm{d} z^{n+1}} \left [  \left ( -e_p z^2 +2z - e_p  \right )^{n-m+1} \left ( \frac{\widetilde{C}_1}{e_p} z^2 + \widetilde{C}_3 z + \frac{\widetilde{C}_2}{e_p} \right )^m \right ]_{z=0}.
\end{align}
Alternatively, one can apply the residue theorem to the integral in equation~(\ref{eq:int_gen1}), leading to the same result. 

For any $n>0$, equation~(\ref{eq:int_gen2}) reduces to a real-valued function of $e_p$, $r_k$, $\ve{e}_p\cdot \ve{r}_k$ and $\ve{j}_p\cdot \ve{r}_k$. In particular, $\widetilde{C}_1$ and $\widetilde{C}_2$ (necessarily) always appear in a real-valued combination, and the plus/minus sign in equations~(\ref{eq:tildeC_def}) always reduces to $+1$. Allowed combinations are, for example,
\begin{align}
\widetilde{C}_1 + \widetilde{C}_2 = 2 \left ( \ve{e}_k \cdot \ve{r}_k \right ); \quad \quad \widetilde{C}_1 \widetilde{C}_2 = \left ( \ve{e}_p \cdot \ve{r}_k \right )^2 + \left(1-e_p^2 \right ) \left [ e_p^2 r_k^2 - \left ( \ve{e}_p \cdot \ve{r}_k \right )^2 - \frac{e_p^2}{1-e_p^2} \left ( \ve{j}_p \cdot \ve{r}_k \right)^2 \right ].
\end{align}
Furthermore, on dimensional grounds, the maximum combined exponent of $r_k$, $\ve{e}_p\cdot \ve{r}_k$ and $\ve{j}_p\cdot \ve{r}_k$ is $m$. Therefore, equation~(\ref{eq:int_gen2}) can be formally written as
\begin{align}
\label{eq:int_gen2b}
\left \langle \left ( \ve{r}_p \cdot \ve{r}_k \right )^m r_p^{n-m} \right \rangle_p &= a_p^n \sum_{\substack{i_1,i_2 \in \, \mathbb{N}^0 \\ i_1+i_2 \leq m}} r_k^{m-i_1-i_2} \left ( \ve{e}_p \cdot \ve{r}_k \right )^{i_1} \left ( \ve{j}_p \cdot \ve{r}_k \right )^{i_2} \mathcal{B}_{i_1,i_2}^{(n,m)} (e_p),
\end{align}
where $\mathbb{N}^0$ are the natural numbers plus zero, and $\mathcal{B}_{i_1,i_2}^{(n,m)} (e_p)$ are polynomial functions of $e_p$ which can be computed from equation~(\ref{eq:int_gen2}).

Subsequently, we average over the `outer' orbit $k$ using the true anomaly $f_k$. From equations~(\ref{eq:S_n_pair_gen_form}) and (\ref{eq:int_gen2b}), it follows that the terms which need to be averaged are of the form (cf. equations~\ref{eq:app:avdef}, \ref{eq:app:r_k_dot_r_u}, \ref{eq:app:r_p} and \ref{eq:app:av_jac})
\begin{align}
\label{eq:int_gen3}
\nonumber &\left \langle \frac{ r_k^{m-i_1-i_2} \left ( \ve{e}_p \cdot \ve{r}_k \right )^{i_1} \left ( \ve{j}_p \cdot \ve{r}_k \right )^{i_2}}{r_k^{n+m+1}} \right \rangle_k = \left \langle \frac{ \left ( \ve{e}_p \cdot \ve{r}_k \right )^{i_1} \left ( \ve{j}_p \cdot \ve{r}_k \right )^{i_2}}{r_k^{n+i_1+i_2+1}} \right \rangle_k \\
\nonumber &= \frac{\left (1-e_k^2 \right )^{\frac{1}{2}-n}}{a_k^{n+1}} \frac{1}{2\pi} \int_0^{2\pi} \mathrm{d} f_k \left [ 1 + e_k \cos(f_k) \right ]^{n-1} \left [ \cos(f_k) \left (\unit{e}_k \cdot \ve{e}_p \right ) + \sin(f_k) \left ( \unit{q}_k \cdot \ve{e}_p \right ) \right ]^{i_1} \left [ \cos(f_k) \left (\unit{e}_k \cdot \ve{j}_p \right ) + \sin(f_k) \left ( \unit{q}_k \cdot \ve{j}_p \right ) \right ]^{i_2} \\
&=  \frac{\left (1-e_k^2 \right )^{\frac{1}{2}-n}}{a_k^{n+1}} \left ( \frac{1}{2} \right )^{n+i_1+i_2-1} \frac{1}{2\pi \complexi} \oint_C \frac{\mathrm{d} z}{z^{n+i_1+i_2}} \left (e_k z^2 + 2z + e_k \right )^{n-1} \left ( \frac{ \widetilde{C}_{2e}}{e_k} z^2 + \frac{\widetilde{C}_{1e}}{e_k} \right )^{i_1} \left ( \frac{\widetilde{C}_{2j}}{e_k} z^2 + \frac{\widetilde{C}_{1j}}{e_k} \right )^{i_2},
\end{align}
where we switched again to complex integration with $z = \exp(\complexi f_k)$, and the constants are given by
\begin{subequations}
\label{eq:tildeCej_def}
\begin{align}
\widetilde{C}_{1e} &\equiv \left ( \ve{e}_p \cdot \ve{e}_k \right ) + \complexi \, e_k \left (\unit{q}_k \cdot \ve{e}_p \right ); \\
\widetilde{C}_{2e} &\equiv \left ( \ve{e}_p \cdot \ve{e}_k \right ) - \complexi \, e_k \left (\unit{q}_k \cdot \ve{e}_p \right ); \\
\widetilde{C}_{1j} &\equiv \left ( \ve{e}_k \cdot \ve{j}_p \right ) + \complexi \, e_k \left (\unit{q}_k \cdot \ve{j}_p \right ); \\
\widetilde{C}_{2j} &\equiv \left ( \ve{e}_k \cdot \ve{j}_p \right ) - \complexi \, e_k \left (\unit{q}_k \cdot \ve{j}_p \right ).
\end{align}
\end{subequations}
As before, the integrand in equation~(\ref{eq:int_gen2}) is a meromorphic function with a single pole at $z=0$. By the same arguments, 
\begin{align}
\label{eq:int_gen4}
\nonumber &\left \langle \frac{ \left ( \ve{e}_p \cdot \ve{r}_k \right )^{i_1} \left ( \ve{j}_p \cdot \ve{r}_k \right )^{i_2}}{r_k^{n+i_1+i_2+1}} \right \rangle_k \\
&= \frac{\left (1-e_k^2 \right )^{\frac{1}{2}-n}}{a_k^{n+1}} \left ( \frac{1}{2} \right )^{n+i_1+i_2-1} \frac{1}{(n+i_1+i_2-1)!} \frac{\mathrm{d}^{n+i_1+i_2-1}}{\mathrm{d} z^{n+i_1+i_2-1}} \left [ \left (e_k z^2 + 2z + e_k \right )^{n-1} \left ( \frac{ \widetilde{C}_{2e}}{e_k} z^2 + \frac{\widetilde{C}_{1e}}{e_k} \right )^{i_1} \left ( \frac{\widetilde{C}_{2j}}{e_k} z^2 + \frac{\widetilde{C}_{1j}}{e_k} \right )^{i_2} \right ]_{z=0}.
\end{align}
The expression in equation~(\ref{eq:int_gen4}) is generally a function of $e_p$, $e_k$, $\ve{e}_p\cdot \ve{e}_k$, $\ve{j}_p\cdot \ve{j}_k$, $\ve{e}_p\cdot \ve{j}_k$ and $\ve{e}_k \cdot \ve{j}_p$. The combined exponent of $\ve{e}_p\cdot \ve{e}_k$ and $\ve{e}_p\cdot \ve{j}_k$ is at most $i_1$, whereas the combined exponent of $\ve{j}_p\cdot \ve{j}_k$ and $\ve{e}_k \cdot \ve{j}_p$ is at most $i_2$. Formally, equation~(\ref{eq:int_gen4}) can be written as
\begin{align}
\left \langle \frac{ \left ( \ve{e}_p \cdot \ve{r}_k \right )^{i_1} \left ( \ve{j}_p \cdot \ve{r}_k \right )^{i_2}}{r_k^{n+i_1+i_2+1}} \right \rangle_k = \frac{\left (1-e_k^2 \right )^{\frac{1}{2}-n}}{a_k^{n+1}} \sum_{\substack{l_1,l_2,l_3,l_4 \in \, \mathbb{N}^0 \\ l_1+l_3 \leq i_1 \\ l_2 + l_4\leq i_2}} \mathcal{C}^{(n,i_1,i_2)}_{l_1,l_2,l_3,l_4}(e_p,e_k) \left ( \ve{e}_p\cdot \ve{e}_k \right )^{l_1} \left ( \ve{j}_p\cdot \ve{j}_k \right )^{l_2} \left ( \ve{e}_p\cdot \ve{j}_k \right )^{l_3} \left ( \ve{e}_k \cdot \ve{j}_p \right )^{l_4},
\end{align}
where $\mathcal{C}^{(n,i_1,i_2)}_{l_1,l_2,l_3,l_4}(e_p,e_k)$ are polynomial functions of $e_p$ and $e_k$ that can be computed from equation~(\ref{eq:int_gen4}) using the vector identity equation~(\ref{eq:vec_id}). 

With these formal expressions for the orbit averages, the averaged pairwise Hamiltonian term to order $n$ is given by
\begin{align}
\label{eq:S_n_pair_gen_final}
\nonumber \overline{S}'_{n;2} &=  (-1)^{n+1} \sum_{k \in \mathrm{B}} \sum_{\substack{p \in \mathrm{B} \\ p \in \{ k.\mathrm{C} \} }} \frac{M_{p.\mathrm{C1}} M_{p.\mathrm{C2}}}{M_p} \frac{ \left | M_{p.\mathrm{C2}}^{n-1} + (-1)^n M_{p.\mathrm{C1}}^{n-1} \right | }{M_p^{n-1}} \frac{GM_{k.\mathrm{CS}(p)}}{a_k}  \left ( \frac{a_p}{a_k} \right )^n \frac{1}{j_k^{2n-1}} \\
&\quad \times \sum_{m=0}^n \sum_{\substack{i_1,i_2 \in \, \mathbb{N}^0 \\ i_1+i_2 \leq m}} \sum_{\substack{l_1,l_2,l_3,l_4 \in \, \mathbb{N}^0 \\ l_1+l_3 \leq i_1 \\ l_2 + l_4\leq i_2}} \mathcal{A}_m^{(n)} \mathcal{B}_{i_1,i_2}^{(n,m)}(e_p) \, \mathcal{C}^{(n,i_1,i_2)}_{l_1,l_2,l_3,l_4}(e_p,e_k) \left ( \ve{e}_p\cdot \ve{e}_k \right )^{l_1} \left ( \ve{j}_p\cdot \ve{j}_k \right )^{l_2} \left ( \ve{e}_p\cdot \ve{j}_k \right )^{l_3} \left ( \ve{e}_k \cdot \ve{j}_p \right )^{l_4},
\end{align}
where $\mathcal{A}_m^{(n)}$ are Legendre polynomial coefficients, given implicitly by (cf. Appendix\,\ref{app:der:pot}, in particular equations~\ref{eq:beta_n} and \ref{eq:A_j})
\begin{align}
\sum_{m=0}^n \mathcal{A}_m^{(n)} x^m = \frac{1}{2^n n!} \frac{\mathrm{d}^n }{\mathrm{d} x^n} \left [ \left (x^2 -1 \right )^n \right ],
\end{align}
and $\mathcal{B}_{i_1,i_2}^{(n,m)}(e_p)$ and $ \mathcal{C}^{(n,m,i_1,i_2)}_{l_1,l_2,l_3,l_4}(e_p,e_k)$ are polynomial functions of $e_p$ and $e_k$, given implicitly by
\begin{subequations}
\begin{align}
&\sum_{\substack{i_1,i_2 \in \, \mathbb{N}^0 \\ i_1+i_2 \leq m}} r_k^{m-i_1-i_2} \left ( \ve{e}_p \cdot \ve{r}_k \right )^{i_1} \left ( \ve{j}_p \cdot \ve{r}_k \right )^{i_2} \mathcal{B}_{i_1,i_2}^{(n,m)} (e_p) =  \left ( \frac{1}{2} \right )^{n+1} \frac{1}{(n+1)!}
\frac{ \mathrm{d}^{n+1}}{\mathrm{d} z^{n+1}} \left [  \left ( -e_p z^2 +2z - e_p  \right )^{n-m+1} \left ( \frac{\widetilde{C}_1}{e_p} z^2 + \widetilde{C}_3 z + \frac{\widetilde{C}_2}{e_p} \right )^m \right ]_{z=0}; \\
&\nonumber \sum_{\substack{l_1,l_2,l_3,l_4 \in \, \mathbb{N}^0 \\ l_1+l_3 \leq i_1 \\ l_2 + l_4\leq i_2}} \mathcal{C}^{(n,i_1,i_2)}_{l_1,l_2,l_3,l_4}(e_p,e_k) \left ( \ve{e}_p\cdot \ve{e}_k \right )^{l_1} \left ( \ve{j}_p\cdot \ve{j}_k \right )^{l_2} \left ( \ve{e}_p\cdot \ve{j}_k \right )^{l_3} \left ( \ve{e}_k \cdot \ve{j}_p \right )^{l_4} \\
&\quad = \left ( \frac{1}{2} \right )^{n+i_1+i_2-1} \frac{1}{(n+i_1+i_2-1)!} \frac{\mathrm{d}^{n+i_1+i_2-1}}{\mathrm{d} z^{n+i_1+i_2-1}} \left [ \left (e_k z^2 + 2z + e_k \right )^{n-1} \left ( \frac{ \widetilde{C}_{2e}}{e_k} z^2 + \frac{\widetilde{C}_{1e}}{e_k} \right )^{i_1} \left ( \frac{\widetilde{C}_{2j}}{e_k} z^2 + \frac{\widetilde{C}_{1j}}{e_k} \right )^{i_2} \right ]_{z=0}.
\end{align}
\end{subequations}
Here, the coefficients $\widetilde{C}_1$ etc. are given by equations~(\ref{eq:tildeC_def}) and (\ref{eq:tildeCej_def}).

\subsection{Estimate of the importance of non-binary terms for nested planetary systems}
\label{app:der:est_nested}
Here, we generalize the estimates of the importance of terms other than pairwise binary terms for nested planetary systems, made in \S\,\ref{sect:der:impl:higher}, to arbitrary order $n$. As in \S\,\ref{sect:der:impl:higher}, we consider three nested binaries $p$, $u$ and $k$, where $p \in \{k.\mathrm{C}\}$, $p \in \{u.\mathrm{C}\}$ and $u \in \{k.\mathrm{C}\}$. First, we consider the importance of the pairwise binary terms compared to the triplet term; for the pairwise binary terms, we make a distinction between the pairwise combinations $(p,u)$ and $(u,k)$. For our purposes of estimating the (orbit-averaged) Hamiltonian, we approximate $\beta_n$ as
\begin{align}
\label{eq:beta_n_est}
\beta_n(\ve{r}_{u_1},...,\ve{r}_{u_n}; \ve{r}_k) \sim \frac{a_{u_1} ... a_{u_n}}{a_k^n},
\end{align}
i.e. we replace the norms of the separation vectors with semimajor axes, and ignore the `angular' dependence, i.e. the dependence on the unit vectors $\unit{r}_{u_l}$ and $\unit{r}_k$ (cf. Appendix\,\ref{app:der:pot}). 

Using equation~(\ref{eq:Sn_gen1}) (or equation~\ref{eq:S_n_pair_gen0}), this gives the following estimate for the (orbit-averaged) pairwise terms, applied to the pair $(p,u)$,
\begin{align}
\left |\overline{S}'_{n;2} \right |_{(p,u)} \sim \frac{G}{a_u} c_n \left ( \frac{a_p}{a_u} \right )^n M_{u.\mathrm{CS}(p)} \frac{M_{p.\mathrm{C1}}M_{p.\mathrm{C2}}}{M_p} \frac{ \left | M_{p.\mathrm{C2}}^{n-1} + (-1)^n M_{p.\mathrm{C1}}^{n-1} \right | }{M_p^{n-1}}.
\end{align}
Applied to the pair $(u,k)$,
\begin{align}
\left |\overline{S}'_{n;2} \right |_{(u,k)} \sim \frac{G}{a_k} c_n \left ( \frac{a_u}{a_k} \right )^n M_{k.\mathrm{CS}(u)} \frac{M_{u.\mathrm{C1}}M_{u.\mathrm{C2}}}{M_u} \frac{ \left | M_{u.\mathrm{C2}}^{n-1} + (-1)^n M_{u.\mathrm{C1}}^{n-1} \right | }{M_u^{n-1}}.
\end{align}
The triplet term at order $n$, applied to $(p,u,k)$, is found in equation~(\ref{eq:Sn_gen1}) by setting (1) $n_1=n$, $n_2=0$, $d_p=n-1$ and $d_q=0$, and (2) $n_1=0$, $n_2=n$, $d_p=n-1$ and $d_q=0$, giving the estimate
\begin{align}
\label{eq:S3_est}
\left |\overline{S}'_{n;3} \right |_{(p,u,k)} \sim \frac{G}{a_k} c_n \frac{a_p^{n-1} a_u}{a_k^n} \frac{n!}{(n-1)!} \frac{M_{u.\mathrm{CS}(p)}}{M_u} \frac{M_{p.\mathrm{C1}}M_{p.\mathrm{C2}}}{M_p} \frac{ \left | M_{p.\mathrm{C2}}^{n-2} + (-1)^{n-1} M_{p.\mathrm{C1}}^{n-2} \right | }{M_p^{n-2}} M_{k.\mathrm{CS}(p)}.
\end{align}
Therefore,
\begin{subequations}
\begin{align}
\frac{ \left |\overline{S}'_{n;3} \right |_{(p,u,k)} }{\left |\overline{S}'_{n;2} \right |_{(p,u)} } &\sim n \frac{M_{k.\mathrm{CS}(p)}}{M_u} \frac{ \left | M_{p.\mathrm{C2}}^{n-2}M_p + (-1)^{n-1} M_{p.\mathrm{C1}}^{n-2}M_p \right | }{ \left | M_{p.\mathrm{C2}}^{n-1}+ (-1)^{n} M_{p.\mathrm{C1}}^{n-1} \right | } \frac{a_u}{a_p} \left ( \frac{a_u}{a_k} \right )^{n+1}; \\
\frac{ \left |\overline{S}'_{n;3} \right |_{(p,u,k)} }{\left |\overline{S}'_{n;2} \right |_{(u,k)} } &\sim n \frac{M_{u.\mathrm{CS}(p)}}{M_p} \frac{M_{p.\mathrm{C1}}M_{p.\mathrm{C2}}}{M_{u.\mathrm{C1}}M_{u.\mathrm{C2}}} \frac{ \left | M_{p.\mathrm{C2}}^{n-2} M_p + (-1)^{n-1} M_{p.\mathrm{C1}}^{n-2} M_p \right | }{ \left | M_{u.\mathrm{C2}}^{n-1} + (-1)^{n} M_{u.\mathrm{C1}}^{n-1} \right | } \left ( \frac{M_u}{M_p} \right )^{n-1} \left ( \frac{a_p}{a_u} \right )^{n-1}.
\end{align}
\end{subequations}
Note that for $n=3$, these estimates, apart from any dependence on the eccentricities, reduce to equations~(\ref{eq:ratio_S32_S33_p_u}) and (\ref{eq:ratio_S32_S33_u_k}). We assume a nested planetary system with stellar mass $M_\star$ and planetary mass $m_\mathrm{p}$ such that $m_\mathrm{p}/M_\star = \tilde{q} \ll 1$, and semimajor axis ratios $a_p/a_u=a_u/a_k = \alpha \ll 1$. In this case, $M_{p.\mathrm{C1}} = M_\star$, $M_{p.\mathrm{C2}} = m_\mathrm{p}$, $M_{u.\mathrm{C1}} \approx M_\star$, $M_{u.\mathrm{C2}} = m_\mathrm{p}$, $M_{k.\mathrm{C1}} \approx M_\star$ and $M_{k.\mathrm{C2}} = m_\mathrm{p}$. This gives
\begin{subequations}
\label{eq:est_nested}
\begin{align}
\frac{ \left |\overline{S}'_{n;3} \right |_{(p,u,k)} }{\left |\overline{S}'_{n;2} \right |_{(p,u)} } &\sim n \tilde{q} \alpha^{n}; \\
\frac{ \left |\overline{S}'_{n;3} \right |_{(p,u,k)} }{\left |\overline{S}'_{n;2} \right |_{(u,k)} } &\sim n \tilde{q} \alpha^{n-1}
\end{align}
\end{subequations}
(note that $| M_{p.\mathrm{C2}}^{n-2}M_p + (-1)^{n-1} M_{p.\mathrm{C1}}^{n-2}M_p | \approx | m_\mathrm{p}^{n-2} M_\star + (-1)^{n-1} M_\star^{n-1} | \approx M_\star^{n-1}$). 

\begin{figure}
\center
\includegraphics[scale = 0.5, trim = 15mm 0mm 0mm 0mm]{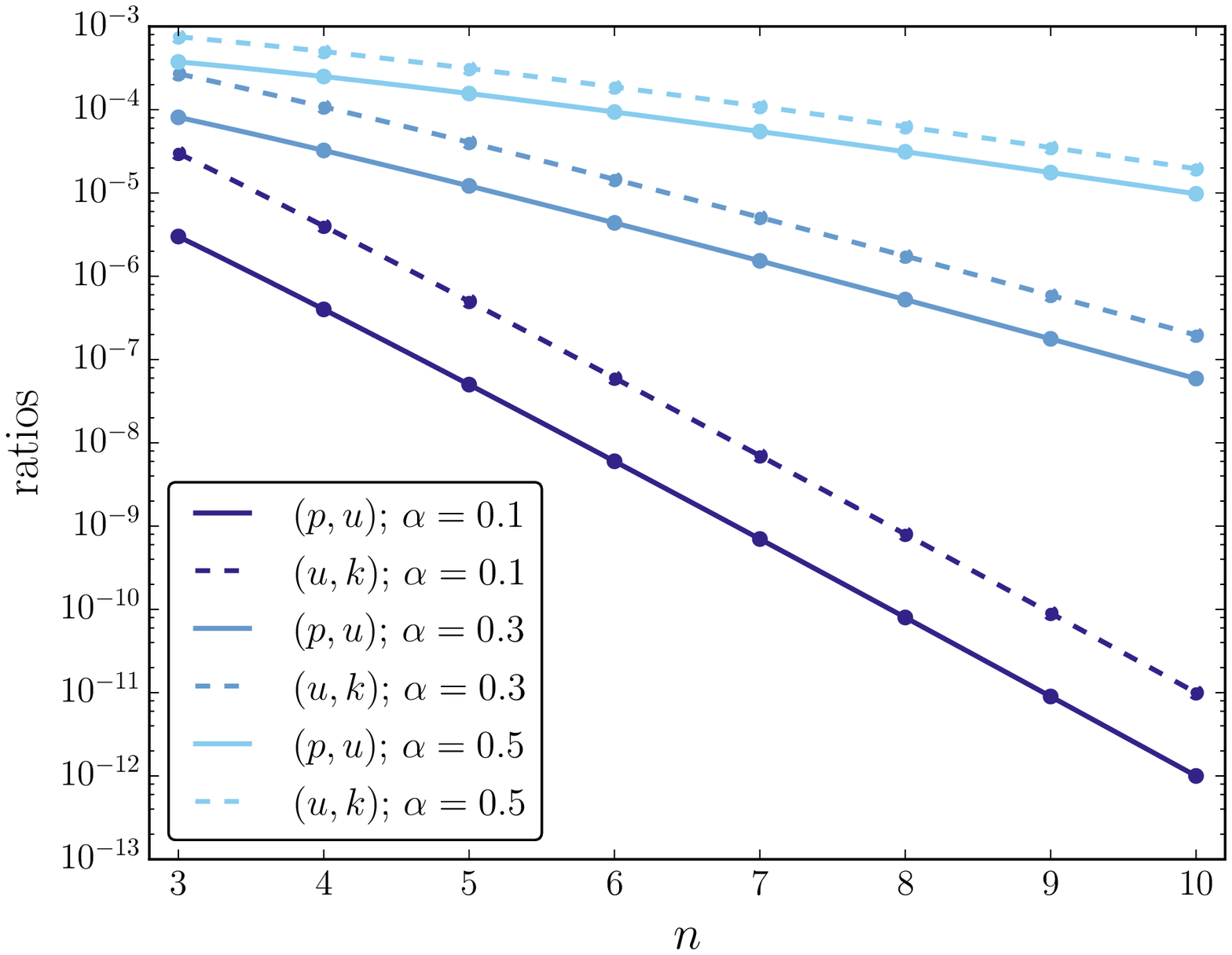}
\caption{\small The ratios of the binary triplet to the pairwise term in the Hamiltonian, applied to both pairs $(p,u)$ and $(u,k)$ (cf. equation~\ref{eq:est_nested}). We assume $\tilde{q} = 10^{-3}$ and three values of $\alpha$, indicated in the legend. }
\label{fig:estimate_nested}
\end{figure}

In \F\,\ref{fig:estimate_nested}, we show the ratios given by equations~(\ref{eq:est_nested}) as a function of $n$, assuming $\tilde{q} = 10^{-3}$ (i.e. order Jupiter-mass planets around a solar-mass star), and various values of $\alpha$. The Hamiltonian ratios decrease strongly for increasing $n$ and decreasing $\alpha$. For the smallest applicable $n$, $n=3$, and $\alpha=0.5$, the ratios are $<10^{-3}$, showing that the triplet terms are small. 

More generally, and less quantitatively, equation~(\ref{eq:Sn_gen1}) shows that for nested planetary systems, the non-binary terms, in contrast to pairwise binary terms, are multiplied by mass ratio factors $[M_{p_l.\mathrm{CS}(p)}/M_{p_l}]^{d_{p_l}} \approx (m_\mathrm{p}/M_\star)^{d_{p_l}} = \tilde{q}^{d_{p_l}} \ll 1$, with $1\leq d_{p_l} \leq n-2$. Also, note that for planetary systems, the quantity $\mathcal{M}_{p;k.\mathrm{CS}(p)}^{(d_p)}$ is approximately the same for pairwise and non-pairwise terms. Furthermore, the function $\beta_n$ (cf. equation~\ref{eq:beta_n_est}) is generally smaller for non-binary pairs compared to adjacent binary pairs, because in addition to ratios of semimajor axes of adjacent pairs, the non-binary pair terms also include ratios of semimajor axes of non-adjacent pairs (e.g. equation~\ref{eq:S3_est} contains the ratio $a_p/a_k \ll a_u/a_k$). We conclude that for nested planetary systems, the non-binary pair terms are generally small compared to the pairwise binary terms applied to adjacent binaries.

\subsection{Ad hoc expression for the 1PN Hamiltonian}
\label{app:der:1PN}
In the post-Newtonian (PN) approximation, corrections to Newtonian gravity due to general relativity are included by adding terms of order $(v/c)^n$, where $v$ is the orbital velocity and $c$ is the speed of light. To lowest order in $(v/c)$, the $N$-body equations of motion are known and have been derived by \citet{lorentz_droste_17} and \citet{1938AnMat..39...65E}. They give rise to a 1PN Hamiltonian that is conserved to order $(v/c)^2$. For the hierarchical three-body problem, the 1PN Hamiltonian was expanded in binary separation ratios and orbit averaged by \citet{2013ApJ...773..187N}. In the latter paper, it was found that, in addition to terms corresponding to separate precession in the inner and outer orbits, there is also a term associated with both inner and outer binaries, i.e. an `interaction' term. 

Here, we do {\it not} attempt to derive, from first principles, the generalized 1PN Hamiltonian $H_\mathrm{1PN}$ for hierarchical $N$-body systems. Rather, we construct an ad hoc expression by assuming that the dominant terms in $H_\mathrm{1PN}$ are given by terms which, individually, depend on only one binary, and which give rise to the well-known rate of precession in the 1PN two-body problem. Extrapolating from e.g. the results of \citet{2013ApJ...773..187N}, these terms for a binary $k$ should have the form
\begin{align}
\label{eq:H1PNk}
H_{\mathrm{1PN},k} = - \frac{3G^2 M_{k.\mathrm{C1}} M_{k.\mathrm{C2}} M_k}{c^2 a_k^2 \left (1-e_k^2 \right)^{1/2}}.
\end{align}
From the equations of motion (cf. equation~\ref{eq:EOM}), it follows that the precession associated with equation~(\ref{eq:H1PNk}), which only depends on $\ve{e}_k$, is
\begin{align}
\frac{ \mathrm{d} \ve{e}_k}{\mathrm{d} t} = \frac{2 \pi}{t_\mathrm{1PN}} \unit{j}_k \times \ve{e}_k,
\end{align}
i.e. $\ve{e}_k$ precesses around $\ve{j}_k$ with time-scale $t_\mathrm{1PN}$. The latter is given by
\begin{align}
\label{eq:t_1PN}
t_\mathrm{1PN} = \frac{1}{3} P_{\mathrm{orb},k} \left (1-e_k^2 \right ) \frac{a_k}{r_{\mathrm{g},k}},
\end{align}
where the orbital period is $P_{\mathrm{orb},k} = 2 \pi \sqrt{a_k^3/(GM_k)}$ and the gravitational radius is $r_{\mathrm{g},k} = GM_k/c^2$. 

Therefore, our assumed 1PN Hamiltonian is
\begin{align}
\label{eq:H1PN}
H_\mathrm{1PN} \approx H - \sum_{k \in \mathrm{B}} \frac{3G^2 M_{k.\mathrm{C1}} M_{k.\mathrm{C2}} M_k}{c^2 a_k^2 \left (1-e_k^2 \right)^{1/2}},
\end{align}
where $H$ is the Newtonian part given by equation~(\ref{eq:H2}). Evidently, in this form, orbit averaging of equation~(\ref{eq:H1PN}) only affects $H$.

\label{lastpage}
\end{document}